\documentclass[usenatbib]{mnras}

\usepackage{multicol}
\usepackage{aas_macros,times,multirow,amsmath,amssymb,longtable,breqn}

\usepackage[T1]{fontenc}
\usepackage[varg]{txfonts}

\usepackage{grffile} 
\usepackage{graphicx}

\usepackage{color}
\usepackage{xcolor}

\def\ltsima{$\; \buildrel < \over \sim \;$}
\def\lsim{\lower.5ex\hbox{\ltsima}}
\def\gtsima{$\; \buildrel > \over \sim \;$}
\def\gsim{\lower.5ex\hbox{\gtsima}}

\newcommand{\be}{\begin{equation}}
\newcommand{\en}{\end{equation}}

\def \DV#1{{\textcolor{orange}{#1}}}

\begin{document}
\label{firstpage}
\pagerange{\pageref{firstpage}--\pageref{lastpage}}

\title[Modelling of the non-thermal emission from pulsars]{Synchro-curvature modelling of the multi-frequency non-thermal emission of pulsars}
\author[D. F. Torres, et al.]
{Diego F. Torres$^{1,2,3}$, Daniele Vigan\`o$^{3,4,5}$, Francesco Coti Zelati$^{1,2}$, \& Jian Li$^{6}$ \\
$^{1}$ Instituci\'o Catalana de Recerca i Estudis Avan\c cats (ICREA), 08010 Barcelona, Spain \\
$^{2}$ Institute of Space Sciences (ICE, CSIC), Campus UAB, Carrer de Can Magrans s/n, 08193 Barcelona, Spain \\
$^{3}$ Institut d'Estudis Espacials de Catalunya (IEEC), Gran Capit\`a 2-4, 08034 Barcelona, Spain \\
$^{4}$ Departament de F\'isica, Universitat de les Illes Balears, Palma de Mallorca, Baleares E-07122, Spain \\
$^{5}$ Institut Aplicacions Computationals (IAC3) Universitat de les Illes Balears, Palma de Mallorca, Baleares E-07122, Spain\\
$^{6}$ Deutsches Elektronen Synchrotron DESY, D-15738 Zeuthen, Germany 
}

\date{}
\maketitle

\pubyear{2019}

\begin{abstract}

We apply a synchro-curvature spectral emission model based on characterizing the dynamics of magnetospheric particles to fit the phase-average spectra of the most extended database for the non-thermal spectra of pulsars. 
We consider 36 pulsars with well-determined non-thermal spectra from X-rays to gamma-rays. 
The sample includes Crab and Crab twin, for which the spectra extends even to the optical/ultraviolet and infrared energies. 
We find that the model --with just three physical parameters and a global scaling-- can fit the observations well across eight orders of magnitude for 18 of the 36 pulsars studied. 
Additionally, we find a set of 8 pulsars for which the model still provides arguably good fits and another set of 10 pulsars for which the model fails in reproducing the spectra. 
We discuss why, propose and provide physical interpretations for a simple model extension (related to the geometry of the accelerating system with regards to the observer) that allows dealing with all such cases, ultimately providing very good fits for all pulsars. 
The extended model is still austere, adding only two additional parameters to the former set, of the same kind of the ones previously used. 
We use these fits to discuss issues going from the observed spectral origin, to the extent of the dominance of synchrotron or curvature regimes, the use of a model as predictor for searching new non-thermal pulsars starting from gamma-ray surveys, and how the model offers a setting where phase shifts between X-ray and gamma-ray light curves would naturally arise.

\end{abstract}


\section{Introduction}
\label{sec:intro}

 \cite{CotiZelati2019} compiled 
a sample (complete up to April 2019)
of the non-thermal X-ray spectral energy distributions (SEDs) of the gamma-ray pulsars reported in the {\it Fermi} Second Pulsar Catalog \citep{2fpc}, significantly enlarging the data availability.
We now have 40 pulsars for which detailed (non-thermal) SEDs covering at least from X-rays to gamma-rays are available.
Here, we embark into trying to model all
such data in conjunction with the corresponding detections at higher energies.
We shall make use of the synchro-curvature, multi-wavelength model that we have recently introduced by \cite{torres18}.
We refer the reader to the Methods and supplementary materials of \cite{torres18} 
for details, formulae, and the applicable numerical implementation, beyond what is explained below.
This model is based on the theoretical works on synchro-curvature radiation by  \cite{paper0} (see also \citet{cheng96}), and extends on 
the applications earlier done to study the gamma-ray data of the pulsars detected by {\it Fermi}-LAT \citep{paper1,paper2,paper3,paper4}.

In this scheme, the emission from a pulsar occurs in one (or two, see the discussion below) accelerating region in the magnetosphere.
A {\it parallel electric field $E_{||}$} is associated to such region.
The region extends from an inner ($x_{in}$, assumed to be 0.5 $R_{lc}$, with $R_{lc}$ representing the light cylinder 
of the pulsar in question, $R_{lc}=cP/2 \pi$) to an outer boundary $x_{out}$, which could in principle exceed the light cylinder itself and is here taken
as 1.5 $R_{lc}$. 
It was earlier shown that 
a precise location and extent of the accelerating region  is not dominating the predicted high-energy spectral shape of the model, justifying fixing these values for all pulsars.
The accelerating region is threaded by a magnetic field assumed to be represented by a power law $B(x) \propto x^{-b}$ (see the discussion in \cite{paper2}), where $x$ is the distance along the field line.
The parameter $b$ is referred to as the {\it magnetic gradient},
and will describe how fast the magnetic field intensity declines along the trajectory of the particles. 
Having defined these two parameters ($E_{||}, b$), for a given particular pulsar --as described by its period and period derivative $(P,\dot P)$--, the model solves the equations of motion of the particle
while moving in the accelerating region, assuming they are injected or enter the region at $x_{in}$ with a (sizeable) pitch angle $\alpha$ (see \cite{torres18} for the explicit formulae of the equations of motion).

The equations of motion for a single particle balance the acceleration and the radiative losses, and are used to compute physical magnitudes such as 
the pitch angle $\alpha$, the Lorentz factor $\Gamma$, the momentum, etc., as a function of time (or position along the trajectory).
Thus, for each position of the trajectory, the physical properties under which radiation is emitted are well-determined.
The radiative losses are computed taking these properties into account and considering the full synchro-curvature formulae, encompassing the transition
of synchrotron-dominated emission at the initial part of the trajectory, associated with the loss of angular momentum, to a more curvature-dominated
emission towards the end, when particles slide along the magnetic field lines.

Once the radiation emitted by a single particle along the trajectory across the accelerating region is known, the total yield by synchro-curvature radiation, which is what we ultimately fit to the observational data, is obtained by convolving ${dP_{\rm sc}}/ {dE_\gamma}$, the single-particle synchro-curvature spectrum \citep{cheng96,paper0}, with the effective particle distribution, i.e., with 
the number of particles per unit of distance that are emitting radiation towards us, $dN_e/dx $.
The explicit formula we use to fit observational data is then
\begin{equation}\label{eq:sed_x}
  \frac{dP_{\rm gap}}{dE_\gamma} =  \int_{x_{\rm in}}^{x_{\rm out}} \left[ \frac{dP_{\rm sc}} {dE_\gamma}\right] \left[\frac{dN_e}{d x}\right] {\rm d}x~.
\end{equation}

Since we lack knowledge of $dN_e/dx $, we can only parameterize it. 
Following \cite{torres18}, we use the particle distribution to define (using a length scale $x_0$) 
the {\it contrast}, the inverse of $x_0/(x_{out}-x_{in})$, as a measure of how uniform is the distribution of particles emitting towards us: 
\begin{equation}
  \frac{dN_e}{dx}= \left [ \frac{N_0}{x_0 (1 - e^{-(x_{\rm out}-x_{\rm in})/x_0}) } \right]
  e^{-(x-x_{\rm in})/x_0},
  \label{eq:distribution}
\end{equation}
where $N_0$ is the {\it normalization}, such that
$
\int_{x_{\rm in}}^{x_{\rm out}} (dN_e/dx) dx=N_0; 
$
its value will not affect the shape of the spectrum but only its absolute scaling. 
For simplicity, and in order to avoid adding degrees of freedom, we assume that all particles are created at the same location and with the same initial pitch angle. As for other parameters, these values should be taken as effective, representative of the averaged magnitudes.
We come back to considerations affecting the assumed form of $dN_e/dx$ below. 

The single-particle synchro-curvature power, ${dP_{\rm sc}}/ {dE_\gamma}$ has a cumbersome expression that is simplified by the use of the synchro-curvature
parameter, $\xi$, making the power 
expression below (Eq. \ref{eq:sed_synchrocurv}) 
to reduce itself to purely synchrotron (when $\xi\gg1$) or curvature radiation (when $\xi\ll 1$),  \citep{paper0}. It can be numerically computed as:
\begin{equation}
\label{eq:sed_synchrocurv}
 \frac{dP_{\rm sc}}{dE} = \frac{\sqrt{3} e^2 \Gamma y}{4\pi \hbar r_{\rm eff} } [ (1 + z) F(y) - (1 - z) K_{2/3}(y)]~,
\end{equation}
where
\begin{eqnarray}
  z &=& (Q_2 r_{\rm eff})^{-2} ~, \label{eq:z}\\
 F(y) &=& \int_y^\infty K_{5/3}(y') dy'~,\label{eq:f_y}\\
  y &=&\frac{E}{E_c} ~, \\
  E_c &=& \frac{3}{2}\hbar cQ_2\Gamma^3~,\label{eq:echar}\\
  r_{\rm eff} &=& \frac{r_c}{\cos^2\alpha}\left(1 + \xi+ \frac{r_{\rm gyr}}{r_c}  \right)^{-1}~,\\
  r_{\rm gyr} &=& \frac{mc^2\Gamma\sin\alpha}{eB}~,\\
  Q_2 &=& \frac{\cos^2\alpha}{r_c}\sqrt{1 + 3\xi  + \xi^2 + \frac{r_{\rm gyr}}{r_c}} \label{eq:q2}~, \\
  \xi &=& \frac{r_c}{r_{\rm gyr}}\frac{\sin^2\alpha}{\cos^2\alpha}~. \label{eq:xi}
  \label{eq:gr}
\end{eqnarray}
Here, $\hbar$ is the reduced Planck constant, $c$ is the speed of light, $K_n$ are the modified Bessel functions of the second kind of index $n$, $m$ and $e$ are the rest mass and charge of the lepton, and $\alpha$, $r_{\rm gyr}$ and $r_c$ are the pitch angle, the Larmor radius, and the radius of curvature of its trajectory, respectively. $E$ is the photon energy, and $E_c$ is the characteristic energy of the emitted radiation. Finally, $B$ is the local strength of the magnetic field at each position. 
We take the following effective parametrization using the magnetic gradient:
\begin{equation}
\label{B}
B(x)=B_s(R_\star/x)^b, 
\end{equation}
with 
$B_s = 6.4 \times 10^{19} ~ \left(P{\rm [s]}\dot{P}\right)^{1/2} {\rm ~G}$, with $R_\star$ being the radius
and $B_s$
being an estimate of the surface magnetic field of the neutron star.
If particles would go in trajectories following radial lines off a dipole, $b$ would be equal to 3.
In realistic cases, we could expect a wider range of values for $b$, due to different magnetic field geometries: smaller than 3 for movements along twisted monopole or dipole lines, or larger than 3 for higher multipoles or turbulent configurations. Here, we keep it general, with the parametric formula above.

The shape of a predicted SED for a given pulsar having timing parameters $(P,\dot P)$ is then fully determined 
by defining the set $(E_{||}, b, x_0/R_{lc})$, and its absolute scale by $N_0$.
Fittings to observational data sets can then proceed by varying the former three parameters, and, for each set of them, searching the value of $N_0$ that minimizes the deviation between the model
and the data. 
Further details about the fitting process (used to quantify the relative goodness-of-fit in each case)
can be found in the Methods section of \cite{torres18}. 
It is interesting to note that the model can then be used to show the contribution of each part of the particles trajectories
to the total predicted yield at each band, as we will show below.

It is also relevant to note that $x_0/R_{lc}$ and $N_0$ are parameters holding relation with the geometry of the system.
For instance, 
the number of particles emitting radiation directed {\it towards the observer} can be different from two accelerating regions located at 
different positions in the magnetosphere, despite their physical parameters $(E_{||}, b)$ being potentially the same (e.g., if we assume symmetric accelerating regions in the two hemispheres, with different Earth's visibility of particle directions).
We come back to this below.

Despite the relative simplicity of the conceptual approach of this model and the reduced number of the physical parameters that are to be varied, it has 
proven successful in describing the X-ray and gamma-ray emission of several 
known pulsars (see the cases shown in \cite{torres18}).
The model was also used to predict pulsars that should be detectable in X-rays out of their detected gamma-ray emission \citep{torres18,li18}. 
Here, we shall confront this model to a larger number of broad-band observational data sets, recalling that prior to the observational analysis effort described 
in \cite{CotiZelati2019}, only a few X-ray to gamma-ray SEDs were available. 
In what follows, then, we apply this spectral model to the whole sample of broad-band SEDs for PSRs and MSPs we have earlier 
compiled.

\section{Modelling non-thermal SEDs}

The sample of certainly non-thermal, X-ray emitting, gamma-ray pulsars 
contains 33 pulsars (PSRs, with $P>10$ ms), and 7 millisecond pulsars (MSPs, with $P<10$ ms) \citep{CotiZelati2019}.
%
%
From this we exclude the three MSPs detailed below, requiring additional {\it Fermi}-LAT analysis in order to have a consistent set of data (we briefly comment on the results we obtain using these refurbished data set below)
and the PSR J2043+2740, due to the current scarcity 
of data points in its SED  (it has only seven data points useful for our fitting summing up the X-ray and gamma-ray domain).
In total, then, we shall consider 32 PSRs and 4 MSPs here.
%
%

\subsection{broad-band SEDs well-describable with a single set of $(E_{||}, b, x_0/R_{lc},N_0)$ parameters}

We found that in 18 out of the 36
pulsars herein considered,  
the model with a single set of parameters  $(E_{||}, b, x_0/R_{lc},N_0)$, as used by \cite{torres18}, is able to qualitatively describe 
well the observational data across seven to eight orders of magnitude in energy.
%
%
This is explicitly shown in Figure~\ref{sed1}.

\begin{figure*}
\begin{center}
\includegraphics[width=0.34\textwidth]{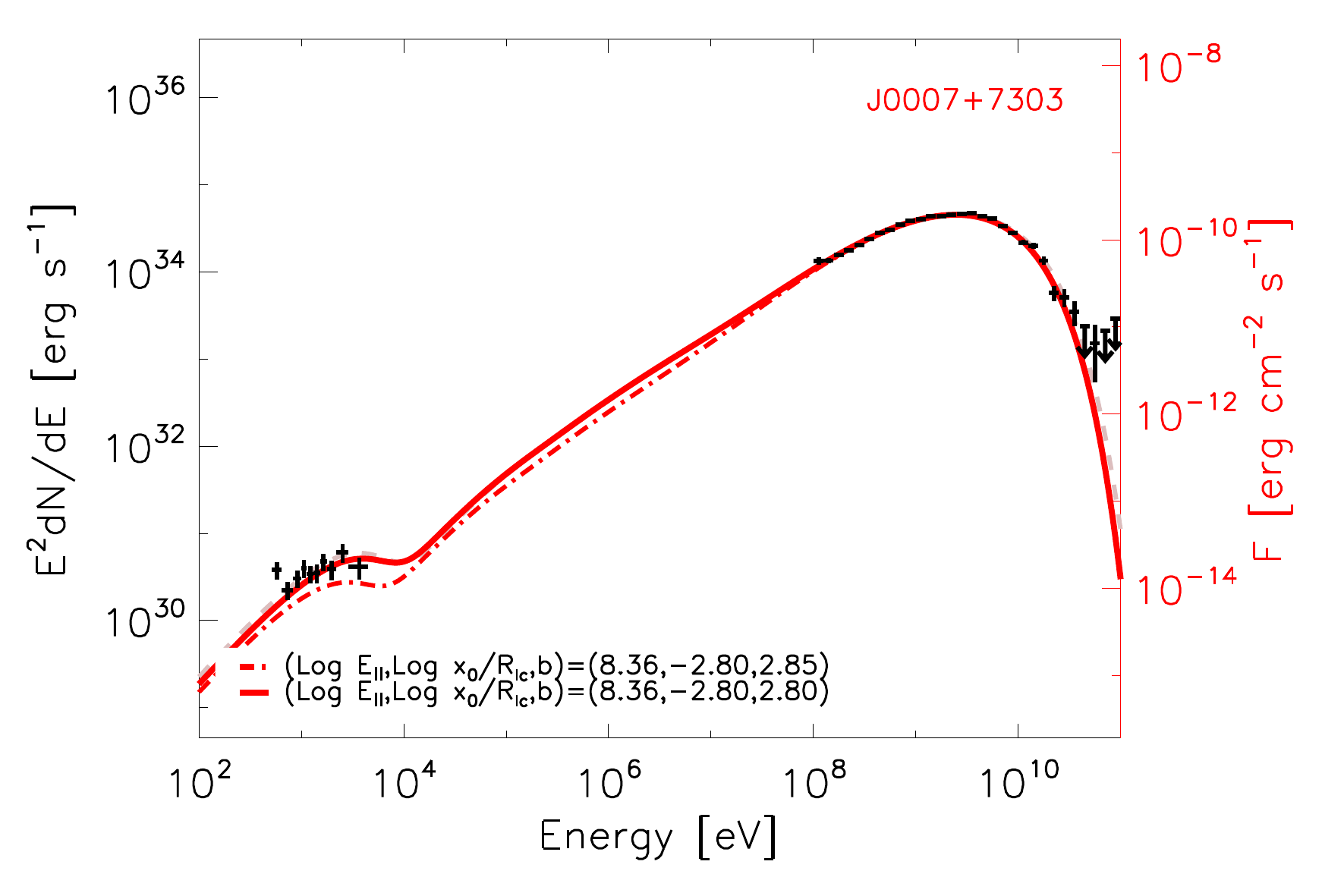}\hspace{-.25cm}
\includegraphics[width=0.34\textwidth]{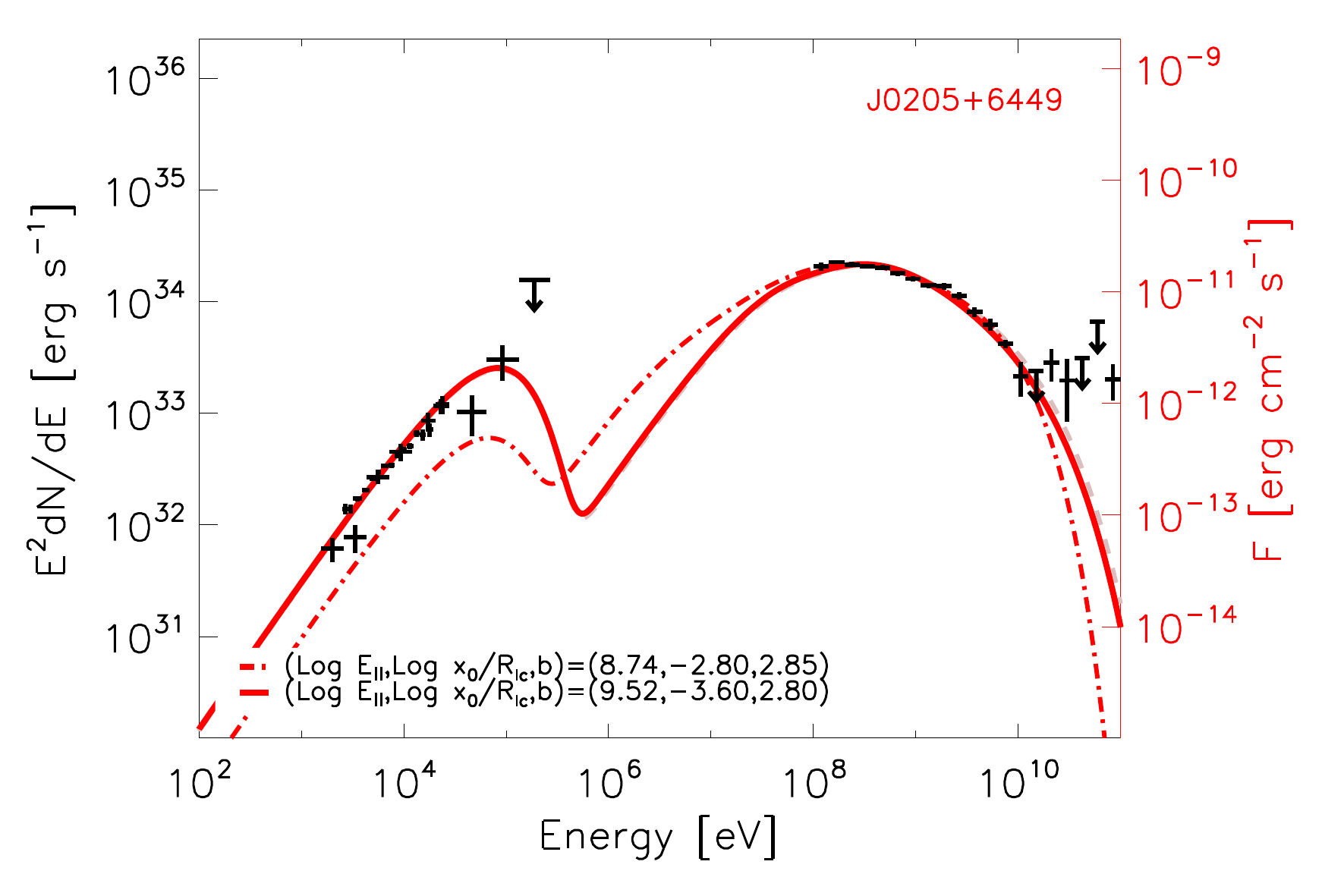}\hspace{-.25cm}
\includegraphics[width=0.34\textwidth]{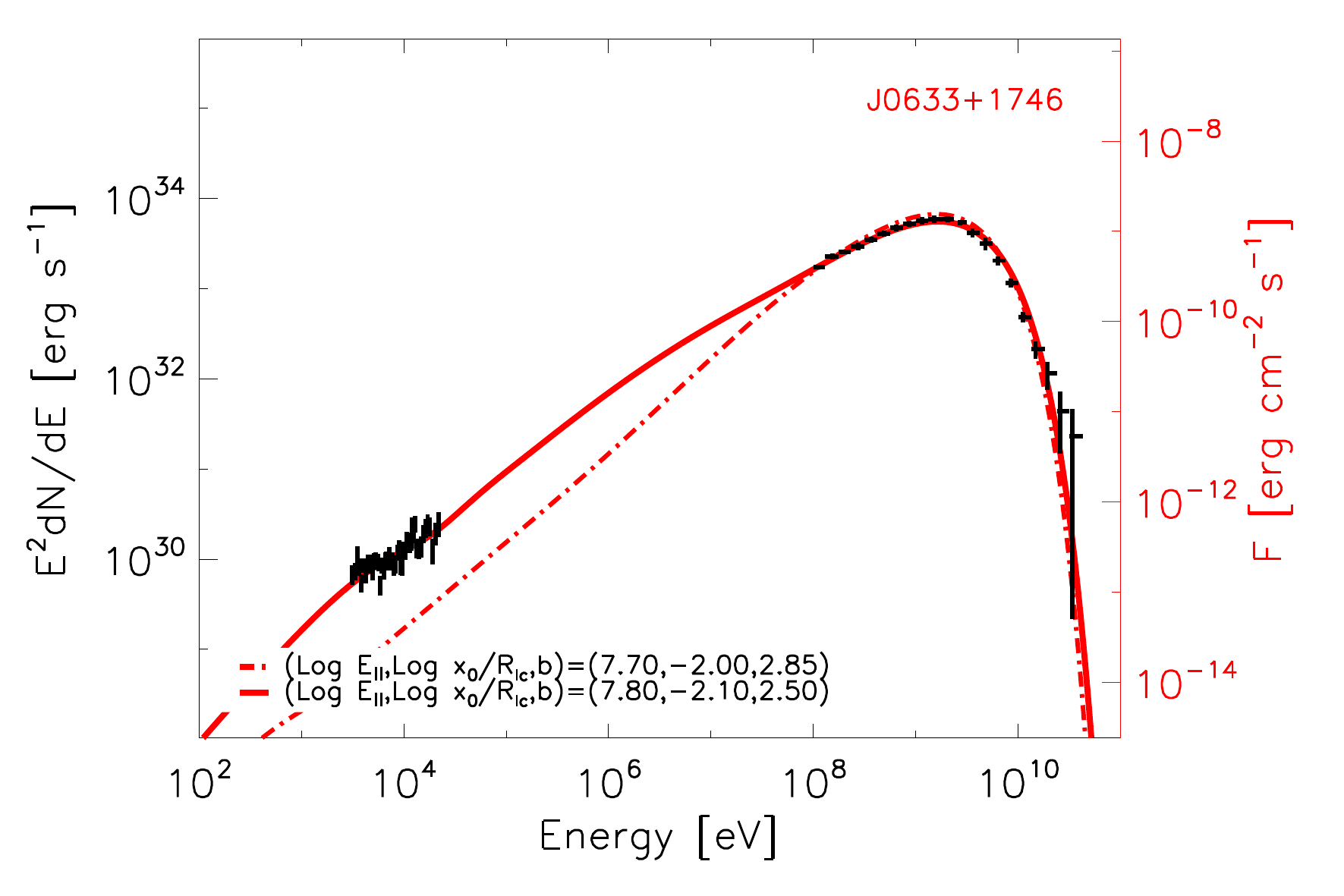}
\includegraphics[width=0.34\textwidth]{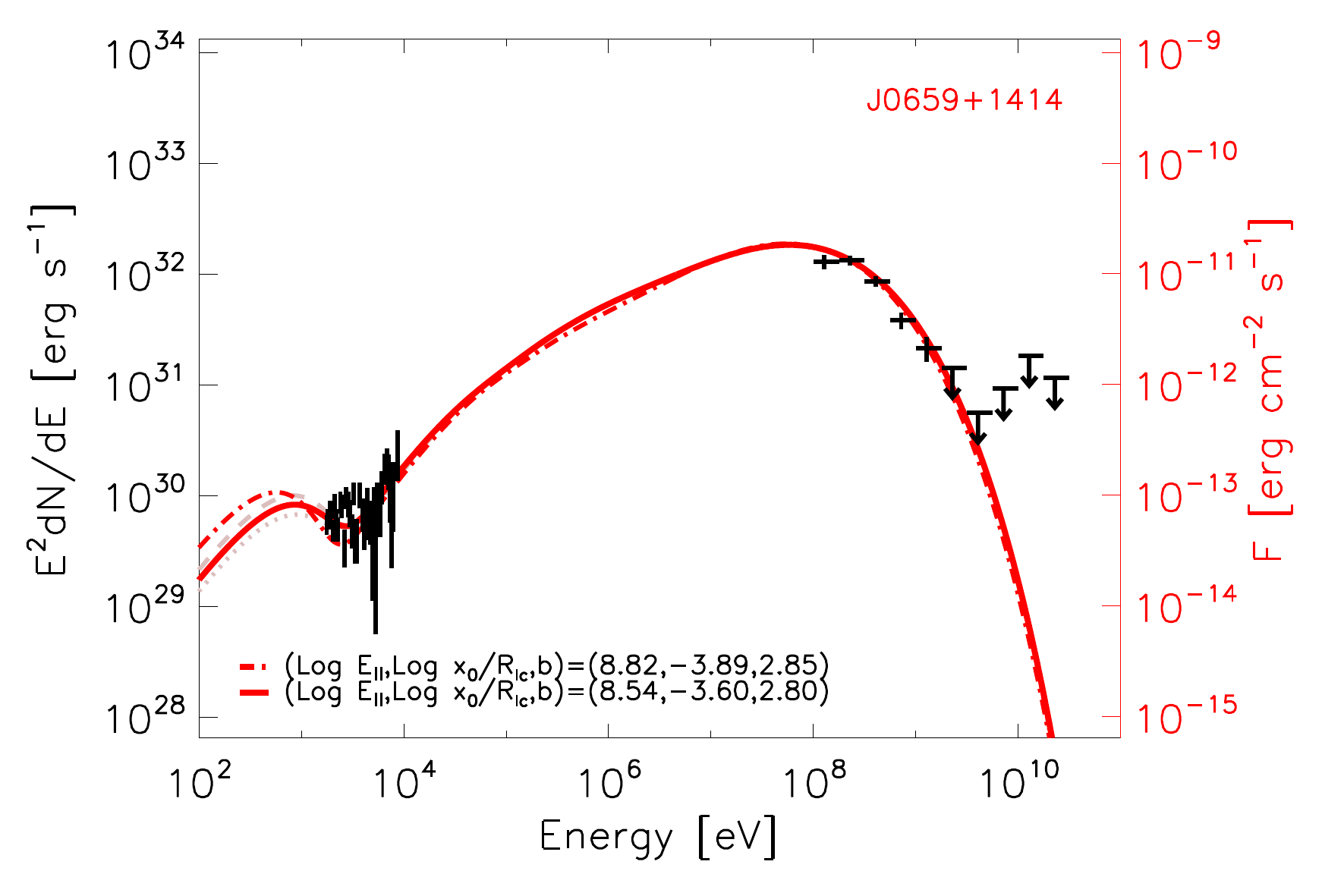}\hspace{-.25cm}
\includegraphics[width=0.34\textwidth]{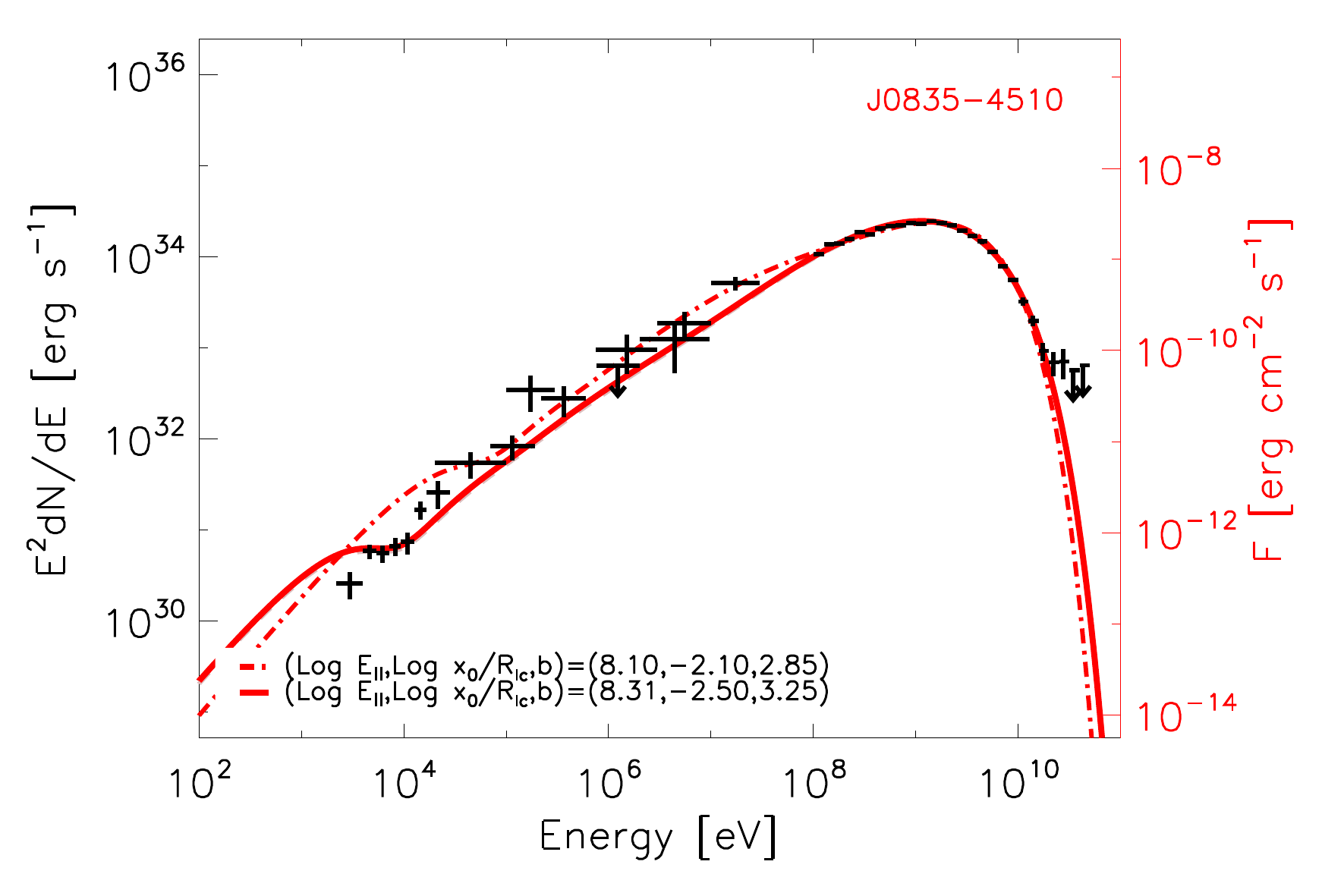}\hspace{-.25cm}
\includegraphics[width=0.34\textwidth]{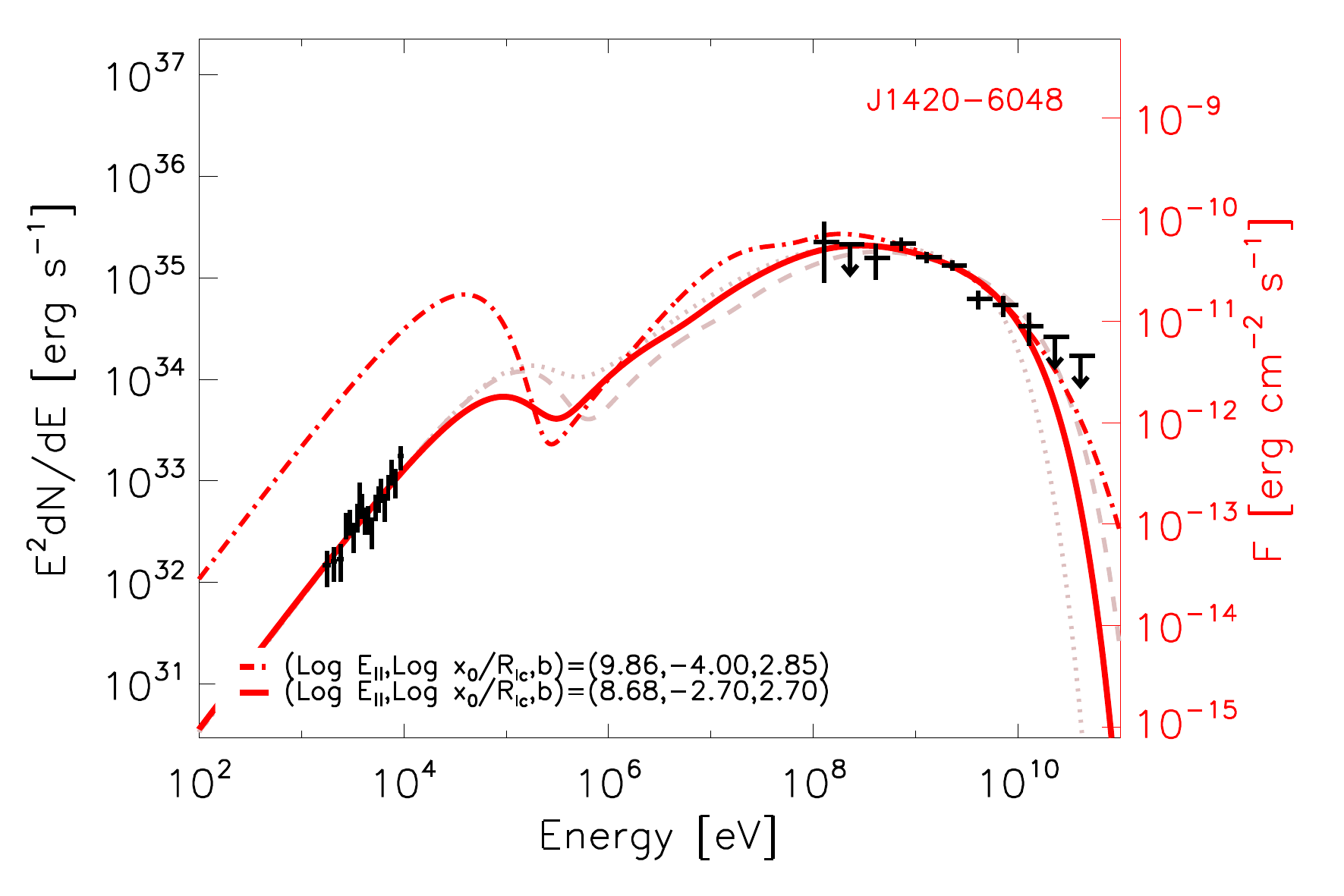}
\includegraphics[width=0.34\textwidth]{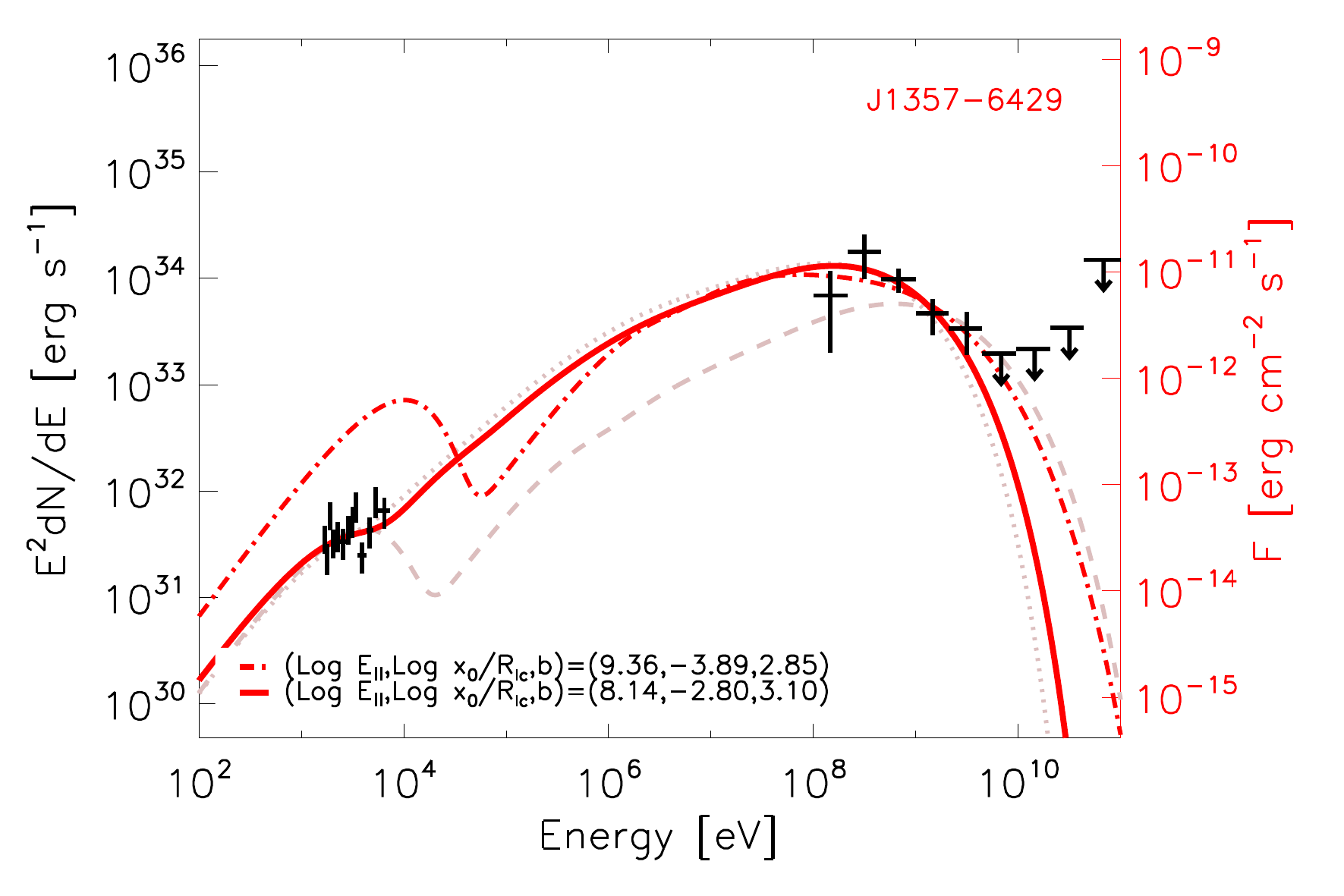}\hspace{-.25cm}
\includegraphics[width=0.34\textwidth]{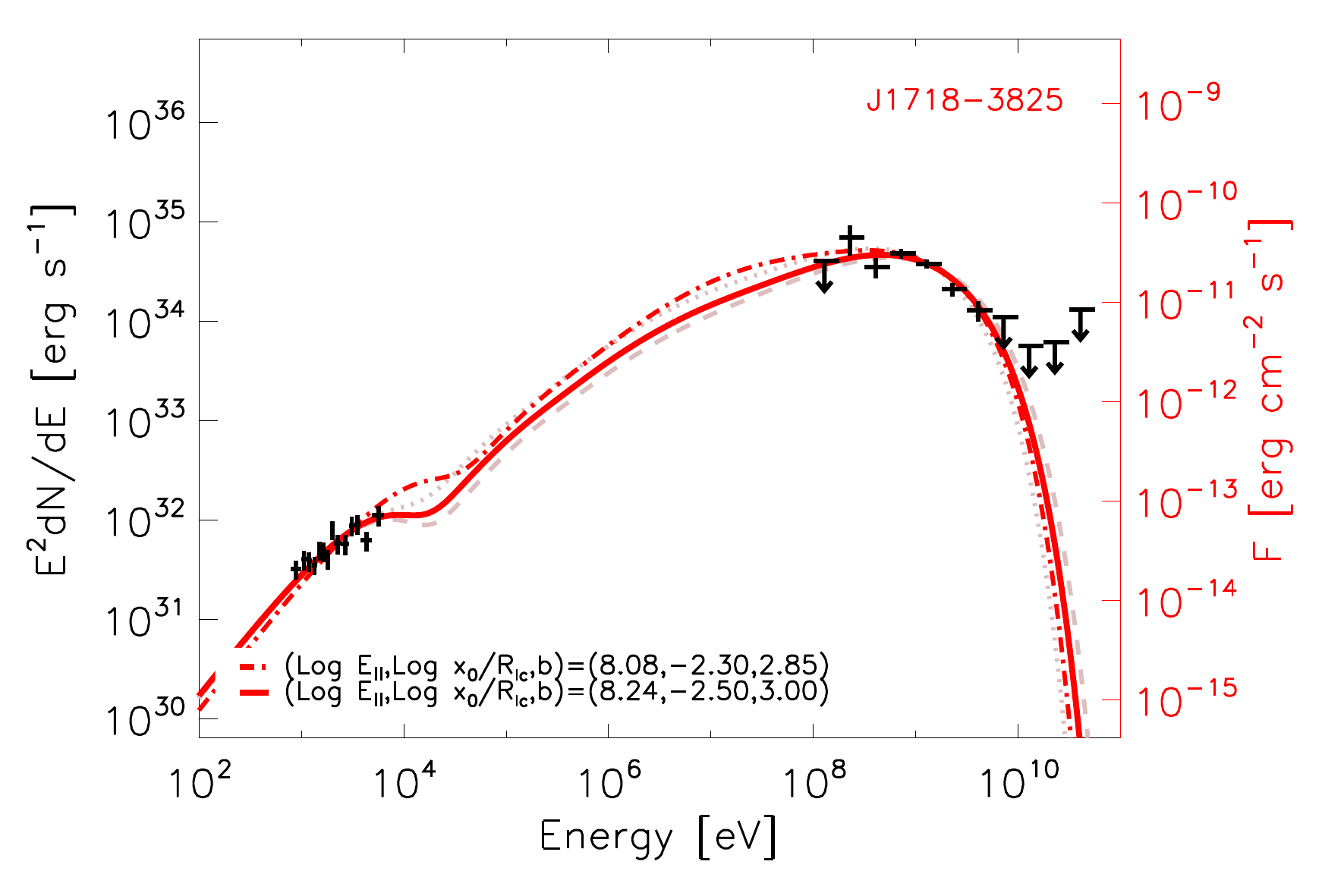}\hspace{-.25cm}
\includegraphics[width=0.34\textwidth]{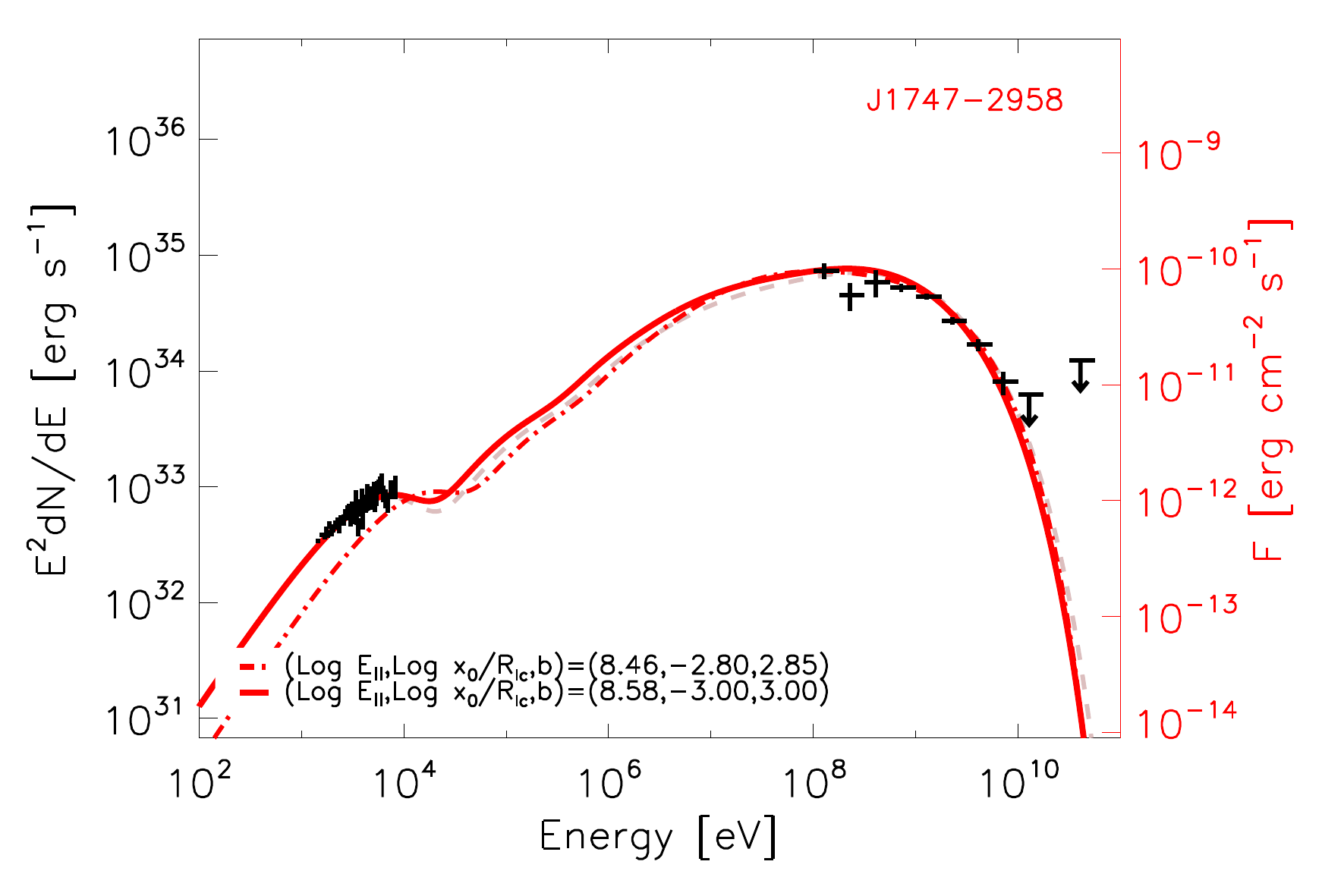}
\caption{Broad-band SEDs well-describable with a single set of $(E_{||}, b, x_0/R_{lc},N_0)$ parameters.
The panels show the corresponding theoretical model fits to the broad-band SEDs of several pulsars using a single set of parameters $(E_{||}, b, x_0/R_{lc},N_0)$ as given in the legend. See the first panel of Table \ref{common-fits} for the uncertainties of the best-fit parameters. The solid red line is the best-fit of our model to the full observational data set (black crosses indicating detections with their statistical 1$\sigma$ errors, and black arrows marking upper limits).
The two milder, grey-colored curves lying close to the solid line correspond to the predicted SEDs 
resulting from varying $E_{||}$ and $x_0$ within their maximum ranges of their 1$\sigma$ uncertainty.
The red dash-dotted lines correspond to the best-fit SEDs obtained when considering the gamma-ray data only. See text for further details.
}
\label{sed1}
\end{center}
\end{figure*}

\begin{figure*}
\begin{center}
\includegraphics[width=0.34\textwidth]{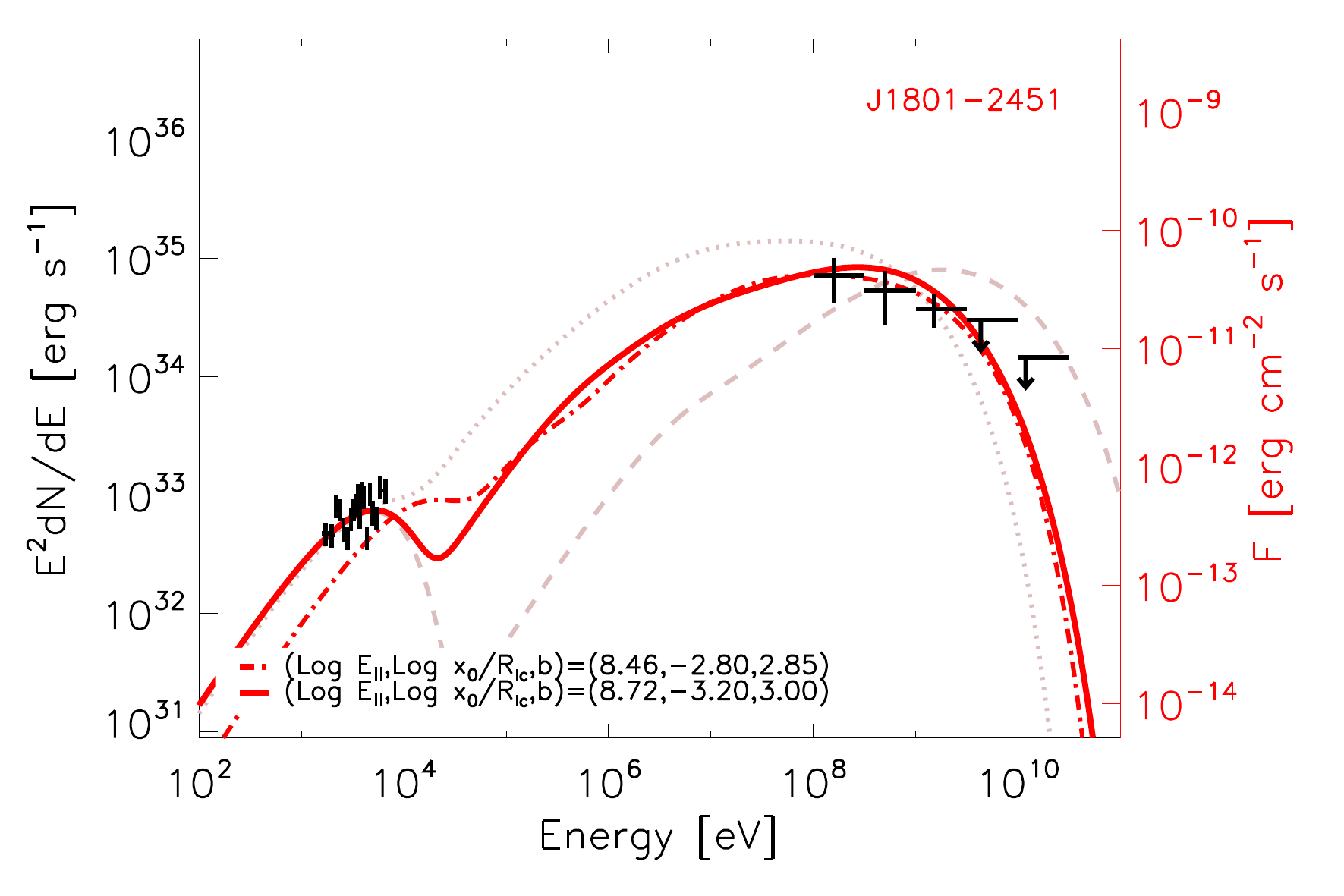}\hspace{-.25cm}
\includegraphics[width=0.34\textwidth]{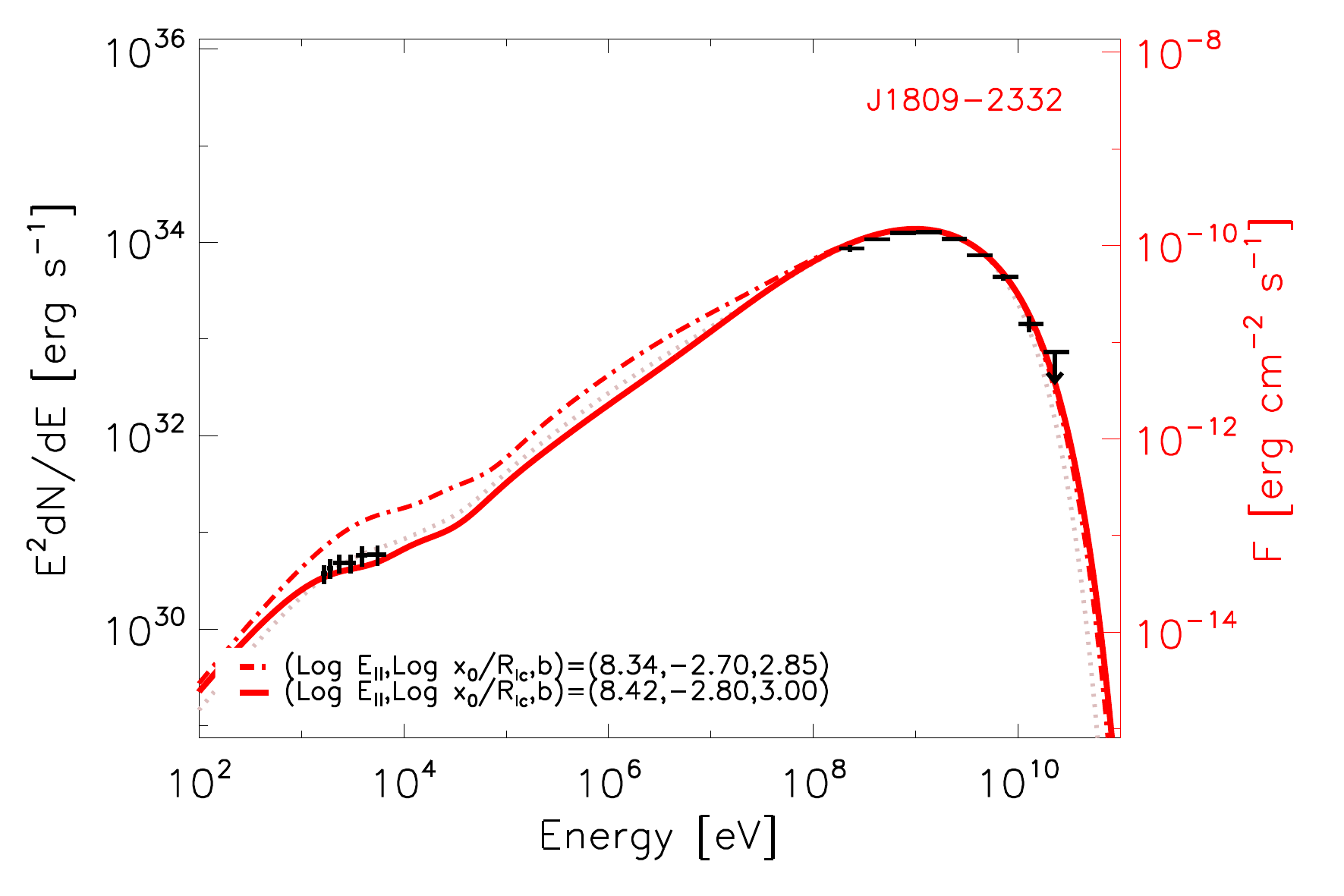}\hspace{-.25cm}
\includegraphics[width=0.34\textwidth]{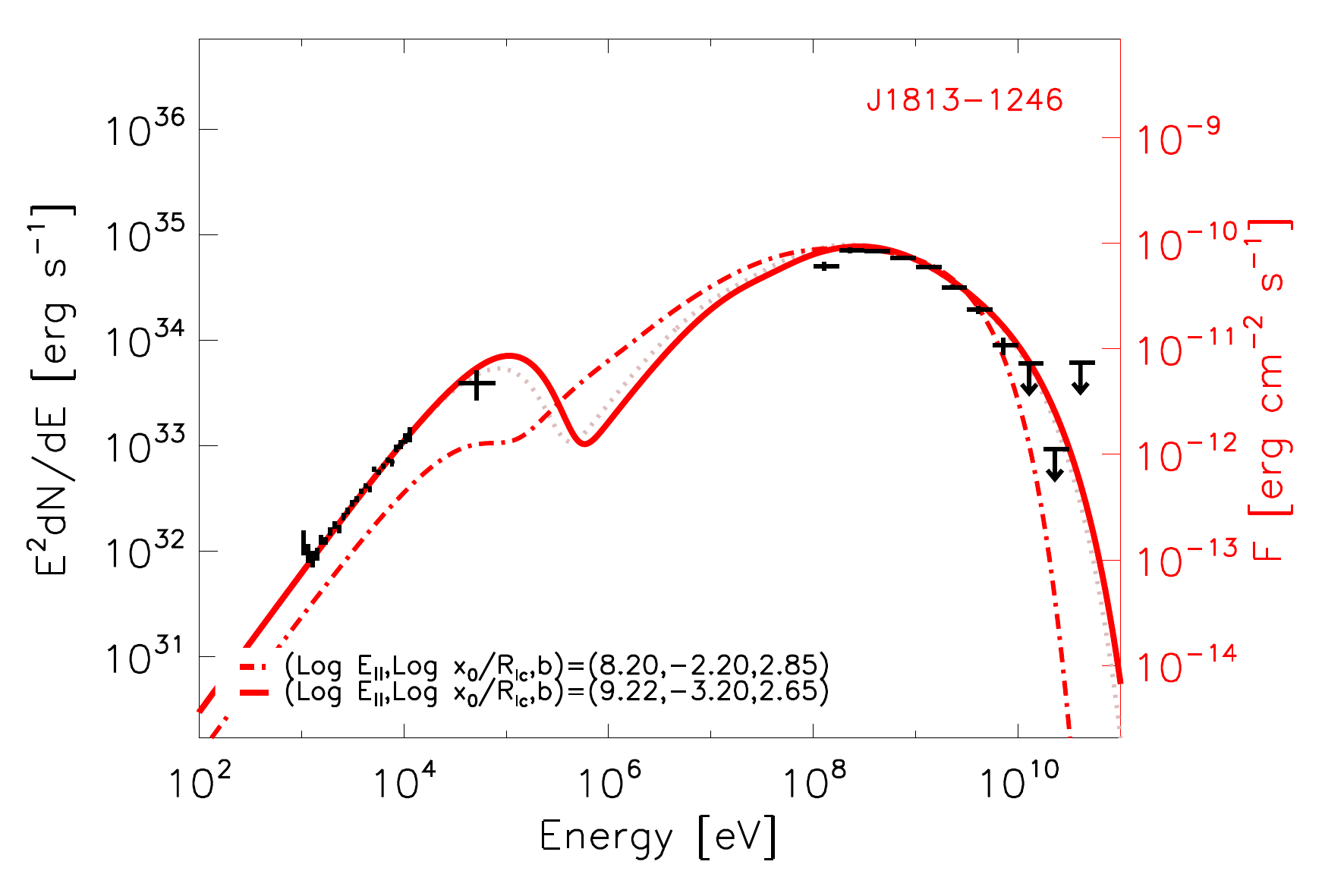}
\includegraphics[width=0.34\textwidth]{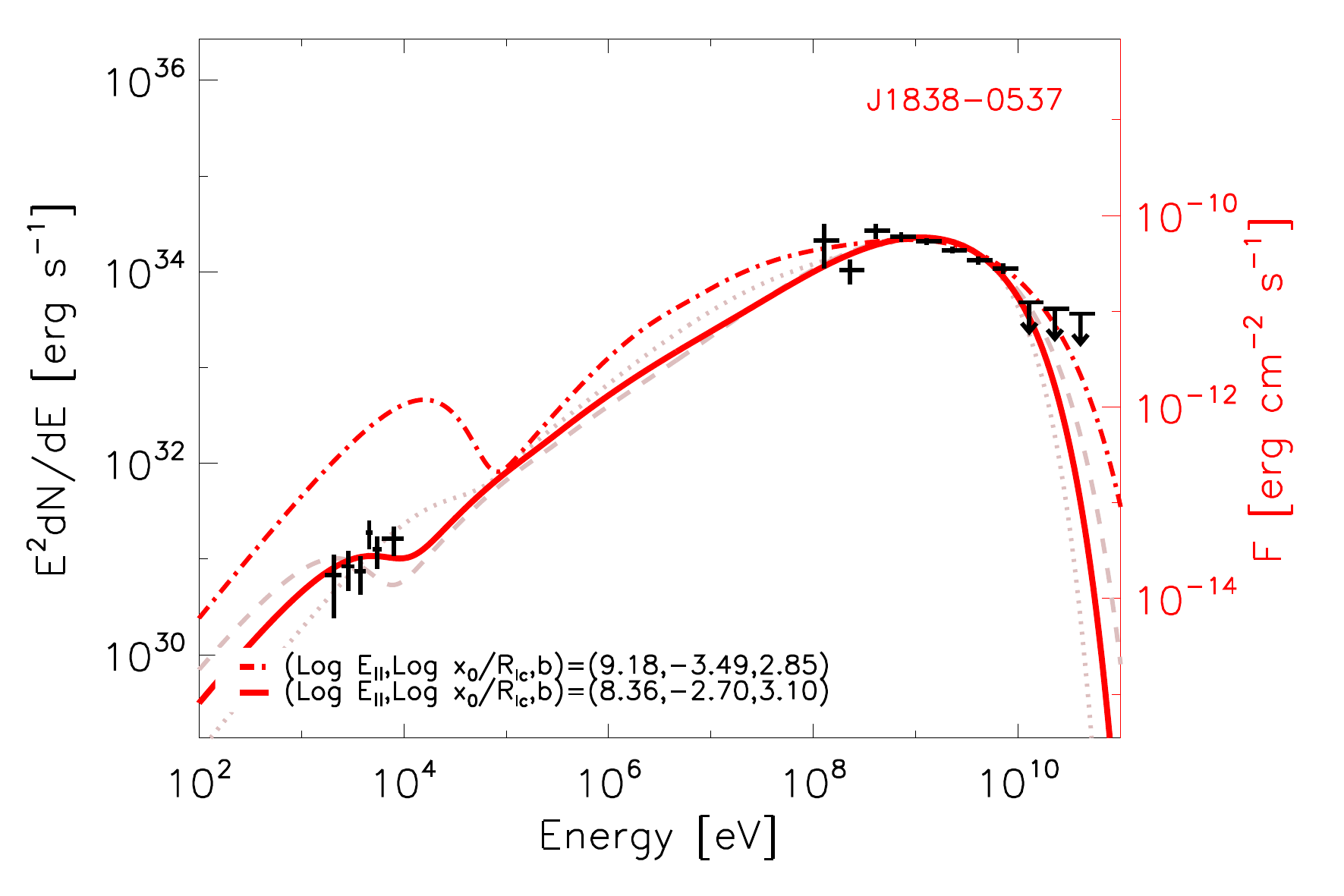}\hspace{-.25cm}
\includegraphics[width=0.34\textwidth]{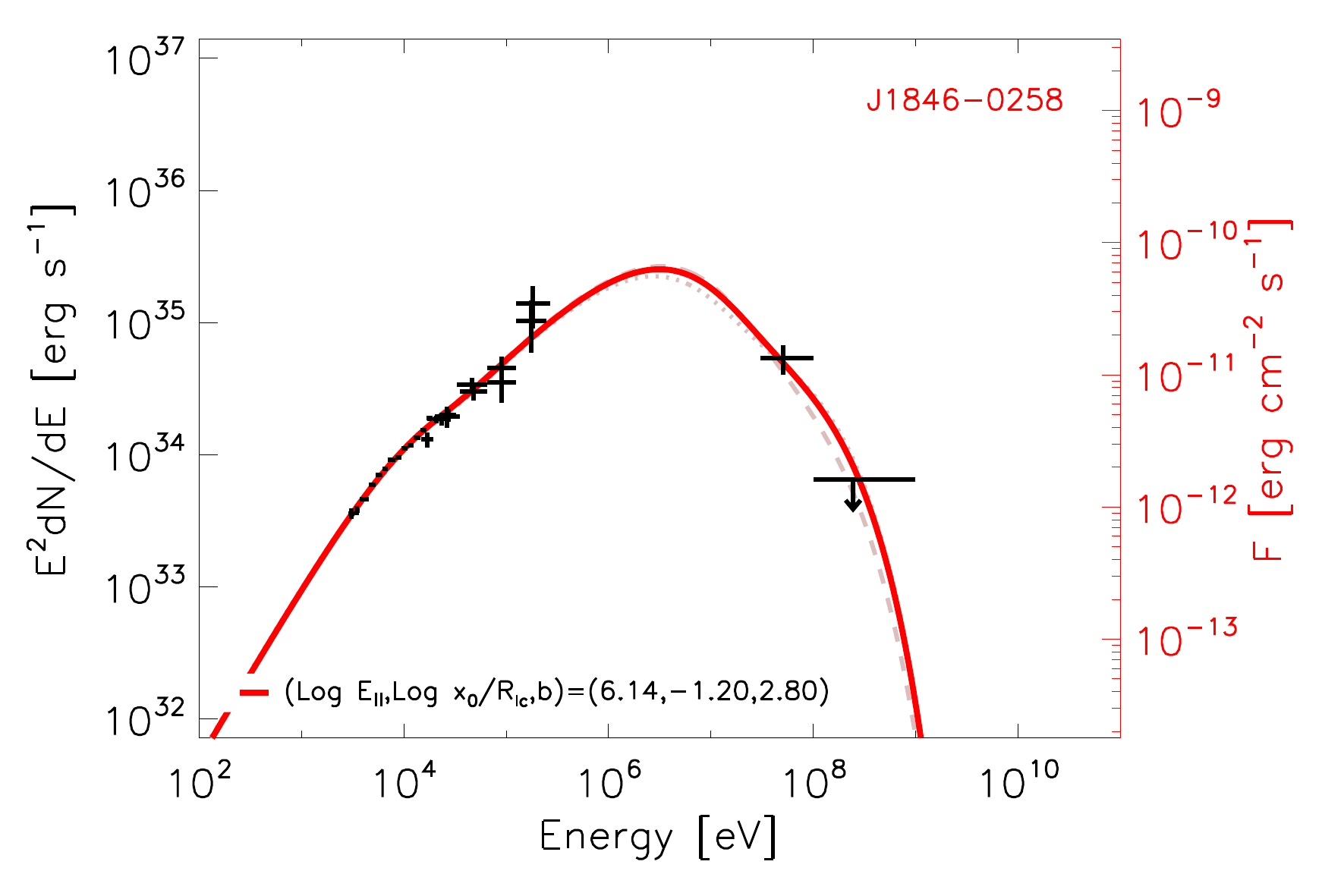}\hspace{-.25cm}
\includegraphics[width=0.34\textwidth]{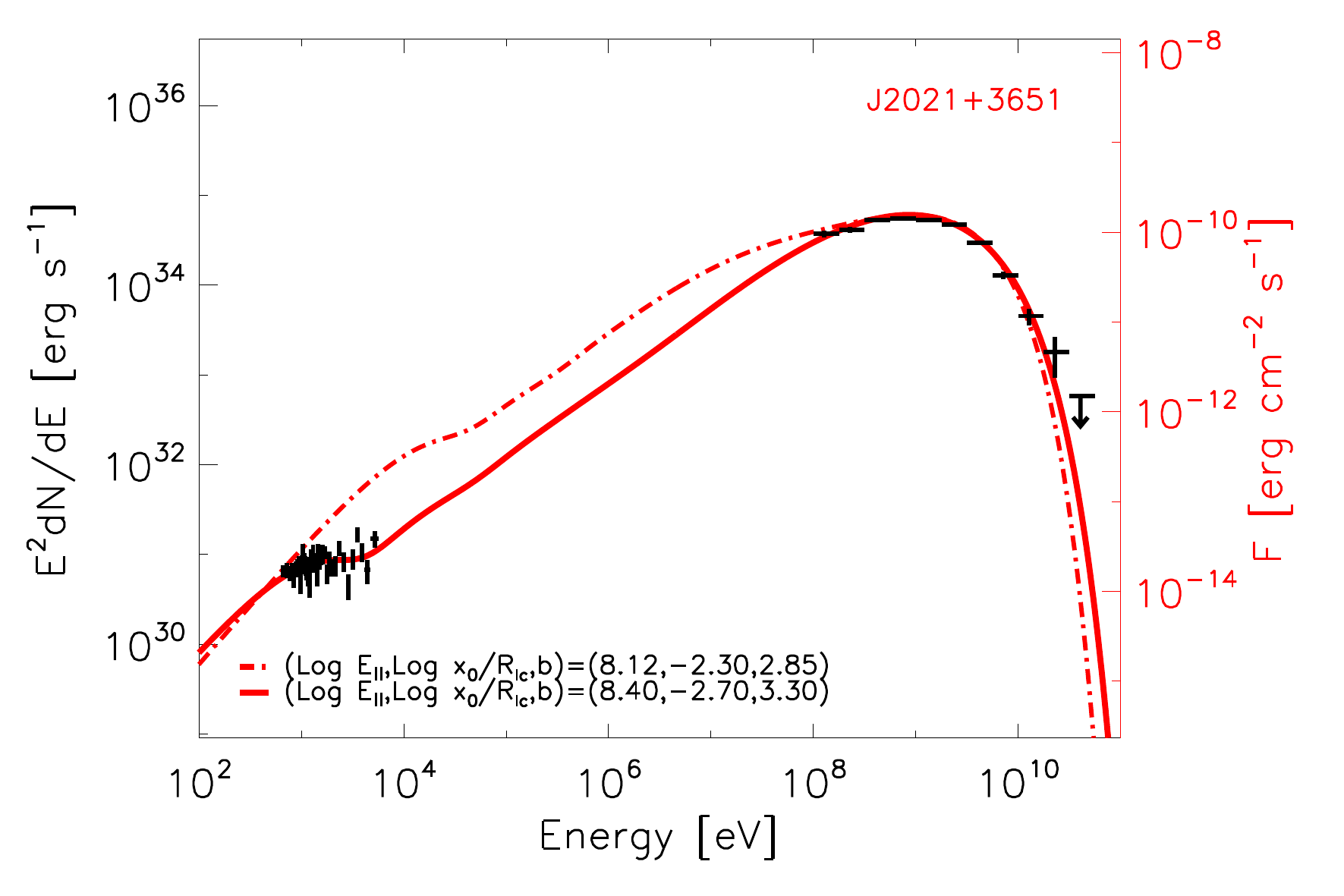}
\includegraphics[width=0.34\textwidth]{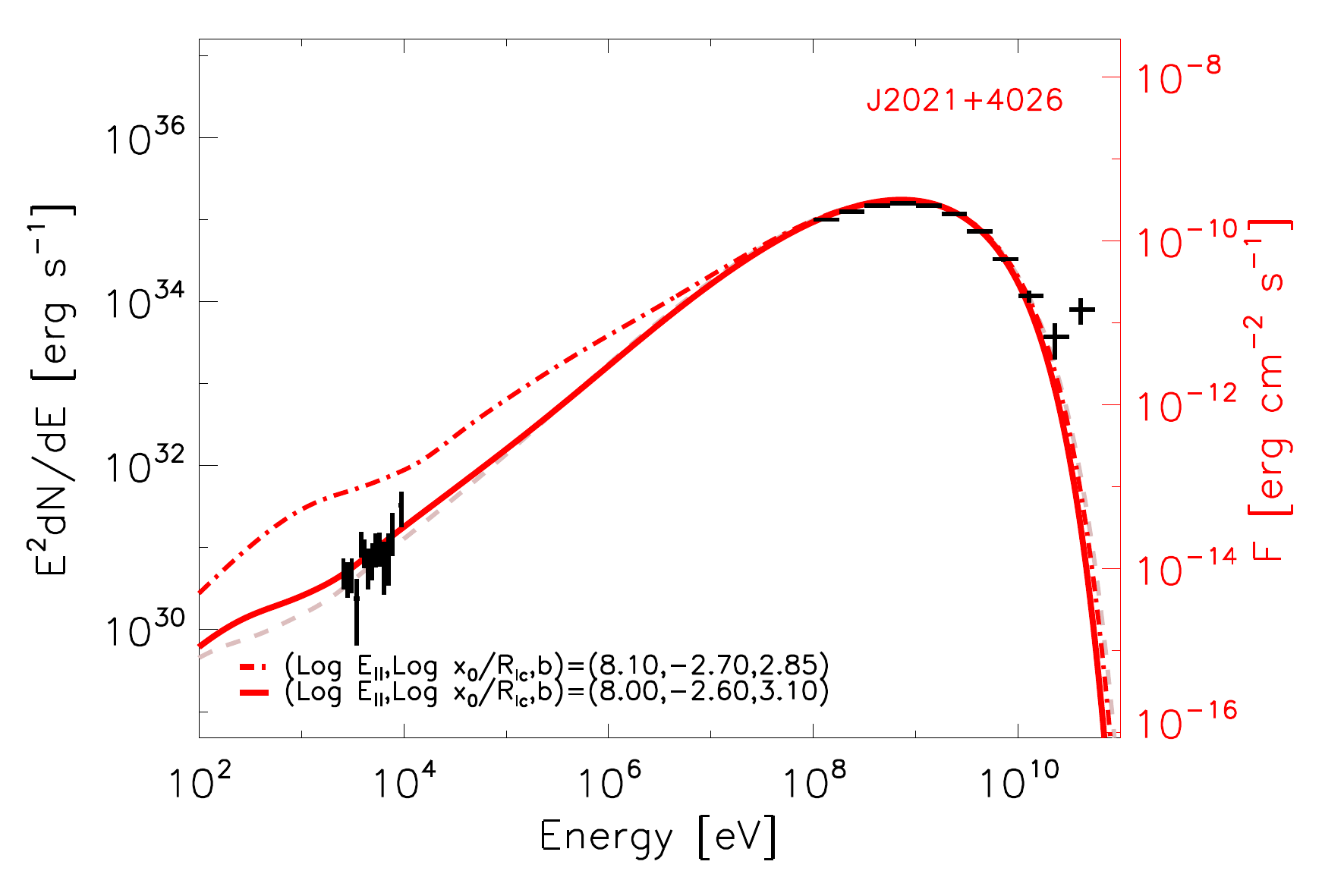}\hspace{-.25cm}
\includegraphics[width=0.34\textwidth]{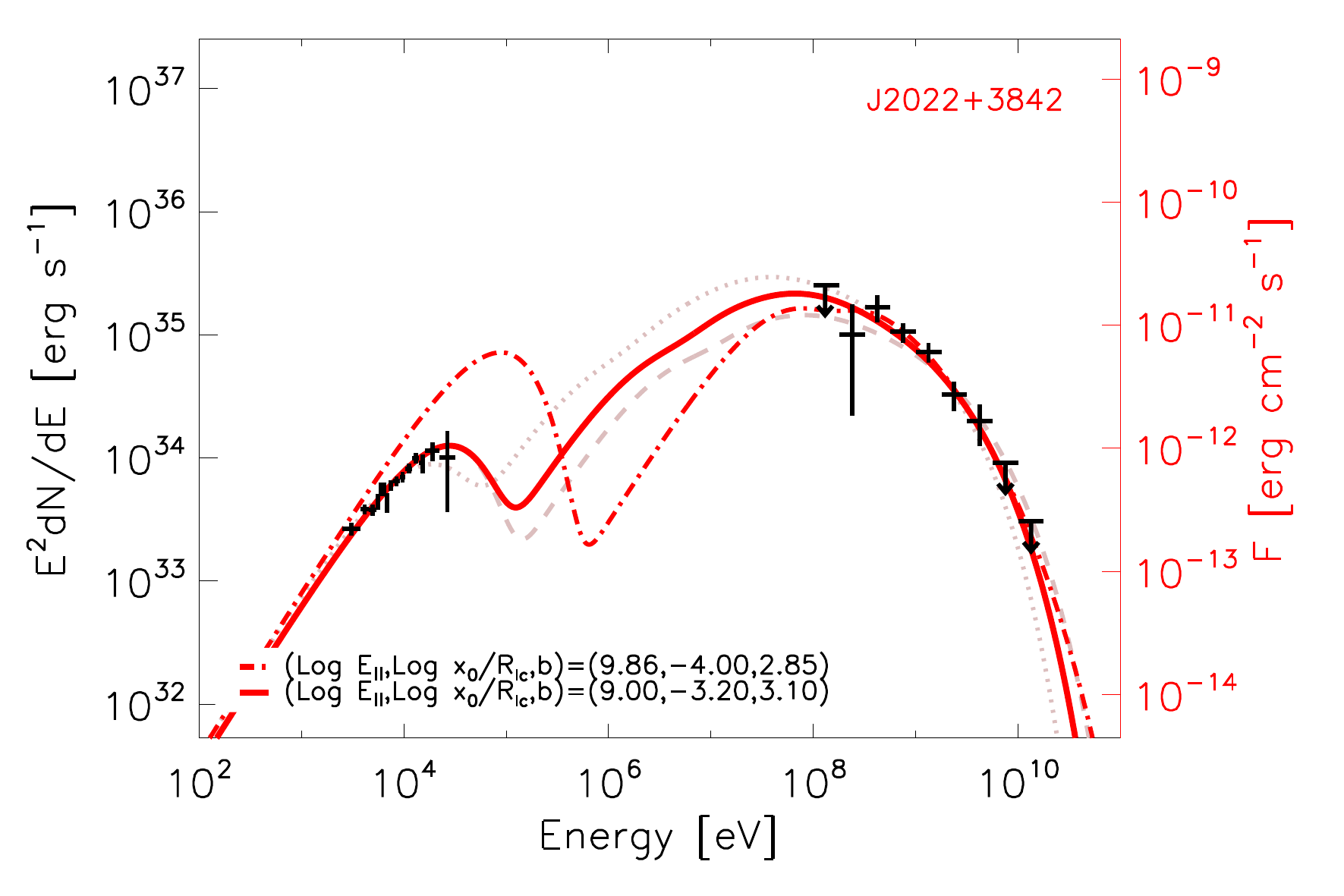}\hspace{-.25cm}
\includegraphics[width=0.34\textwidth]{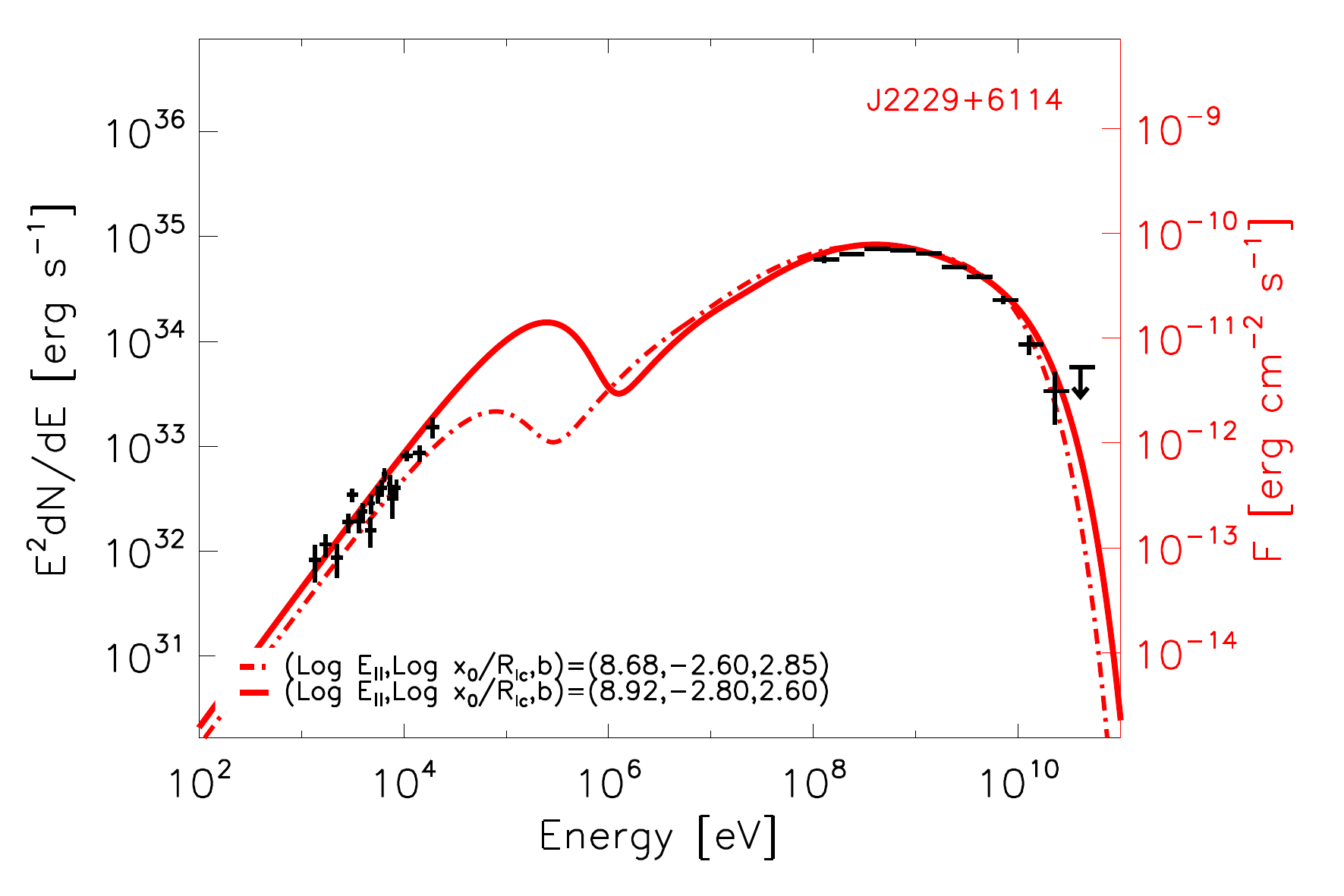}
\contcaption{
}
\label{sed2}
\end{center}
\end{figure*}

Each panel of Figure~\ref{sed1}  show several curves: the solid red line is the best-fit of our model to the full observational data set.
For each case, the two milder, grey-colored curves lying close to the solid line correspond to the predicted SEDs 
resulting from varying $E_{||}$ and $x_0$ within their maximum ranges of their 1$\sigma$ uncertainty.
In particular, we plot with a dotted line the minimum (maximum) in the range of $E_{||}$ ($x_0/R_{lc}$) and the corresponding value of $b$
(usually the minimum in the range of $b$);
and with a dashed line 
the maximum (minimum) in the range of $E_{||}$ ($x_0/R_{lc}$) and  the corresponding value of $b$
(usually the maximum in the range of $b$). 
They give an idea of the overall uncertainty of the predicted SED.
When the  uncertainties in the fitted parameters are low, these curves overlap in the scale plotted. 
%

%
%
%
%

In the same Figure~\ref{sed1} we also plot the best-fit SEDs obtained when considering the gamma-ray data only, indicated by red dash-dotted lines. In these fits, we fix $b=2.85$ in all cases, in order to reduce the degrees of freedom. This choice is motivated by the fact that, as it was shown in previous studies (see \cite{paper2,torres18}) the high-energy gamma-ray data can barely distinguish the value of $b$, especially if there is a lack of well determined points at low and high gamma-ray energies. 
This is also confirmed by these figures.
Physically, this happens because the $b$-parameter (i.e., the magnetic gradient of the field intensity) affects mostly the synchrotron-dominated synchro-curvature radiation, whereas the gamma-ray emission occurs in full synchro-curvature, and sometimes, in curvature-dominated regimes.
%
In the fitting to the gamma-ray data only, then, the dotted-line model of Figure~\ref{sed1} is based on the exploration of only two parameters ($E_{||}, x_0$) and a global scaling (the normalization $N_0$).
The central values of all the best-fit parameters are quoted within each panel, whereas the fitted parameters and uncertainties of the X-ray to gamma-ray fit are given also in the first panel of Table \ref{common-fits}. 
The latter table also shows the pulsar type, name, the distance used in the conversion of flux to luminosities, the best fit values, and their 1$\sigma$ uncertainty. 
%
%

\begin{figure*}
\begin{center}
\includegraphics[width=0.34\textwidth]{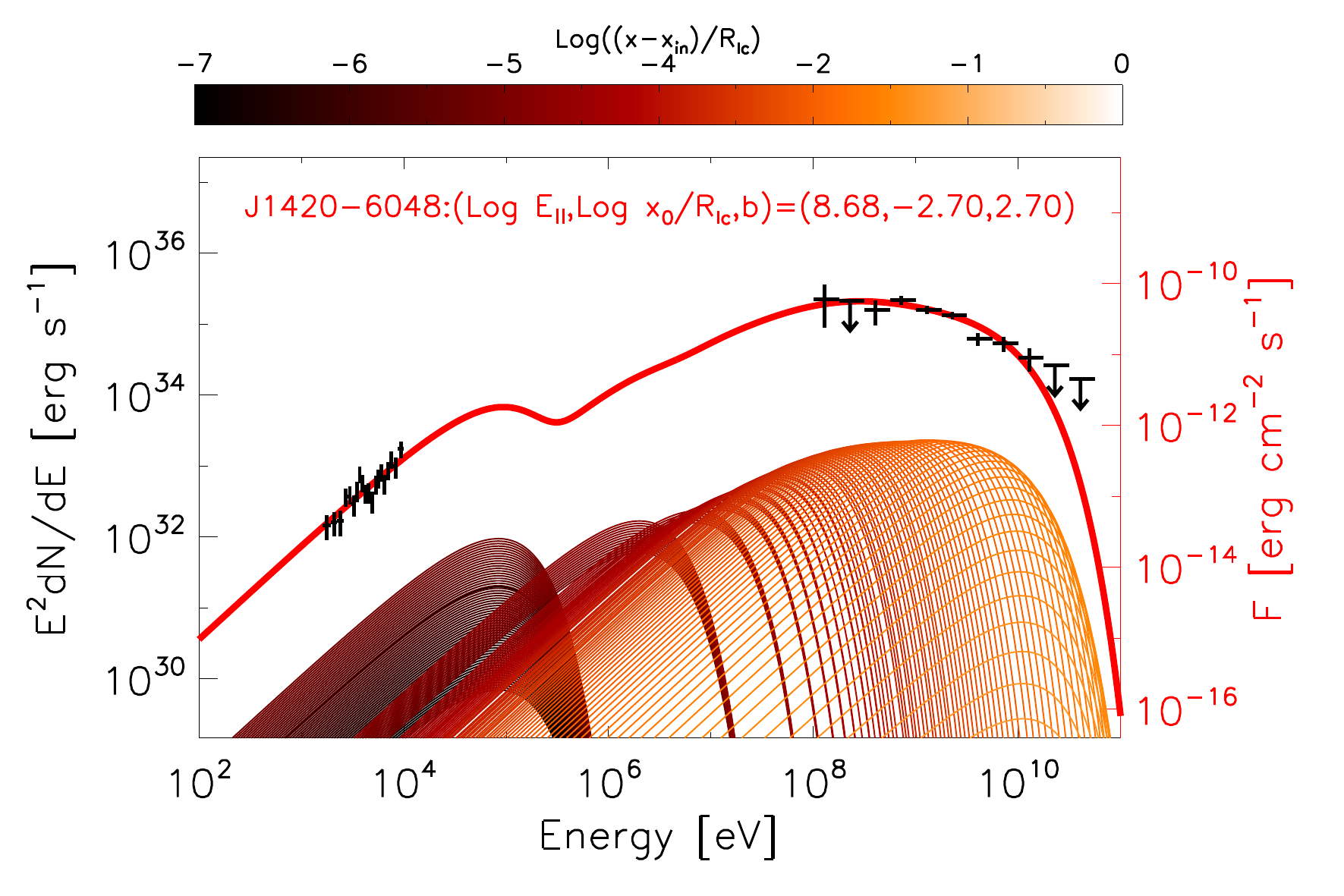}\hspace{-.25cm}
\includegraphics[width=0.34\textwidth]{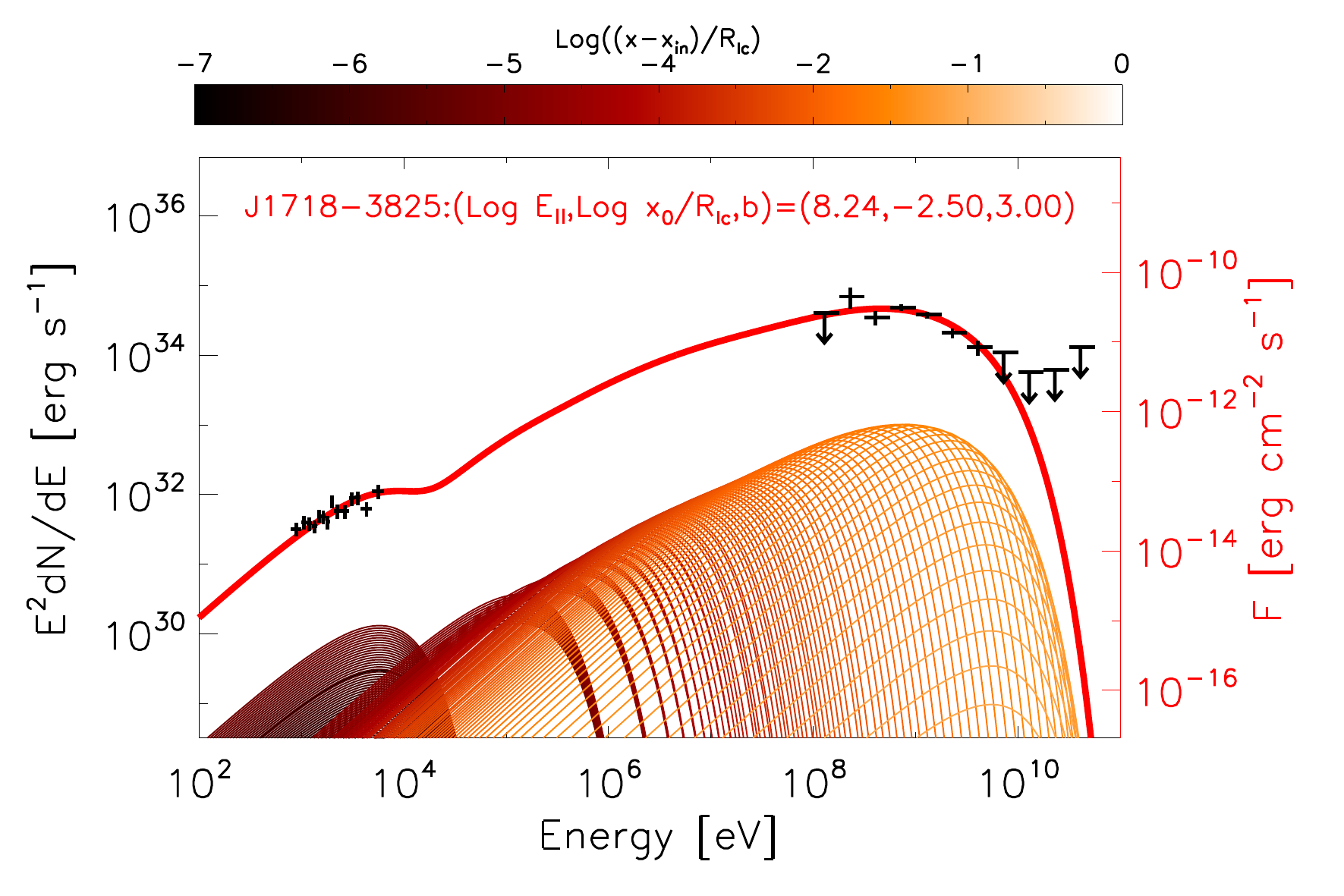}\hspace{-.25cm}
\includegraphics[width=0.34\textwidth]{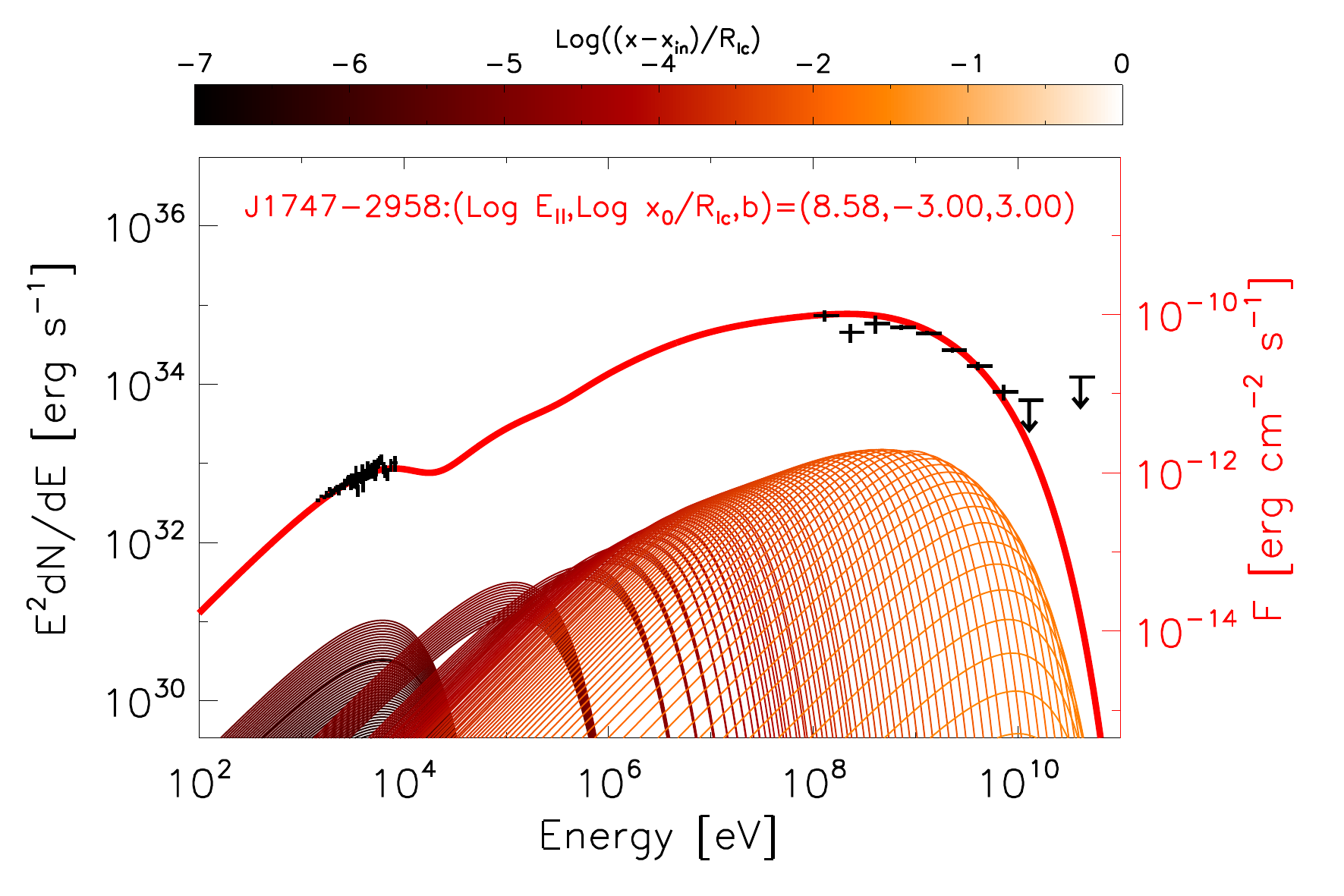}
\includegraphics[width=0.34\textwidth]{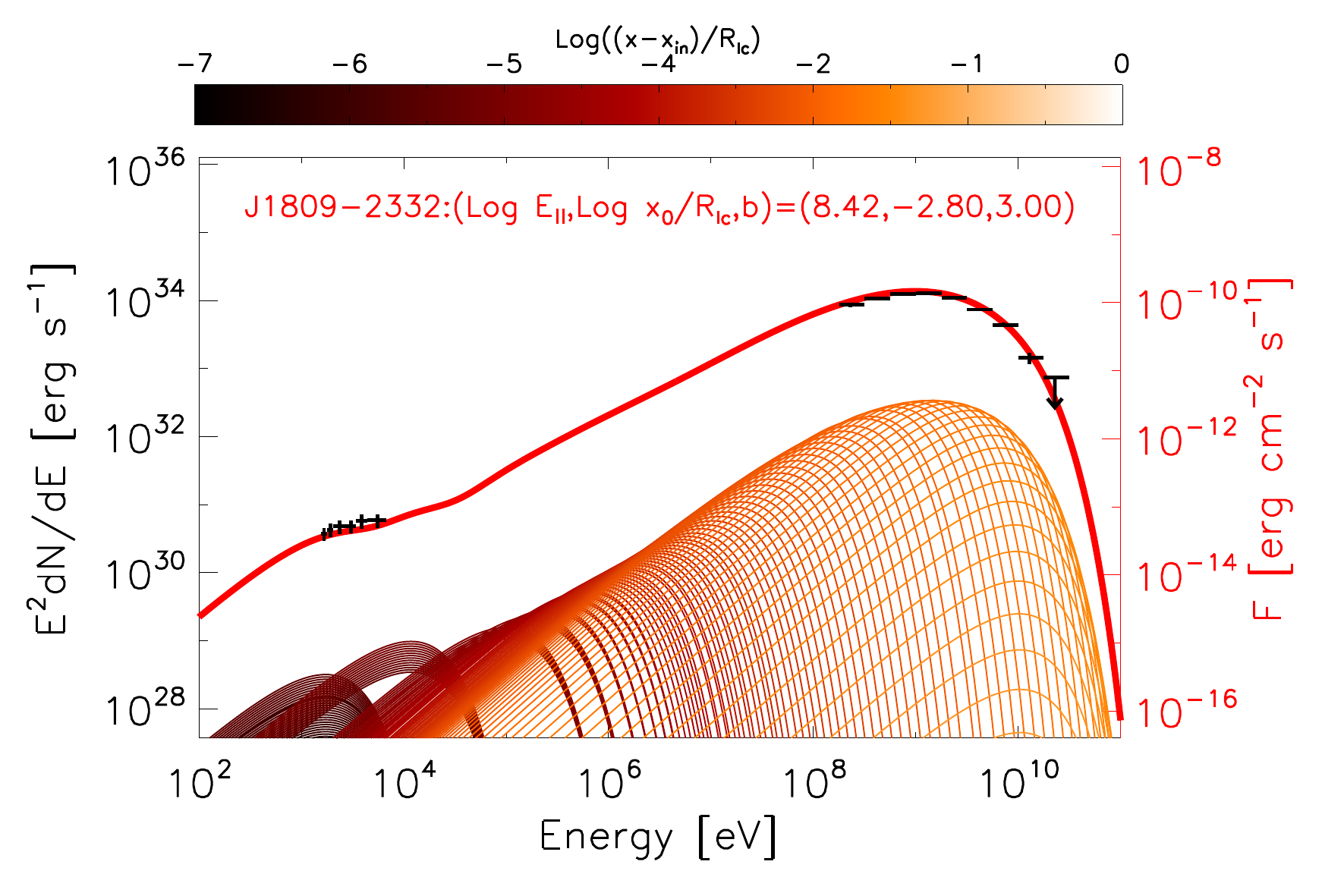}\hspace{-.25cm}
\includegraphics[width=0.34\textwidth]{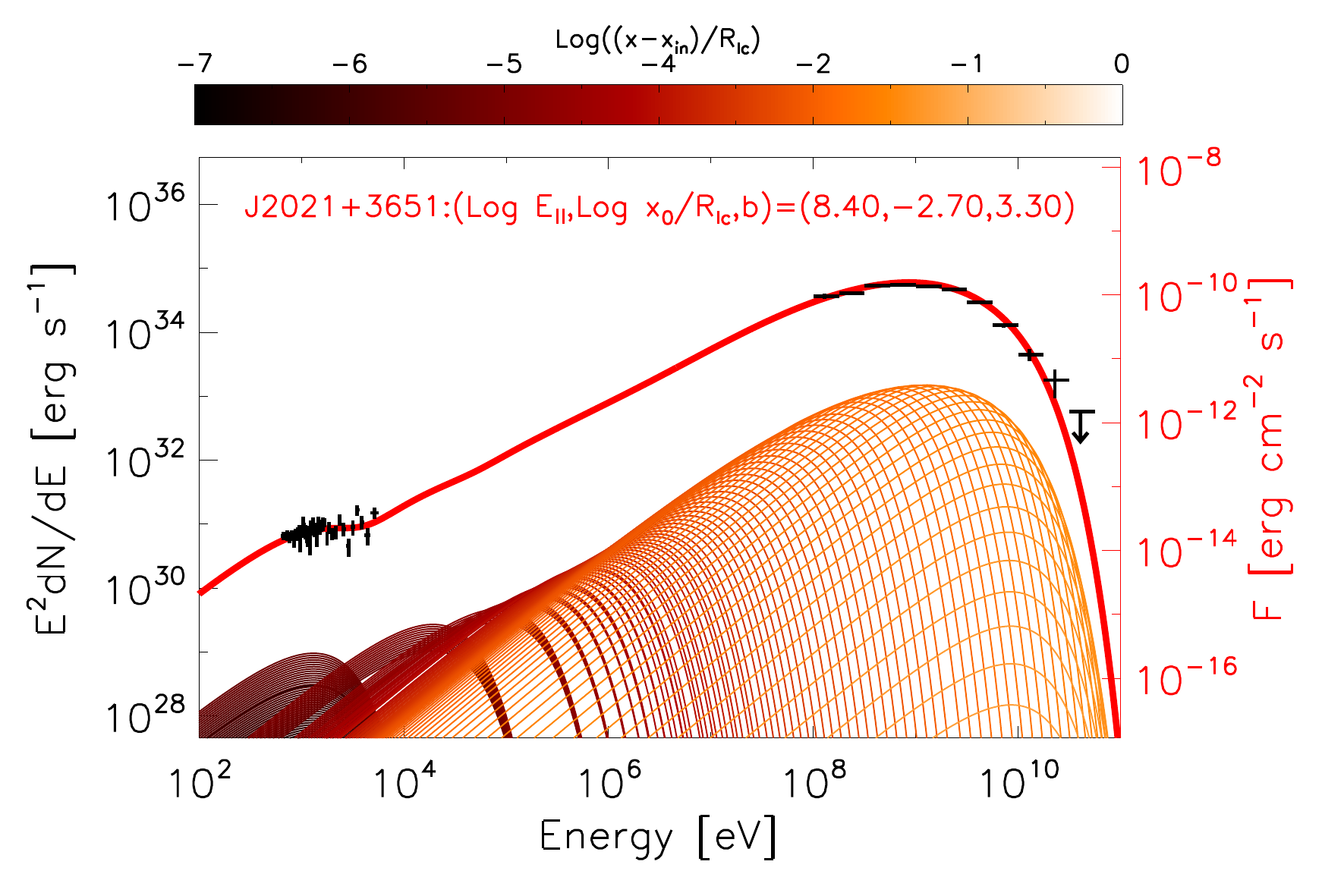}\hspace{-.25cm}
\includegraphics[width=0.34\textwidth]{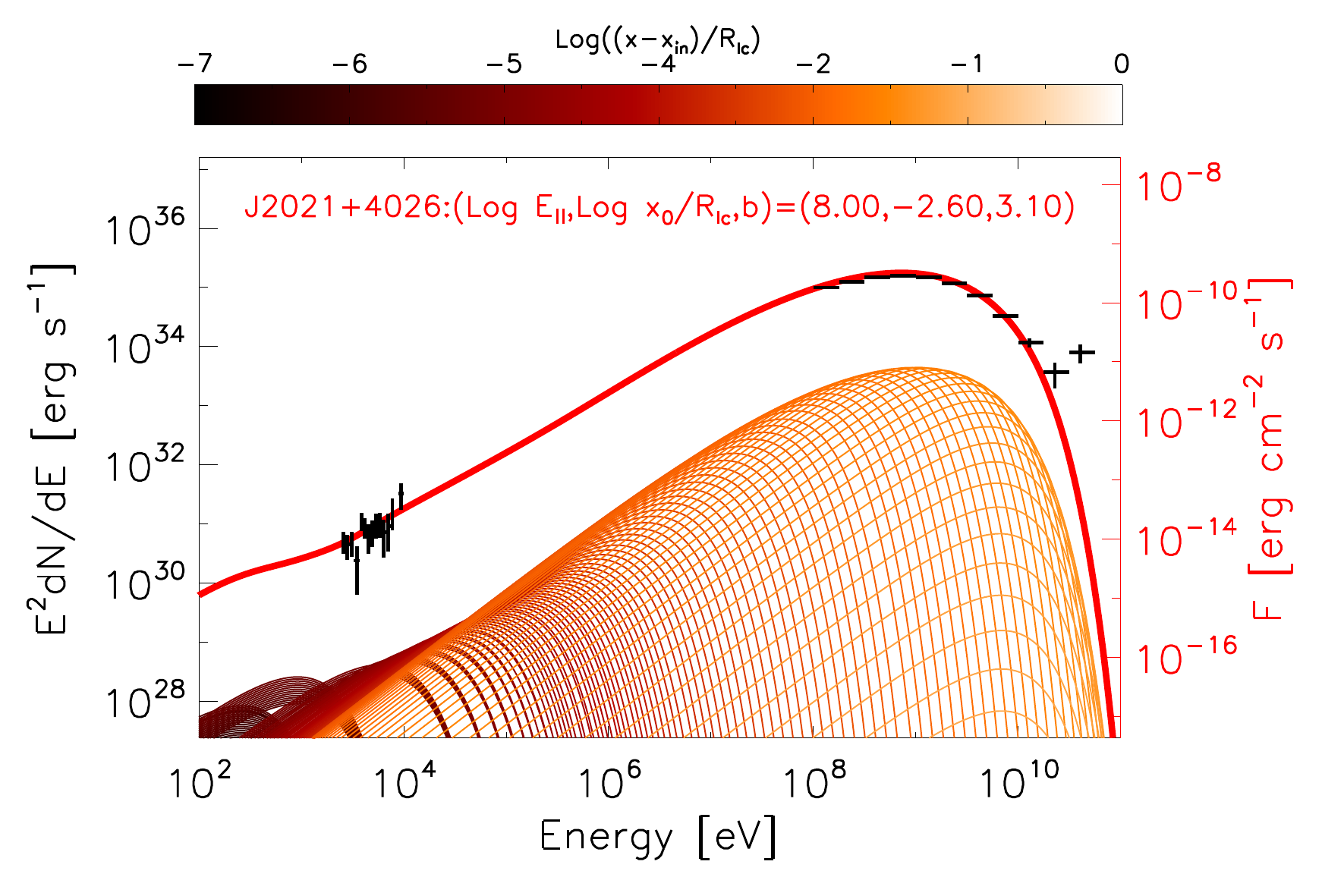}
\caption{
Contribution of each part of the particle trajectories along the accelerating regions (color-coded at the top) to the total fit in six examples of the pulsars studied in Figure~\ref{sed1}.
See text for discussion.
}
\label{formation}
\end{center}
\end{figure*}

Figure~\ref{sed1} shows that the gamma-ray-based best-fits SEDs can be used to predict the X-rays, where they result to be close 
(typically within a factor of a few) to the observed SED in that energy range.
This extended sample confirms what was advanced by \cite{torres18}: using the model to fit the gamma-ray data only (available from the {\it Fermi}-LAT or other future surveys) could predict which pulsars could be observable in X-rays.
%


Figure~\ref{formation} provides examples of the total SED formation for some of the pulsars of our sample, dissecting how the model works. 
Each curve plotted in the figure represents the contribution of a given portion of the particles's trajectory along the accelerating region, 
whose position is color-coded at the top. In order to make the figure legible, not all spatial bins actually computed are shown. 
The sum of all these contributions constitutes the final fit, depicted by the red curve (which corresponds to the fit shown in Figure~\ref{sed1}). 
Figure~\ref{formation} is then similar to Figure 2 of \cite{torres18}, and shows how the same mechanism applies for all the new cases studied in Figure~\ref{sed1}.
We see that the X-ray emission is produced as a result of the emission at the initial part of the particle trajectories, when there is a large loss of the momentum component perpendicular to the magnetic field, and a consequent fast decrease of the pitch angle.
In turn, the gamma-ray emission is produced further out in the particle's evolution, as a result of synchro-curvature emission that starts (but in most cases not yet dramatically) being dominated 
by curvature.

\begin{figure*}
\begin{center}
\includegraphics[width=0.34\textwidth]{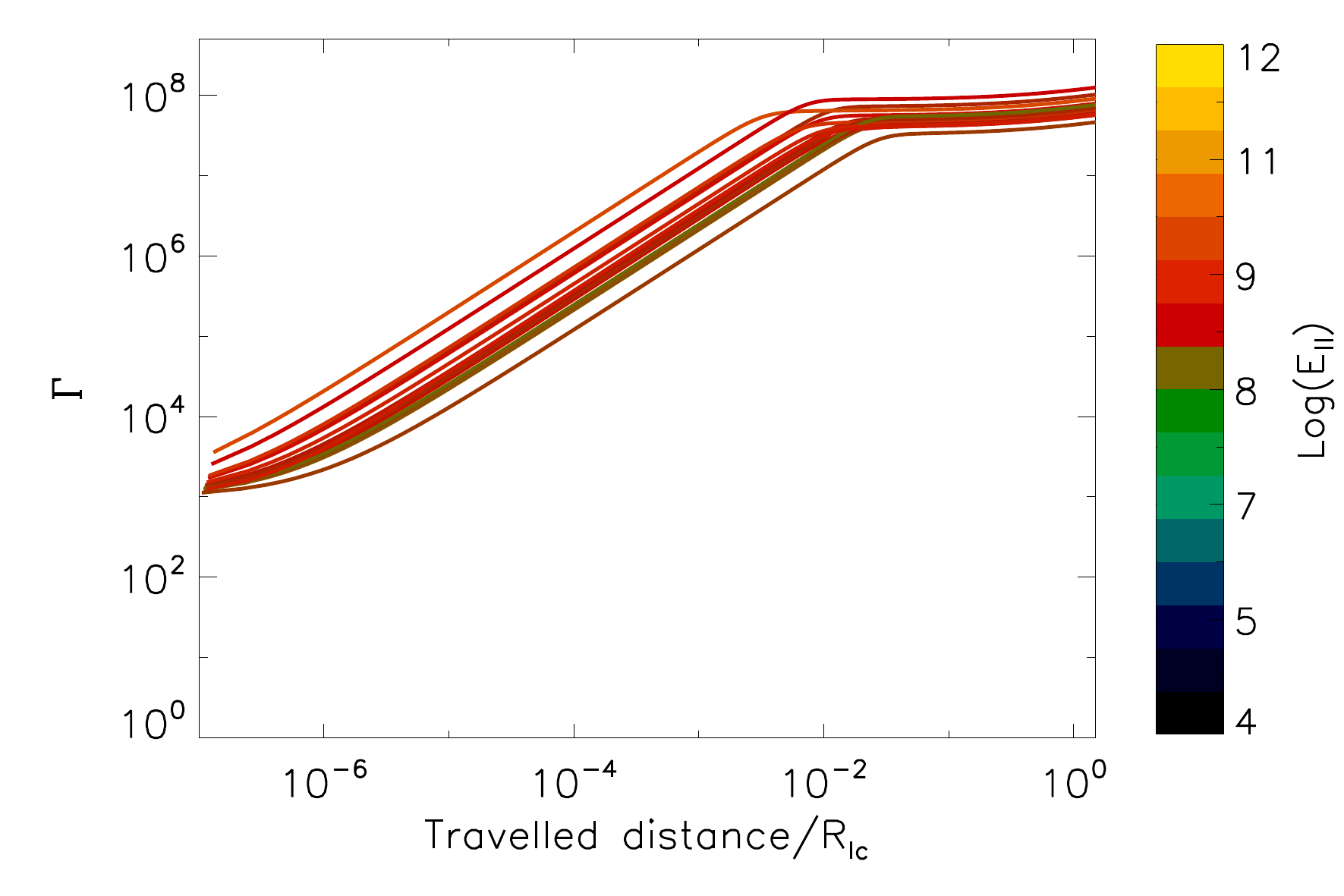} \hspace{0.3cm}
\includegraphics[width=0.34\textwidth]{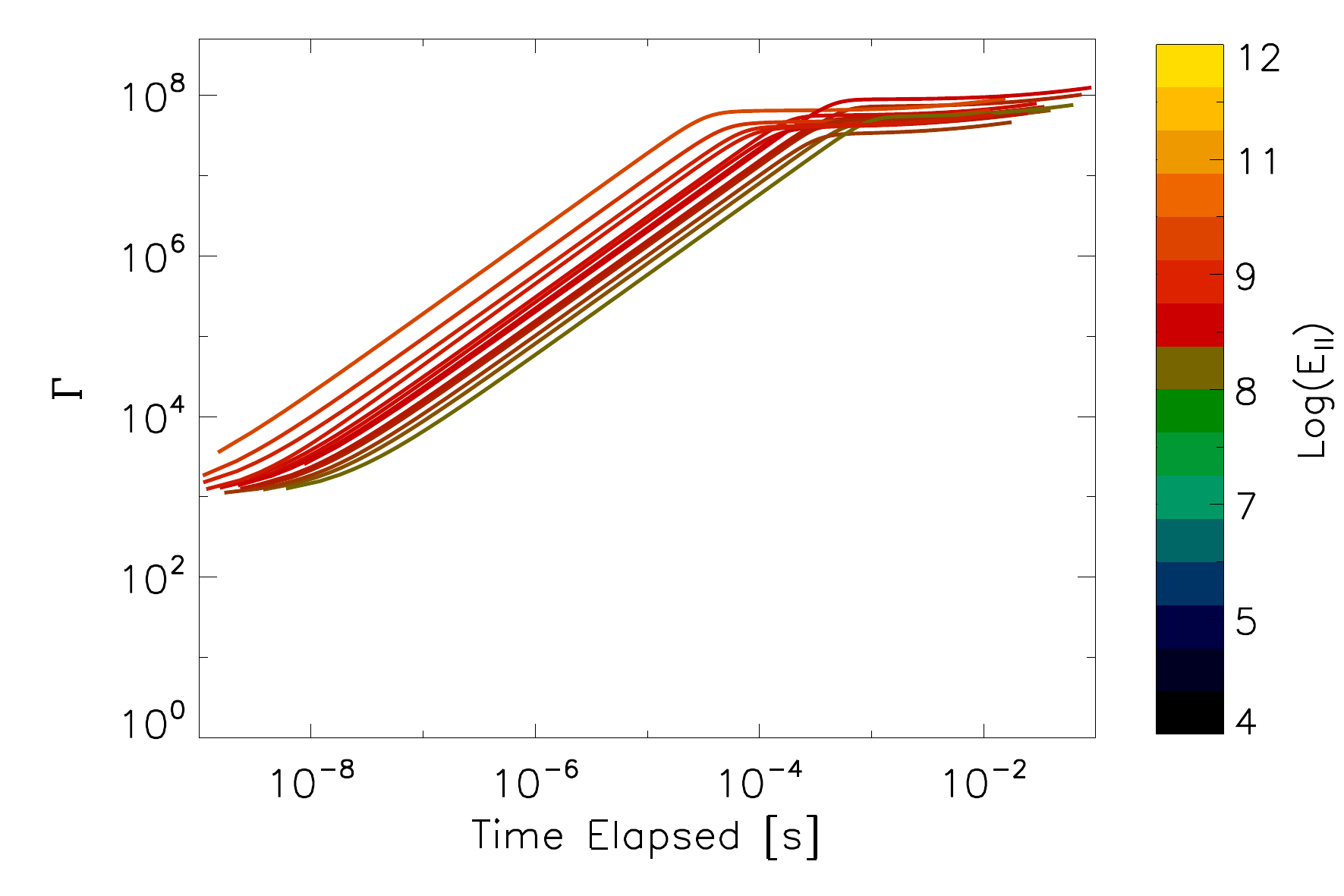}
\includegraphics[width=0.34\textwidth]{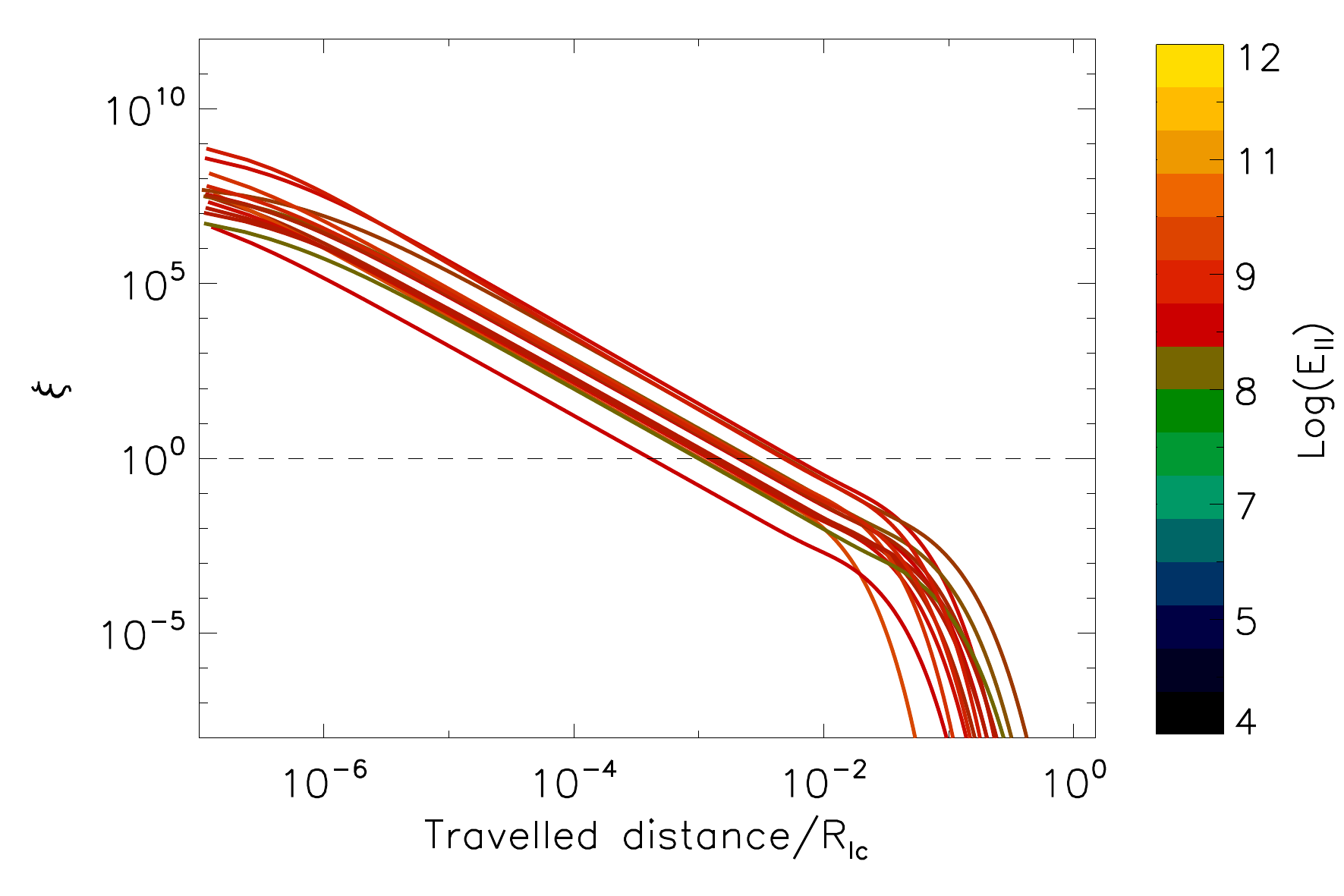}\hspace{0.3cm}
\includegraphics[width=0.34\textwidth]{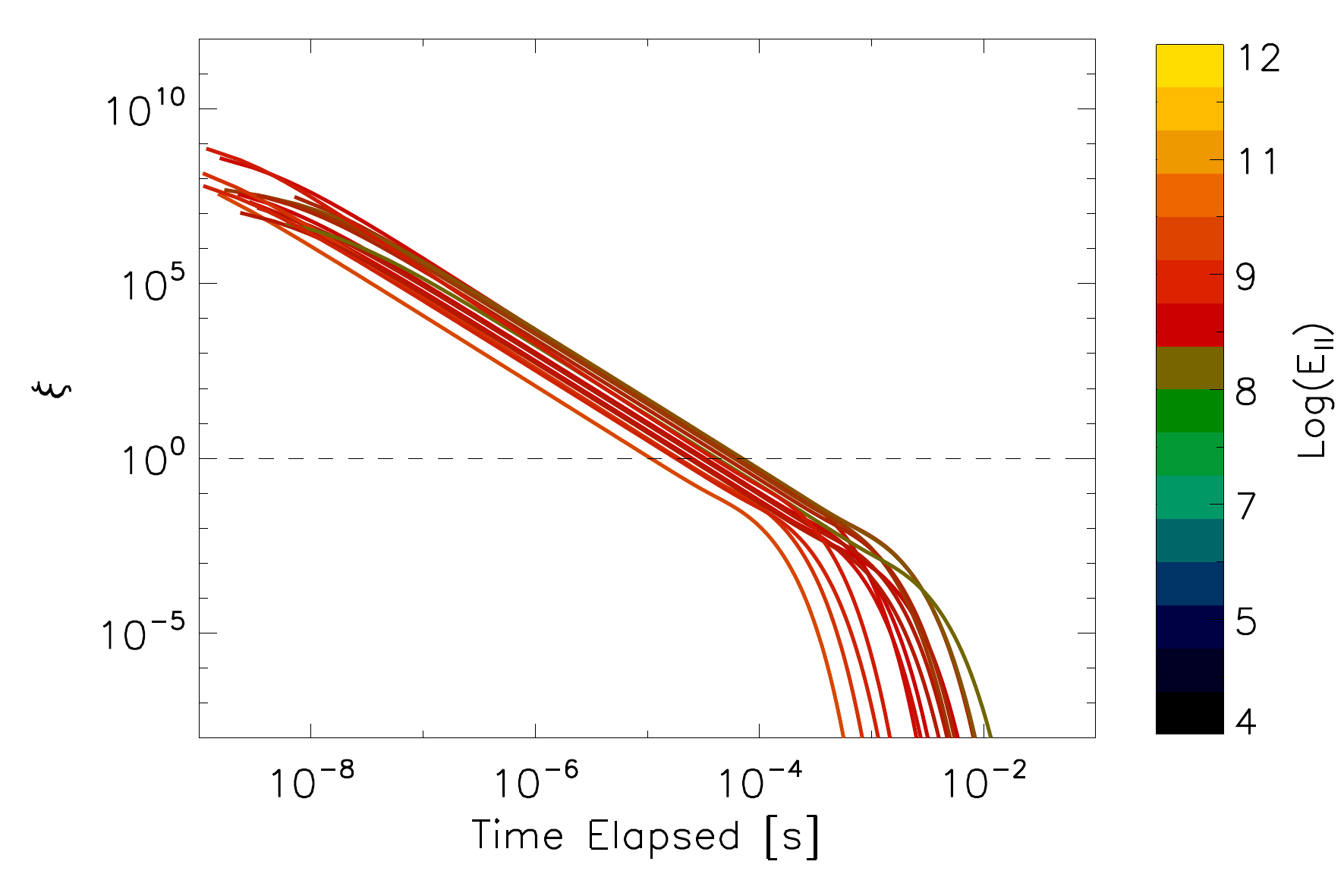}
\includegraphics[width=0.34\textwidth]{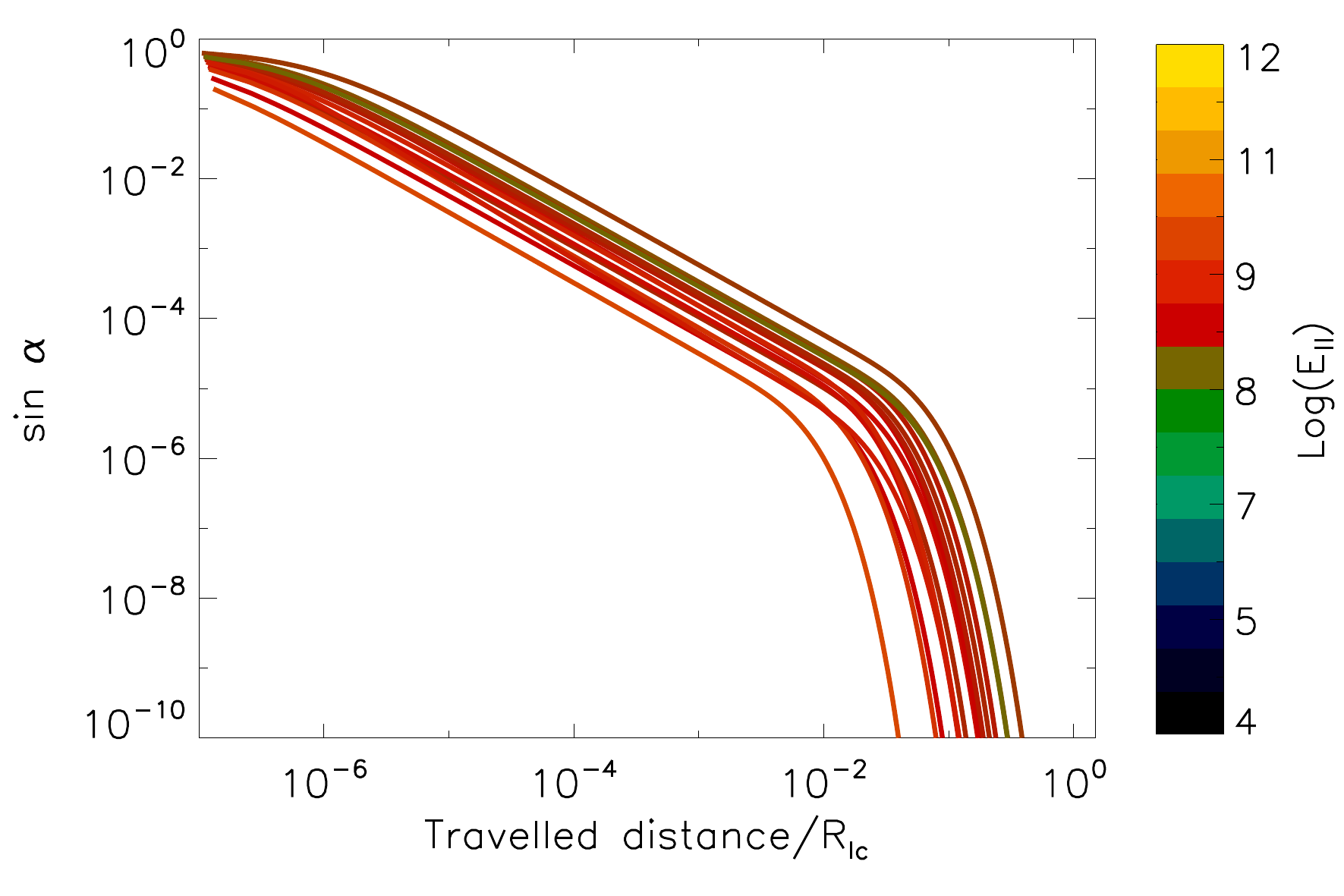}\hspace{0.3cm}
\includegraphics[width=0.34\textwidth]{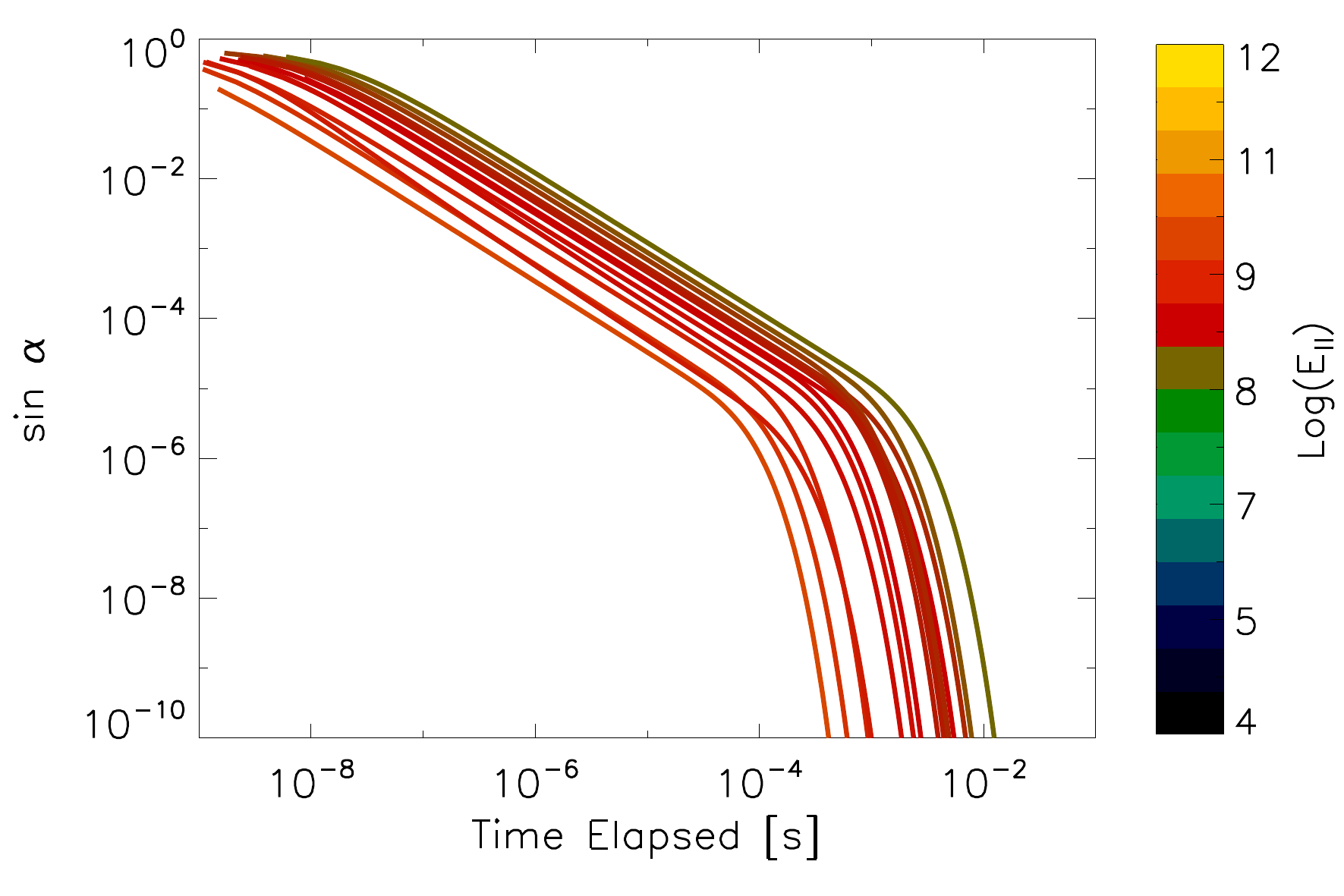}
\caption{Main properties of the trajectories of particles that generate the SED fittings of Figure~\ref{sed1}, both as a function of travelled distance (left) and time elapsed since injection (right panels). 
The color coding is used to represent the value of the accelerating electric field, in units of V/m.
}
\label{traj1}
\end{center}
\end{figure*}

This can be seen better in conjunction with the analysis of the properties of the trajectories themselves, output of the equations of motion.
Figure~\ref{traj1} shows three of the most important quantities entering into the computation of the SED: the Lorentz factor, the synchro-curvature parameter (recall that $\xi=1$ formally marks the transition
from synchrotron ($\xi \ll 1$) to curvature ($\xi \gg 1$) dominated emission) and the pitch angle.
Note how synchrotron emission is very relevant in the initial and less pronounced drop of the pitch angle, and how the Lorentz factor grows linearly when synchro-curvature dominates, until it saturates when the curvature-dominated radiation is large enough to balance the electric acceleration.

\begin{figure*}
\begin{center}
\includegraphics[width=0.34\textwidth]{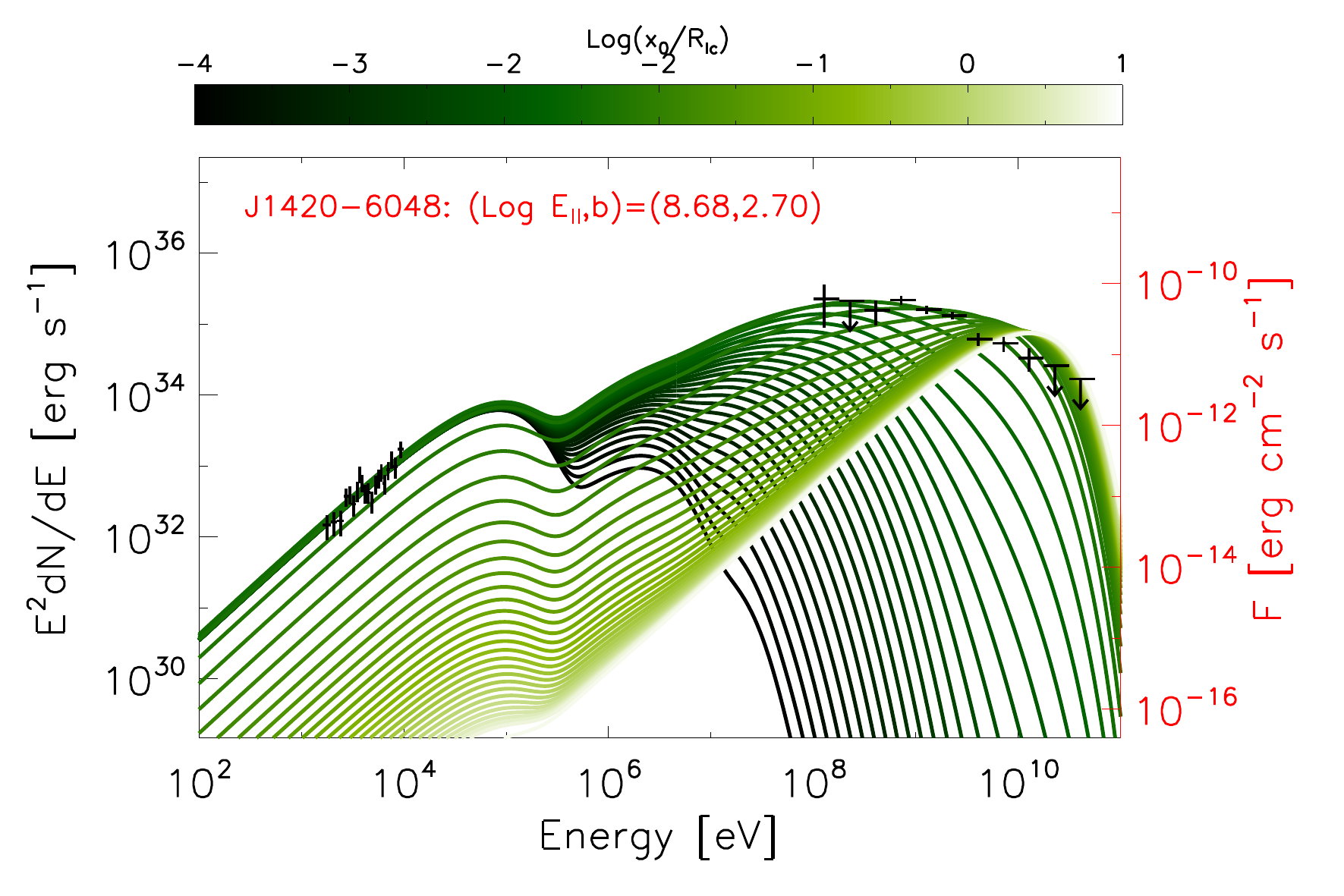}\hspace{-.25cm}
\includegraphics[width=0.34\textwidth]{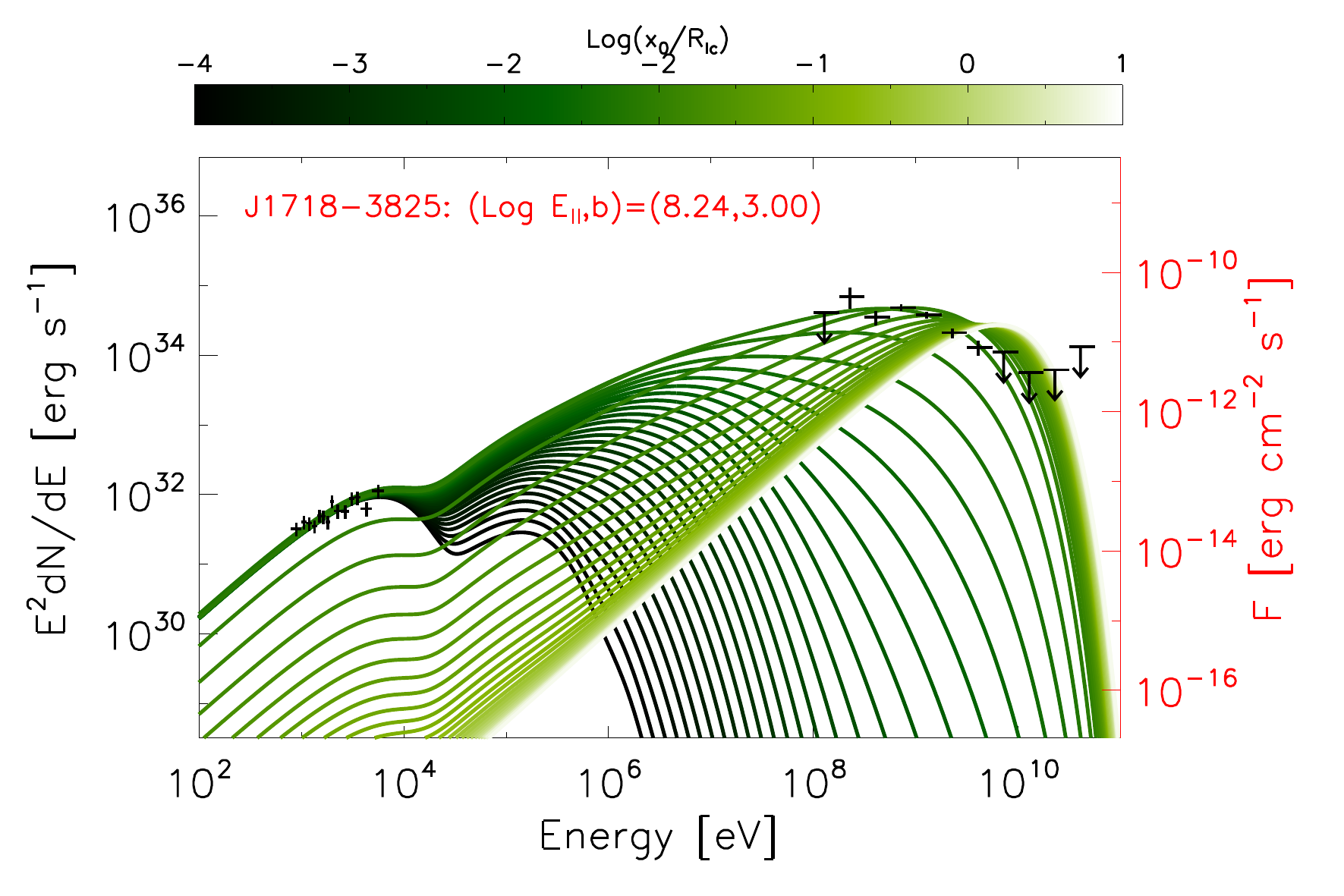}\hspace{-.25cm}
\includegraphics[width=0.34\textwidth]{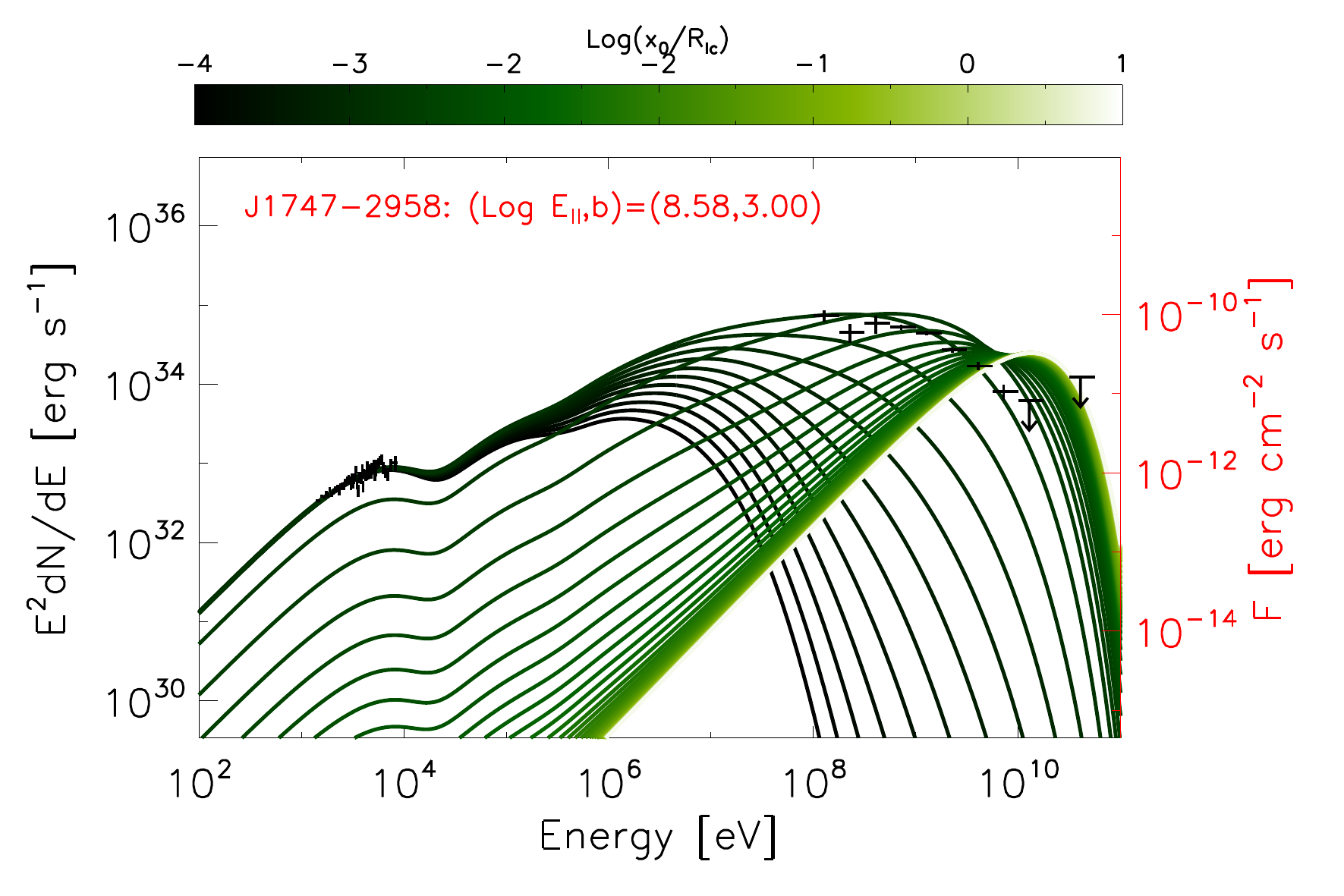}
\includegraphics[width=0.34\textwidth]{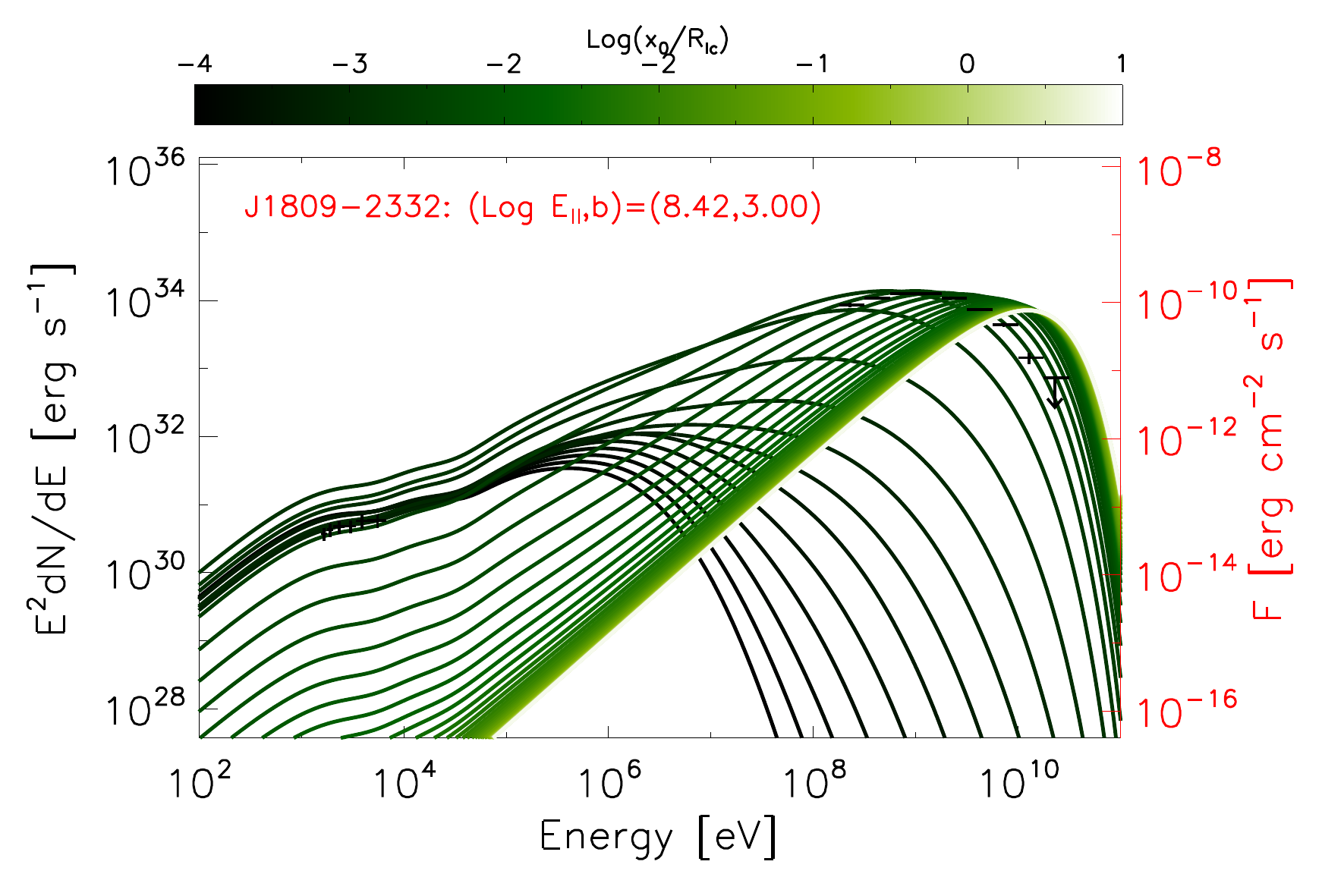}\hspace{-.25cm}
\includegraphics[width=0.34\textwidth]{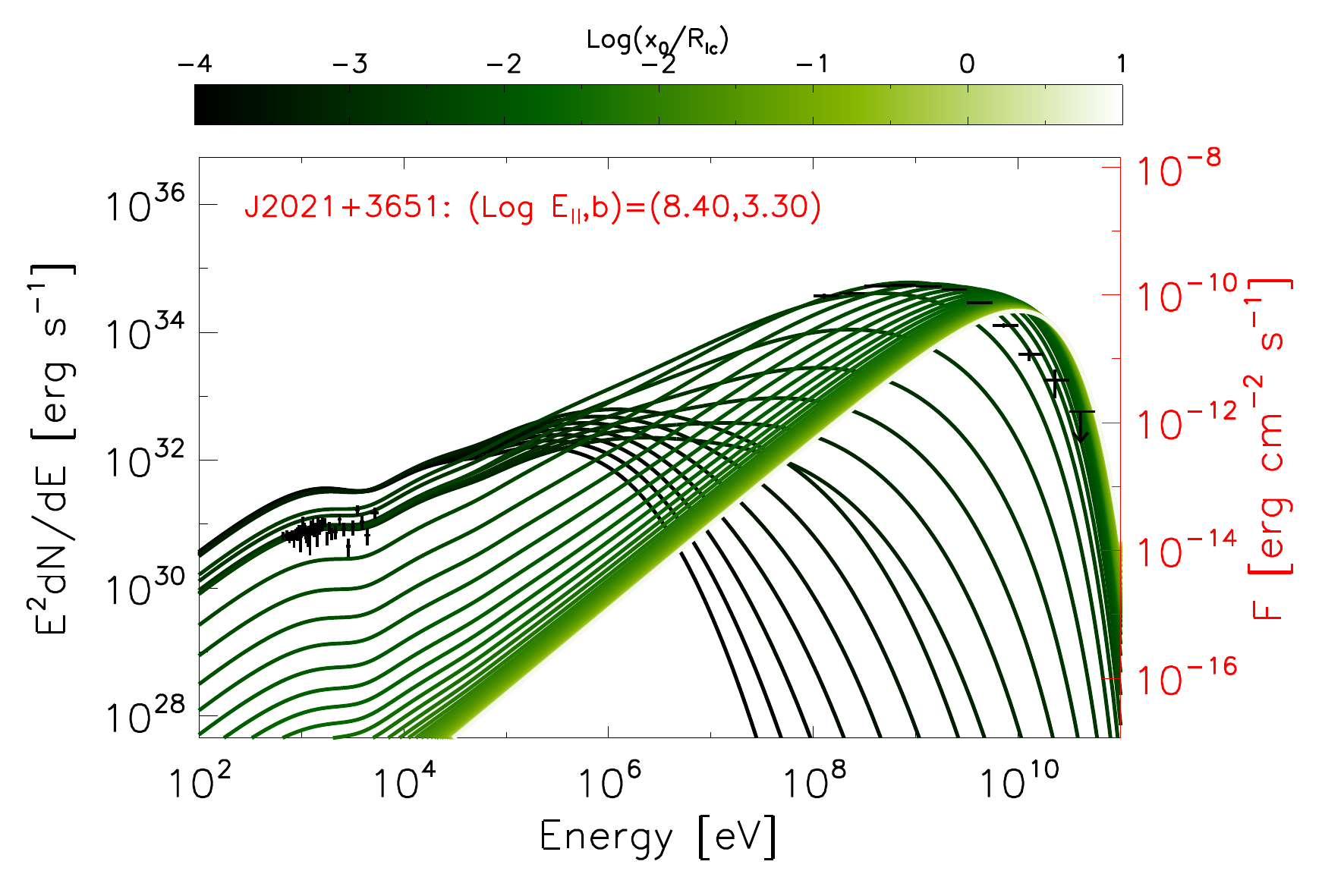}\hspace{-.25cm}
\includegraphics[width=0.34\textwidth]{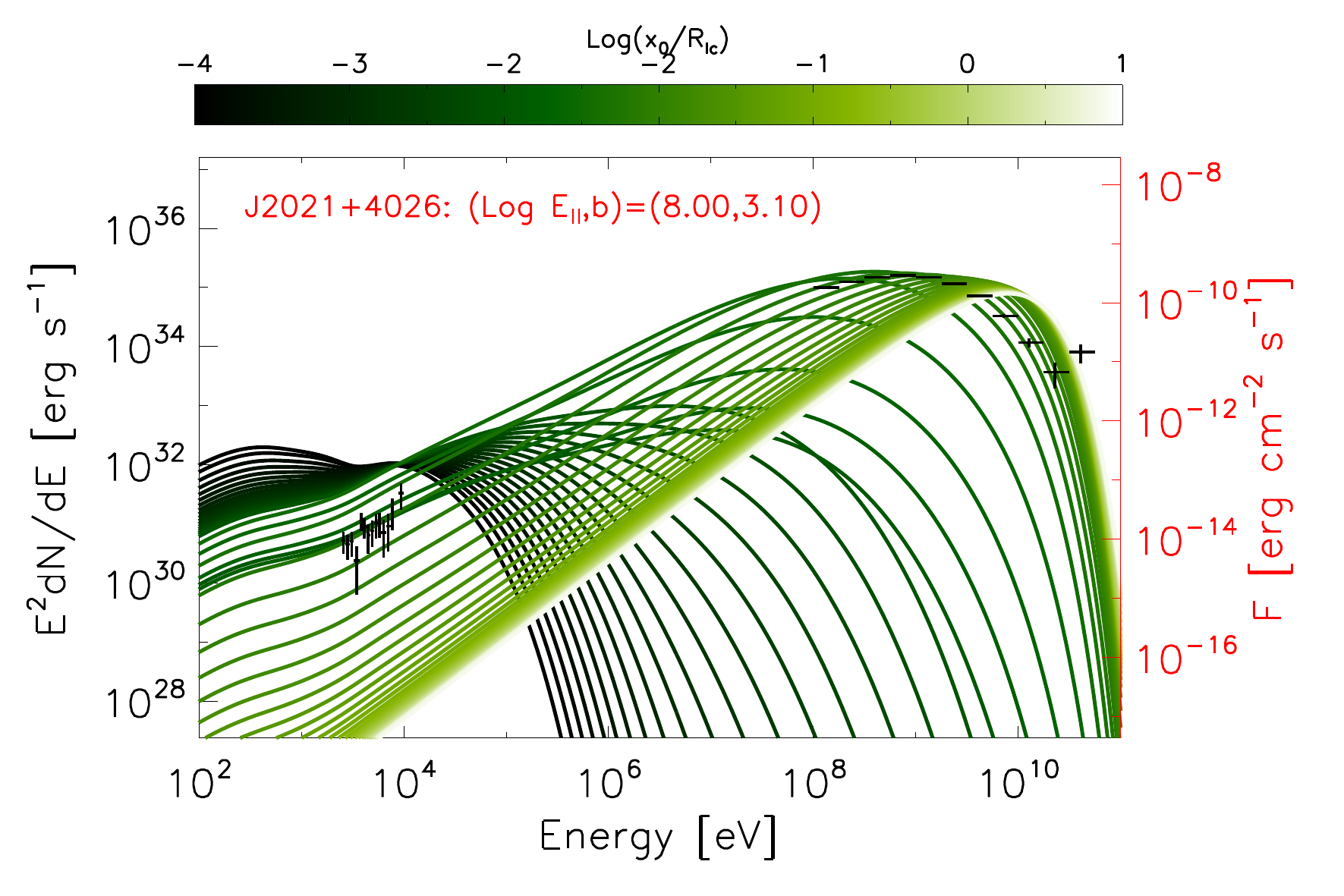}
\caption{The same six examples of Figure~\ref{formation} used to show the influence of the contrast parameter. Each of the panels has the electric field and magnetic gradient fixed at the values
found to be the best fit to the data in Figure~\ref{sed1}. The contrast is color-coded from uniform (formally, $x_0/R_{lc}\rightarrow \infty$, white color), to finite decreasing values $x_0/R_{lc} \in [10^{-1}-10^{-5}]$ (light to dark color).
}
\label{x0-influence}
\end{center}
\end{figure*}

Examples of how the SEDs are affected by the contrast parameter are shown in Figure~\ref{x0-influence}, and again we find similarity
with the cases studied earlier by \cite{torres18}.
A large variety of spectral shapes can be obtained from the same pulsar, the same synchro-curvature process, and the same accelerating region (whose physical properties are fixed by the 
accelerating field and magnetic field gradient), just as a result of the variations in the contrast.
%

Figure~\ref{x0-influence} already poses the issue of how difficult may be to predict the whole SED if having access to X-ray data only. 
Unless the X-ray data uncertainty is very low, and its energy coverage large, there might be a qualitatively degeneracy in the contrast values in such cases. 
As an example, focus on the middle panel of the top row of Figure~\ref{x0-influence}, the case of PSR J1709-3825. 
If only the X-ray data set is available, a fit with low $x_0/R_{lc}$ (dark curves) can be (at least qualitatively) as good as one with larger values (lighter curves).
Such degeneracy can of course be broken by high quality X-ray data, and even better, by harder X-ray or eventually by MeV data.
Note however that this degeneracy is much less severe starting from gamma-ray data (i.e., gamma-rays are a better predictor for the total SED): a good sampling in this energy domain leave little freedom to the value
of $x_0/R_{lc}$ in all the cases where a single set of parameters is already enough to represent the whole SED well. 
%

\subsection{Additional broad-band SEDs arguably well-describable with a single set of $(E_{||}, b, x_0/R_{lc},N_0)$ parameters}

\begin{figure*}
\begin{center}
\includegraphics[width=0.34\textwidth]{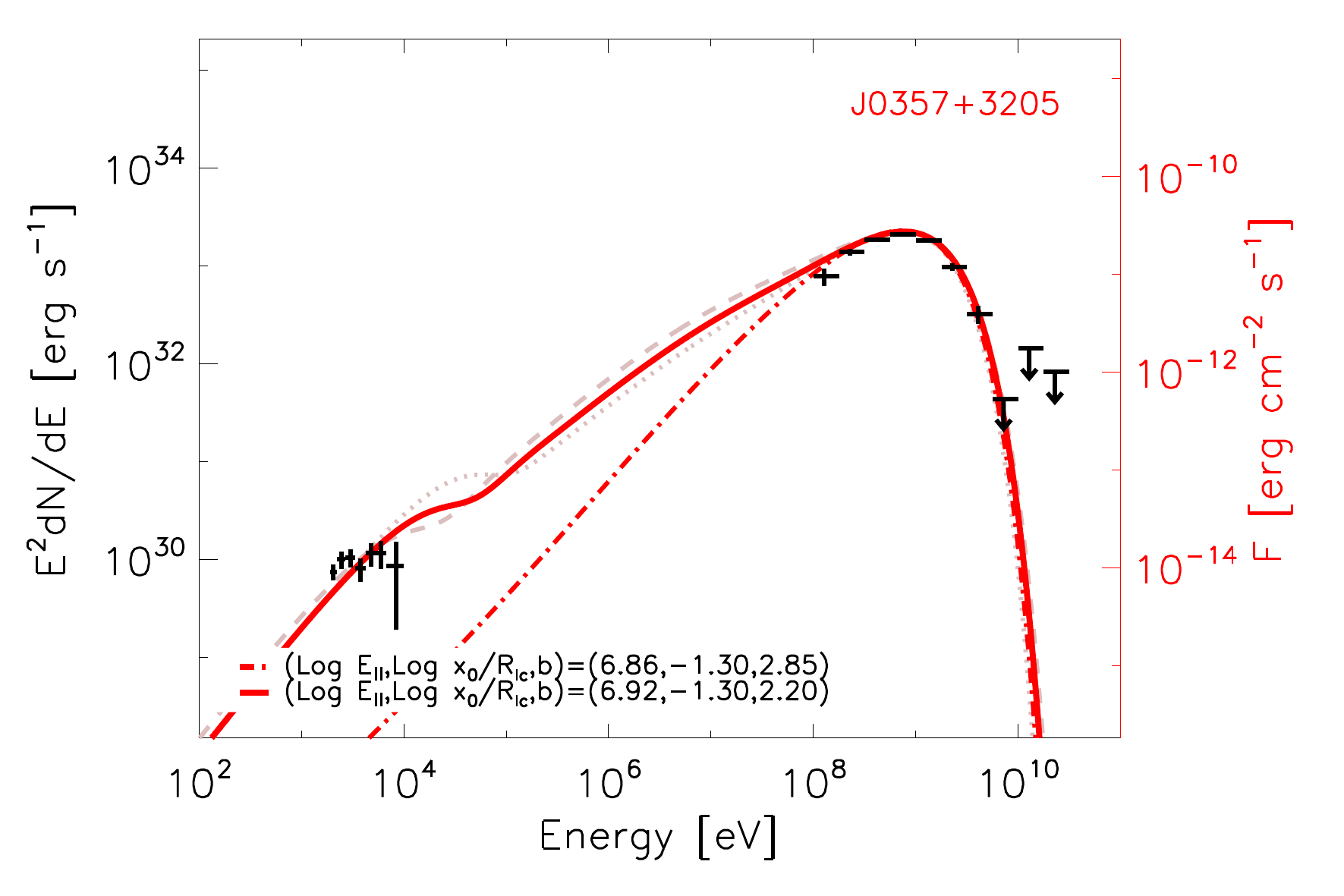}\hspace{-.25cm}
\includegraphics[width=0.34\textwidth]{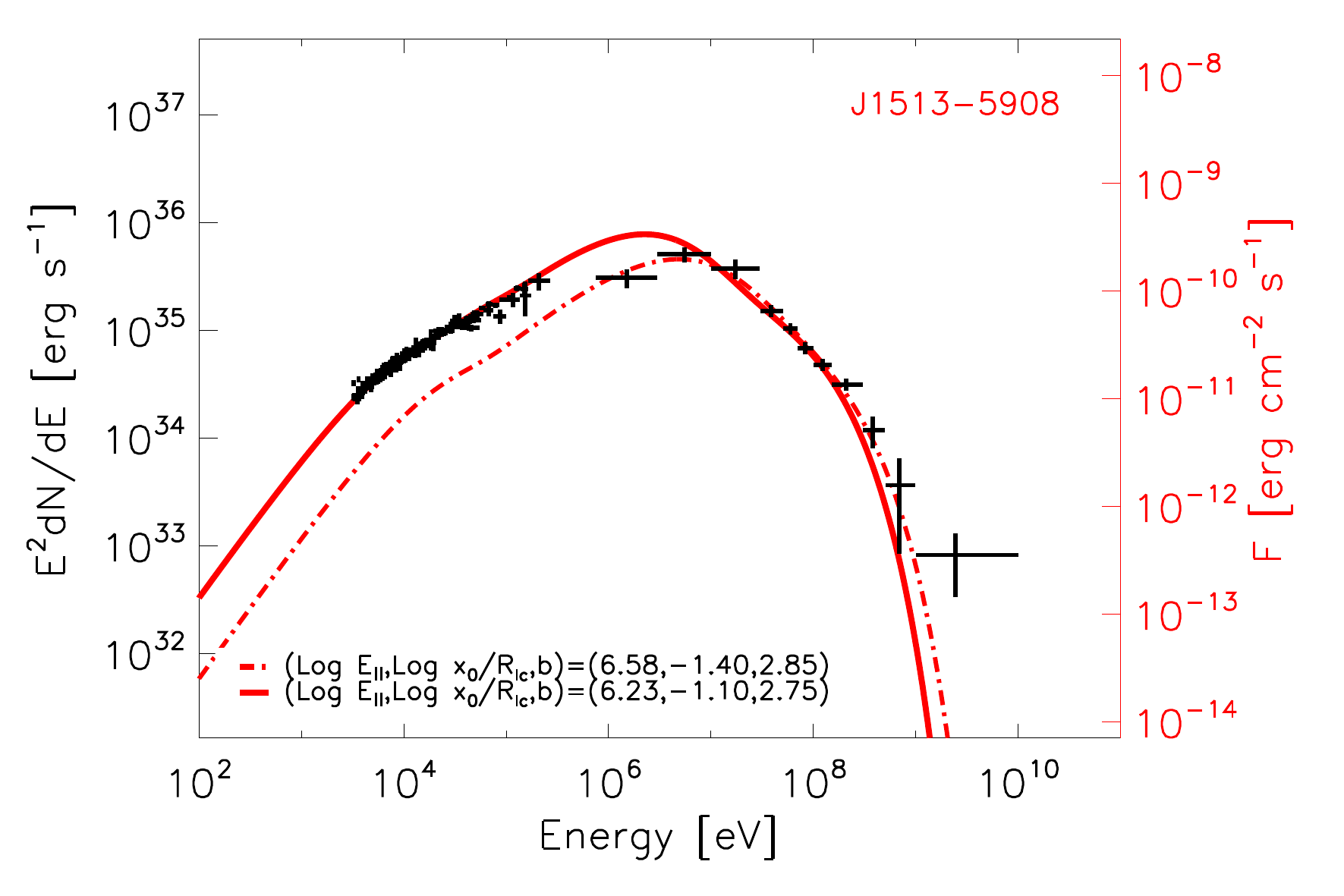}\hspace{-.25cm}
\includegraphics[width=0.34\textwidth]{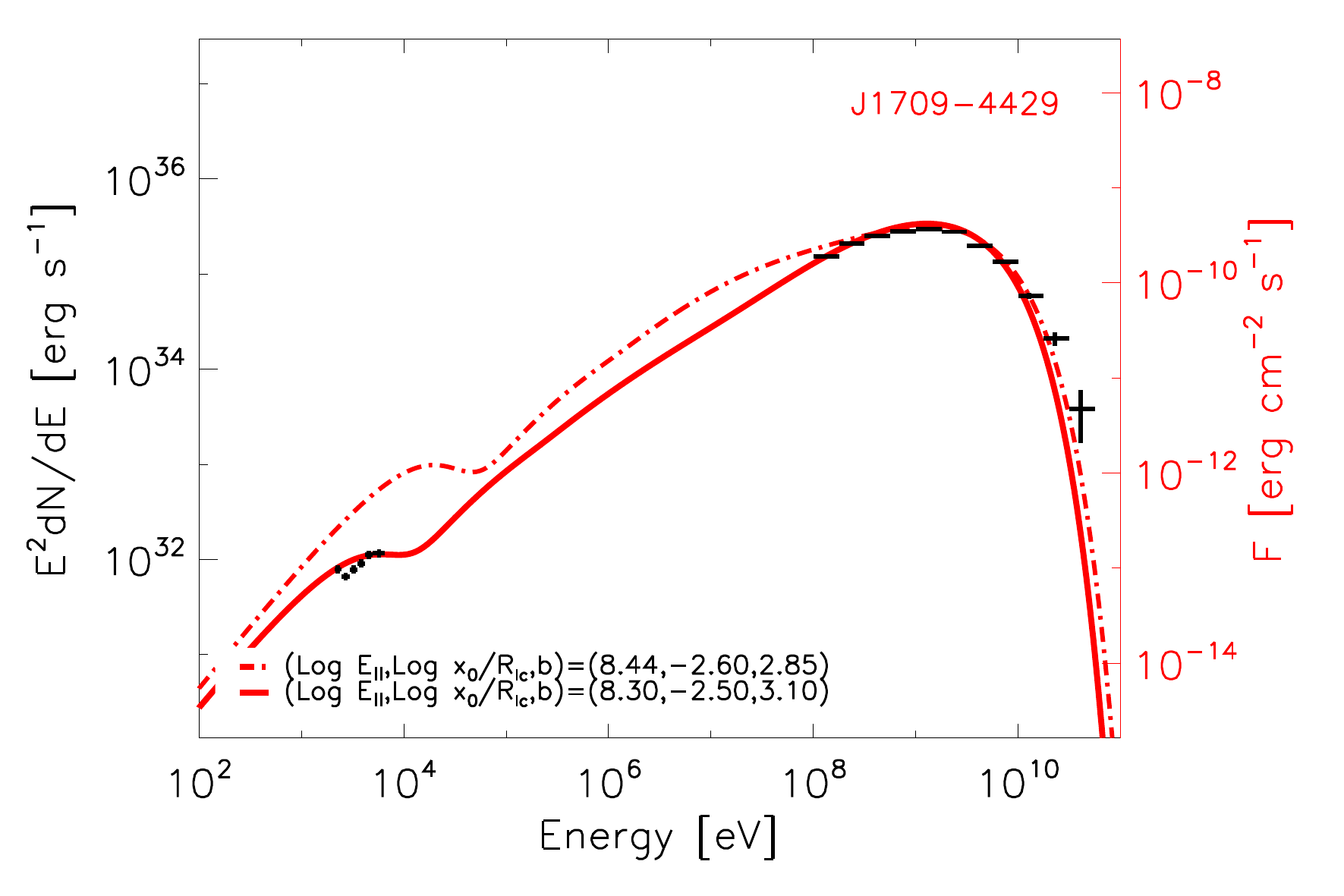}
\includegraphics[width=0.34\textwidth]{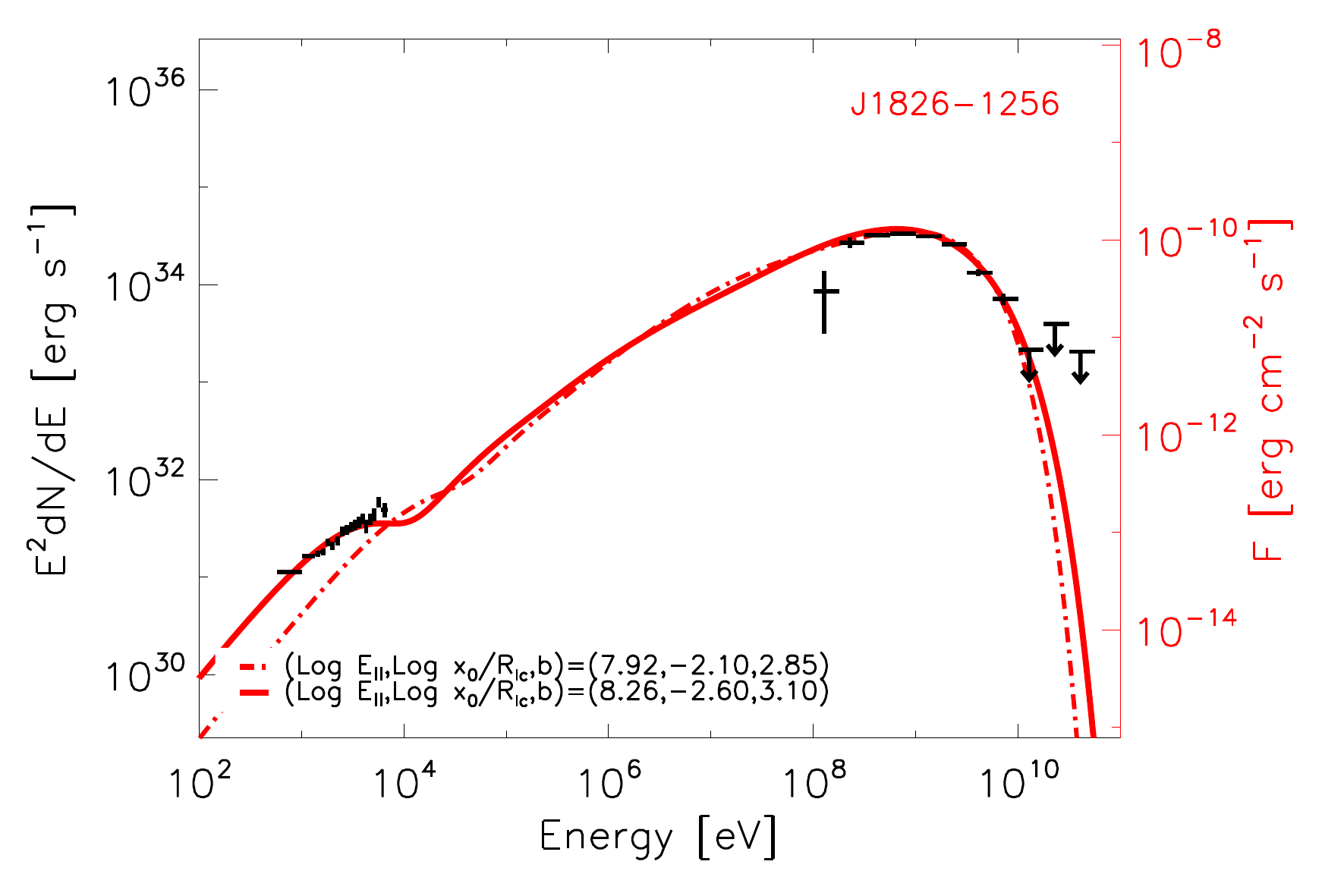}\hspace{-.25cm}
\includegraphics[width=0.34\textwidth]{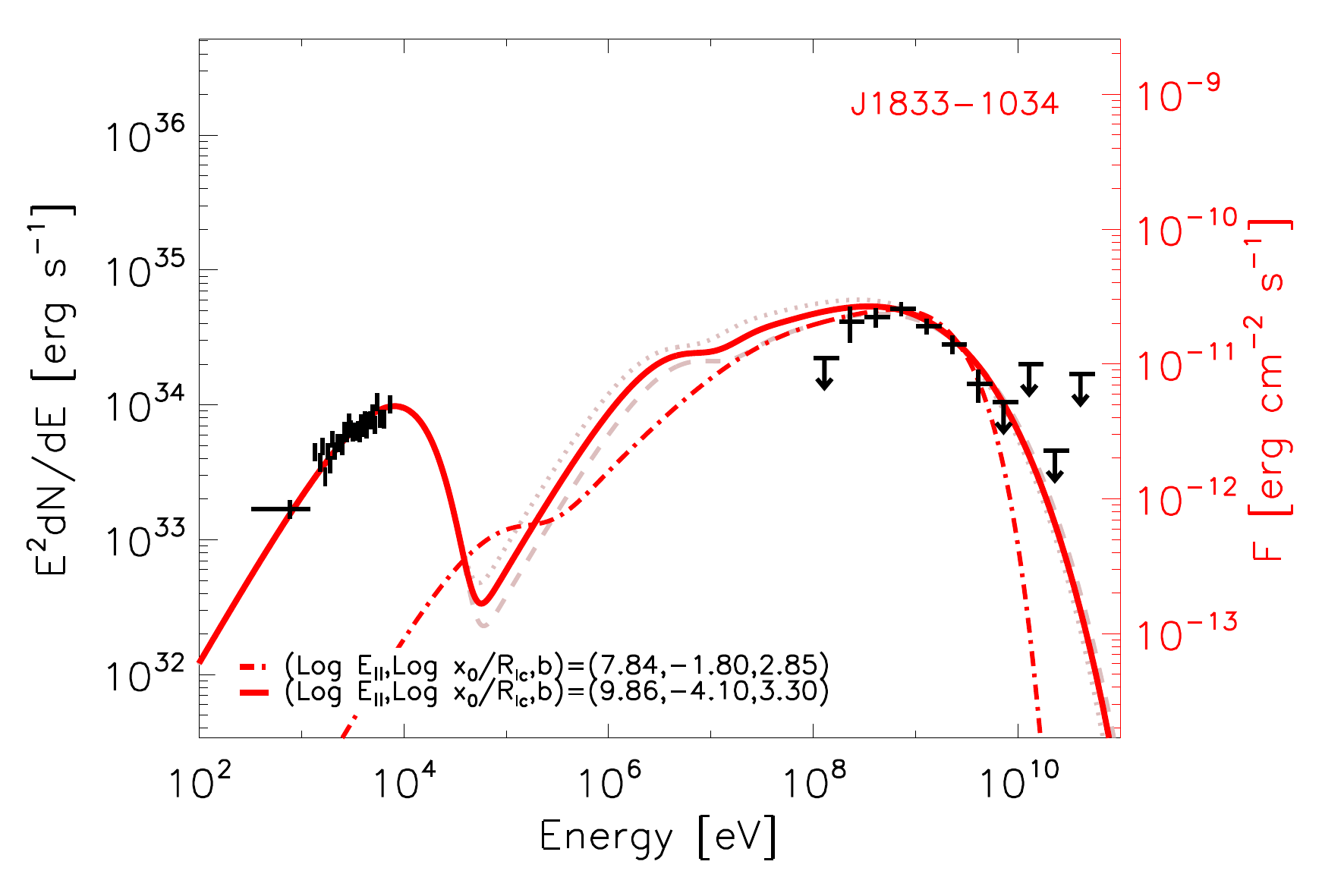}\hspace{-.25cm}
\includegraphics[width=0.34\textwidth]{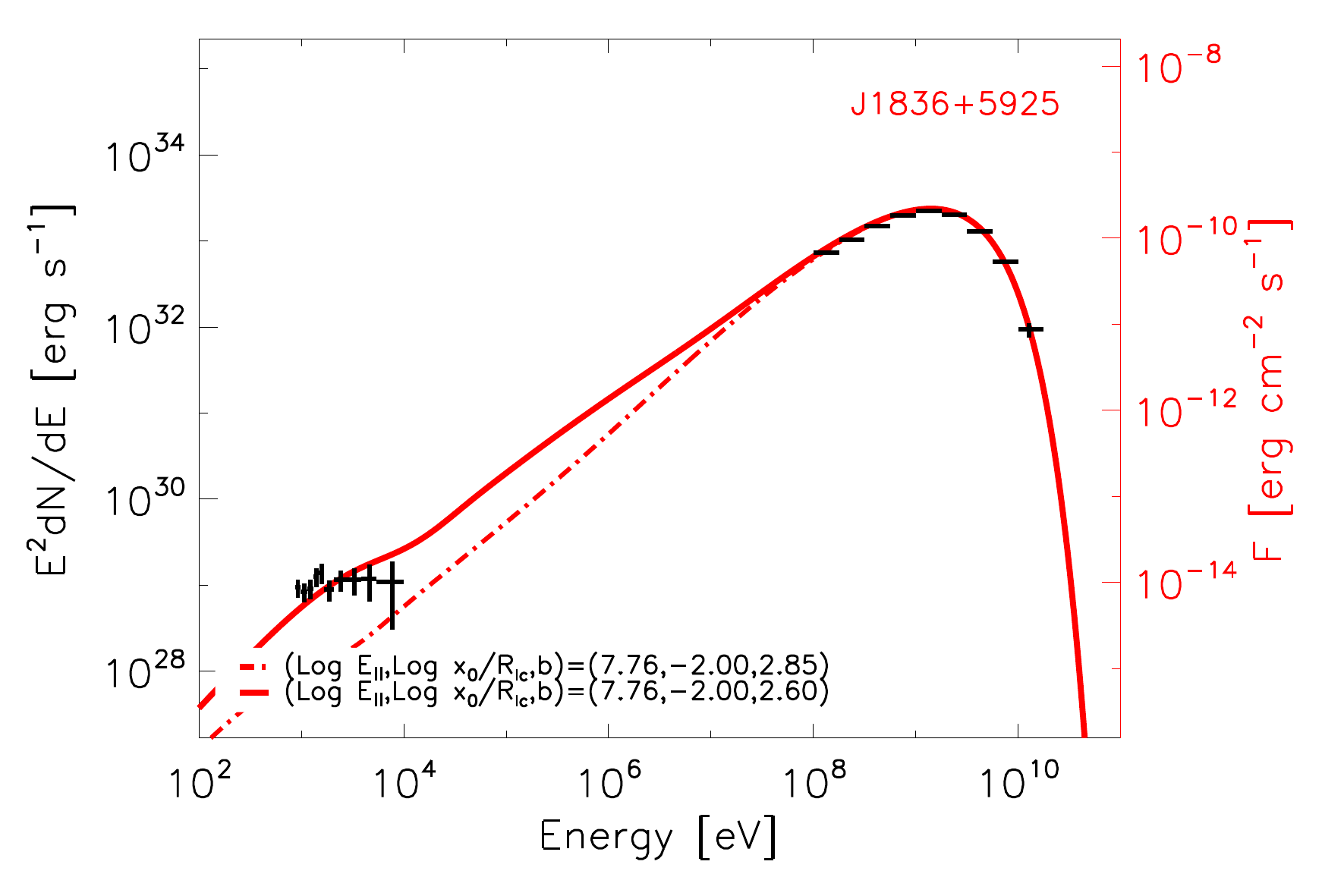}
\includegraphics[width=0.34\textwidth]{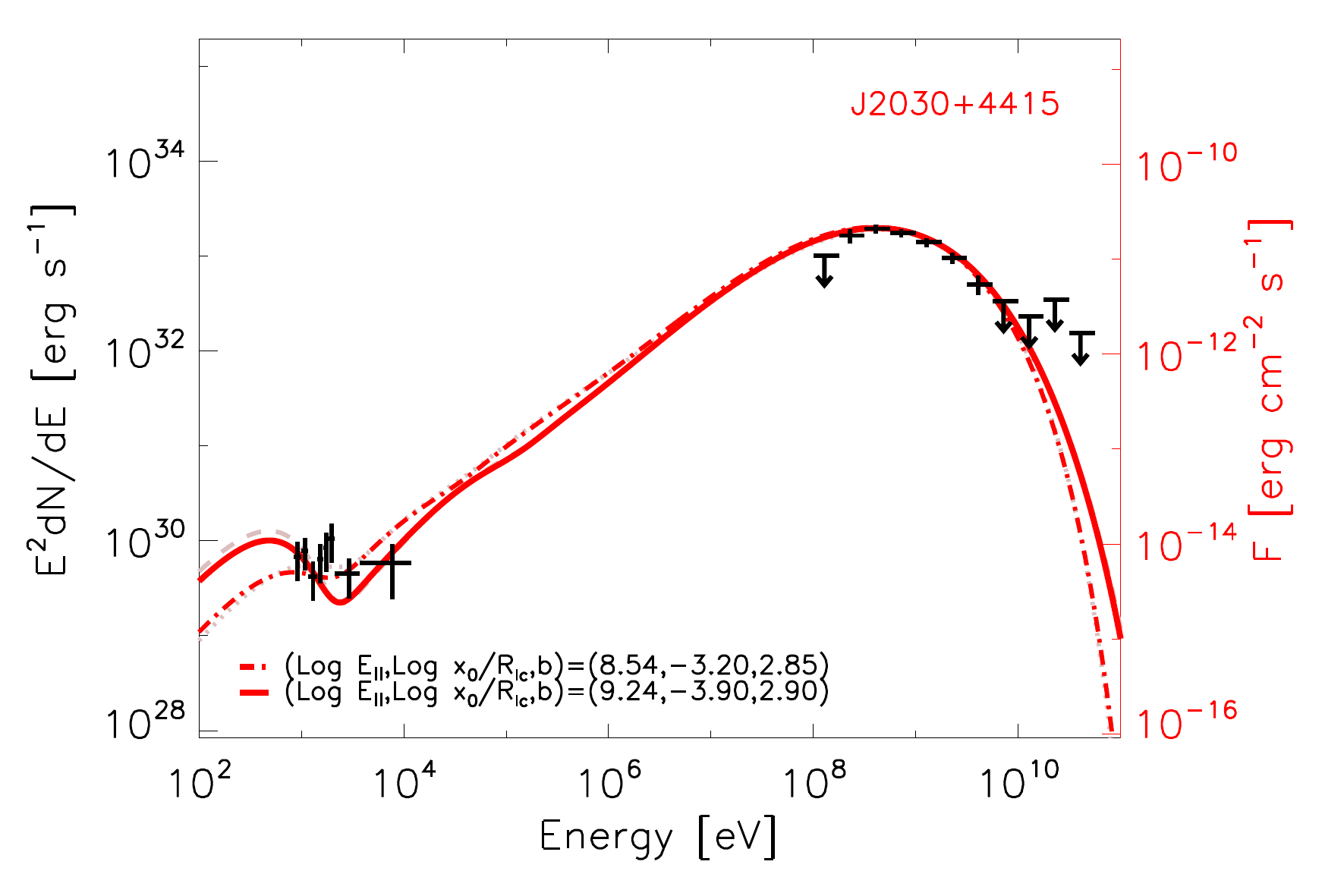}\hspace{-.25cm}
\includegraphics[width=0.34\textwidth]{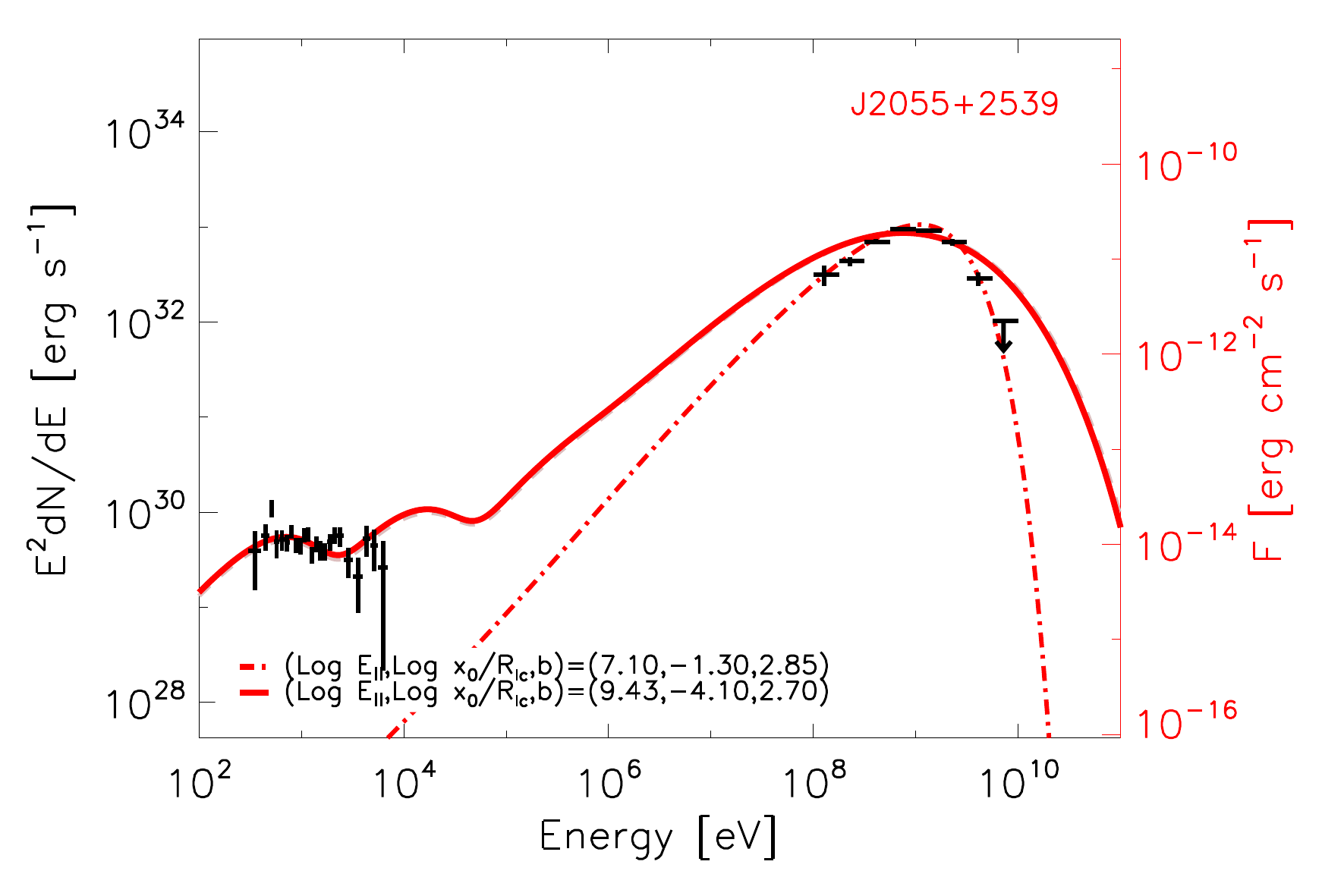}\hspace{-.25cm}
\caption{Broad-band SEDs arguably well-describable with a single set of $(E_{||}, b, x_0/R_{lc},N_0)$ parameters.
The panels show the fits to the SEDs of several pulsars (according to the corresponding label) using a single set of parameters $(E_{||}, b, x_0/R_{lc},N_0)$
in our theoretical model. In these cases, however, one can see that the model is failing in describing some specific properties of the data, see text. Curves shown are as in Figure~\ref{sed1}.
}
\label{arguably-good}
\end{center}
\end{figure*}

The panels of Figure~\ref{arguably-good} show a second set of 8 pulsars in our sample where the best-fitting models, while maintaining the general broad-band trend shown by the available data, misses a few of the observational points or a given peculiarity of the spectrum. 
The parameters of the common X-ray and gamma-ray fittings of Figure~\ref{arguably-good} are given in the second panel of Table \ref{common-fits}.
In particular, we notice the following mismatches:
\begin{description}
\item {\it J0357+3205:} the model is not describing well the flatness of the X-ray dataset.

\item {\it J1513-5908:} the model seems to over-predict the 1~MeV flux. 

\item {\it J1709-4429:} the model does not correctly describe the peculiar  X-ray spectrum, described with data featuring small error bars. 

\item {\it J826-1256:} the model is missing the lowest energy gamma-ray point (which, however, has a large error bar) and the highest energy X-ray data.

\item {\it J1833-1034:} the model is incompatible with the  upper limit of the smallest-energy gamma-ray bin. 

\item {\it J1836-5925:} the model is missing the flatness of the X-ray SED. 

\item {\it J2030+4415:} the model is incompatible with the  upper limit of the smallest-energy gamma-ray bin.

\item {\it J2055+2539:} the model is missing the highest energy X-ray data (too flat for the best fit) and the slope of the gamma-ray spectrum.

\end{description}

Despite these mismatches, the overall look of all fits of Figure~\ref{arguably-good} can be considered reasonable.
This conclusion is emphasized  recalling that the theoretical spectral shape only depend on 3 parameters 
and that the lowest gamma-ray point may be the most affected by the treatment of the diffuse emission and/or the removal of any pulsar wind nebula component, thus 
and care must be exercised if accepting or rejecting a model only based on such data point only.

However, perhaps the inability of capturing the spectral flatness in X-rays and/or the lowest or highest energy data points in X-rays or gamma-rays may indicate an over simplification of our model.
%
%
We shall come back to these cases below, after considering a few pulsars for which the model is indeed undoubtedly unable to find a good SED fitting with a single set of $(E_{||}, b, x_0/R_{lc},N_0)$ parameters.

\subsection{broad-band SEDs not describable with a single set of $(E_{||}, b, x_0/R_{lc},N_0)$  parameters}
\label{doubletern}

\begin{figure*}
\begin{center}
\includegraphics[width=0.34\textwidth]{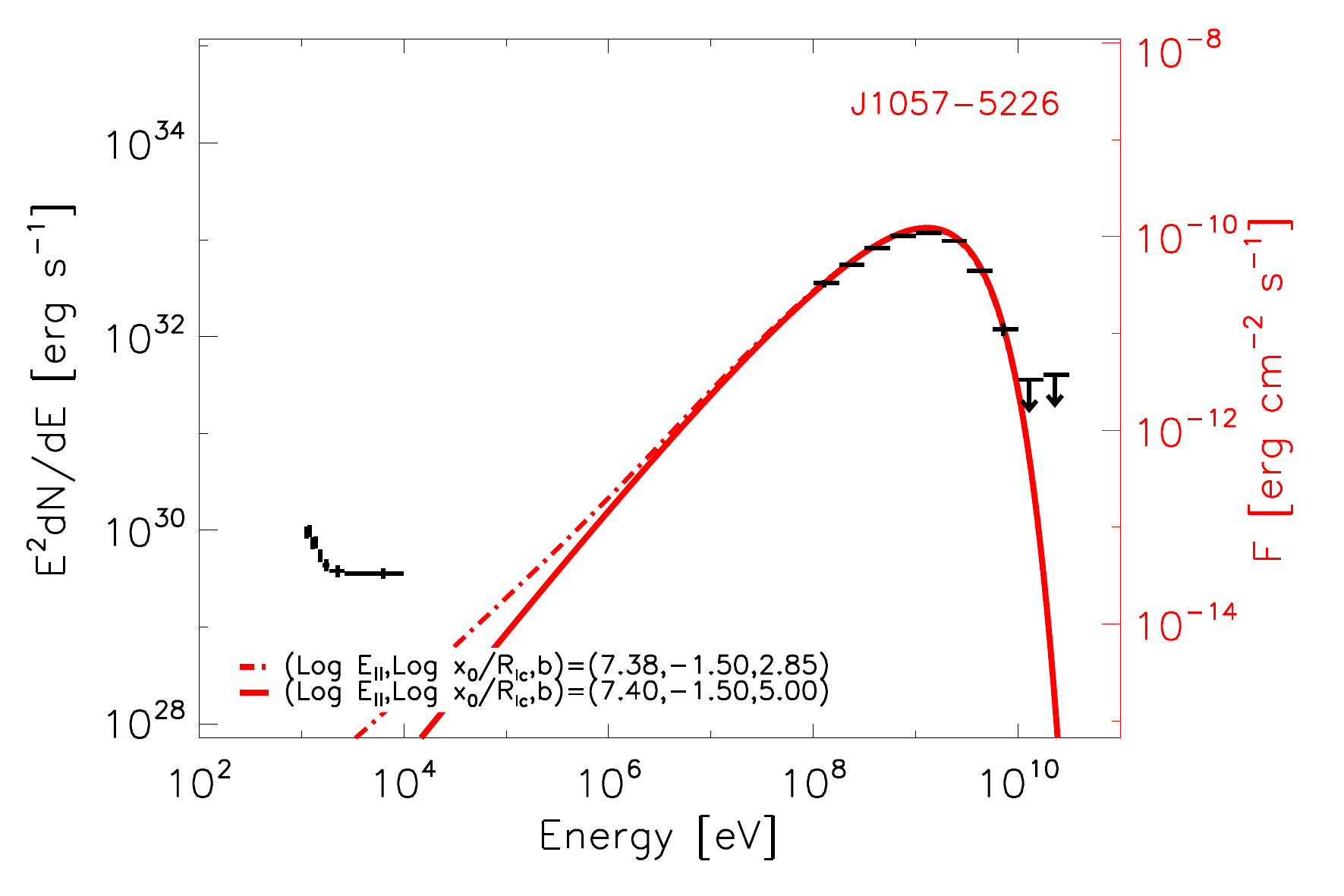}\hspace{-.25cm}
\includegraphics[width=0.34\textwidth]{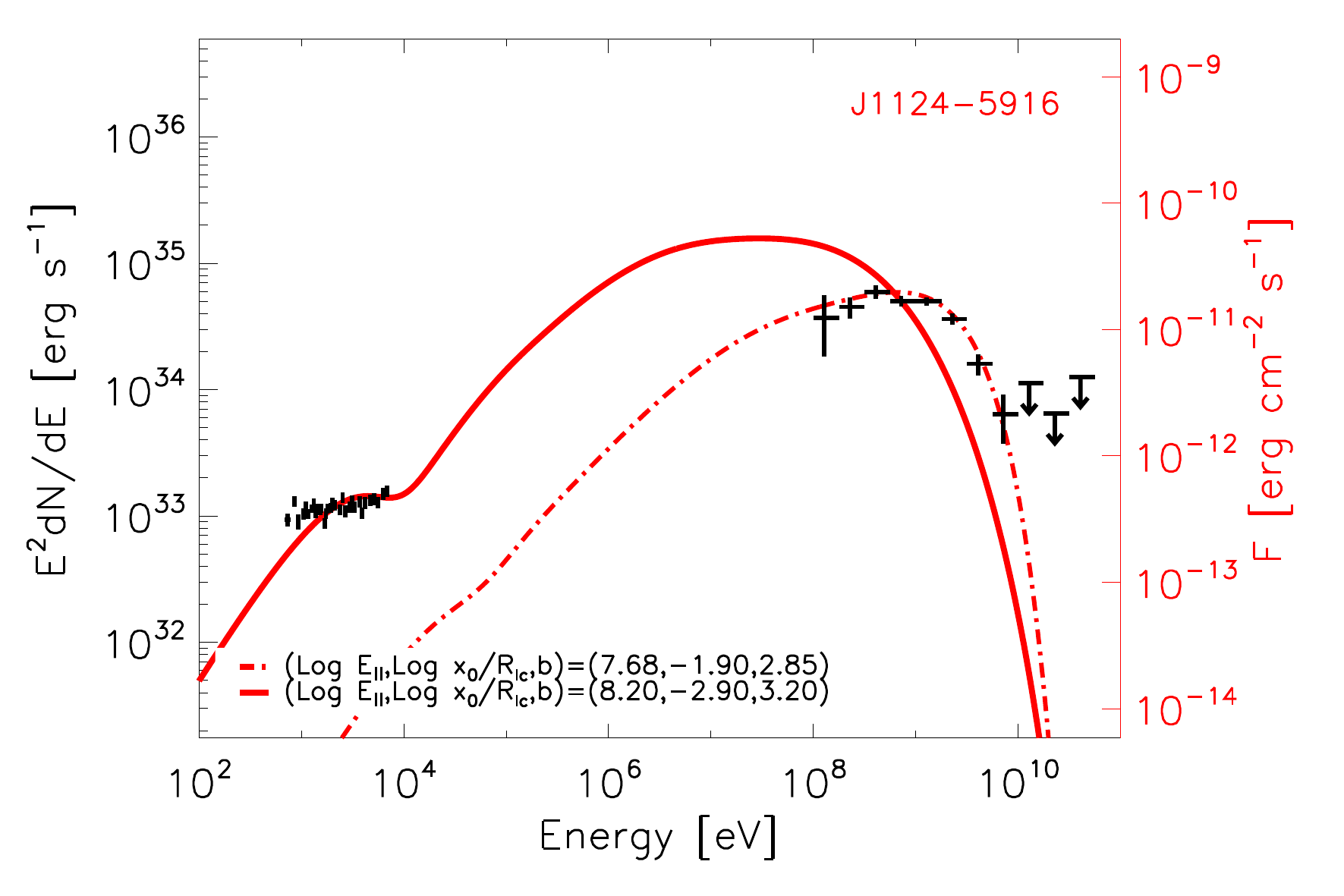}\hspace{-.25cm}
\includegraphics[width=0.34\textwidth]{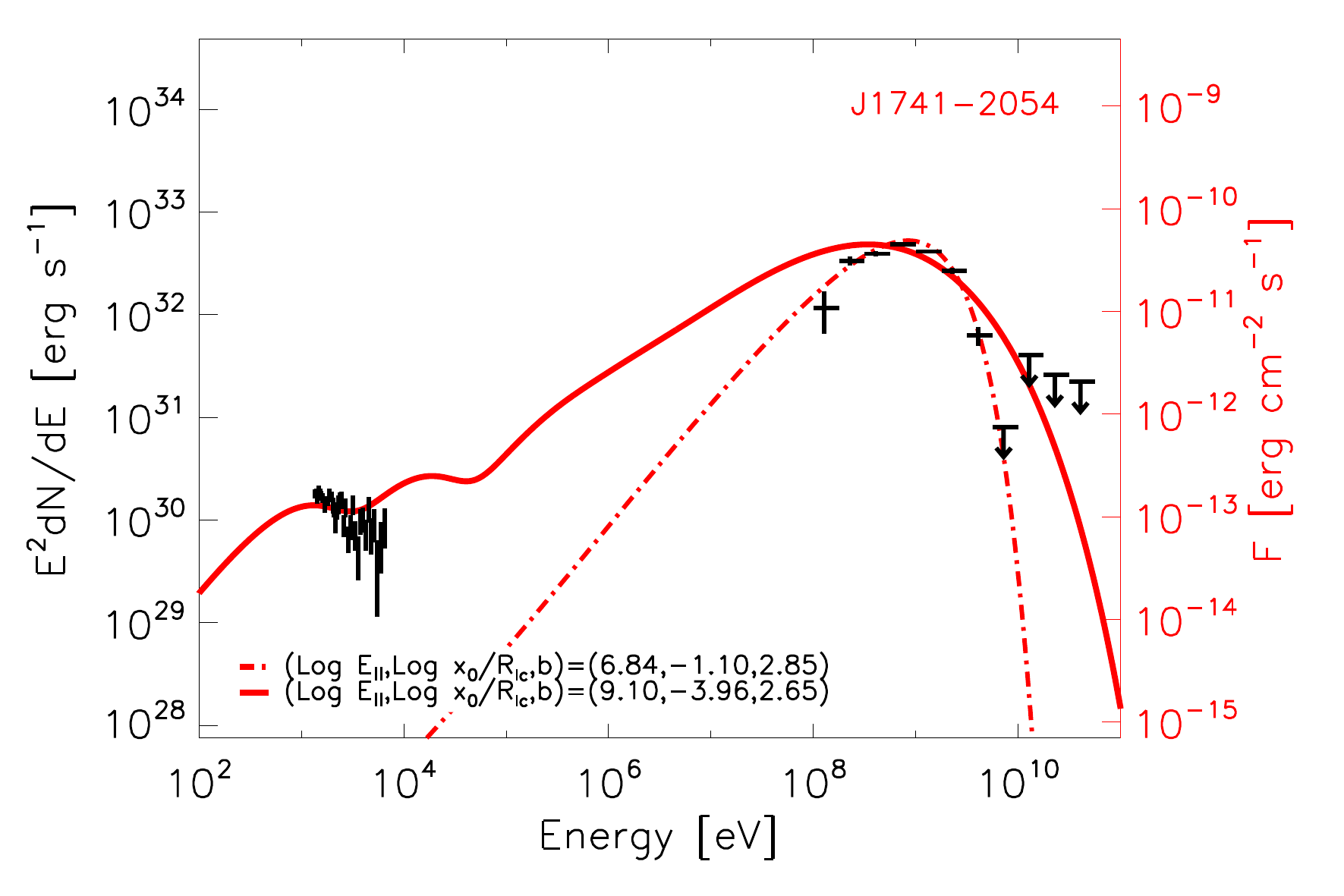}
\includegraphics[width=0.34\textwidth]{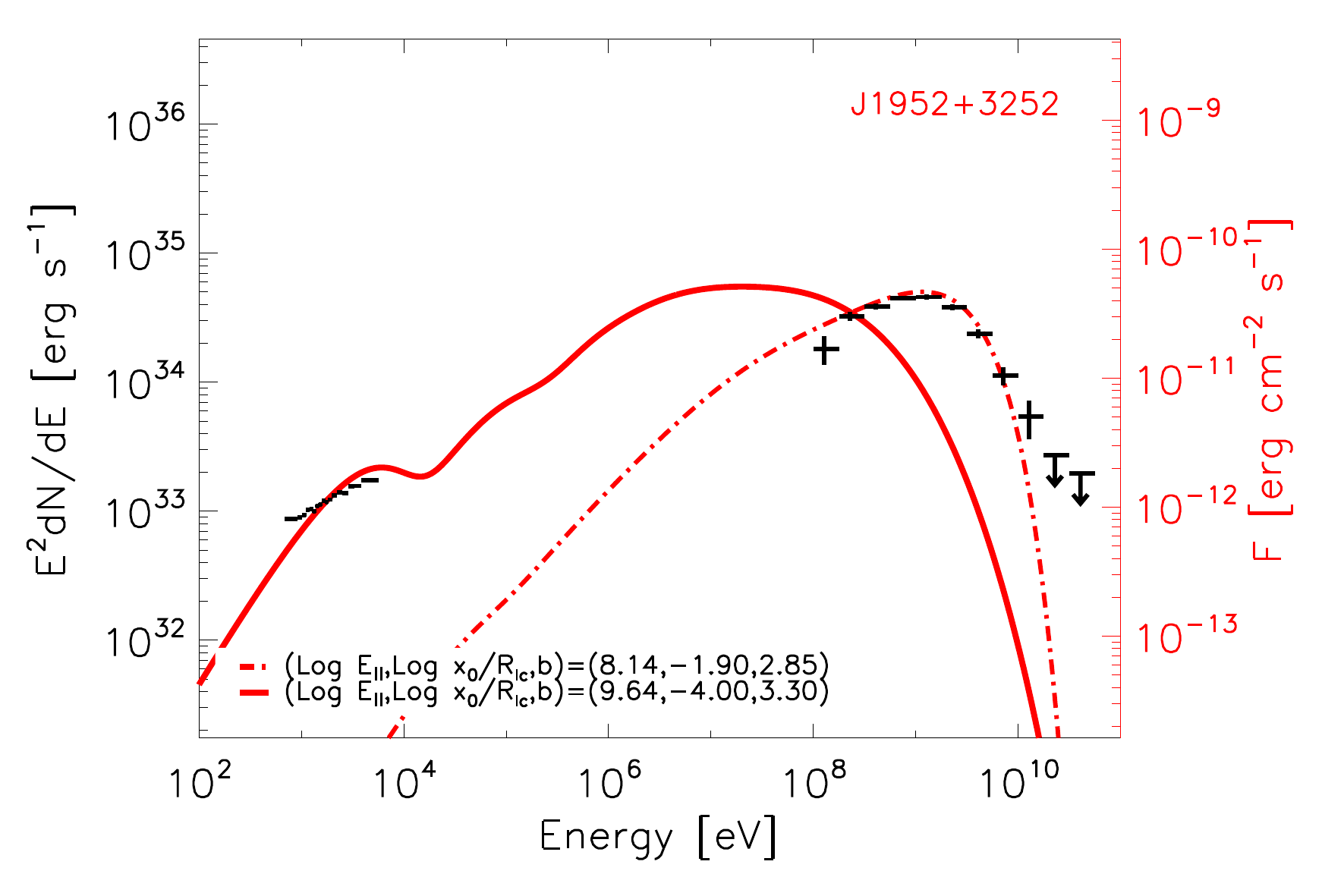}\hspace{-.25cm}
\includegraphics[width=0.34\textwidth]{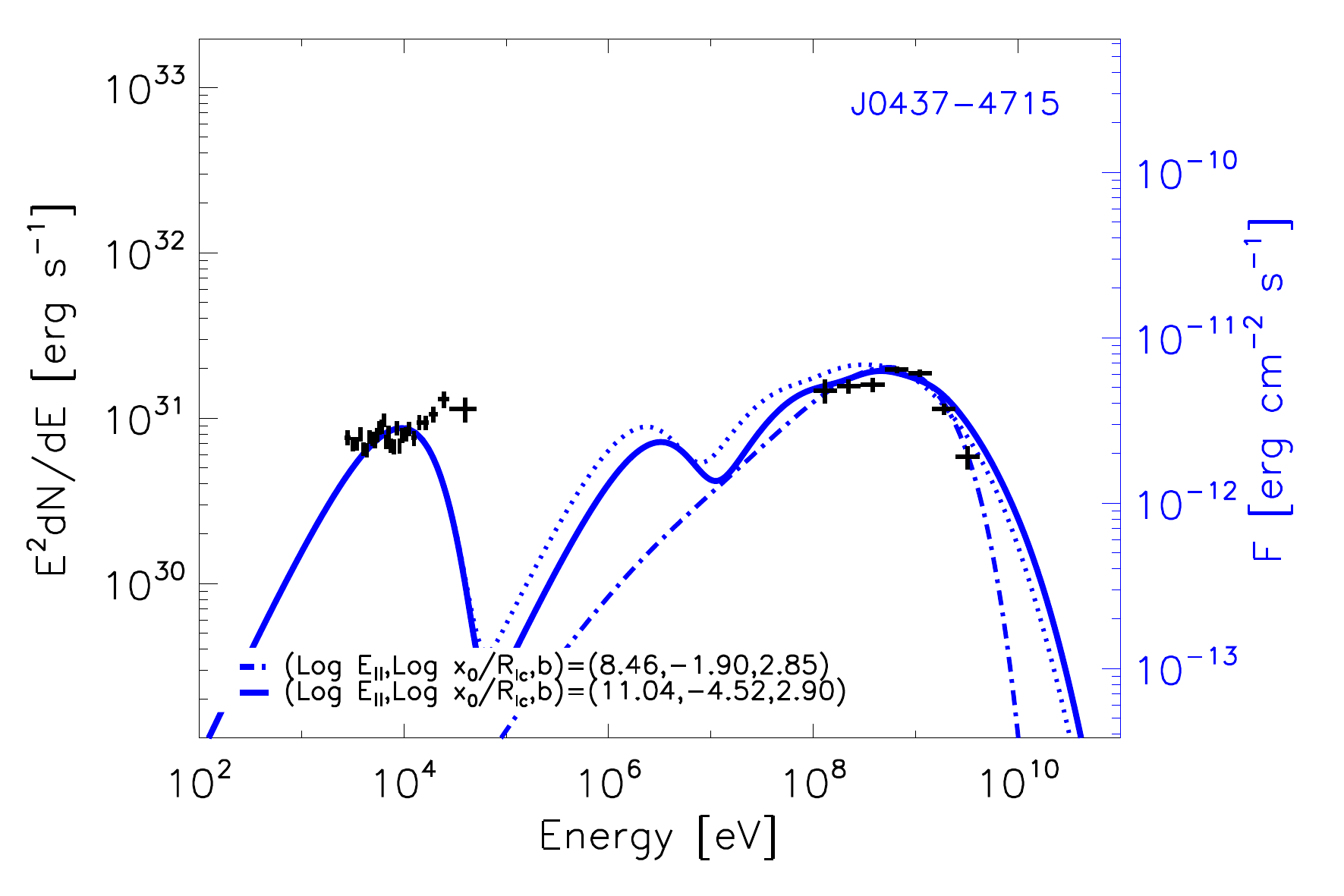}\hspace{-.25cm}
\includegraphics[width=0.34\textwidth]{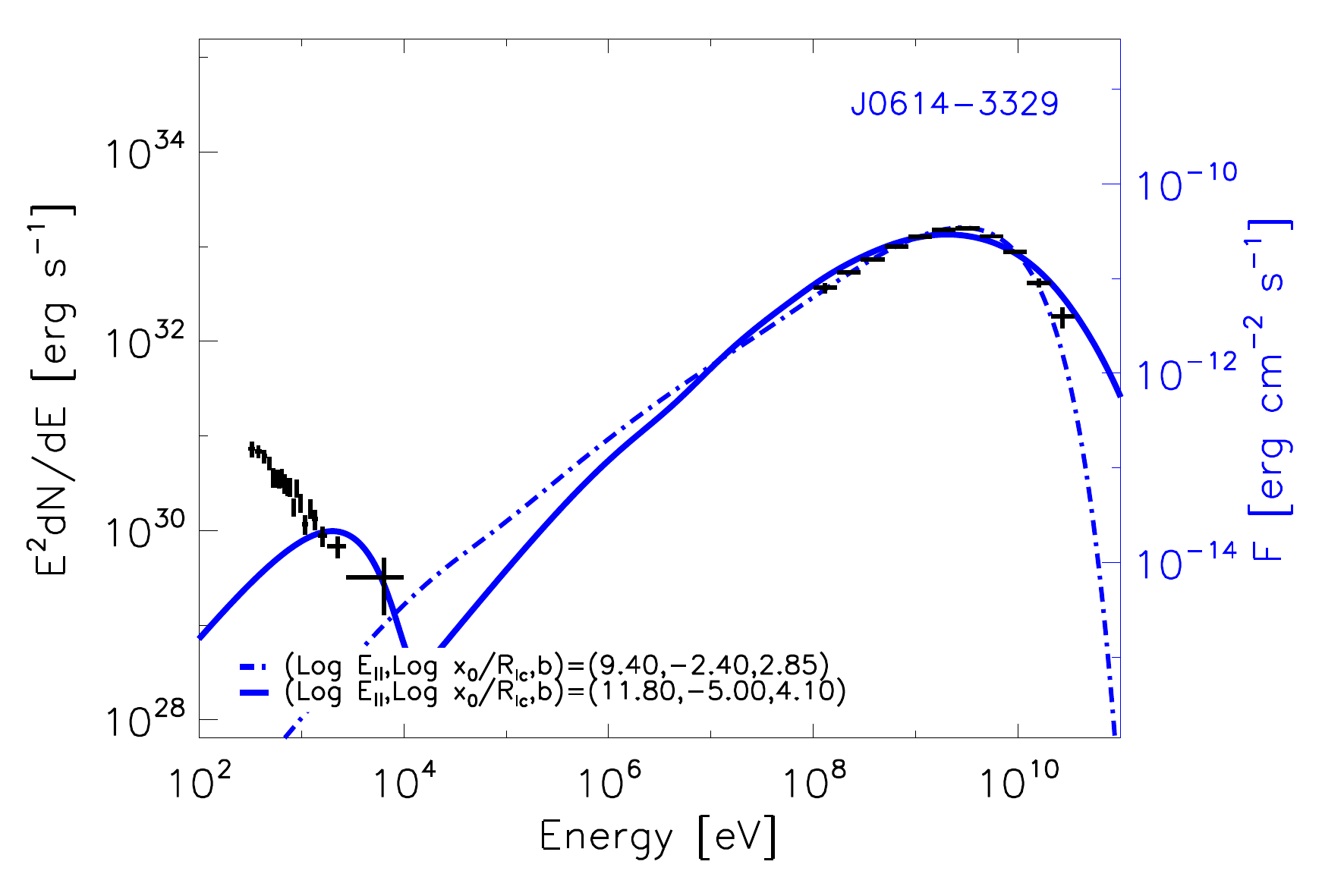}
\includegraphics[width=0.34\textwidth]{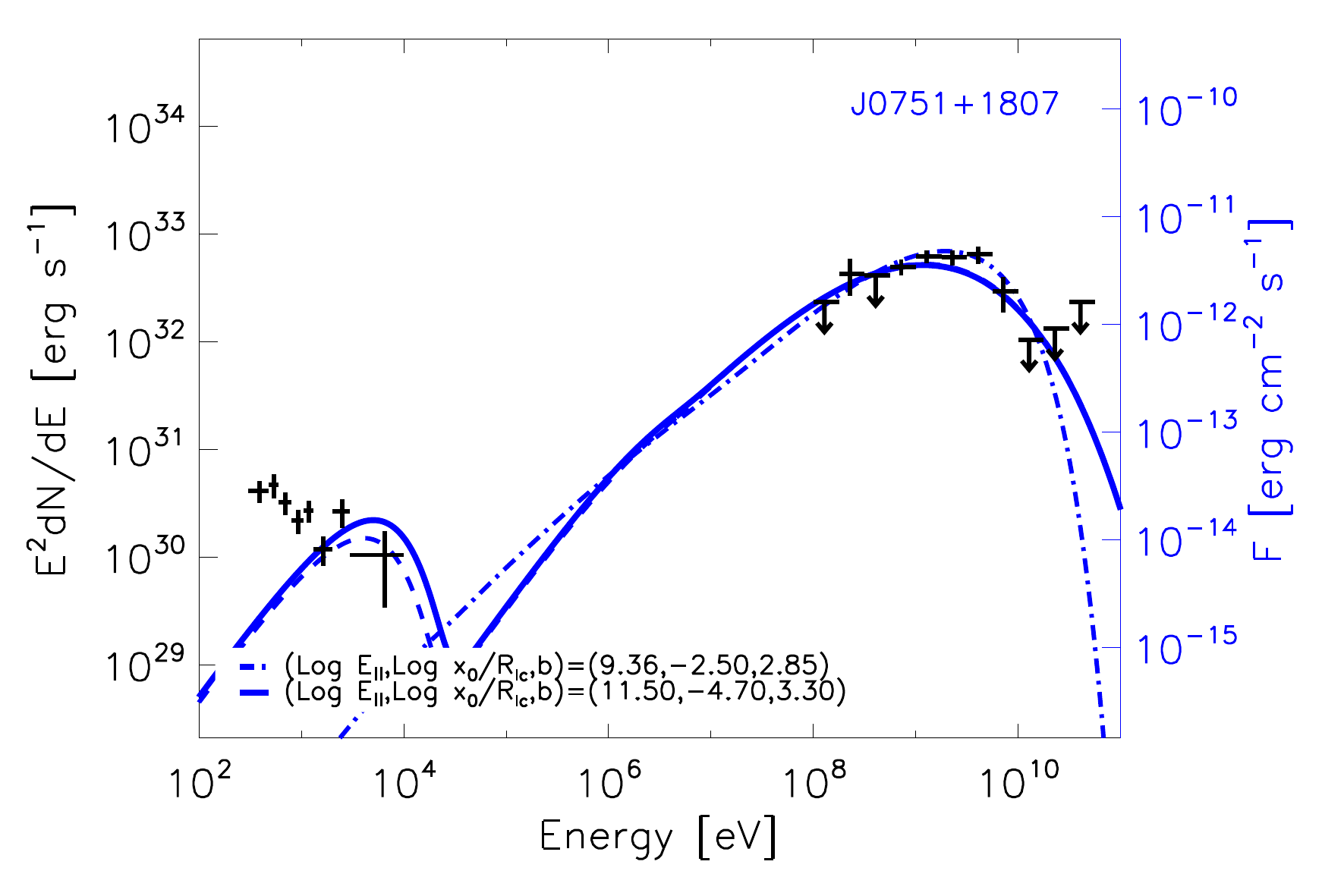}\hspace{-.25cm}
\includegraphics[width=0.34\textwidth]{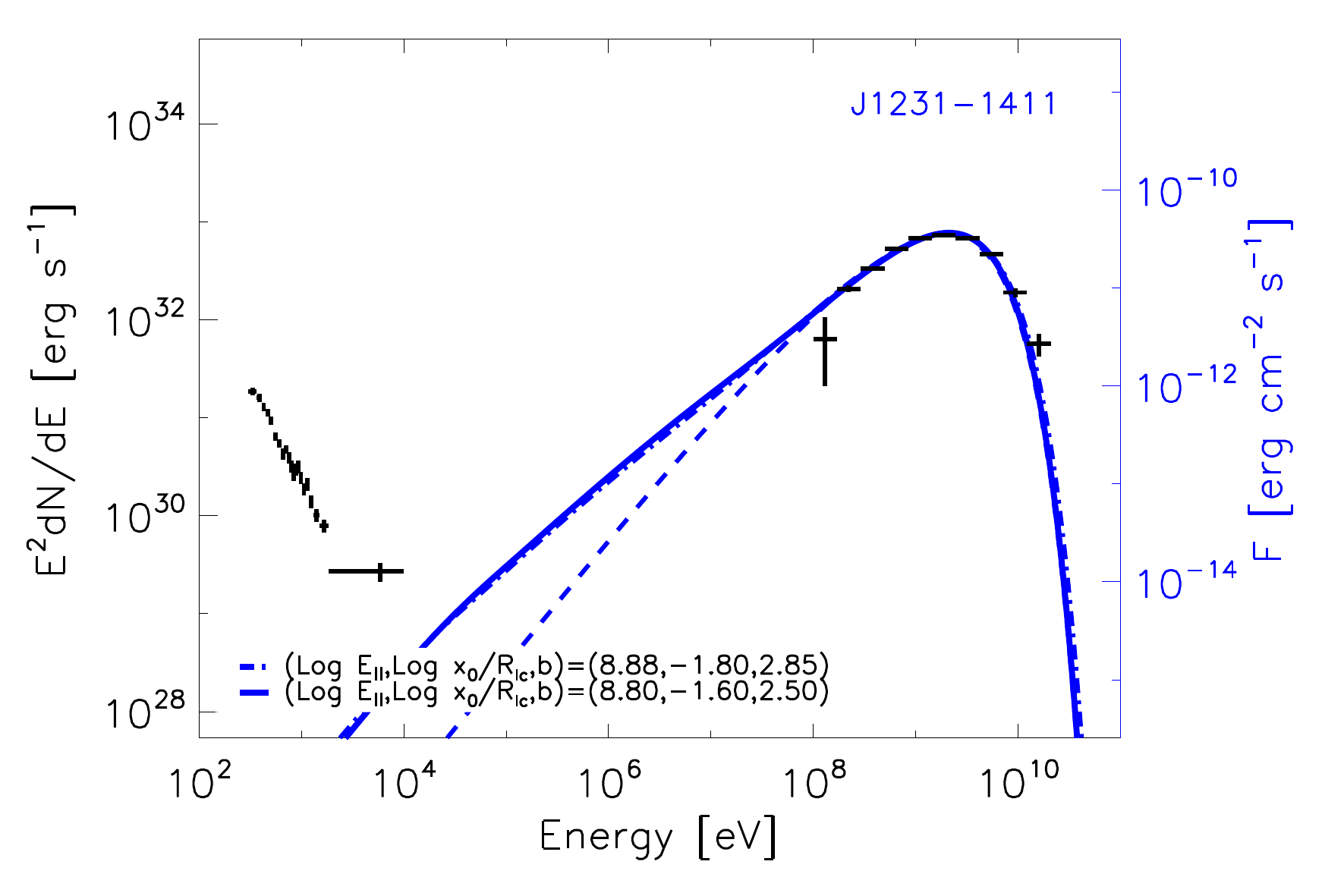}\hspace{-.25cm}
\caption{Broad-band SEDs not describable with a single set of $(E_{||}, b, x_0/R_{lc},N_0)$  parameters.
The panels show the SEDs of pulsars in our sample for which an acceptable fit using a single set of parameters $(E_{||}, b, x_0/R_{lc}, N_0)$
 is not found. 
 We use red for pulsars with $P>10 ms$ and blue for MSPs. Curves shown are as in Figure~\ref{sed1}.
}
\label{no-good-fits}
\end{center}
\end{figure*}

The third group on which we focus is formed by 8 sources for which
the use of our model with a single set of $(E_{||}, b, x_0/R_{lc},N_0)$ parameters
indeed fails to produce a good fit (see Figure~\ref{no-good-fits}).
For reference, the values of these {\it fitted} parameters are also separately quoted in the final panel of Table \ref{common-fits} in all these cases.

Why  is the model unable to fit the data of these pulsars?
First of all, one should recall that the parameters $(P,\dot P)$ defining the specific pulsar are taken into account when fitting.
Thus, fitting a specific SED could be possible for a specific pulsar, but maybe not for another; comparing just the similarity of the spectra does not show the whole picture.

On the other hand, some cases show an {\it inverted} X-ray spectra \cite{CotiZelati2019}, for which the X-ray SED is decreasing with energy (e.g., as PSR 
J1057-5227, J1741-2054, and the MSPs).
This could in principle be fitted in the decaying part of the synchrotron-dominated radiation.

The consideration of the X-ray dataset for the pulsars in Figures \ref{arguably-good} and \ref{no-good-fits} 
has the result of worsening how the model
describes the gamma-ray data (see e.g., J1057-5226, J1741-2054, J1836+5925, J2055+2539),
up to the point of completely missing the high-energy data points when  approaching the X-ray data in turn (see e.g., J1952+3252, or J1124-5916).
A similar situation occurs for the MSPs studied here.
In other words, for these pulsars we find that the best-fit model tends to minimize the deviations from observed data by approaching either the X-ray data or the gamma-ray data; coping with both is impossible.

\begin{figure*}
	\begin{center}
		\includegraphics[width=0.34\textwidth]{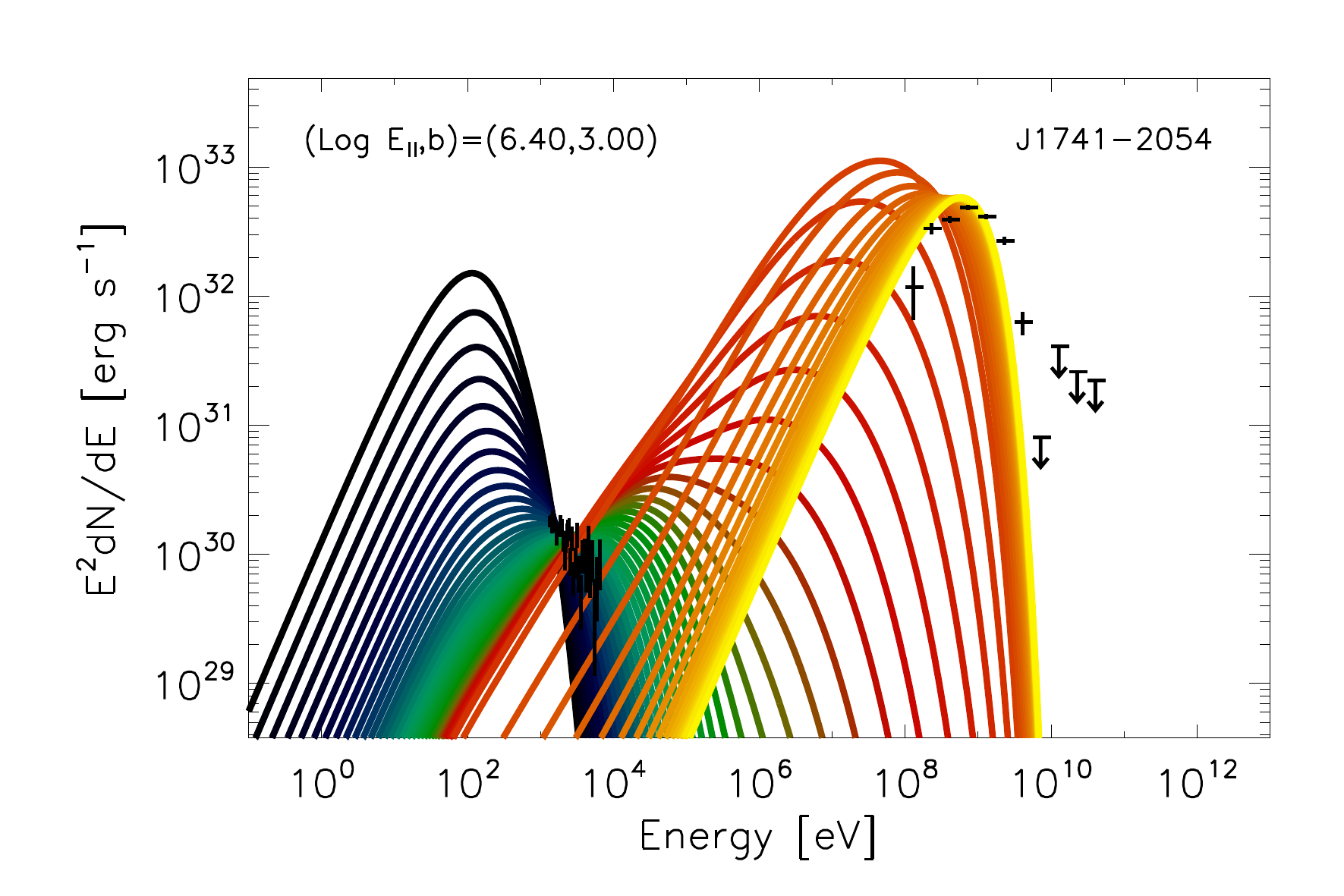}\hspace{-.25cm}
		\includegraphics[width=0.34\textwidth]{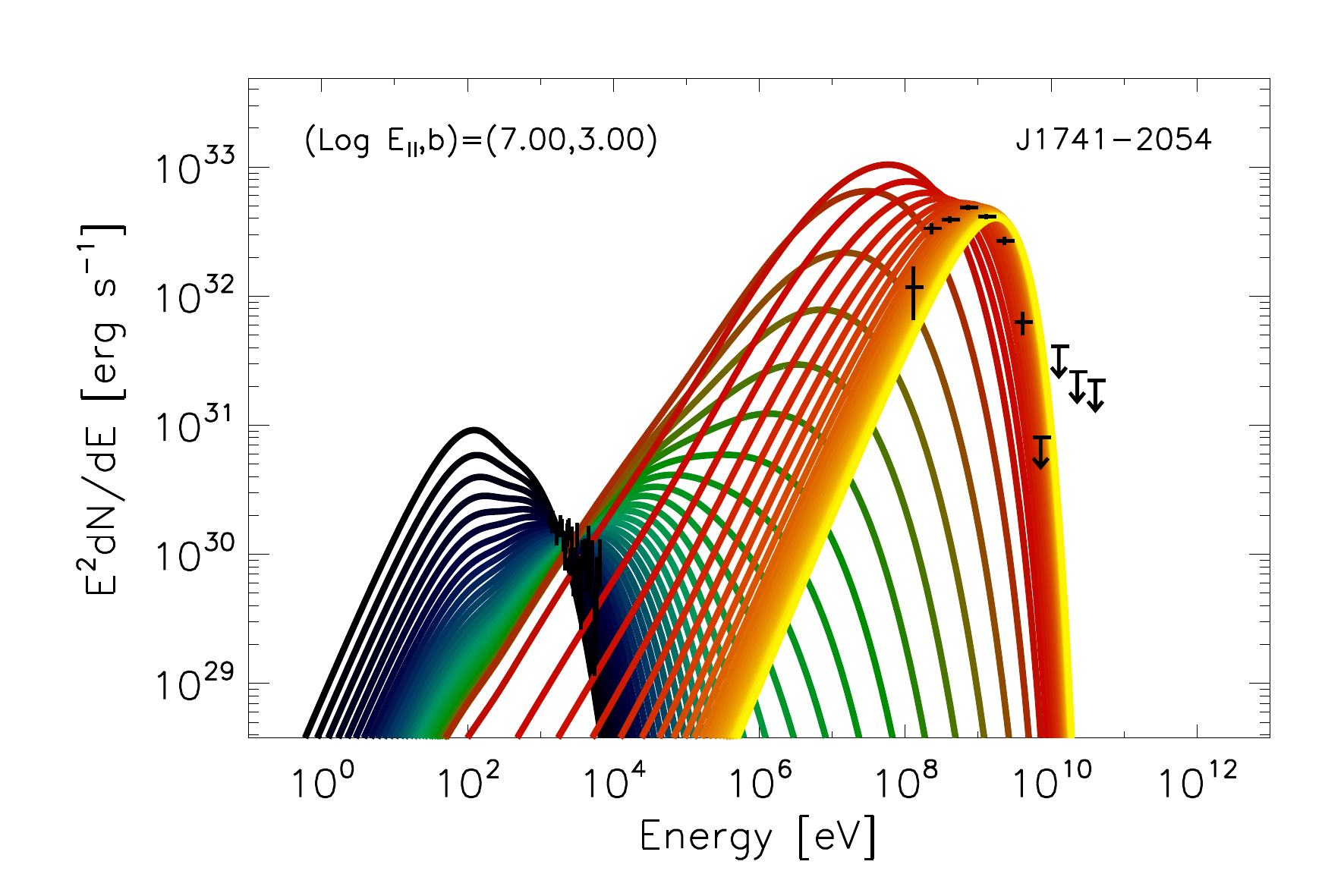}\hspace{-.25cm}
		\includegraphics[width=0.34\textwidth]{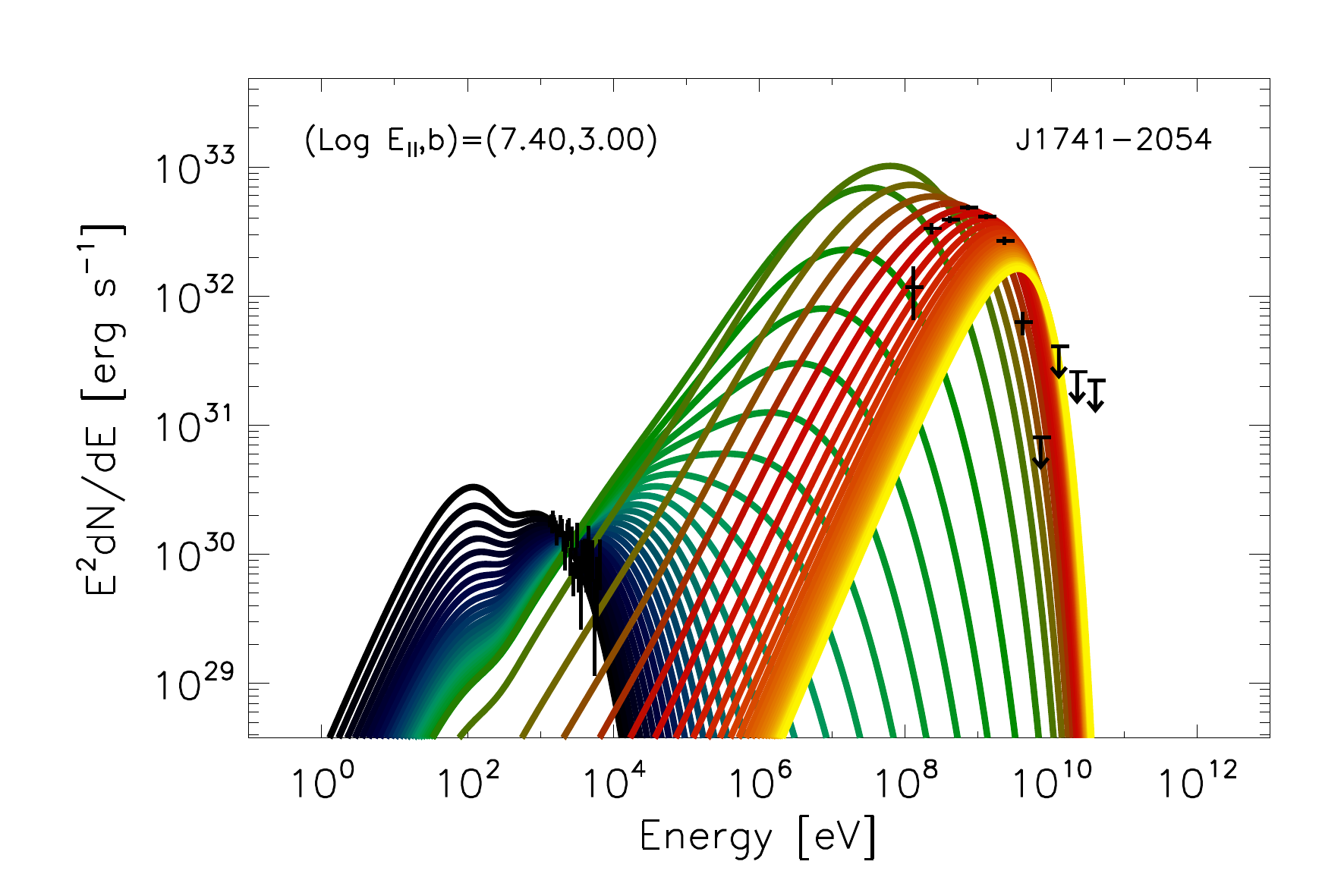}
		\includegraphics[width=0.34\textwidth]{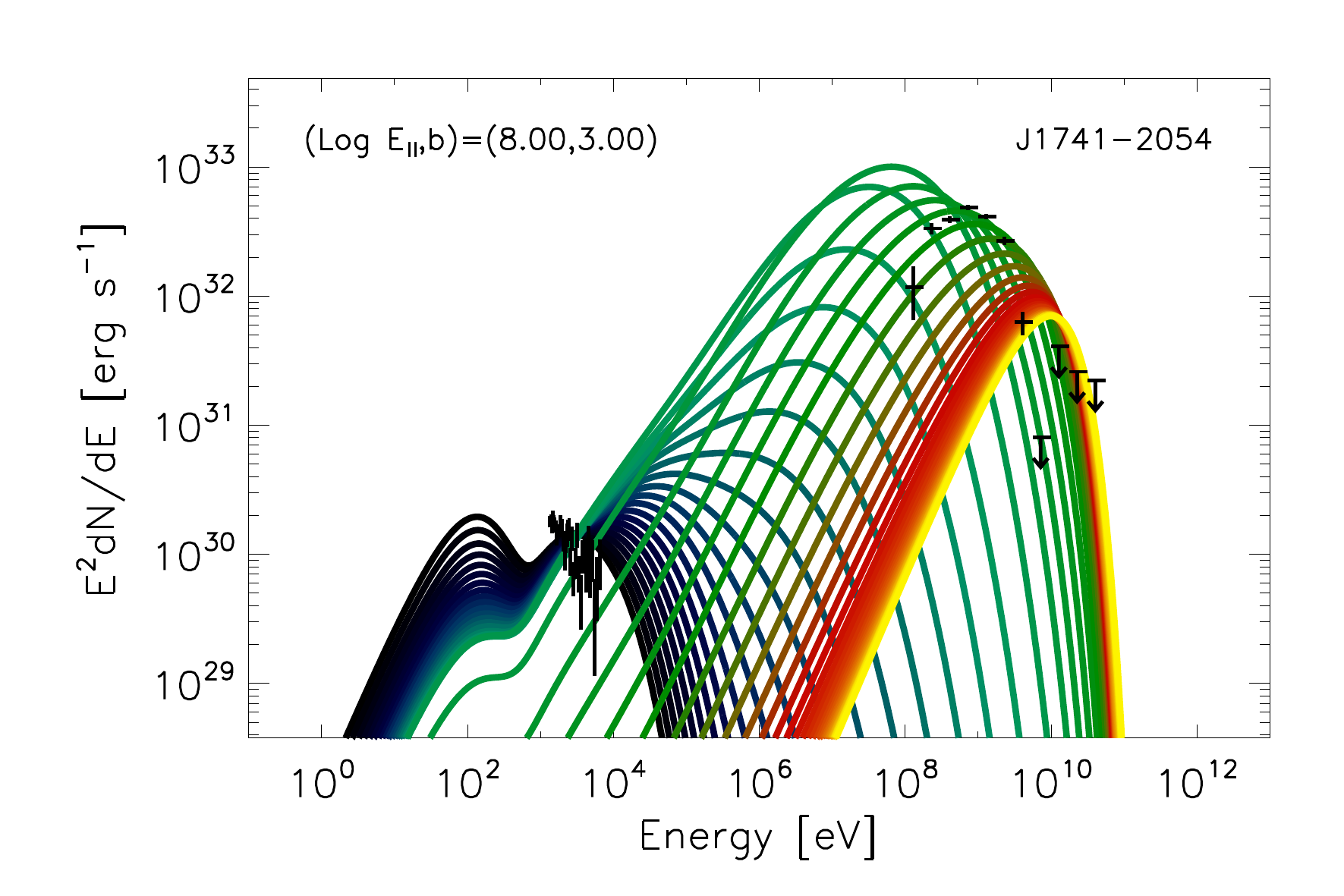}\hspace{-.25cm}
		\includegraphics[width=0.34\textwidth]{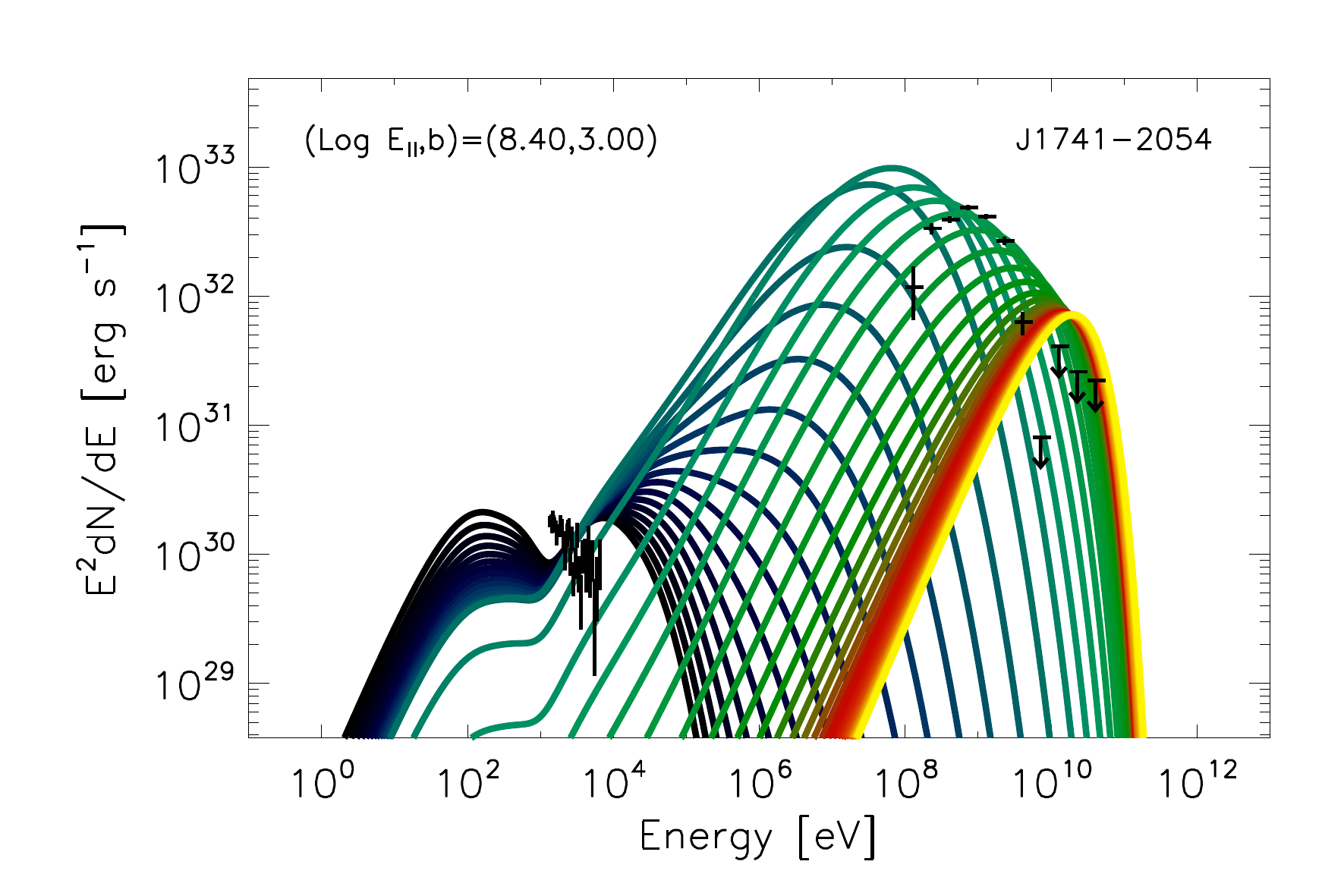}\hspace{-.25cm}
		\includegraphics[width=0.34\textwidth]{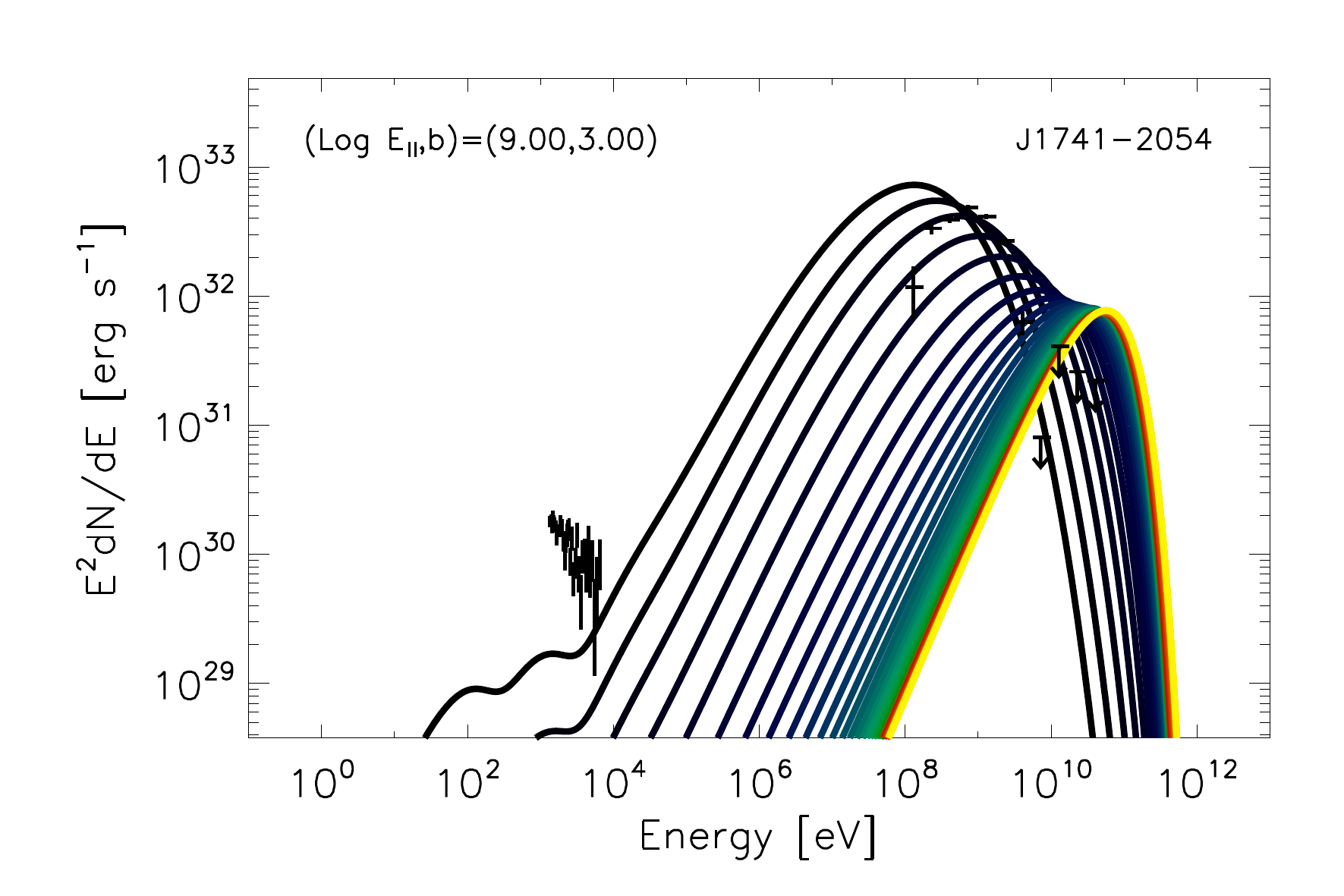}
		\caption{Example of model results for J1741-2054, searching for a common fit to whole spectrum from X-rays to gamma-rays. In these examples we \DV{vary} $E_{||}$ (different in each panel, according to the legend) and $x_0/R_{lc}$ (in the color scale, see text for details), for a fixed value of $b$ as given in the legend.}
		\label{no_single_fit_1}
	\end{center}
\end{figure*}

\begin{figure*}
\begin{center}
\includegraphics[width=0.34\textwidth]{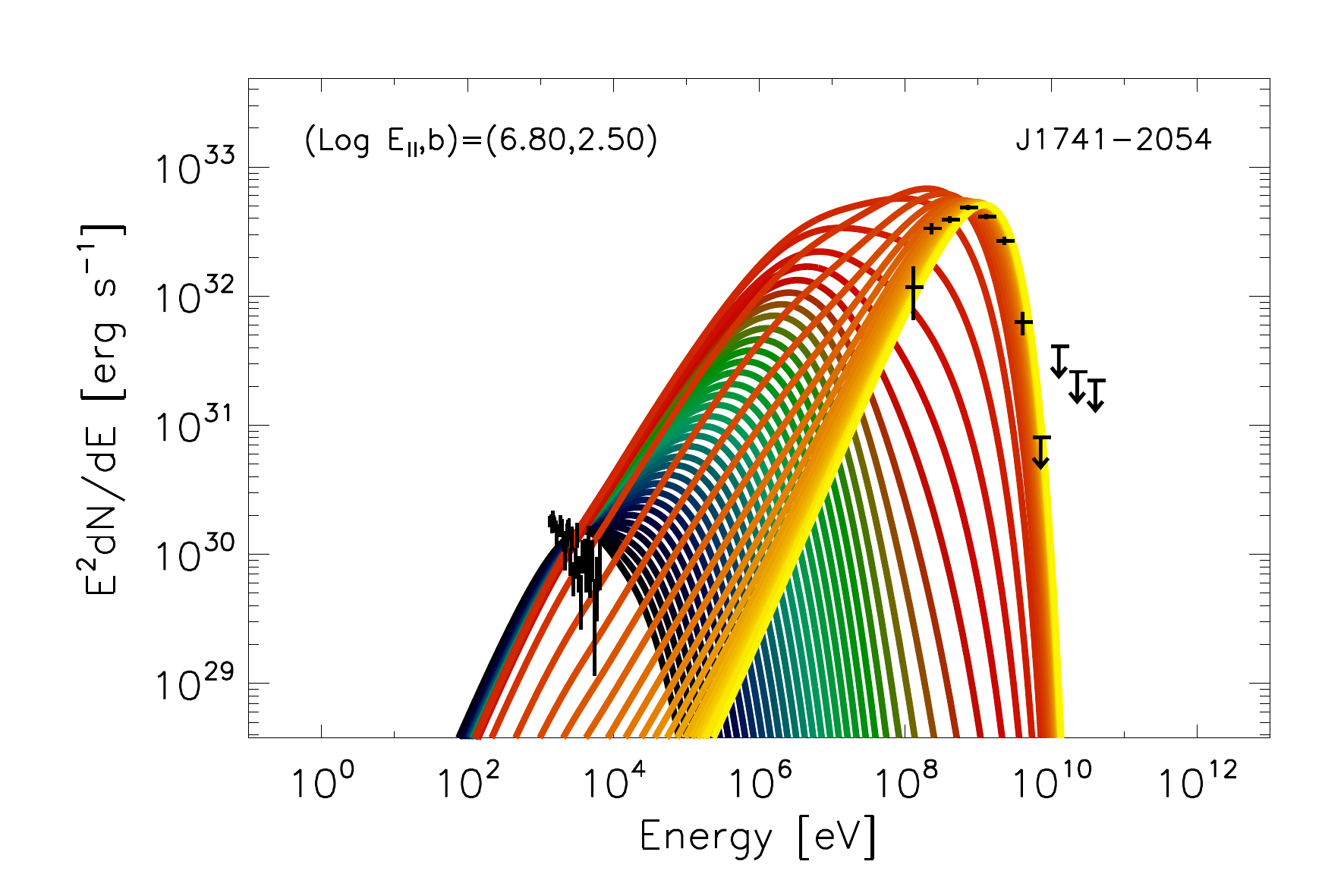}\hspace{-.25cm}
\includegraphics[width=0.34\textwidth]{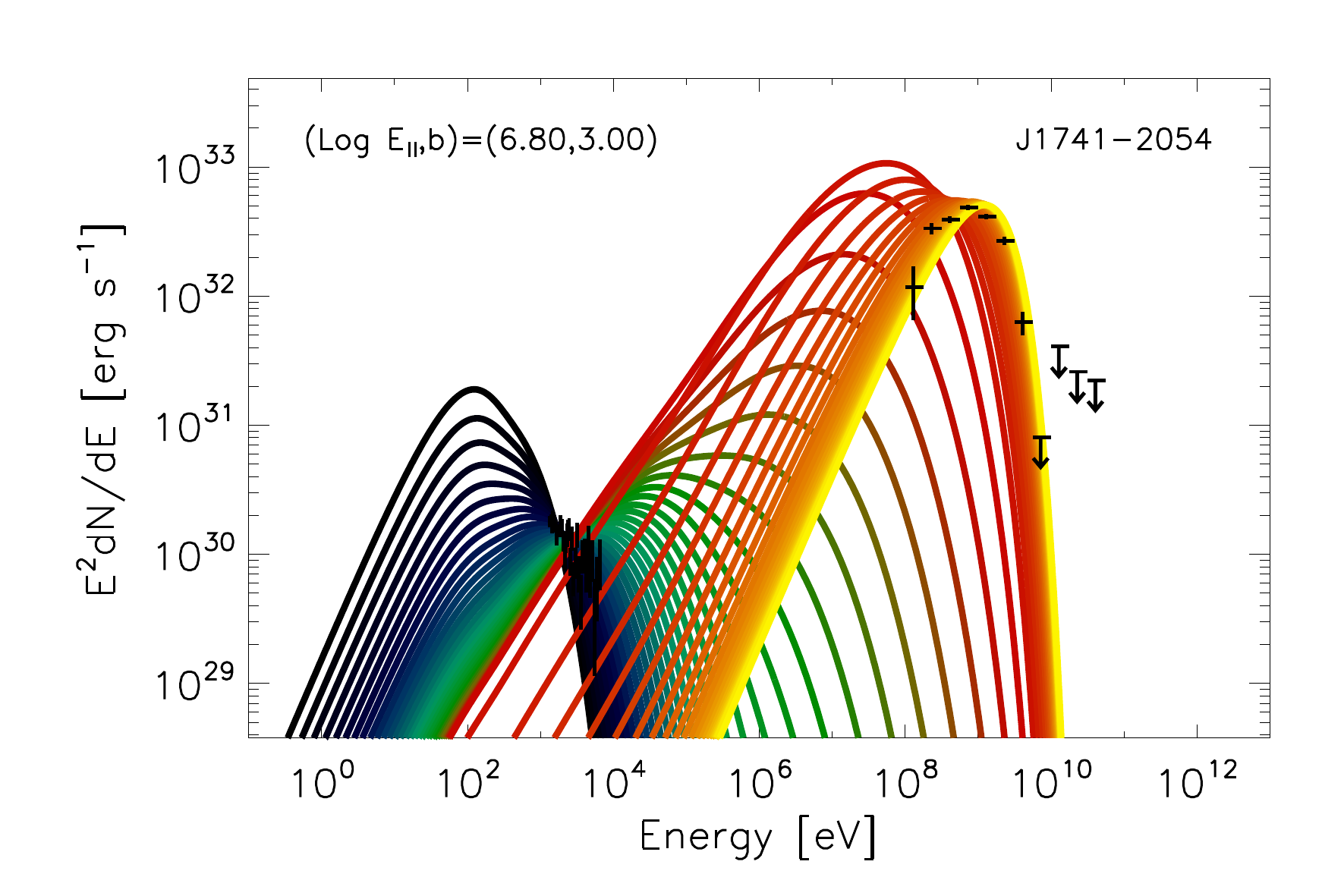}\hspace{-.25cm}
\includegraphics[width=0.34\textwidth]{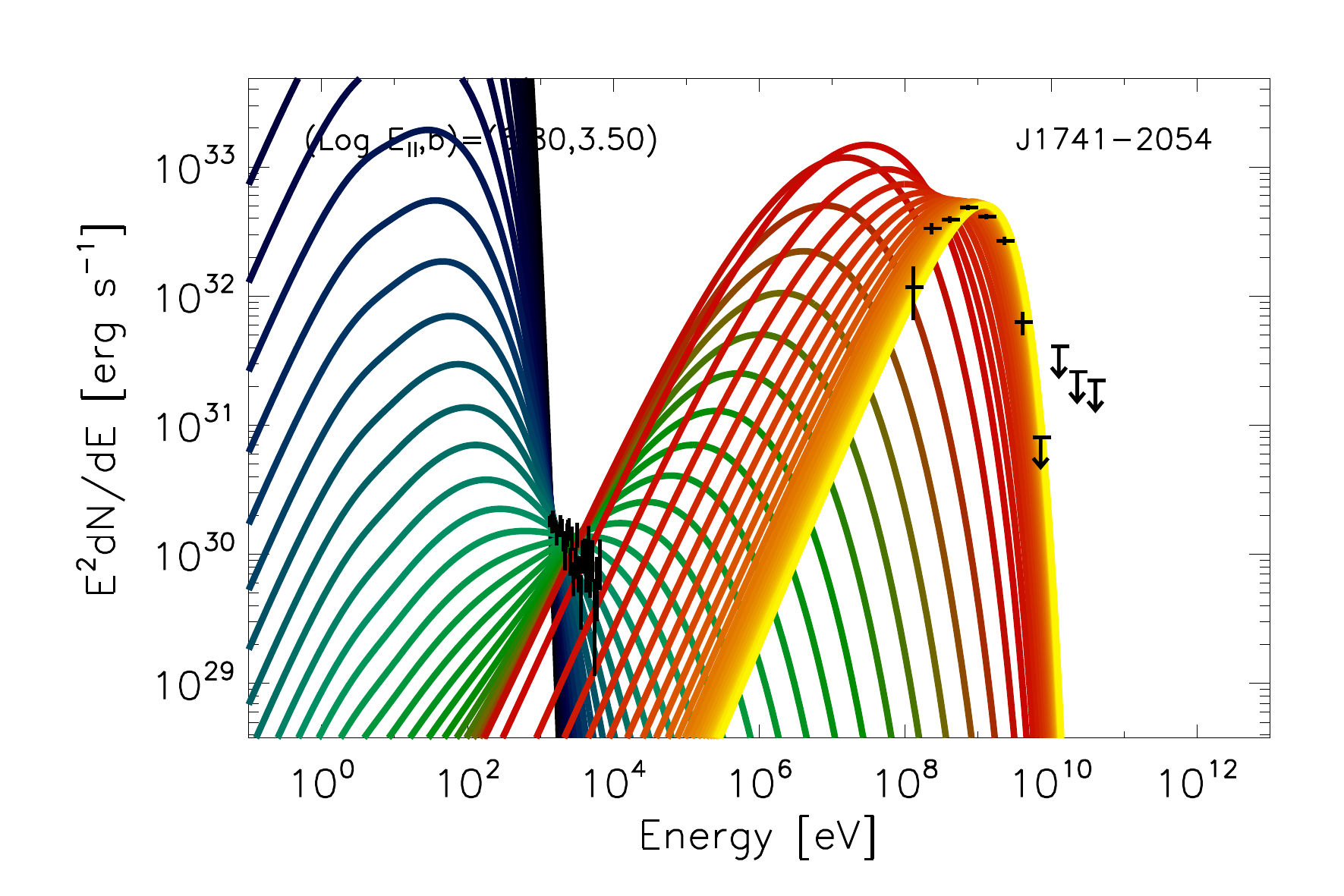}
\caption{Same as Figure~\ref{no_single_fit_1}, but showing different values of $b$ (different in each panel, according to the legend) instead of $E_{||}$, which is fixed.}
\label{no_single_fit_2}
\end{center}
\end{figure*}

To understand better how this happens we show examples of our search for a best fit in 
Figure~\ref{no_single_fit_1}  and \ref{no_single_fit_2}.  
We take J1741-2054 as a study case, but the general scenario is similar for the other pulsars too.
Each panel of Figure~\ref{no_single_fit_1}  shows 
changes of $E_{||}$ for a given value of $b$ (fixed at $b=3$ in this example), varying $x_0/R_{lc}$ from 10$^{-5}$ to $1$, shown with the color scale (blue-green-red-yellow scale, with the addition of the uniform case, $x_0\rightarrow \infty$, the brightest curve).
%
%
%
At relatively low values of $E_{||}$ we find that the model can minimize the deviations from data 
either trying to fit the X-ray observations (at low  $x_0/R_{lc}$ values, the blue curves), or the gamma-ray observations (at larger  $x_0/R_{lc}$ values, between green and red curves).
No single curve is close to both datasets.

A similar effect is shown when varying the magnetic gradient~$b$ for a fixed value of $E_{||}$. 
Figure~\ref{no_single_fit_2} shows a few such examples, using $b=2.5,3.0,$ and 3.5 for a
fixed value of $E_{||}$ and with the same ranging of the $x_0/R_{lc}$ values as commented above, represented 
in the color coding.
Depending on the relative values of $b$ and $x_0/R_{lc}$, again the fit prefers to minimize the deviation either approaching the X-ray or the gamma-ray data, but it cannot do it for both at once either.

Figures \ref{no_single_fit_1} and \ref{no_single_fit_2} show only a set of the examples we explored.
When searching for a common fit,  we systematically vary $b$ and $E_{||}$ in a wide range, in addition to the exploration of $x_0/R_{lc}$ for each case, substantiating that no good fit using a single set of parameters can be found in these cases.

We note that the inability to fit the broad-band non-thermal spectra as commented is not common to all MSPs known.
It just so happens that the four analyzed here are in this set, but our preliminary results using a reanalysis of the gamma-ray data for the remaining three MSPs with broad-band spectra,
B1821$-$24, B1937+21, and J0218+4232, for which non-thermal X-ray emission has been discovered by \cite{Gotthelf2017}, 
show that a single set of parameters in our model can be enough to achieve a good fit.
These MSPs are left out of the sample analyzed in this work, since additional dedicated analysis of the gamma-ray data is needed to have a consistent data set.
This will be presented elsewhere.

\section{Solution and interpretations of the fitting mismatches}

We shall first consider what kind of extension should the model be subject to in order to deal with the cases of Figure~\ref{no-good-fits}.
Two different interpretations for this extension are discussed next.

\subsection{The same $(E_{||}, b$) but different ($x_0/R_{lc}$, $N_0$)?}
\label{sdoublefit}

\begin{figure*}
\begin{center}
\includegraphics[width=0.598\textwidth]{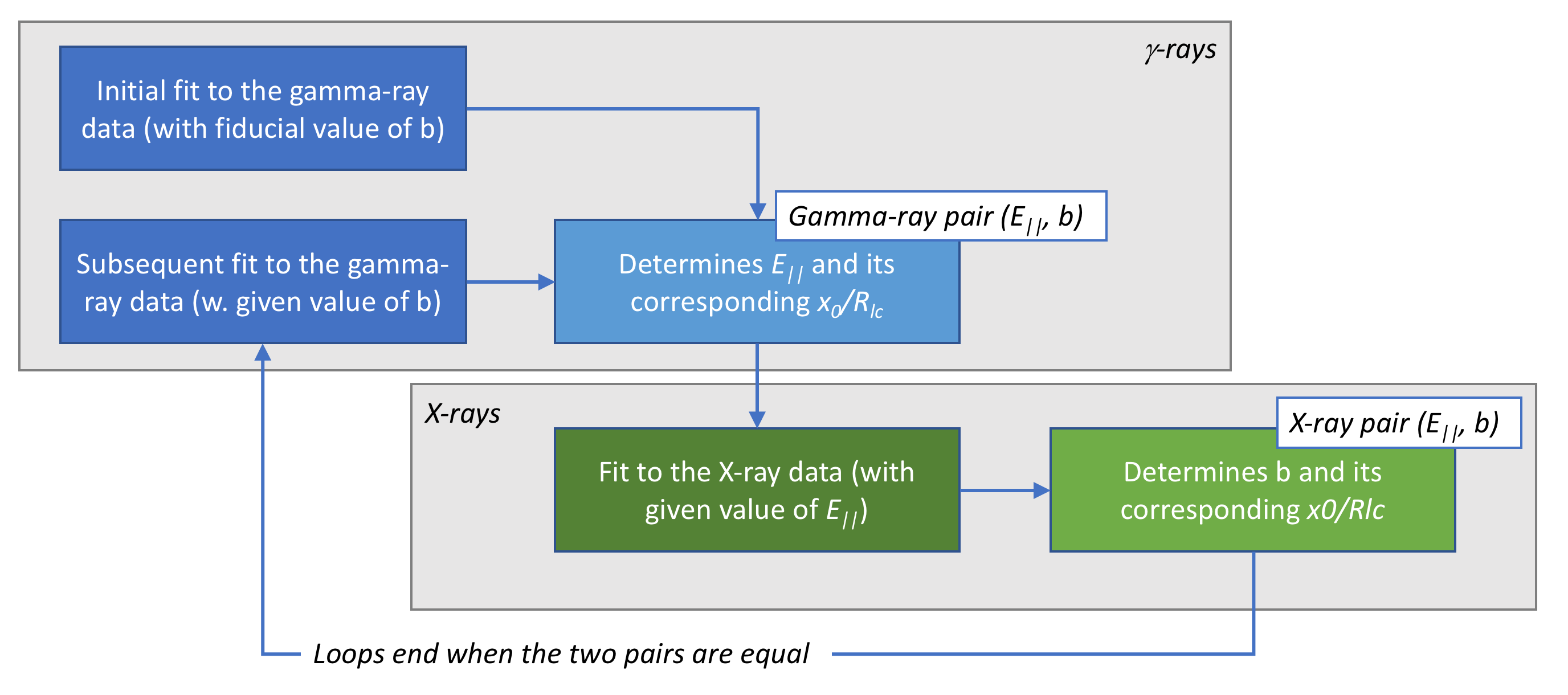}
\caption{Conceptual representation of the fitting loop to determine the physical parameters of the accelerating
region.
}
\label{loop}
\end{center}
\end{figure*}

As commented above, the pair $(E_{||}, b$) represents physical parameters related to the region of acceleration, whereas the pair ($x_0/R_{lc},N_0$) represents 
values related to the geometry: how uniform is the distribution of particles emitting radiation directed towards the observer, and how many particles does this distribution contain.
As the gamma-ray data alone does not in general impose strong constraints on $b$,
it can be used to define an $E_{||}$ range.
That is, one can fix $b$ to a fiducial value (even the same for all pulsars) and get an approximate
range and best value of $E_{||}$ that will be useful for fitting the gamma-ray data.
Then, for that given value of $E_{||}$
one can find a best-fitting pair of $b$ and $x_0/R_{lc}$ that better describes the X-ray data. 
The overall fitting can then be done in a loop, as conceptually represented in Figure~\ref{loop}: 
with the best value of $E_{||}$ obtained from the gamma-ray fitting, we can span a range of $b$-values 
and fit the X-ray data, obtaining a best-fit value of $b$, with the corresponding 
$x_0/R_{lc}$ and $N_0$. 
With such value of $b$ one can then come back to the gamma-ray data set and fit it, to gather a new $E_{||}$  and the corresponding $x_0/R_{lc}$ and $N_0$.
The iterative process continues until convergence of both $E_{||}$ and $b$ values.
For well-determined X-ray spectra, the loop can also be started from a fit to X-ray energies, leading to the same solution.

\begin{figure*}
\begin{center}
\includegraphics[width=0.34\textwidth]{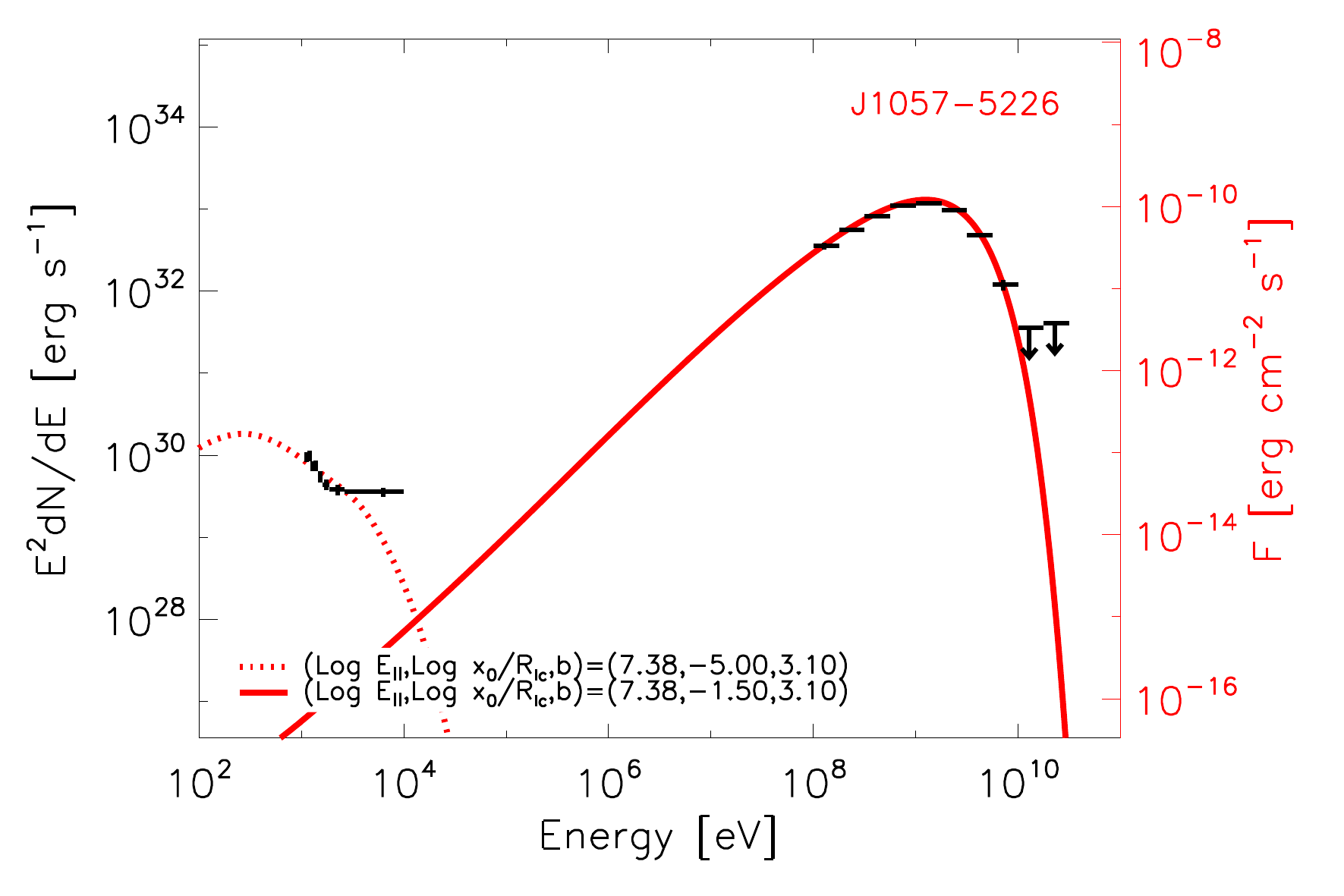}\hspace{-.25cm}
\includegraphics[width=0.34\textwidth]{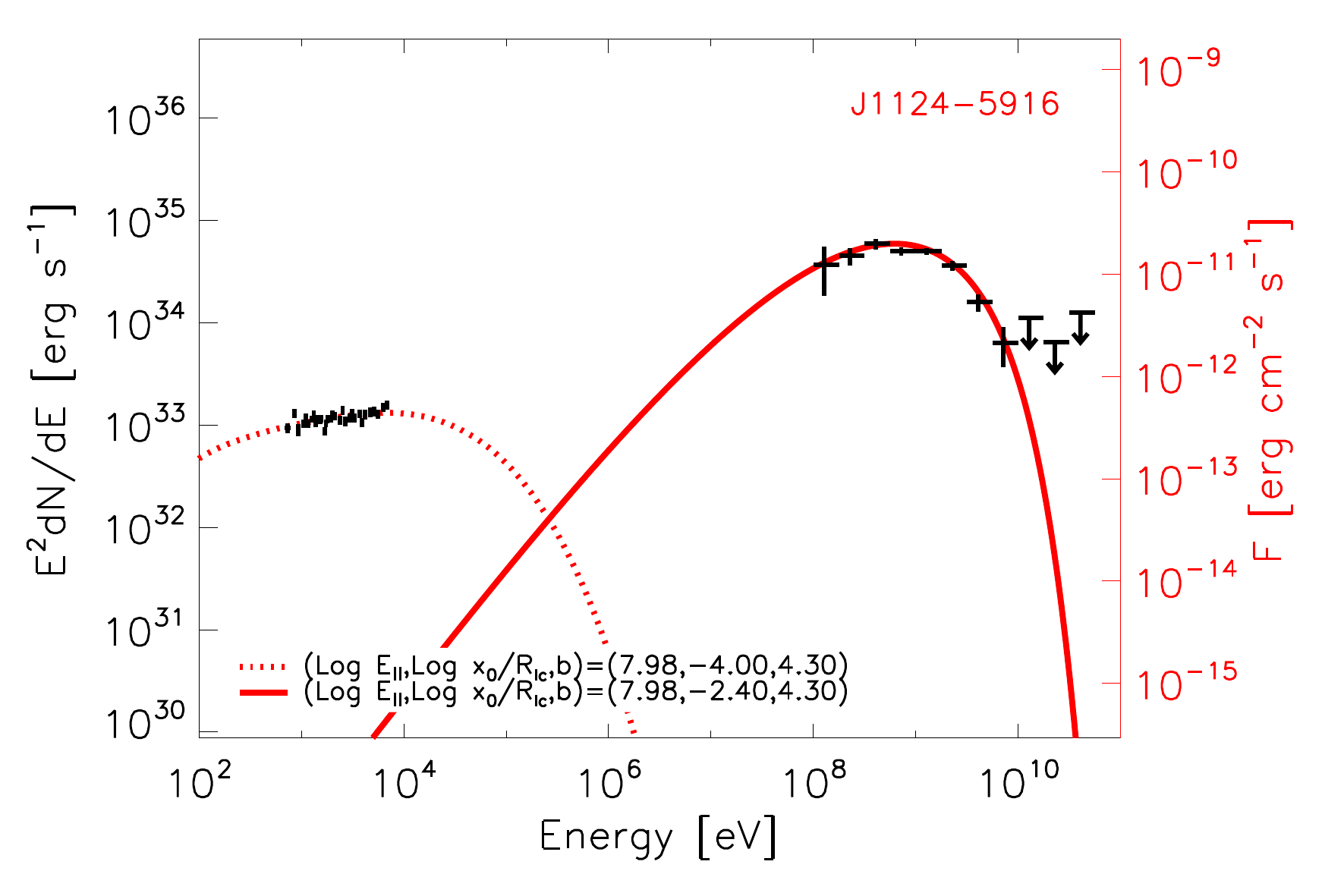}\hspace{-.25cm}
\includegraphics[width=0.34\textwidth]{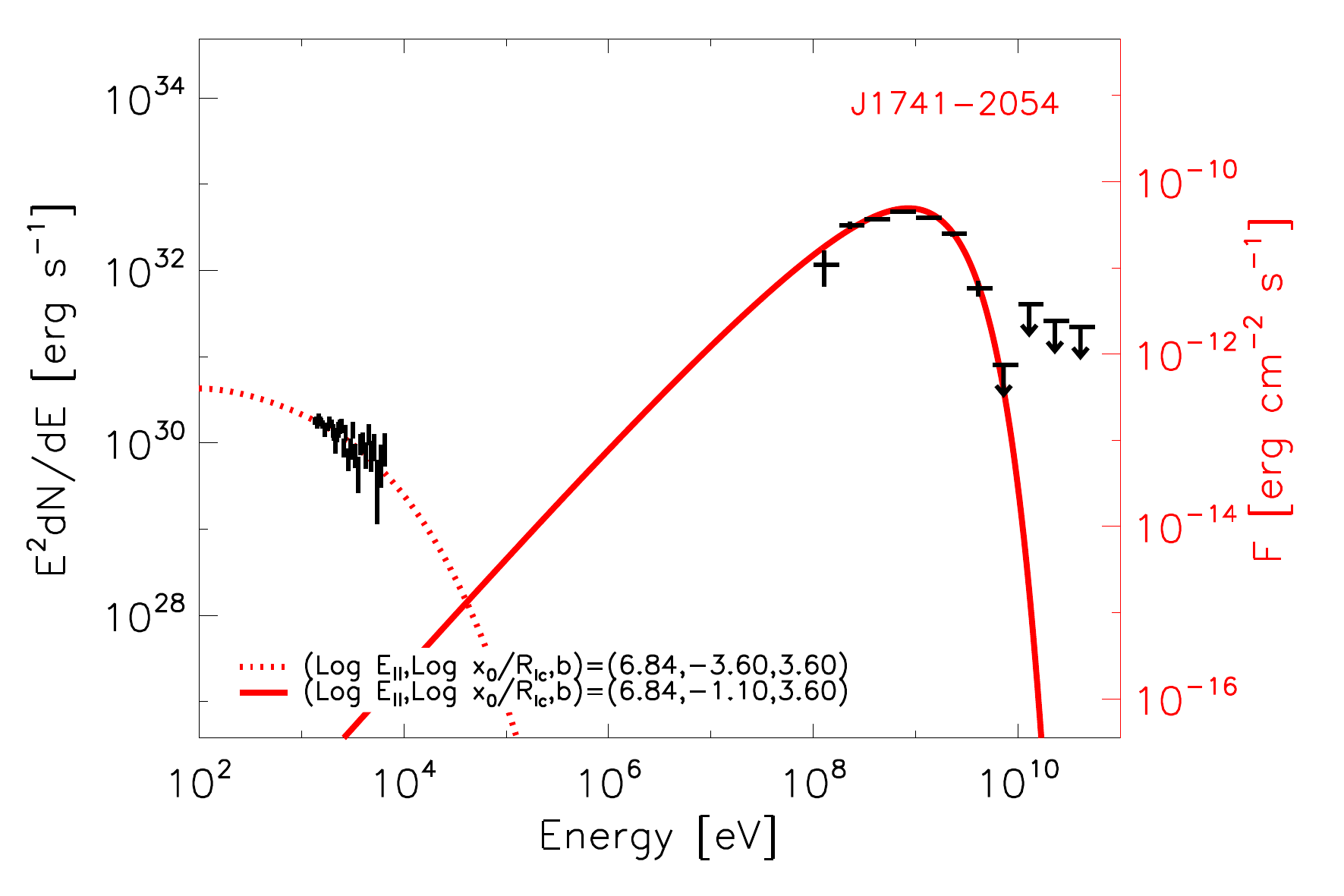}
\includegraphics[width=0.34\textwidth]{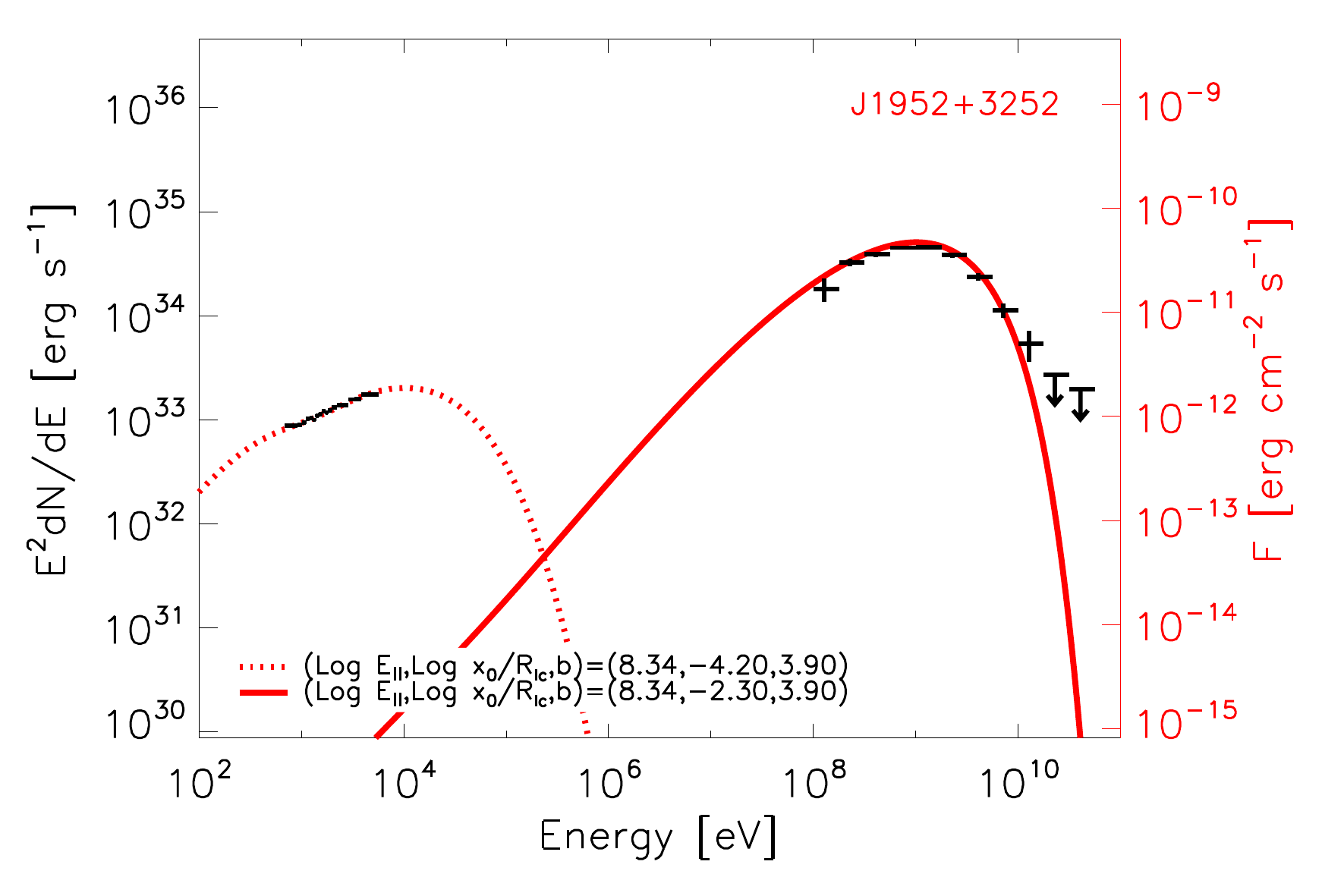}\hspace{-.25cm}
\includegraphics[width=0.34\textwidth]{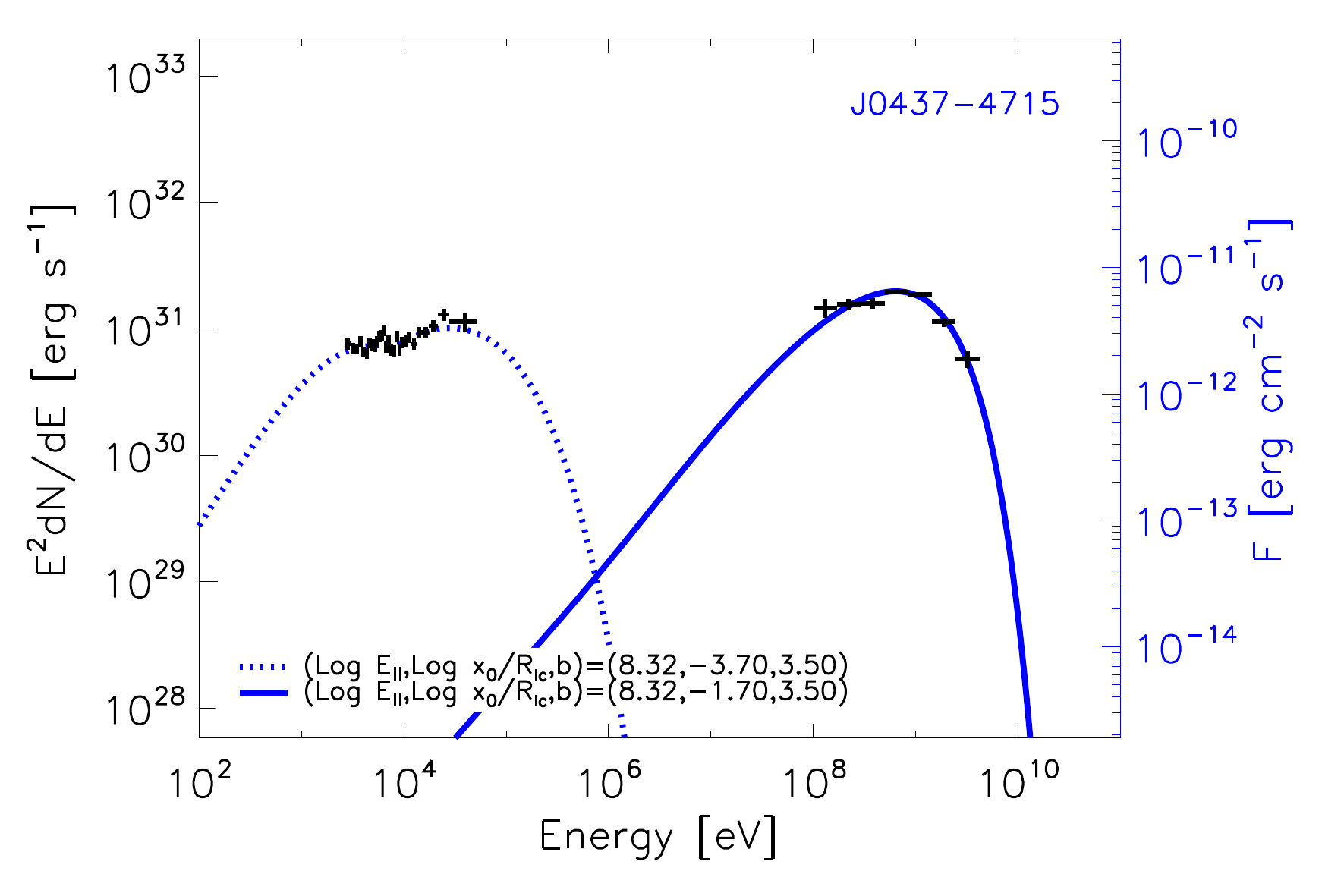}\hspace{-.25cm}
\includegraphics[width=0.34\textwidth]{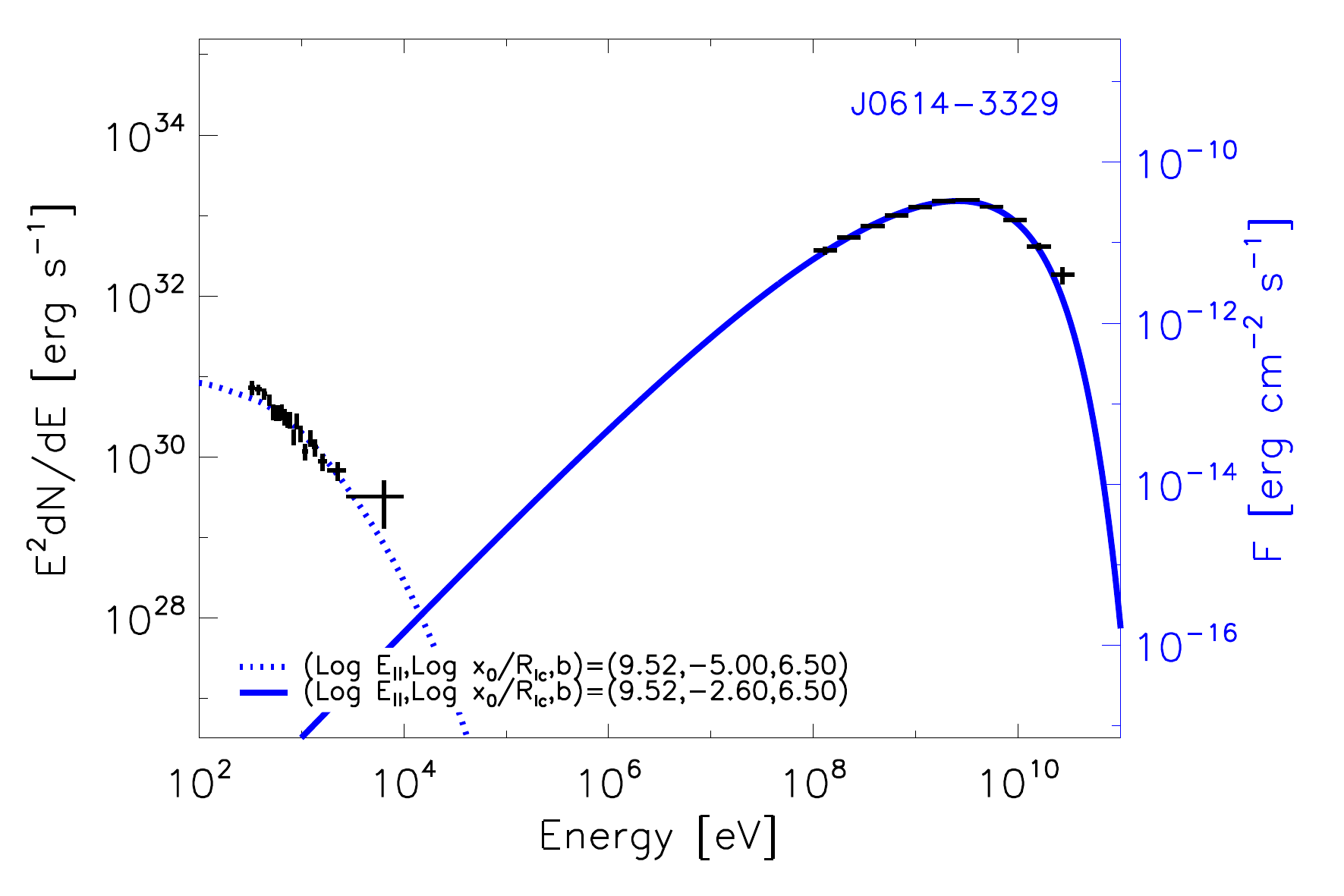}
\includegraphics[width=0.34\textwidth]{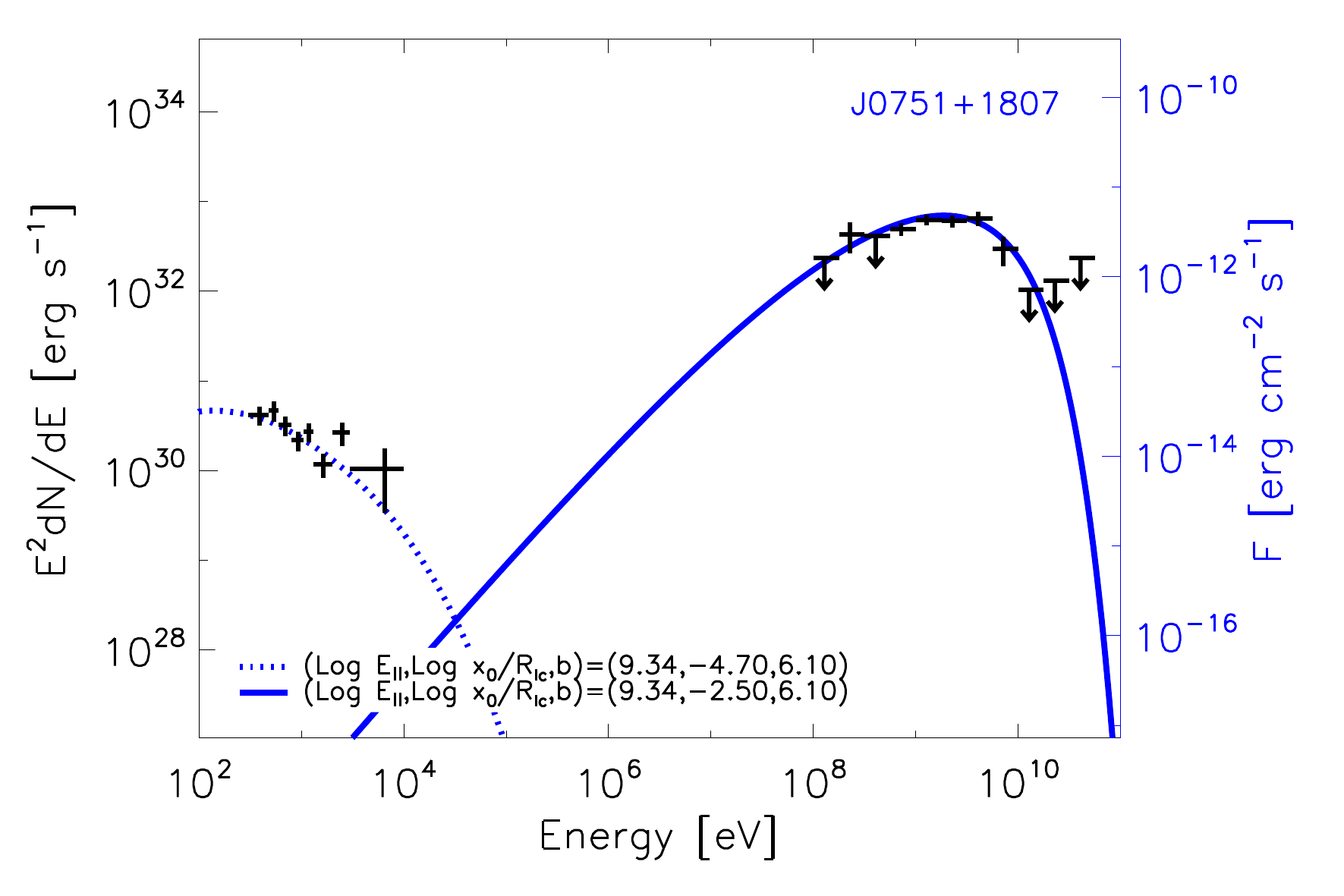}\hspace{-.25cm}
\includegraphics[width=0.34\textwidth]{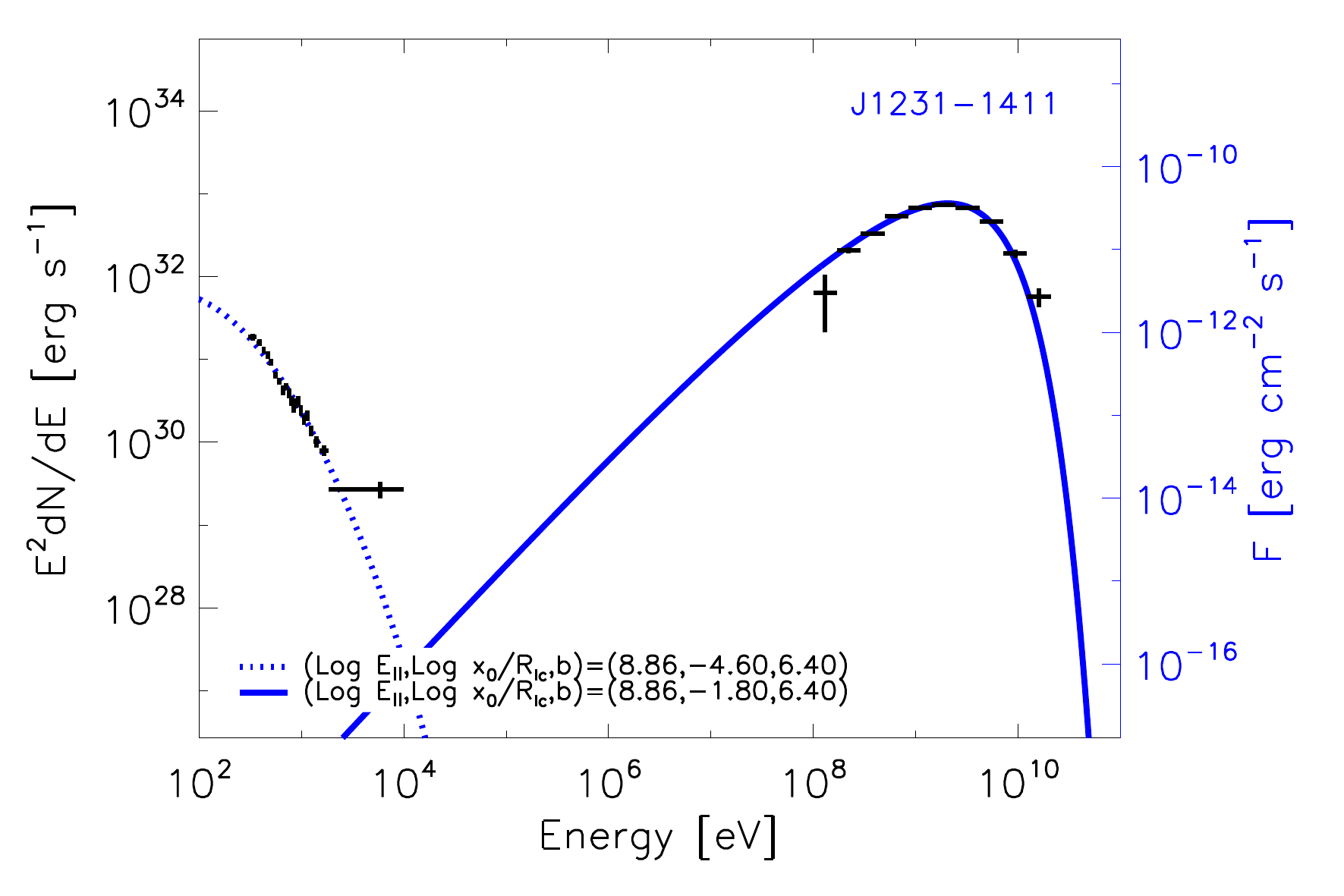}\hspace{-.25cm}
\caption{SEDs of PSRs (red) and MSP (blue) of Figure~\ref{no-good-fits} fitted when assuming two sets of parameters ($x_0/R_{lc}$, $N_0$), as could 
plausibly come from two regions of acceleration being visible, or from a more complex description of the particle distribution that emit radiation in direction to the observer.
As noted in the legend
the same value of $(E_{||}, b$) can be used to fit both parts of the SEDs. See text for discussion.}
\label{composite_fits}
\end{center}
\end{figure*}

We note that in all of the cases of \S \ref{doubletern}, as are shown in Figure~\ref{no-good-fits},
with the same pair of values $(E_{||}, b$) and a different pair of values ($x_0/R_{lc}$ and $N_0$),
we can fit both the X-ray and the gamma-ray data. 
These combined fittings are shown in Figure~\ref{composite_fits}.
The value of the parameters in these (separate) fits are given in first panel of Table~\ref{uncommon-fits}. 

\begin{figure*}
\begin{center}
\includegraphics[width=0.34\textwidth]{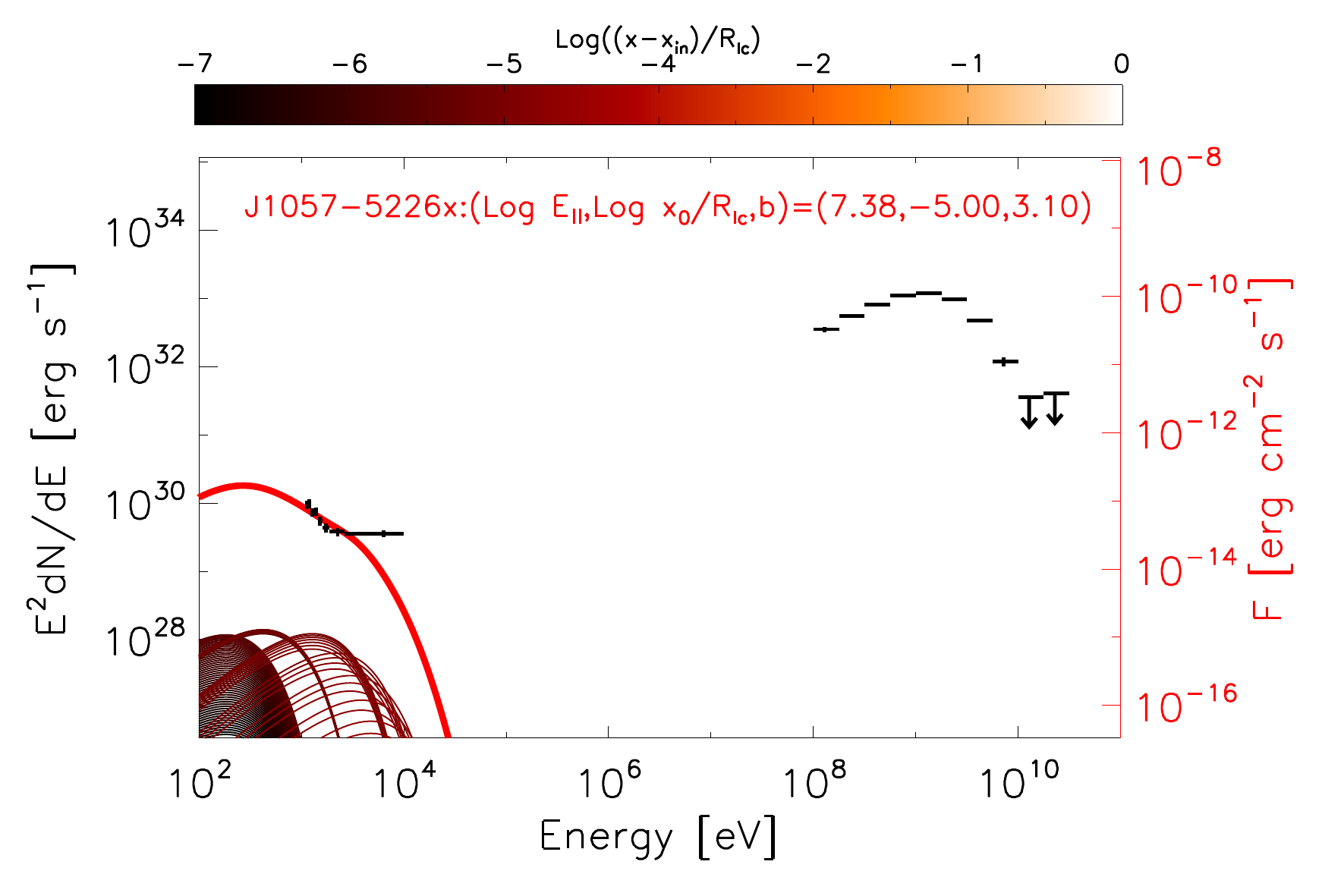}\hspace{-.25cm}
\includegraphics[width=0.34\textwidth]{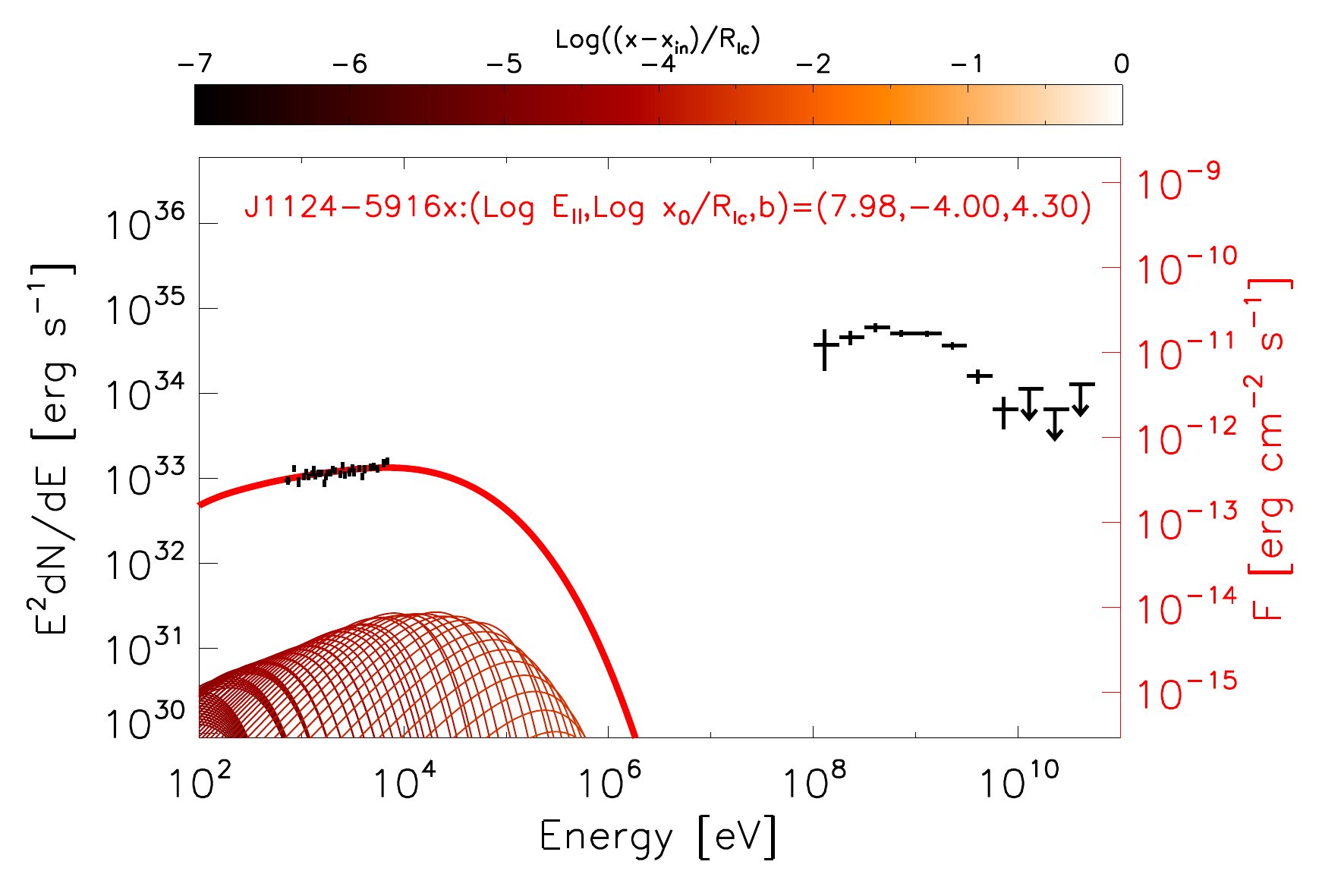}\hspace{-.25cm}
\includegraphics[width=0.34\textwidth]{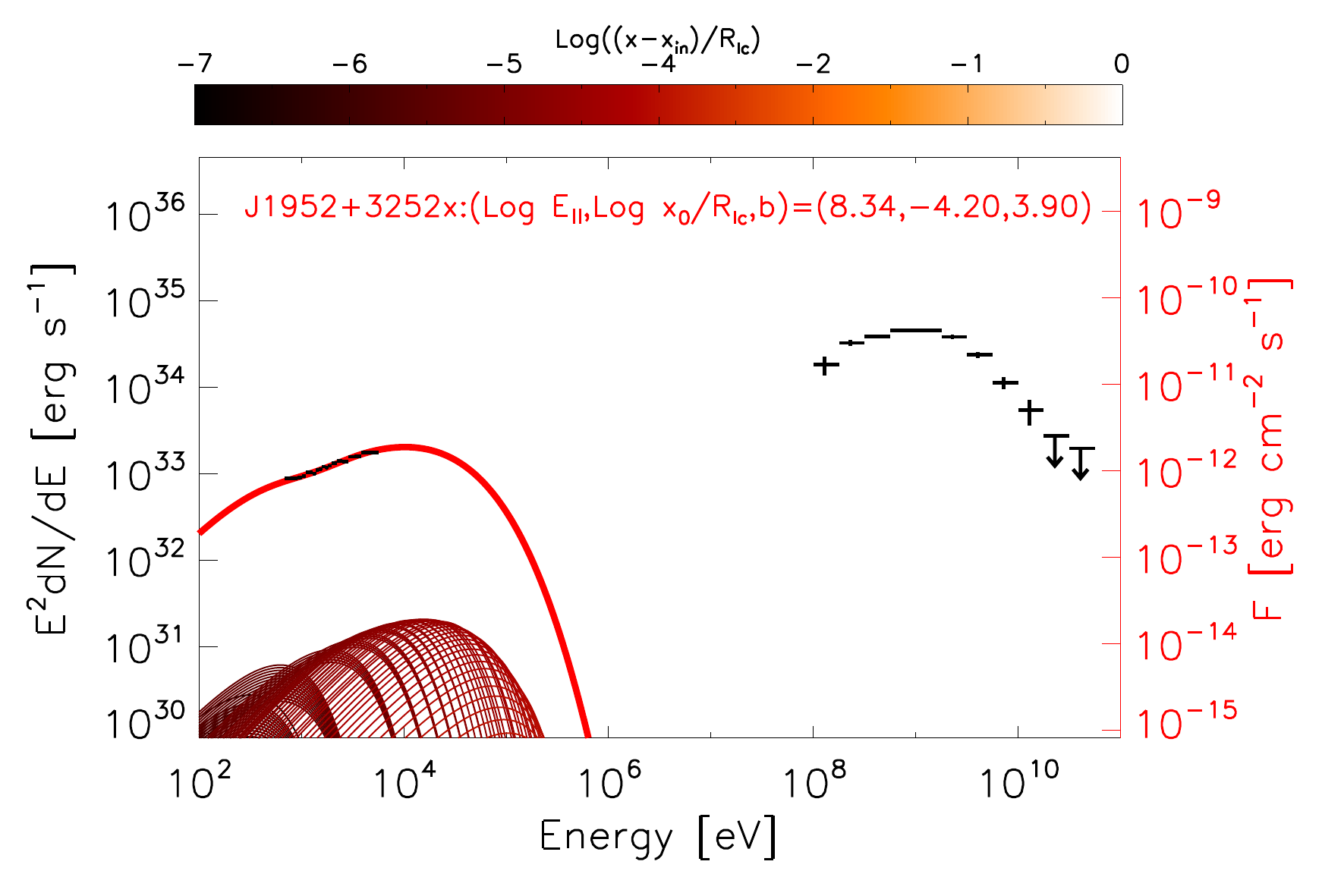}\\
\includegraphics[width=0.34\textwidth]{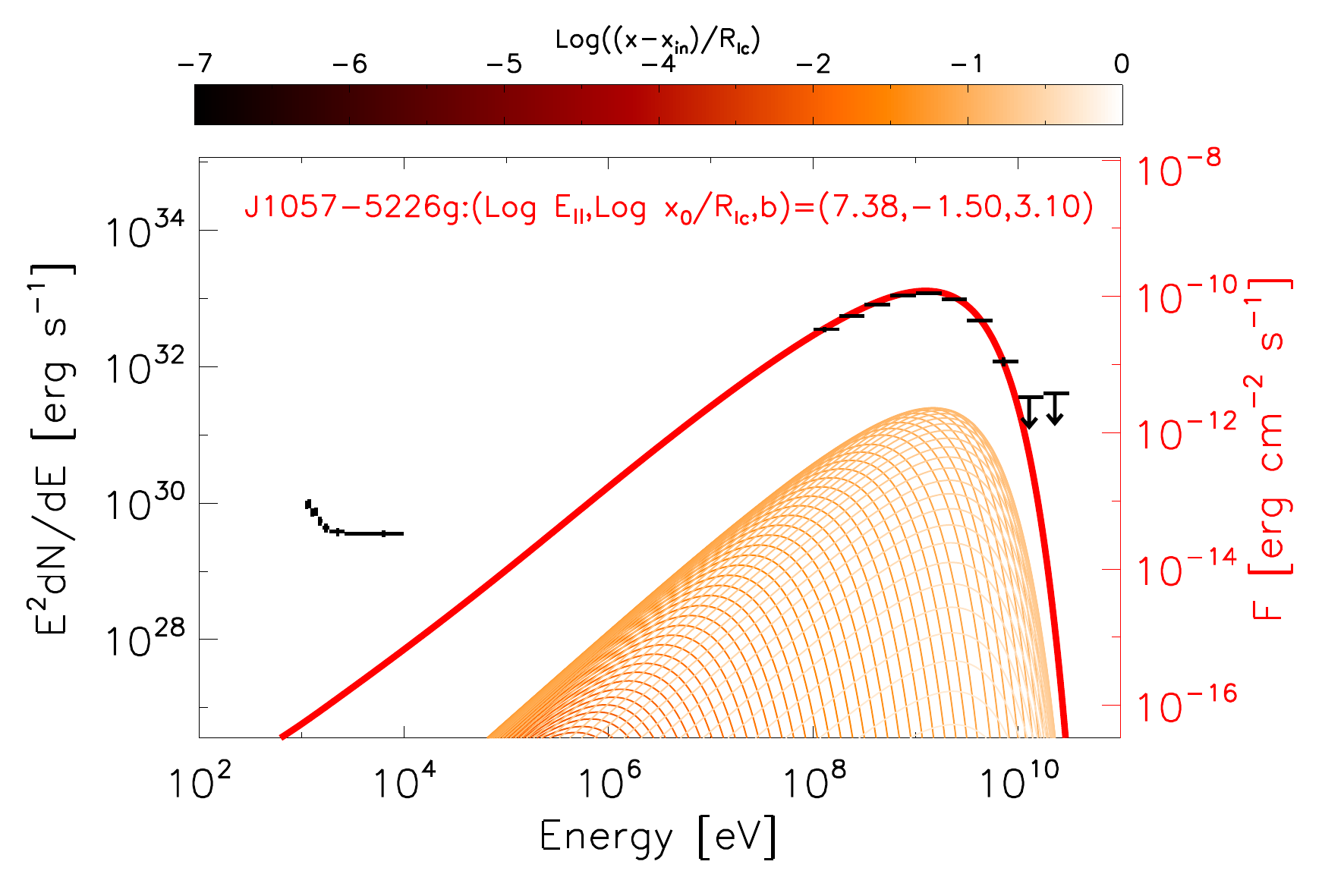}\hspace{-.25cm}
\includegraphics[width=0.34\textwidth]{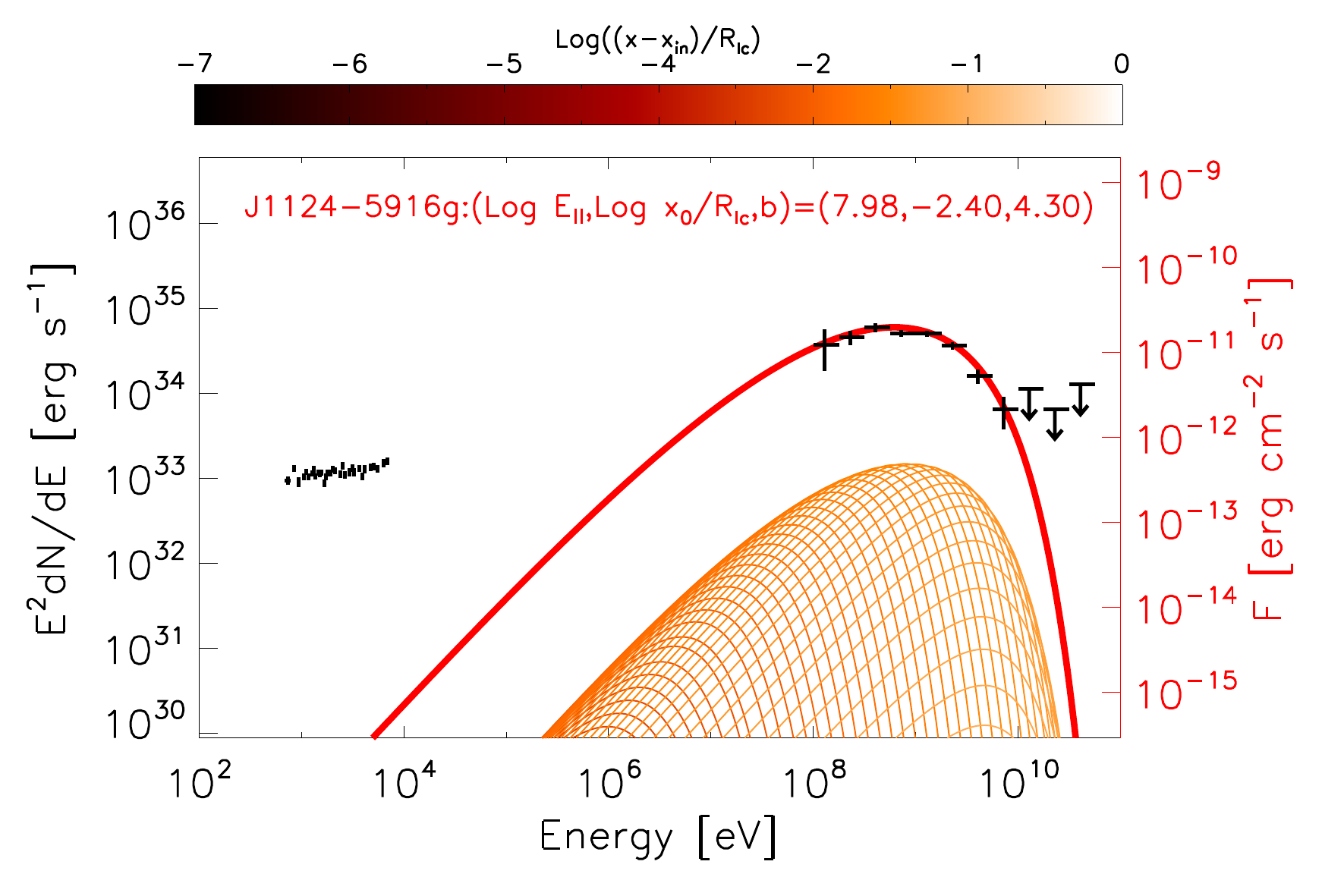}\hspace{-.25cm}
\includegraphics[width=0.34\textwidth]{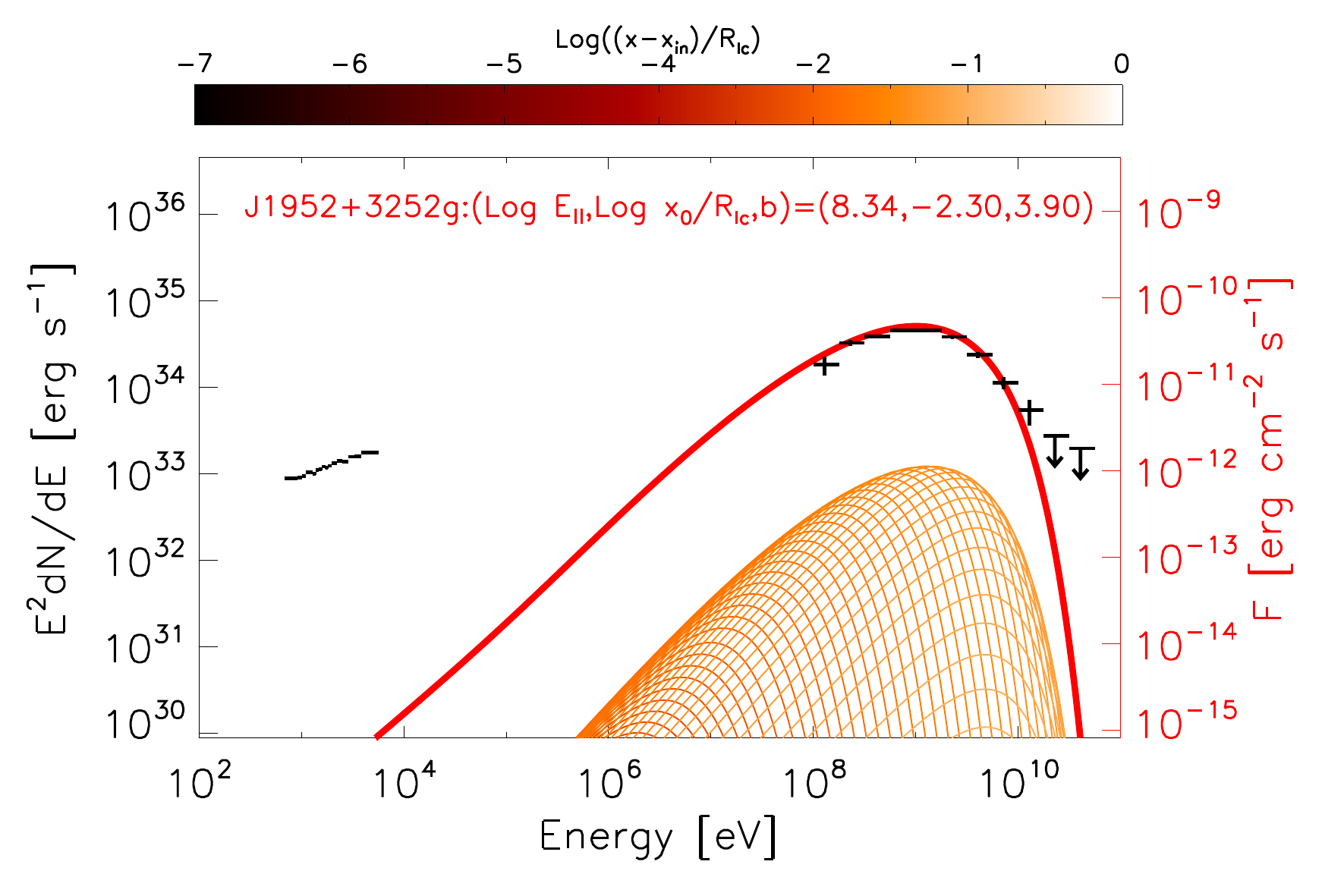}\\
\includegraphics[width=0.34\textwidth]{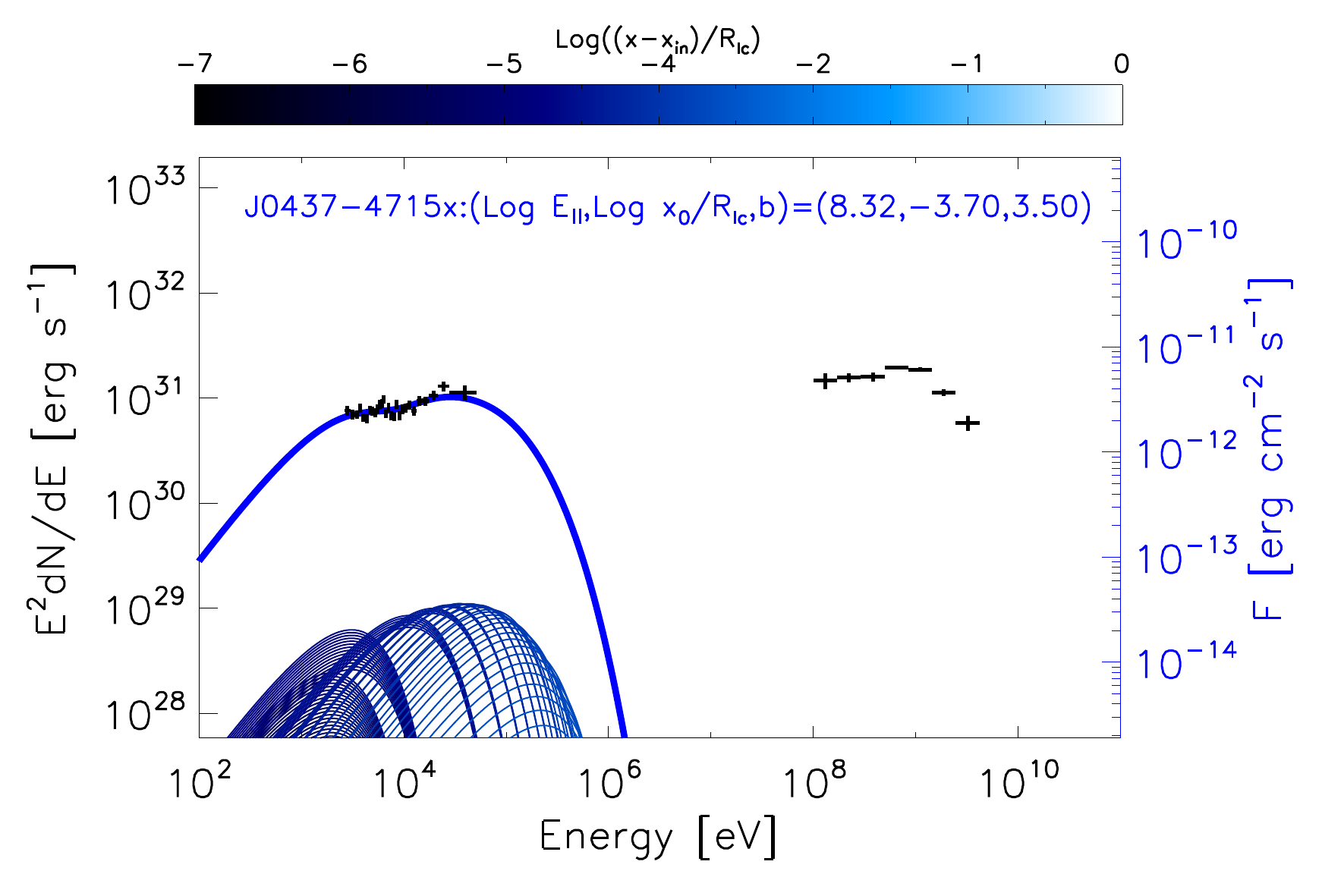}\hspace{-.25cm}
\includegraphics[width=0.34\textwidth]{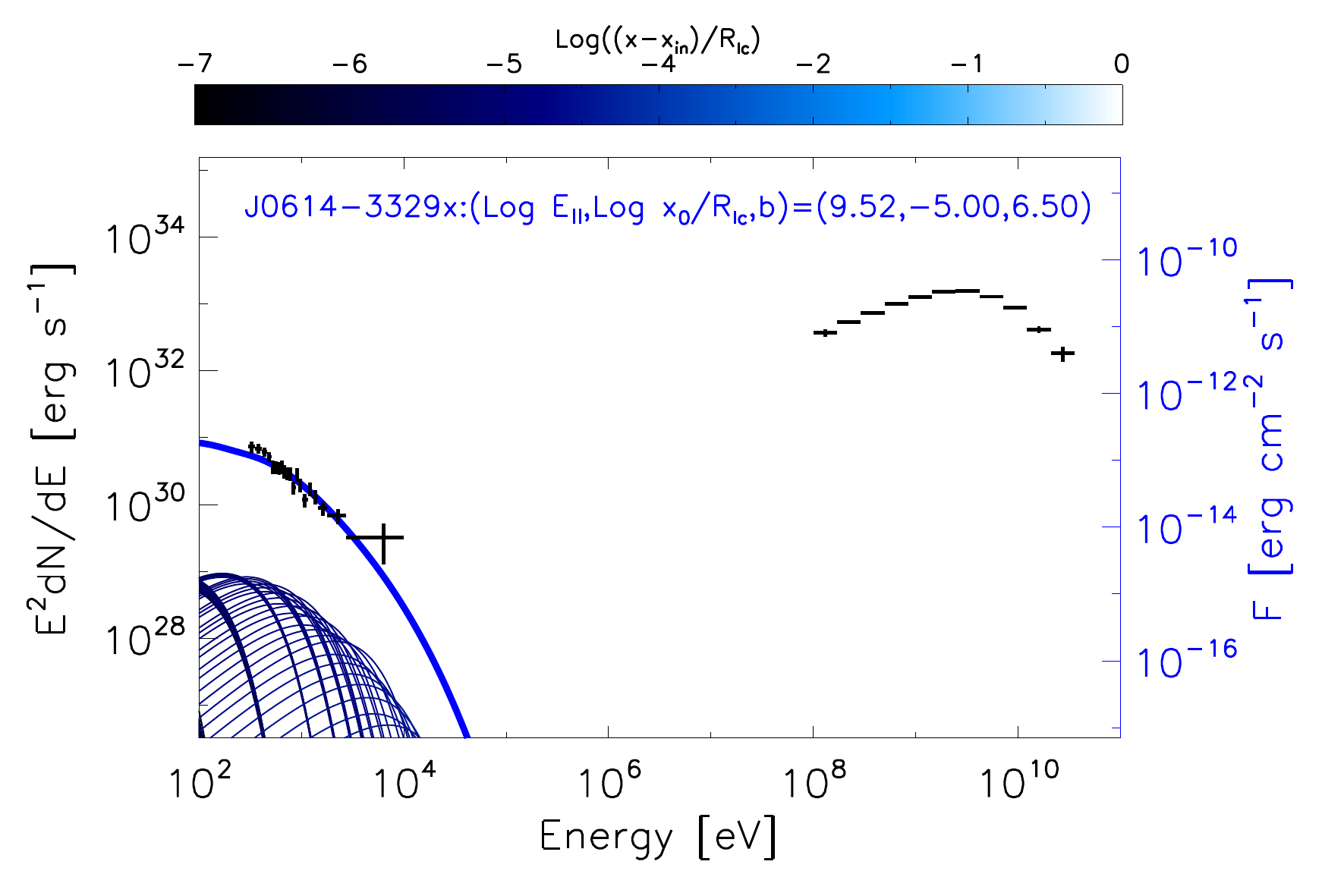}\hspace{-.25cm}
\includegraphics[width=0.34\textwidth]{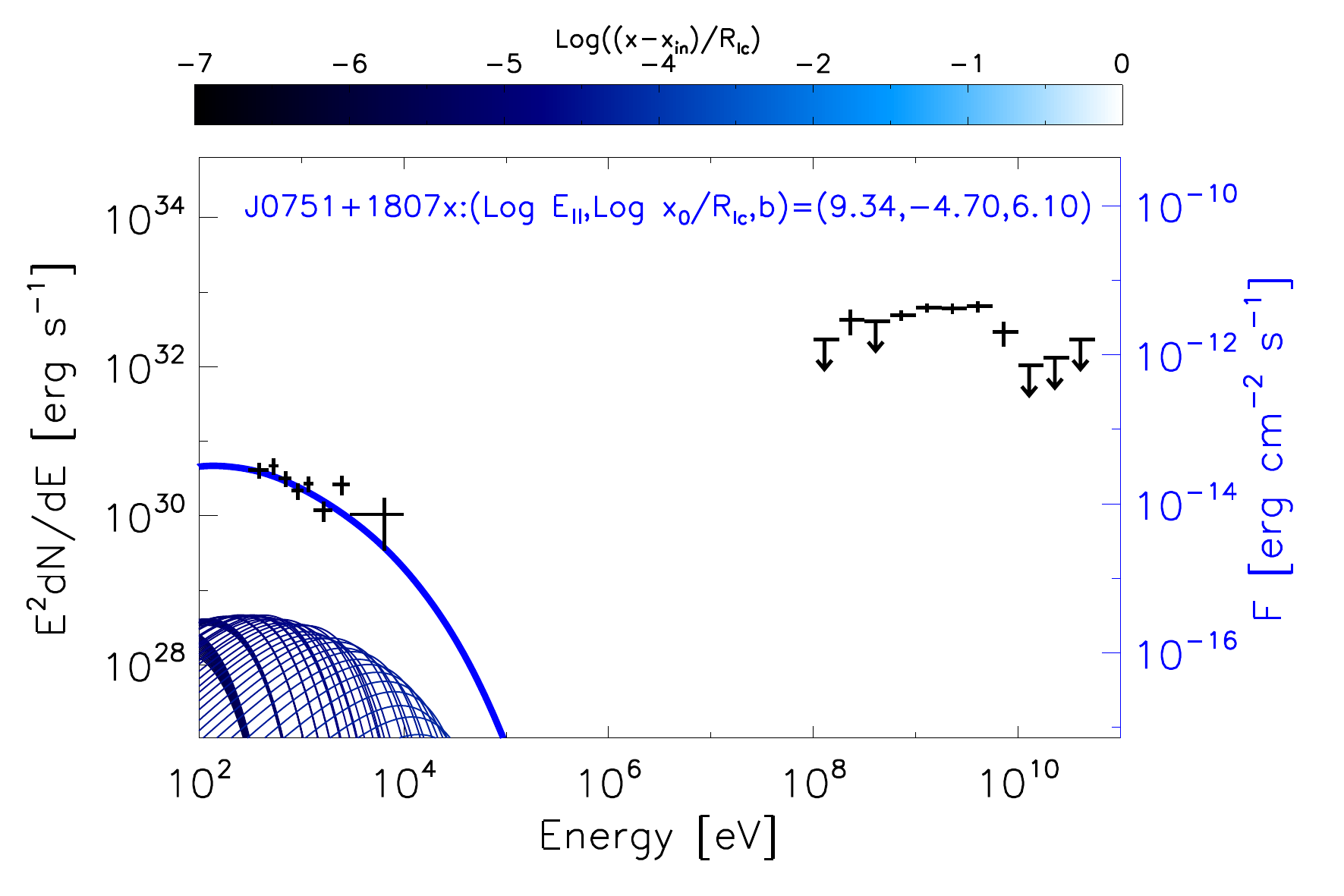}\\
\includegraphics[width=0.34\textwidth]{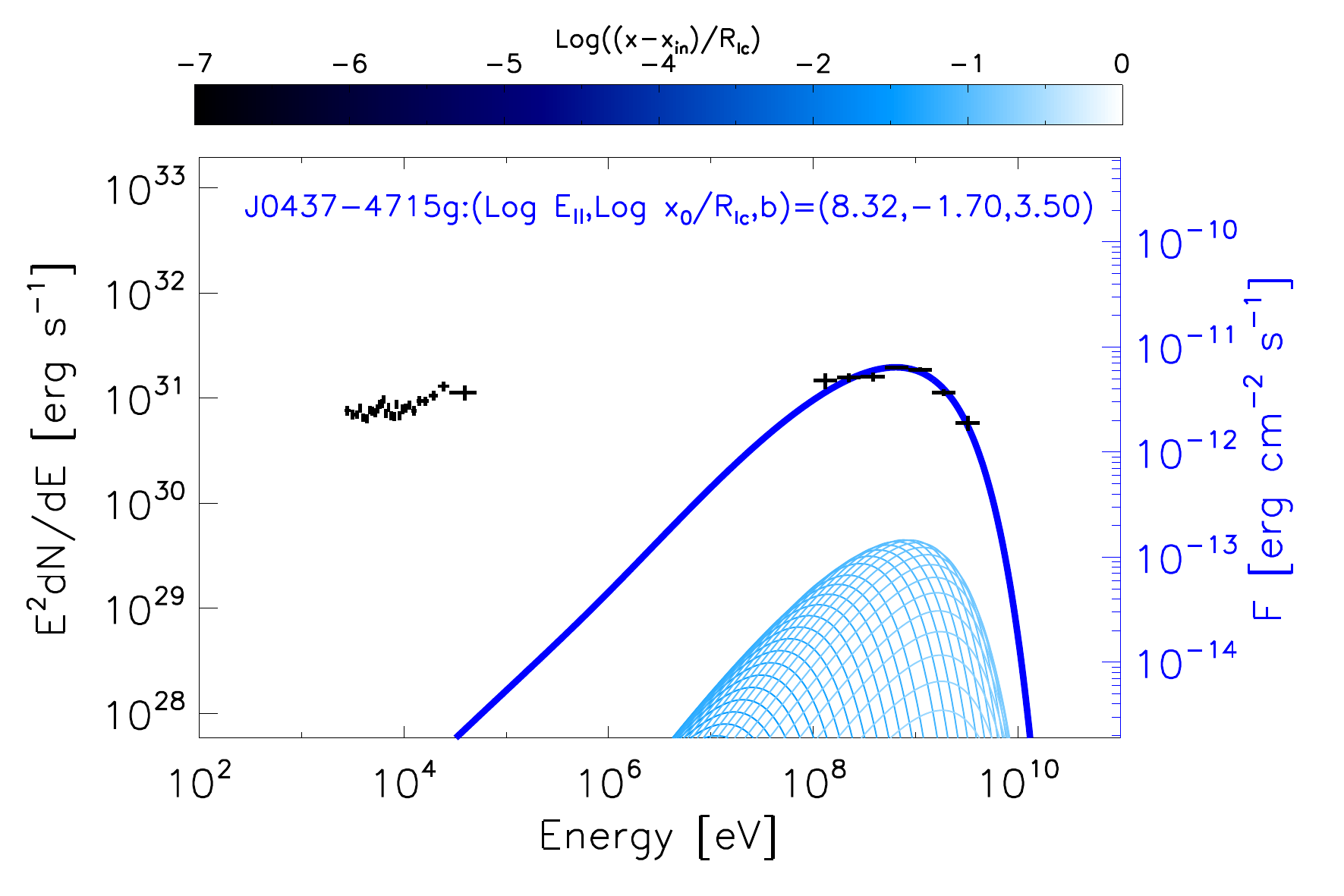}\hspace{-.25cm}
\includegraphics[width=0.34\textwidth]{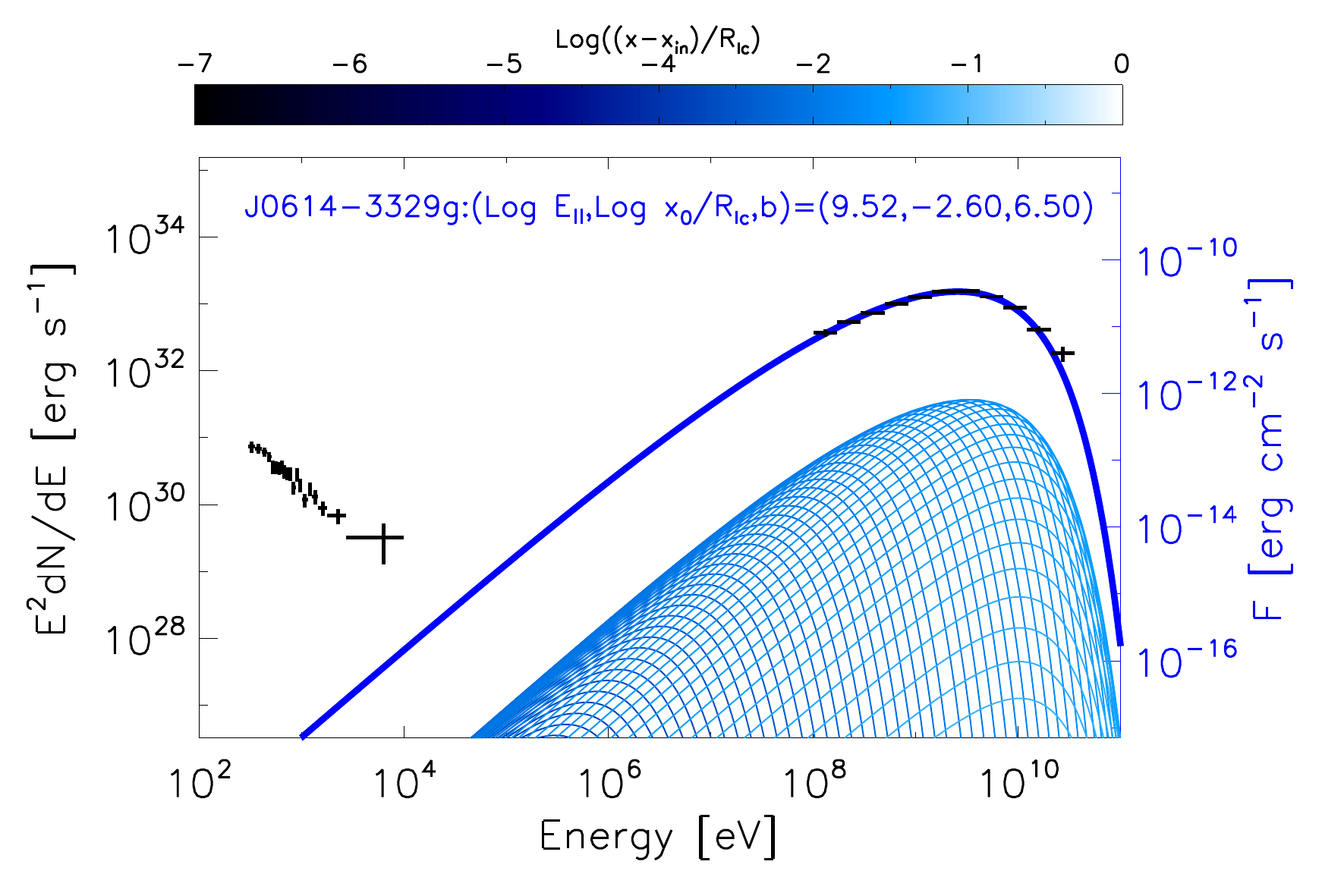}\hspace{-.25cm}
\includegraphics[width=0.34\textwidth]{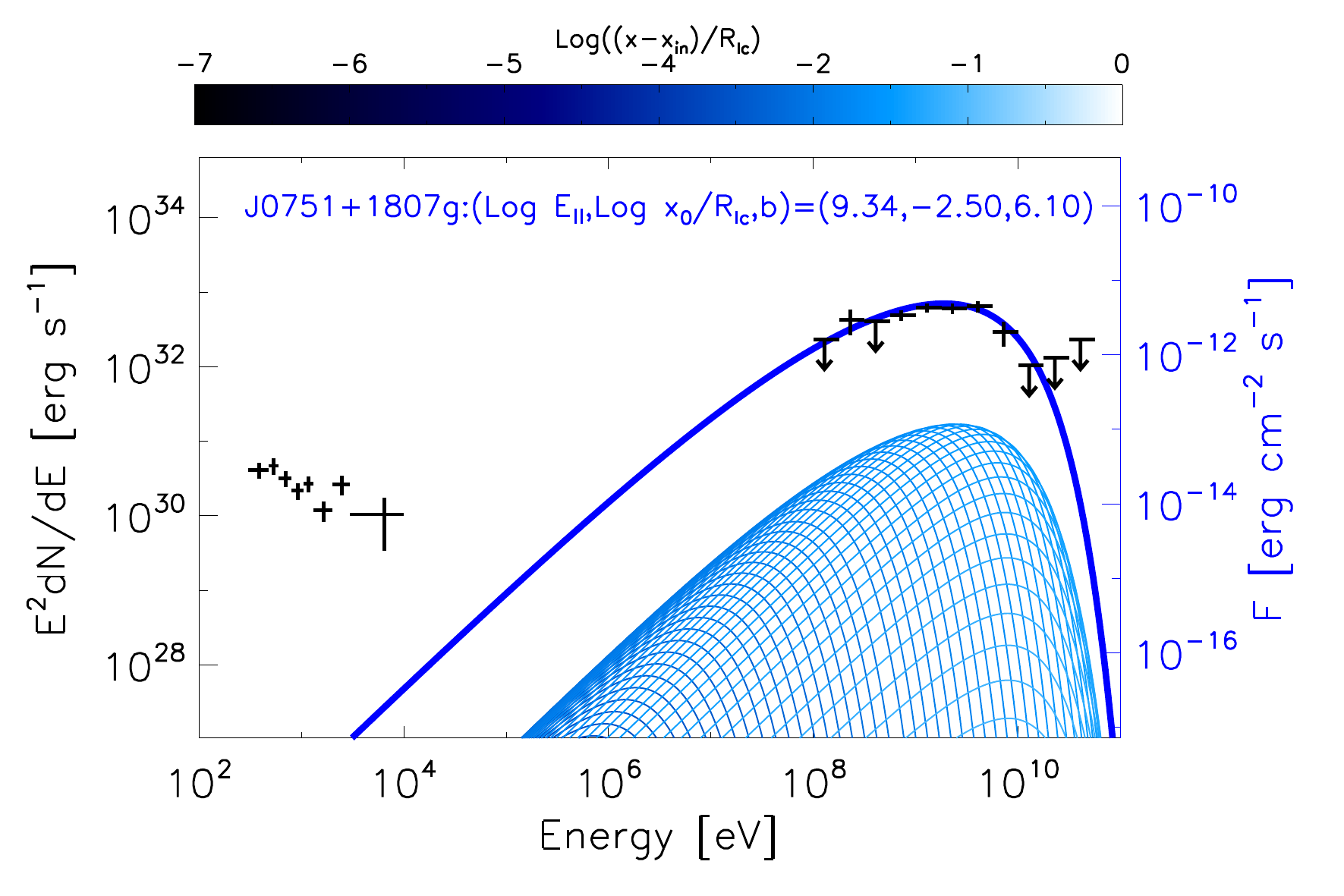}
\caption{
Examples for the formation of the SEDs in cases where two sets of $(x_0/R_{lc}, N_0$) are needed to represent 
the full SED. Two panels are shown for each pulsar, associated to the X-ray and gamma-ray fits.
}
\label{formation2}
\end{center}
\end{figure*}

The strong(er) dependence of each part spectrum on a particular parameter ($b$ for X-rays, and $E_{||}$ for gamma-rays), especially visible when the model
is confronted with data featuring a large energy coverage in these energy ranges and small error bars, does not imply at all 
that these separate fits are devoid of physical information.
On the contrary, since here we fit the gamma-ray data in a completely independent way from the X-rays, 
there is nothing securing a-priori that this fit will not be above the observed X-ray flux level, being incompatible.
As an example, one can look
at all the gamma-ray fits shown in Figure~\ref{sed1} -- dash-dotted lines, to see that, in most cases, the fit to the gamma-ray data already introduces a significant contribution in the X-ray band.
This is not the case here: all fits to the gamma-ray data of Figure~\ref{composite_fits} produce a very low yield in X-ray energies, under-predicting the observed X-ray SED
 and not interfering with the X-ray fit obtained next. 
This is better shown in the spectral formation plots for these cases, which are exemplified in Figure~\ref{formation2}.
It is there shown how the X-ray emission is again in all cases produced by synchrotron emission at the initial part of the particle's trajectories, and how the gamma-ray emission is in turn produced by the farther contributions.
The extrapolation of each contribution into the other range (gamma-ray best-fit to X-ray, and viceversa) is negligible.

It is also interesting to note that the $x_0/R_{lc}$ value selected in the fit to the gamma-ray data is always larger
than that selected in the fit to the X-ray data.

Note that if we used the model fitting only to the gamma-ray part of the SEDs of these pulsars (assuming that we do not yet know the X-ray part, and thus using a single set of parameters to describe the full SED) to infer which pulsars should be visible in X-rays, we would predict that the pulsars in Figures \ref{composite_fits} are not detectable. 
This implies that a selected sample of X-ray emitting pulsars starting from a gamma-ray model fitting 
maybe incomplete, but we shall not waste X-ray satellites observation time
searching for gamma-ray selected pulsars:
those pulsars that may ultimately not be fitted well by the gamma-ray derived parameters, despite being 
noted by the gamma-ray fitting as detectable in X-rays, can be more (not less) luminous.

\subsection{Two regions of acceleration visible?}

\begin{figure*}
\begin{center}
\includegraphics[width=0.45\textwidth]{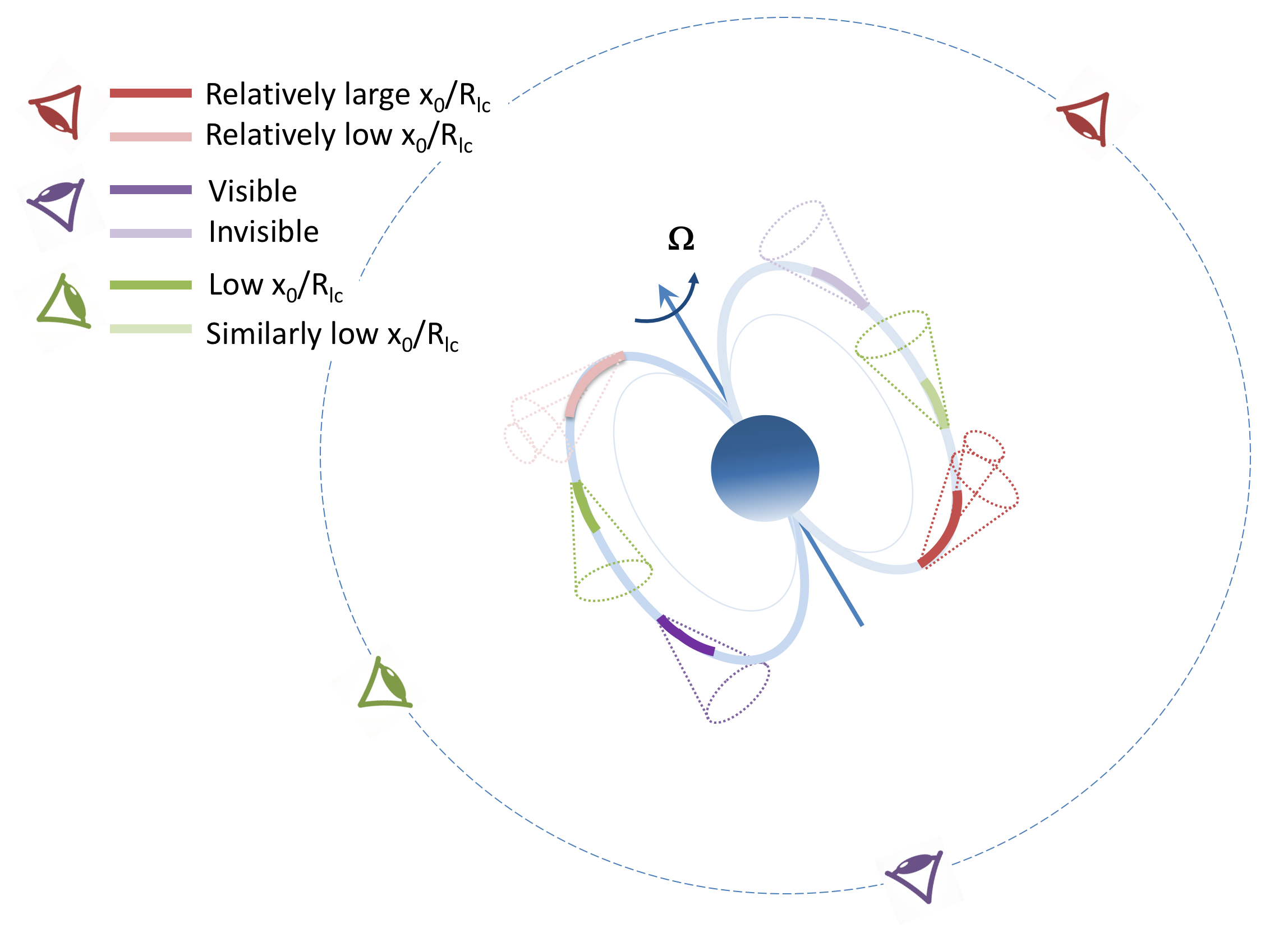} \hspace{0.5cm}
\includegraphics[width=0.45\textwidth]{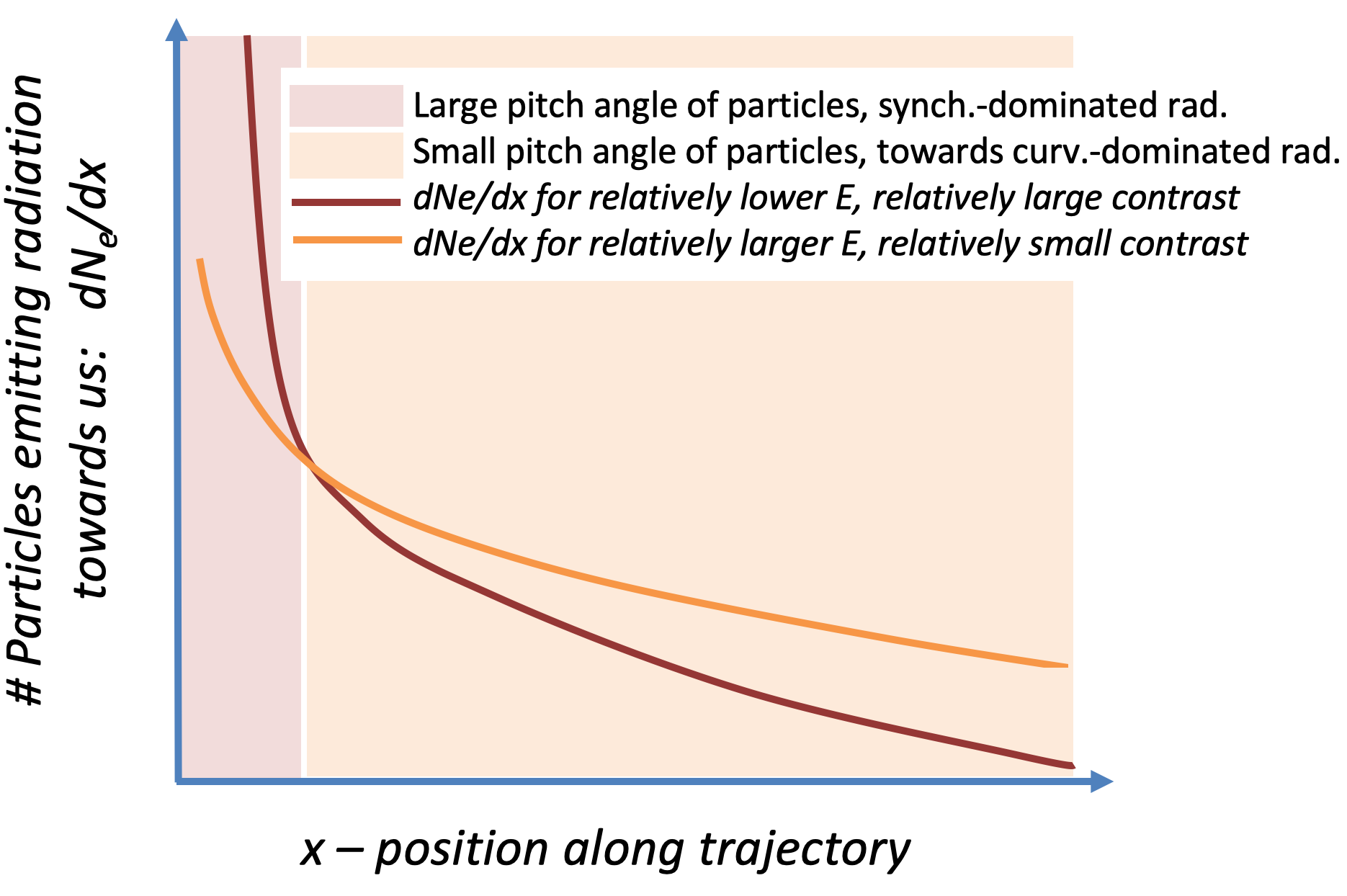}
\caption{Left: Conceptual representation of the effect of the orientation to the observer with regards to the parameter $x_0/R_{lc}$. We use (for the sake of keeping the plot simple) locations of the acceleration regions along a closed field line, to convey the idea only. The positions of the two accelerating regions, and the corresponding observer, are color-coded. 
Right: Conceptual representation of the possible inability of a single set of $x_0/R_{lc}, N_0$ parameters to describe the population of particles emitting towards the observer
at all energies. The total number of particles would be the sum of the two curves.
See text for further discussion.
}
\label{concept}
\end{center}
\end{figure*}

One obvious possibility for interpreting how two populations of particles subject to  the same $(E_{||}, b$) but featuring a different ($x_0/R_{lc}$, $N_0$) can intervene in the generation of the SED 
is to entertain that there are two acceleration regions visible to the observer, and that they sustain a different geometrical orientation.

Indeed, if there are two regions of acceleration, it is reasonable to suppose that they shall have equal physical properties (i.e., the same pair of intrinsic parameters, $(E_{||}, b$)), corresponding to the same pulsar, 
and located in symmetrical positions in the pulsar magnetosphere.
However, if both are visible, they will likely present a different pair of ($x_0/R_{lc}, N_0$) for describing the relevant population of particles emitting radiation towards the observer,
given that the latter is in a different direction with respect to each accelerating region.

Some observers would see the emission mostly coming from one acceleration region, the geometry of which would be more favourable: during a larger part of the trajectories there would be particles moving 
(and emitting) in the direction of the observer. 
This would lead to a (relatively) more uniform distribution, and thus to a larger  $x_0/R_{lc}$. 
The geometry of the other accelerating region may be less favourable, such that the distribution of particles 
emitting towards the observer is more skewed, leading to a smaller $x_0/R_{lc}$. 
The fact that the $x_0/R_{lc}$ value selected in the fit to the gamma-ray data is larger
than that selected in the the fit to the X-ray data is consistent with this geometrical interpretation: 
the gamma-ray emission is produced further along the trajectories of particles, thus from the more uniform population, 
and we see particles reaching to a higher energies. 
Because of that, the $x_0/R_{lc}$ value fitting the gamma-ray observations is larger (the contrast is smaller) than the one chosen by the X-ray data fit.

This concept is represented in the left panel of  Figure~\ref{concept}. 
In that figure, the location of several possible acceleration regions 
are noted with different colors, assuming that two symmetric locations appear. 
Several different observers (color coded) are noted as an example with respect to which assess the geometrical configuration.
For instance, in  Figure~\ref{concept}, the dark and light red represent pairs of regions of acceleration whose geometrical orientation may be leading to 
different values of $x_0/R_{lc}$ and $N_0$, as noted in the legend. 
The cones are representing the direction of emission at different parts of the trajectories of particles: they are wider at the initial part where synchrotron 
domination occurs via the loss of perpendicular angular momentum, and smaller in regions where the trajectory is more curvature-dominated.
Note that even when existing,  geometry would not alway allows seeing the two regions of acceleration (see the violet case in 
Figure~\ref{concept}, where the represented geometry would lead to only one region of acceleration being visible). 
Finally, in cases in which both accelerating regions are similarly oriented (e.g., the green case of the conceptual left panel of Figure~\ref{concept}), or in cases 
in which only one region is visible, a single set of $(E_{||}, b$,$x_0/R_{lc}$, $N_0$) parameters would suffice to represent the SED.
This interpretation seems a proper explanation for all cases, but has additional consequences on the expected light curves that we put in context below, after discussing
an alternative.

\subsection{One region of acceleration with a more complex particle distribution?}
\label{DD}

There is a second possible interpretation for a double set of ($x_0, N_0$) parameters.
The distribution of the particles emitting towards us, $(dN_e/dx)$, may be different at different energies,
even when only one region of emission is visible, and therefore may not be well-describable by a single set of parameters ($x_0, N_0$) as used in Eq.~\ref{eq:distribution}. 

For the relativistic particles we are considering ($\Gamma >10^3$), the opening of the emission cone centered around the direction of motion
(at which boundaries the radiation emitted peaks and 
along which the angular distribution of the radiation is spread), despite being energy-dependent, is vanishingly small at all energies. 
This implies that photons are always emitted in the instantaneous direction of motion of the particles.  
When the pitch angle tends to zero, which  is valid in all cases where curvature dominates synchro-curvature emission \citep{paper0}, all radiation points in the tangent 
direction. But this is not the case if synchrotron
emission dominates synchro-curvature, when large pitch angles are found.
The gyro-averaged emission distribution, for a given position along the line, describes
a circle centered around the tangent direction, with a radius given by the particle pitch angle.
Thus, particles that are not moving in the direction to the observer, may still radiate photons that do when the pitch angle is large.
To correctly describe that such a relatively large number of particles are able to emit in the direction to the observer in the synchrotron-dominated synchro-curvature regime,
a large contrast is needed in Eq. \ref{eq:distribution}. 
However, this contrast may incorrectly underestimate the number of particles emitting towards the observer at larger energies.
This is conceptually represented in the right panel of Figure~\ref{concept}.
Given that the sum of two populations, $p_1$ and $p_2$, each with parameters $(x_0^{p_1}/R_{lc},N_0^{p_1})$ and $(x_0^{p_2}/R_{lc},N_0^{p_2})$ can not be encompassed in a
single set of new parameters ($x_0, N_0$), the use of single distribution may be a bad approximation to the relevant number of particles.

We note that in this scheme, the use of two sets of values ($x_0, N_0$) to describe a high (and more uniform) 
and low (and more skewed) intervening particle population is also an approximation, of course.
It is better than using just one set, but still 
an approximation to a more complex situation where ultimately 
$(dN_e/dx)$ is a function of energy; there is only one particle population and one accelerating region in this interpretation.
We come back to this concept when discussing Crab below.

\subsection{Relation of these  interpretations with the pulsar light curves}

Light curves cannot be directly modelled by our SED approach without further significant
extensions.
Several effects related to the emission process itself and to where this emission happens along the particles trajectories 
concur to shape the light curves beyond the location of the acceleration regions. 
However, we can already note that the two interpretations suggested above should be qualitatively reflected in different light curves properties.

For instance, the X-ray and gamma-ray emission, if coming predominantly from different regions of acceleration as in the left panel of  Figure~\ref{concept}, should appear phase-shifted.
For at least some cases (but not all) where simultaneous light curves have been measured, this indeed appears to occur:
such are for instance the cases of J1057-5226 or J1741-2054, where the gamma-ray and X-ray peaks are phase-offset by roughly a half rotation, e.g., see Figure 3 of \cite{marelli14}.

Instead, if both the X-ray and the gamma-ray emission come from the same emission region, as would be the case in the right panel of  Figure~\ref{concept} where
a more complex particle distribution 
is needed to describe the population emitting towards us, there should be a smaller phase shift between the low and high energy light curve.

Whereas we are not yet offering a first-principle computation of the light curves predicted by our model, 
we note that as commented by \cite{marelli14}, no model has been able to account for offsets between gamma-ray and X-ray peaks  before. 
It seems that this fact can be qualitatively explained here: we offer a plausible interpretation by which 
not always the X-ray and gamma-ray emission come from the same place.

We also note that even in the case of pulsars that can be fitted by a single set of parameters, 
if particles are all injected at the same location there should be a delay between the emission of X-rays
and that of gamma-rays (see the right set of panels of Figure~\ref{traj1} for an explicit computation). 
This delay could be a significant fraction of the the period of the pulsar, what would ultimately introduce a phase shift.

\section{Another look to the better-fitted SEDs}

\subsection{A double fitting approach for the arguably-good SED fittings}

Using the fitting approach described above for all the pulsars signaled in
Figure   \ref{arguably-good}, we also note that the model is able to fit better both the X-ray and gamma-ray data.
The addition of only these two extra parameters in the model (the extra set of $x_0,N_0$) indeed promotes qualitative improvements in all these cases.

\begin{figure*}
\begin{center}
\includegraphics[width=0.34\textwidth]{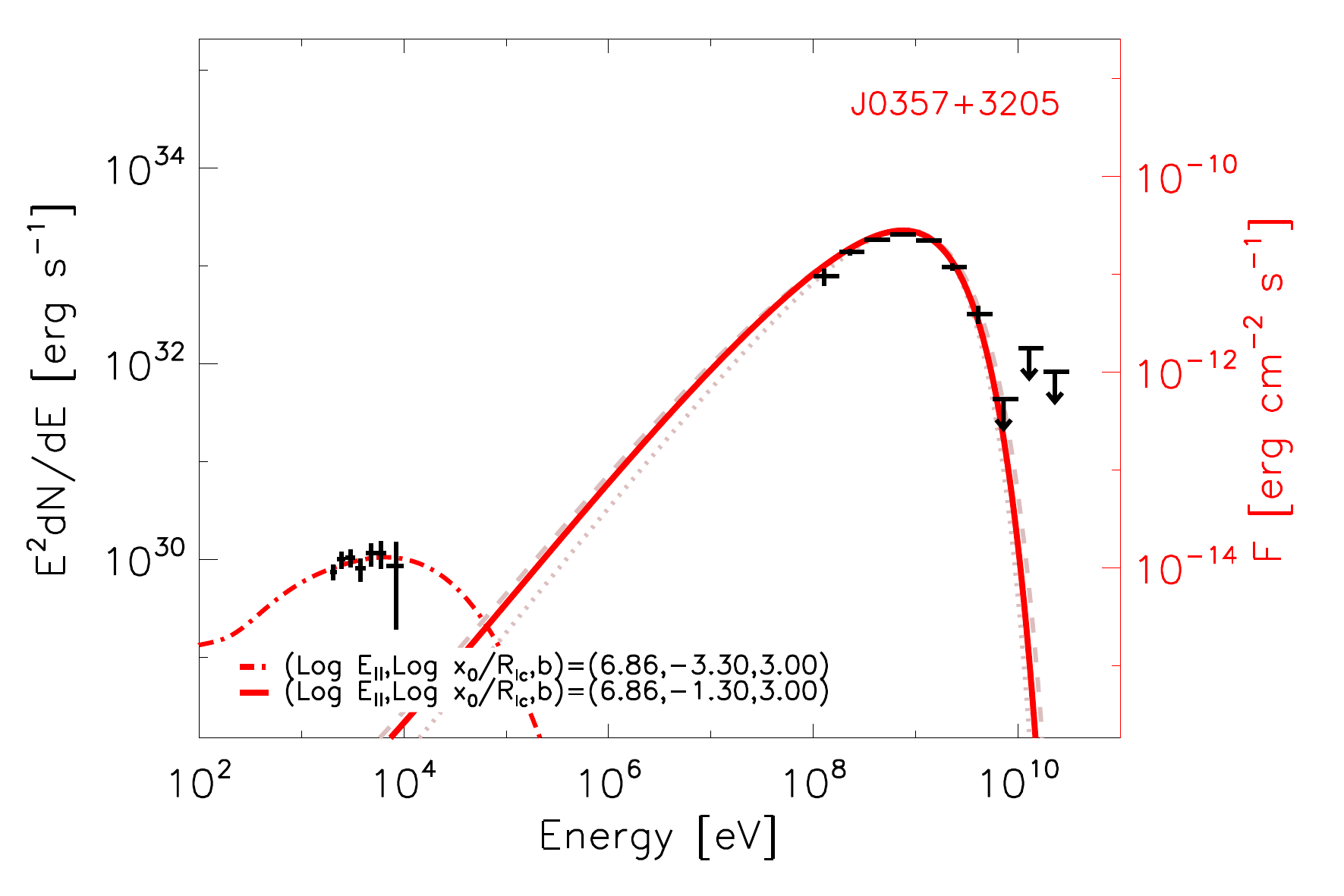}\hspace{-.25cm}
\includegraphics[width=0.34\textwidth]{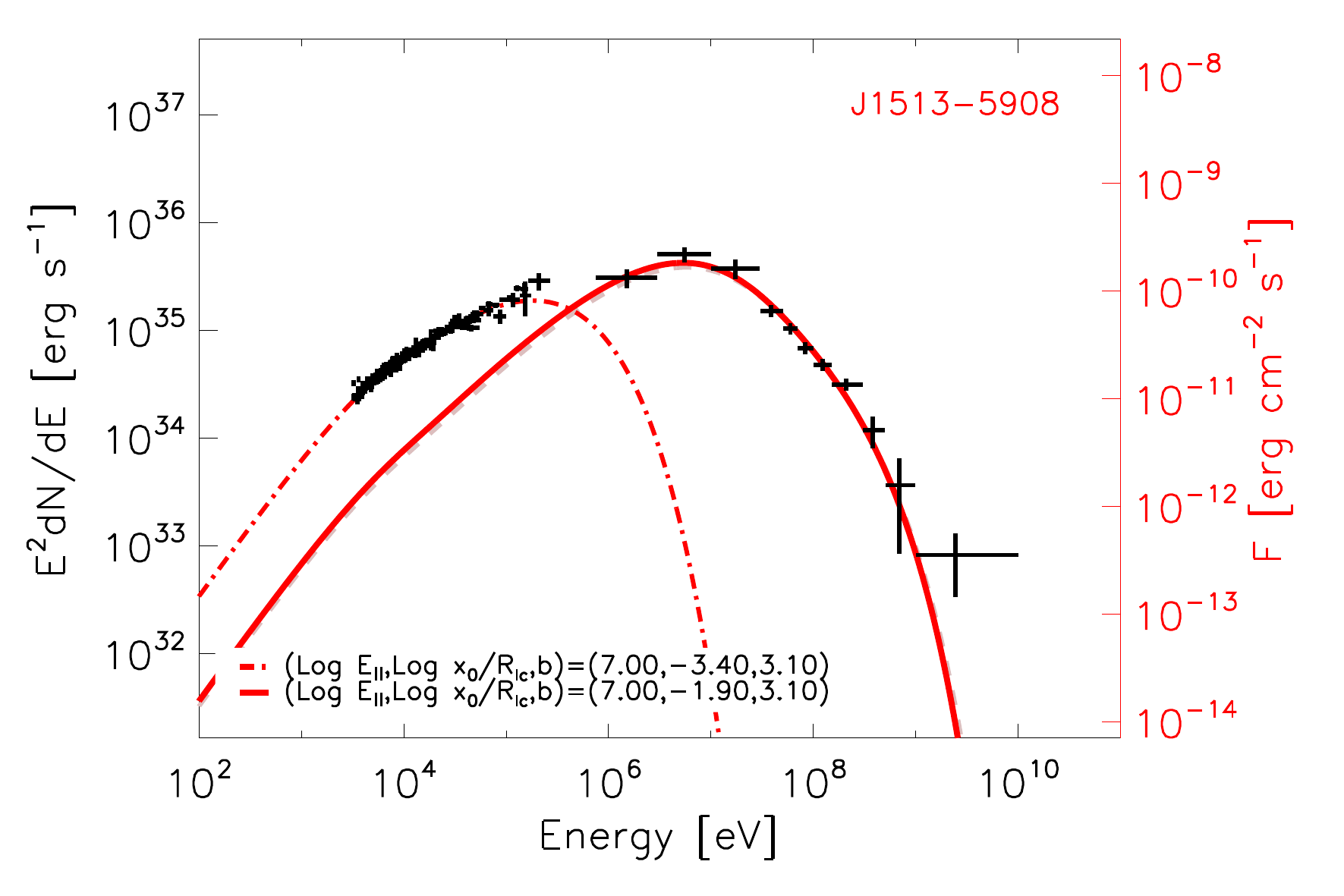}\hspace{-.25cm}
\includegraphics[width=0.34\textwidth]{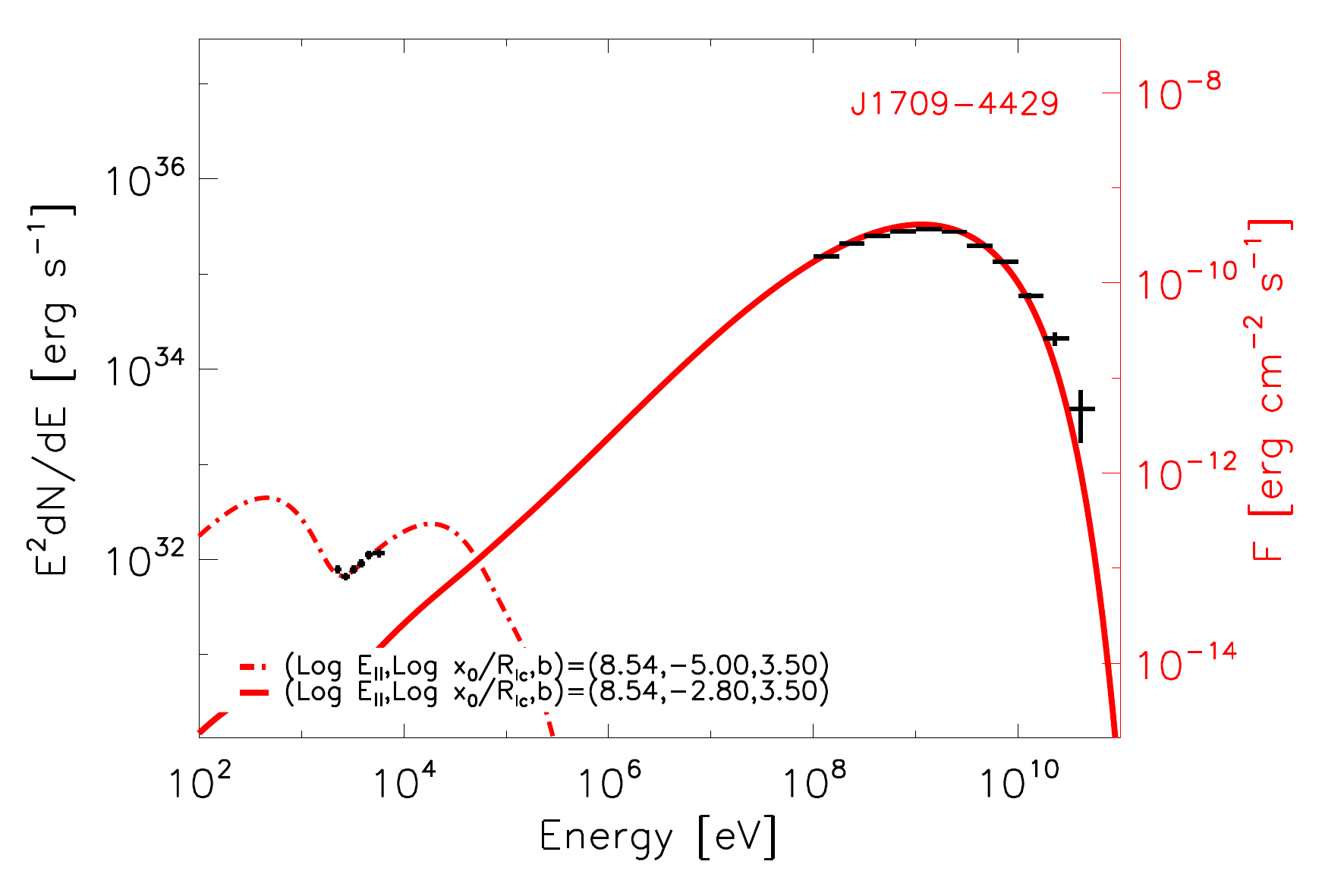}
\includegraphics[width=0.34\textwidth]{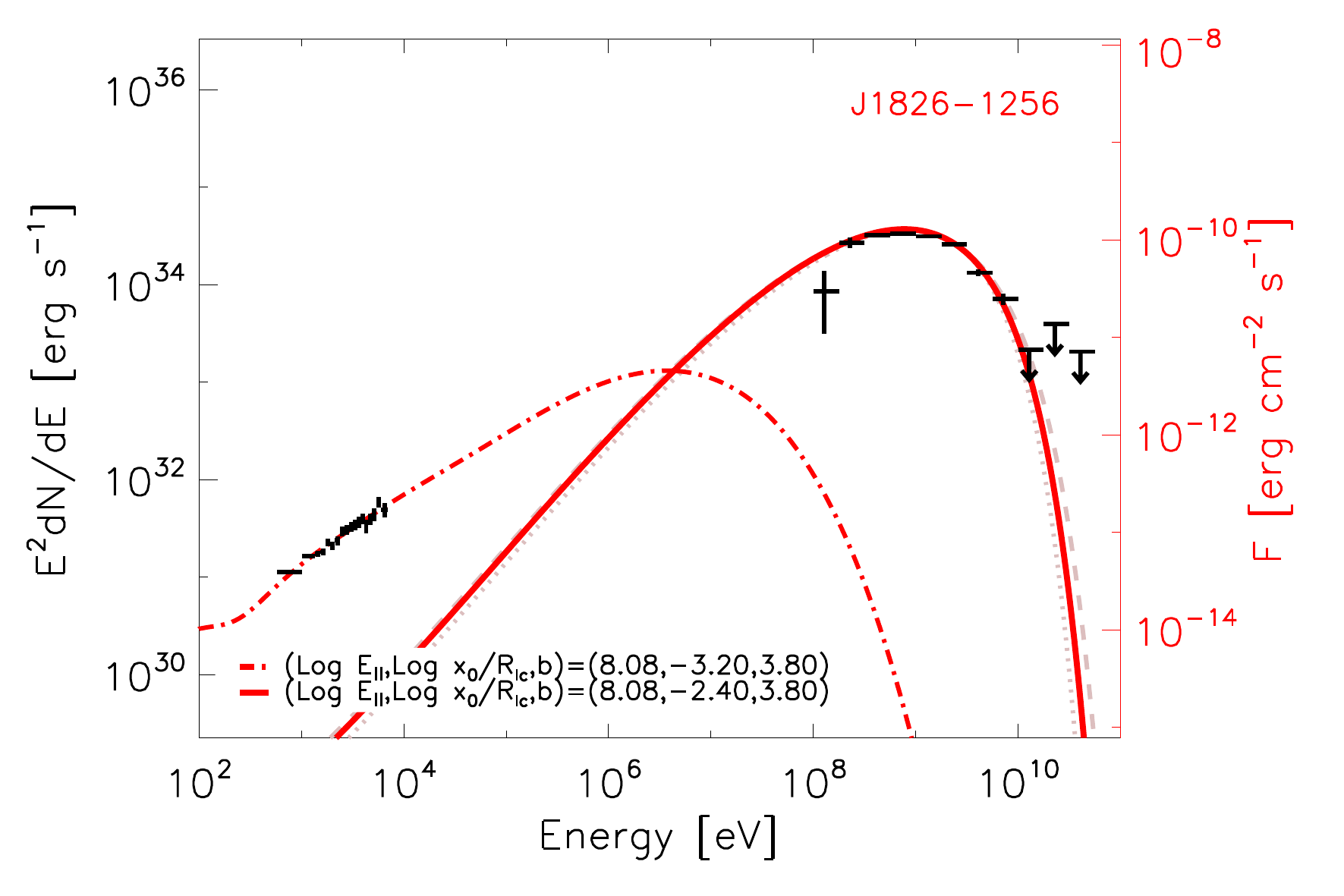}\hspace{-.25cm}
\includegraphics[width=0.34\textwidth]{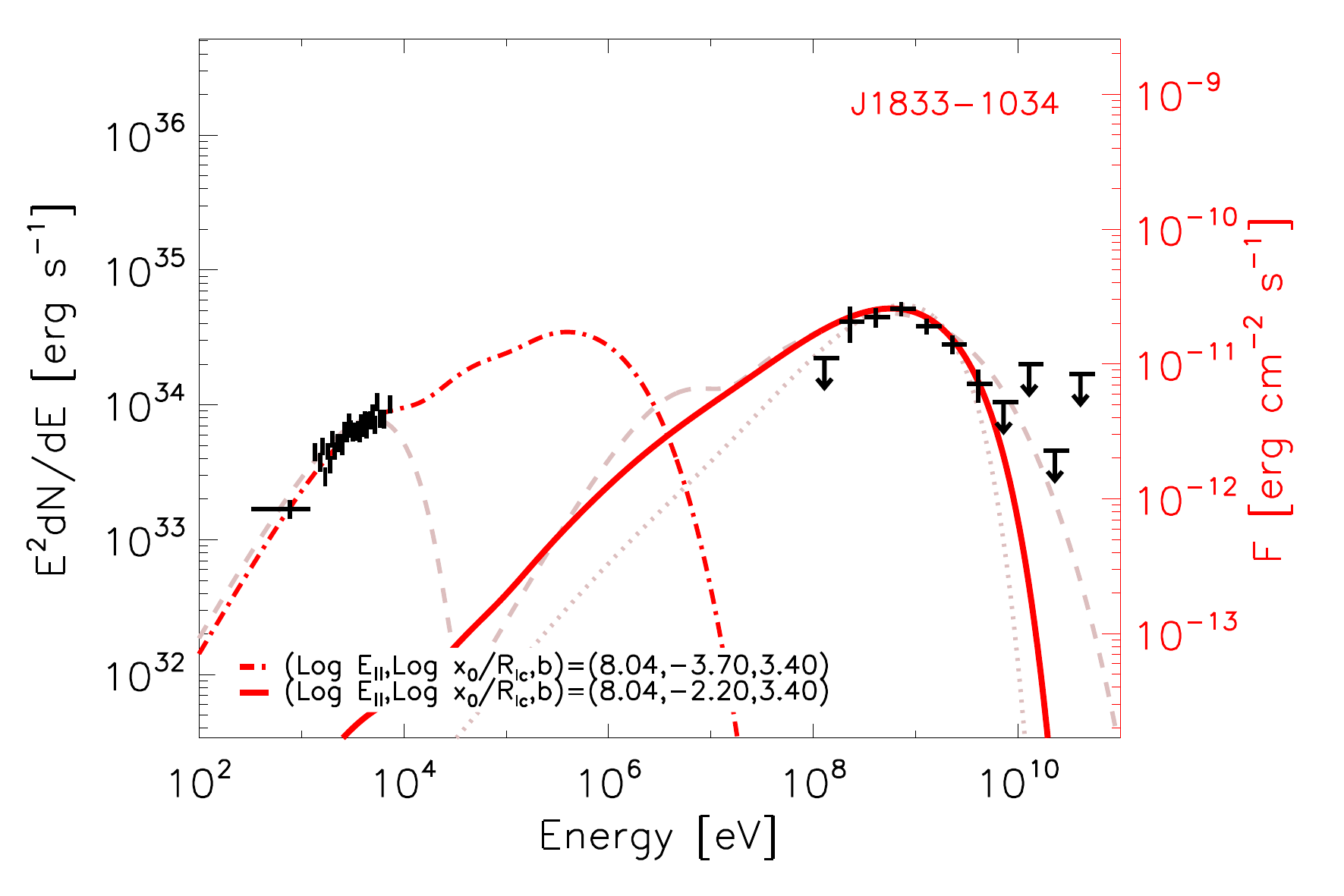}\hspace{-.25cm}
\includegraphics[width=0.34\textwidth]{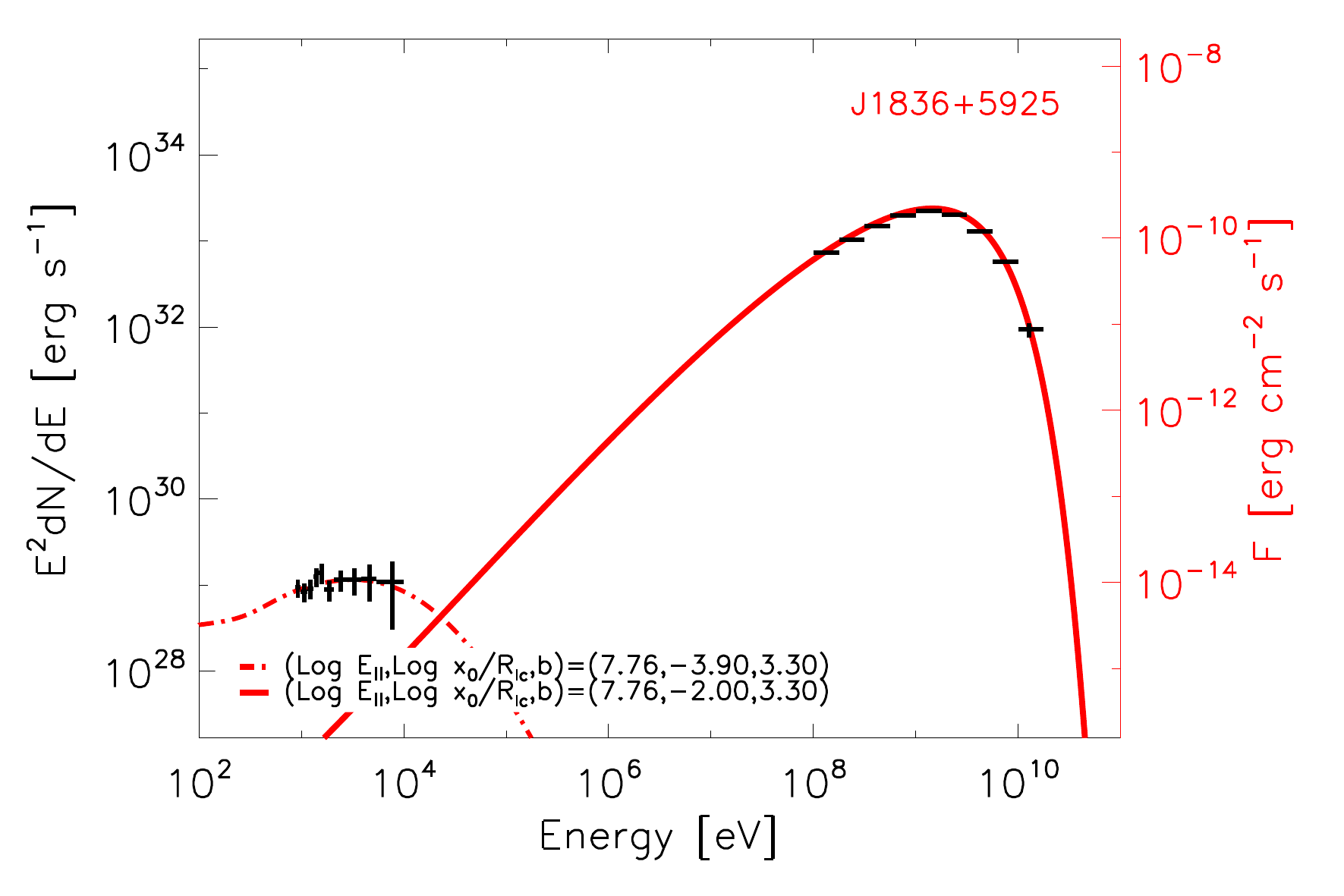}
\includegraphics[width=0.34\textwidth]{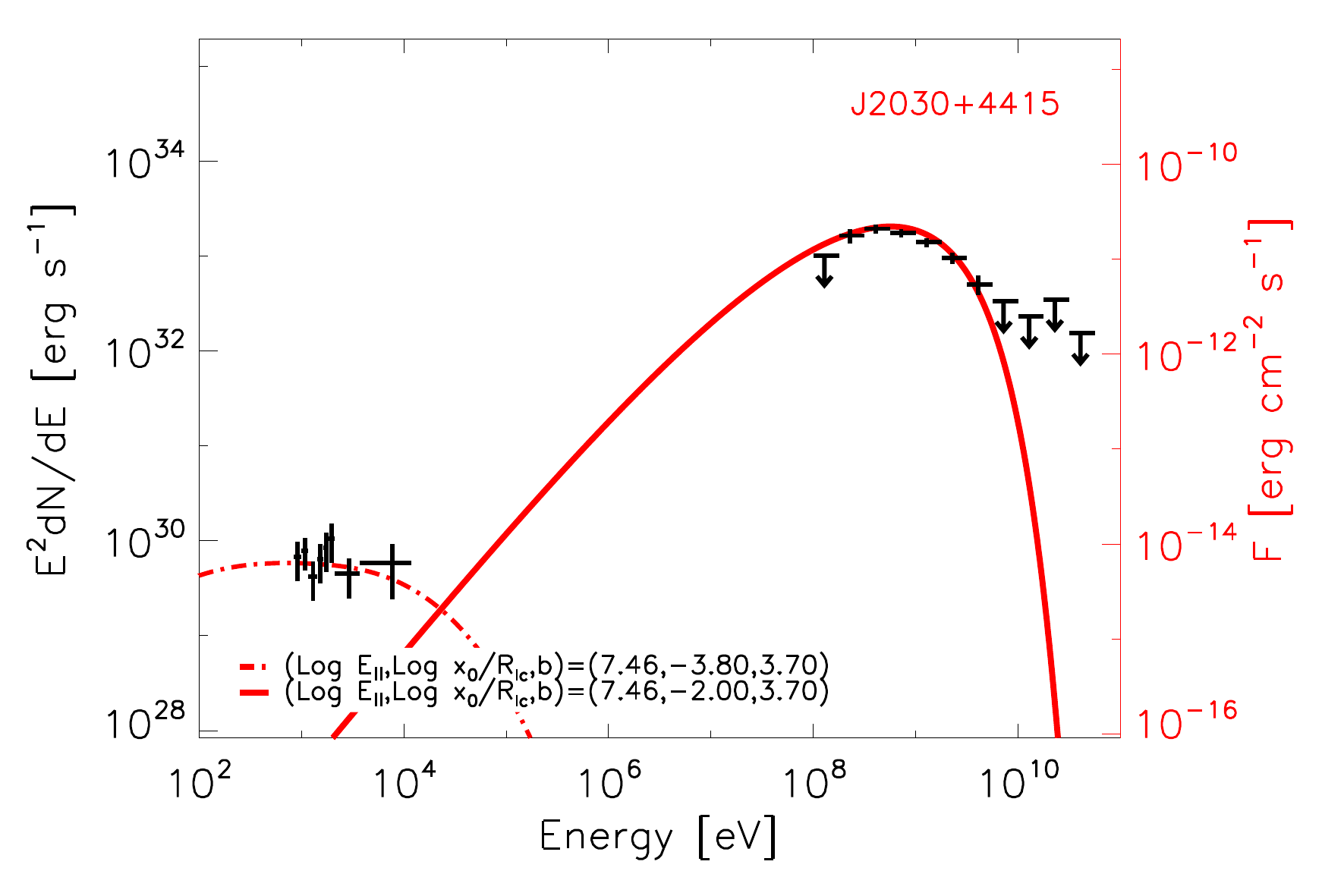}\hspace{-.25cm}
\includegraphics[width=0.34\textwidth]{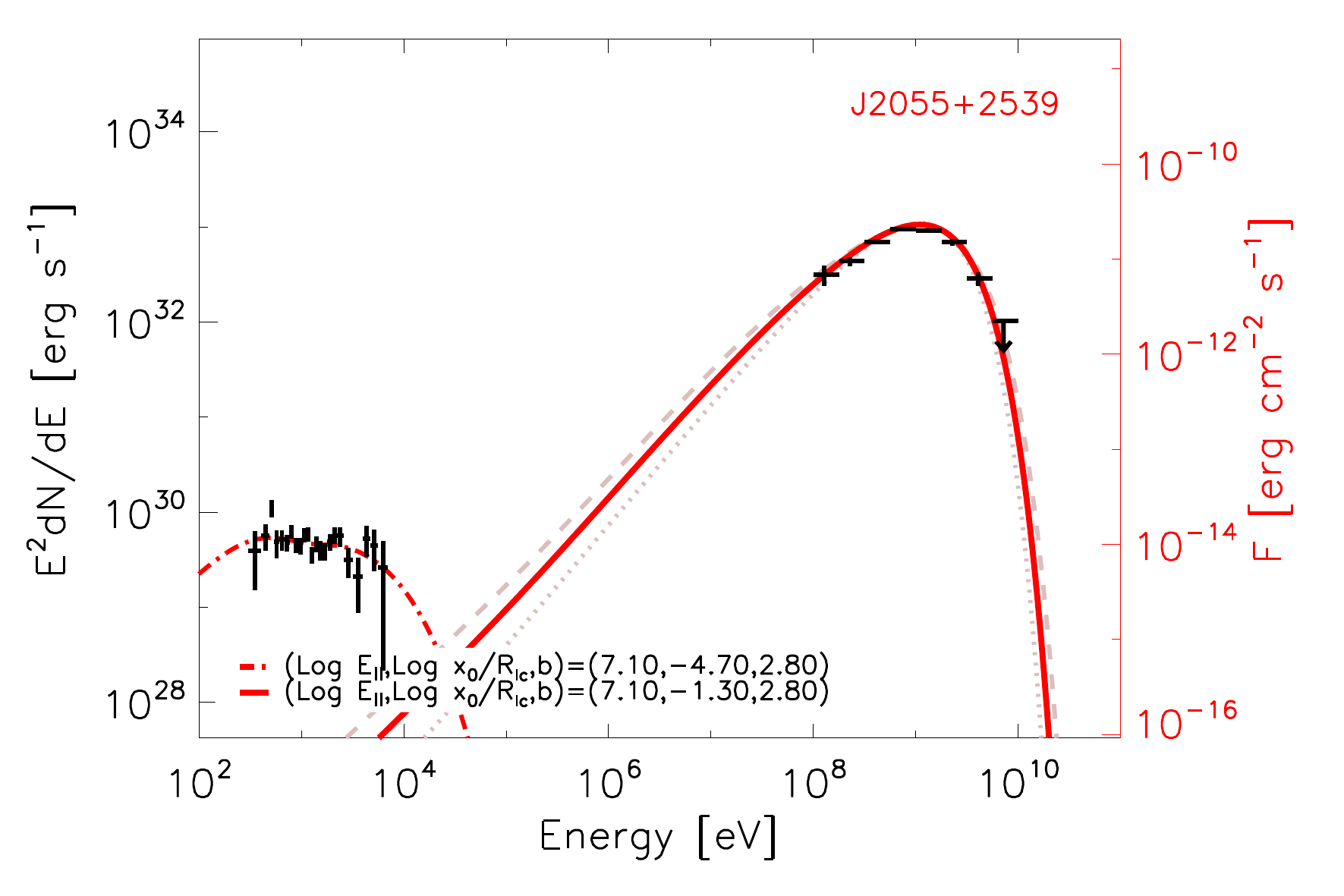}\hspace{-.25cm}
\caption{SEDs of the pulsars of Figure~\ref{arguably-good} analyzed under the perspective of a separate fitting to the X-ray and gamma-ray part of the spectrum. Also here, and as noted in the legend, the same value of the pair $(E_{||}, b$) can be used to fit both parts of the SEDs. See text for discussion.  } 
\label{arguablydouble}
\end{center}
\end{figure*}

The double fits are shown in Figure~\ref{arguablydouble}, and the parameters for these fits are stated in the second panel of Table \ref{uncommon-fits}.
A few comments may be in order to clarify the values there. 
Especially due to the energy coverage, shape, and error bars of the current X-ray data set,
in some of the cases, e.g., PSR J0357+3205, J1826-1256, or J2055+2539, neither the X-ray nor the gamma-ray data sets by themselves
promote a significant constraint on $b$, i.e., essentially all $b$-values explored are admissible within 1$\sigma$. 
This would of course change in case additional data are available. 
In such cases, then we fixed a fiducial value: $b=3.00$ for J0357+3205, $b=3.80$ for J1826-1256, and $b=2.80$ for J2055+2539 (that are selected as the best solutions)
respectively, to show that indeed the same possible solution of a double fitting can occur.
This is why there are no errors in the $b$ values of the X-ray fittings in Table \ref{uncommon-fits} for these cases.
A hard X-ray observation in these (and essentially in all) cases would be extremely relevant to fix the model and the physics it contains better.
These cases are similar to that of J1741-2054 treated before (see Table \ref{uncommon-fits}), where the admissible range of $b$ was also large, and we assumed the best-fitting solution 
to gather the $b$-value used in the gamma-ray fitting.
Finally, in the case of PSR J2030+4415, the gamma-ray fitting only provides a lower limit on $E_{||}$
which is what we use to represent the common solution used in both X and gamma-ray data.

The spectral formation in both components is similar in all of these cases to that exemplified in Figure~\ref{formation2}.

\subsection{Is there degeneracy in the fitting of well-determined SEDs?}

In principle, if for a given SED,  a single set of parameters ($x_0,N_0$) does fit qualitatively well, this is better (less parameters, more elegant solution) 
than having to use two sets.
But can one have e.g., two accelerating regions, with different $x_0/R_{lc}$ values 
and still have good (and different) spectral fits in the cases treated in Figure~\ref{sed1}?

We studied these issues with the best measured cases, among them PSR J0007+7303, Vela (J0835-4510), Geminga (0633+1746), and J1813-1246.
We found that in all of these cases, the iterative loop for fitting used above for the less adequate single-set cases  (see Figure~\ref{loop}) leads
to a solution that is compatible within 1$\sigma$ to the common X-ray and gamma-ray fit of Figure~\ref{sed1}.
For instance, in the case of J0007+7303, the X-ray fit using the $E_{||}$ value obtained from a gamma-ray fit would put no constraint on $b$, providing equally good solutions in the range explored.
For one of these values of $b$, then, we would reproduce the common X-ray / gamma-ray fit.
In the case of Geminga, fixing the $E_{||}$ value as obtained from the gamma-ray only fit, the best solution for the X-ray data already chooses $b=2.5$, and thus the iteration 
would exactly reproduce the common X-ray / gamma-ray fit.
Vela and J1813-1246 are cases for which one can argue a priori that a separate X-ray and gamma-ray fit may improve the global appearance.
For instance, for Vela 
the single set model correctly reproduces the general trend and the flatness of a well-determined low energy X-ray spectrum, but misses 
the lowest energy X-ray data point and seems to under-predict the rising trend in the hard X-ray domain.
However, a separate fit to the two components does no better, with the loop also converging to the same (single) solution.
For J1813-1246, one can also argue that the model is missing the lowest energy gamma-ray point (but see the comment above regarding the caution needed to reject a model based on this point only).
In addition, for J1813-1246 the difference between the gamma-ray only fit (using the fixed value of $b=2.85$) and the common X-ray / gamma-ray fit is relatively large (providing electric fields from $E_{||}$=8.20 to 9.22), which may promote hopes that a separate fitting would find a better alternative solution.
We tried the separate fitting starting from both values of $E_{||}$.
However, the iterative loop would either converge to a solution where the common X-ray / gamma-ray fit would be admissible within the uncertainty (for the larger value of $E_{||}$) or do not find a good fit for both parts of the spectrum at once.
This same situation also happens in other cases with a lesser quality of observational data, like J1357-6429.
For J1813-1246, we have also tried to fit the gamma-ray data only ranging on the $b$ value, to find that the best parameter is already close (or the same within errors) to that in the common X-ray and gamma-ray fit,
thus providing a very good representation of the X-ray data already (although this fit is not optimized since the X-ray data are not taken into account  and thus the fit is not preferred in comparison o the common one).

These detailed fitting studies led us to conclude that there are no obvious separate fitting solutions to the best-determined cases in Figure~\ref{sed1} that qualitatively improve 
those shown there, minimizing the possible degeneracy issues in these cases.

\section{Crab and the Crab twin}

The Crab and the Crab twin present a relatively flat spectral energy  distribution across several orders of magnitude in energy in the X-ray regime, at a level comparable to the gamma-ray yield.
This is a distinctive feature.
As Figure 2 of \cite{CotiZelati2019} shows, most other pulsars present SEDs that are either clearly rising or clearly decreasing with energy.
Those few cases for which the X-ray part of the SED is flatter, present a much lower yield there in comparison with gamma-rays, except perhaps for a few MSPs/
For Crab, the X-ray yield actually exceeds that in gamma-rays.

In our model, we find that it is impossible that a single set of parameters $(E_{||}, b, x_0, N_0$) produce a 
qualitatively good fit across the full energy range for Crab or the Crab Twin, similar to the other cases of Figure~\ref{no-good-fits}.

Given the quality of the data set we profit Crab as a testbed to see whether variations in the parameters appearing in Eq. \ref{eq:sed_x} but
kept fixed in our modelling would be of any help in providing a common fit.
In short, the answer is no, as expected. Details about some of these tests are given next.

\subsection{Crab}

The data we use here, apart from the gamma-ray measurements from \cite{2fpc}, comes from 
the compilation of \cite{Buhler2014}, the MeV data from  \cite{Kuiper2001}, and the 
 infrared (IR) to ultraviolet (UV) measurements reported in 
\cite{Sollerman2000} and \cite{Tziamtzis2009}.
For the TeV data see \cite{Aliu2008,Aleksic2011,Aliu2011,Ansoldi2016}.
This is the most complete sample for the phase averaged spectrum
of a pulsar, and fitting them in a model has proven difficult. 

\begin{figure*}
\begin{center}
\includegraphics[width=0.34\textwidth]{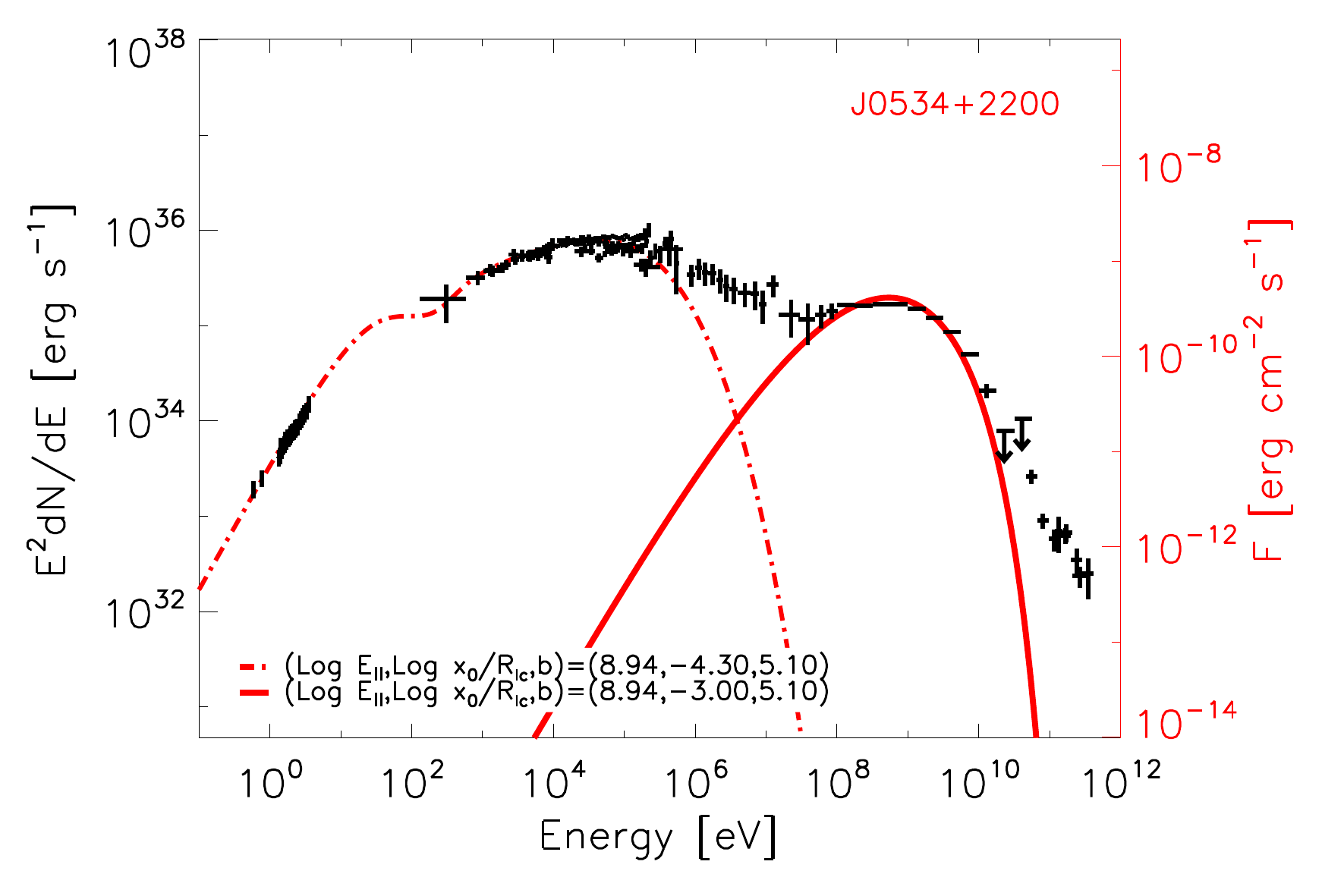}
\includegraphics[width=0.34\textwidth]{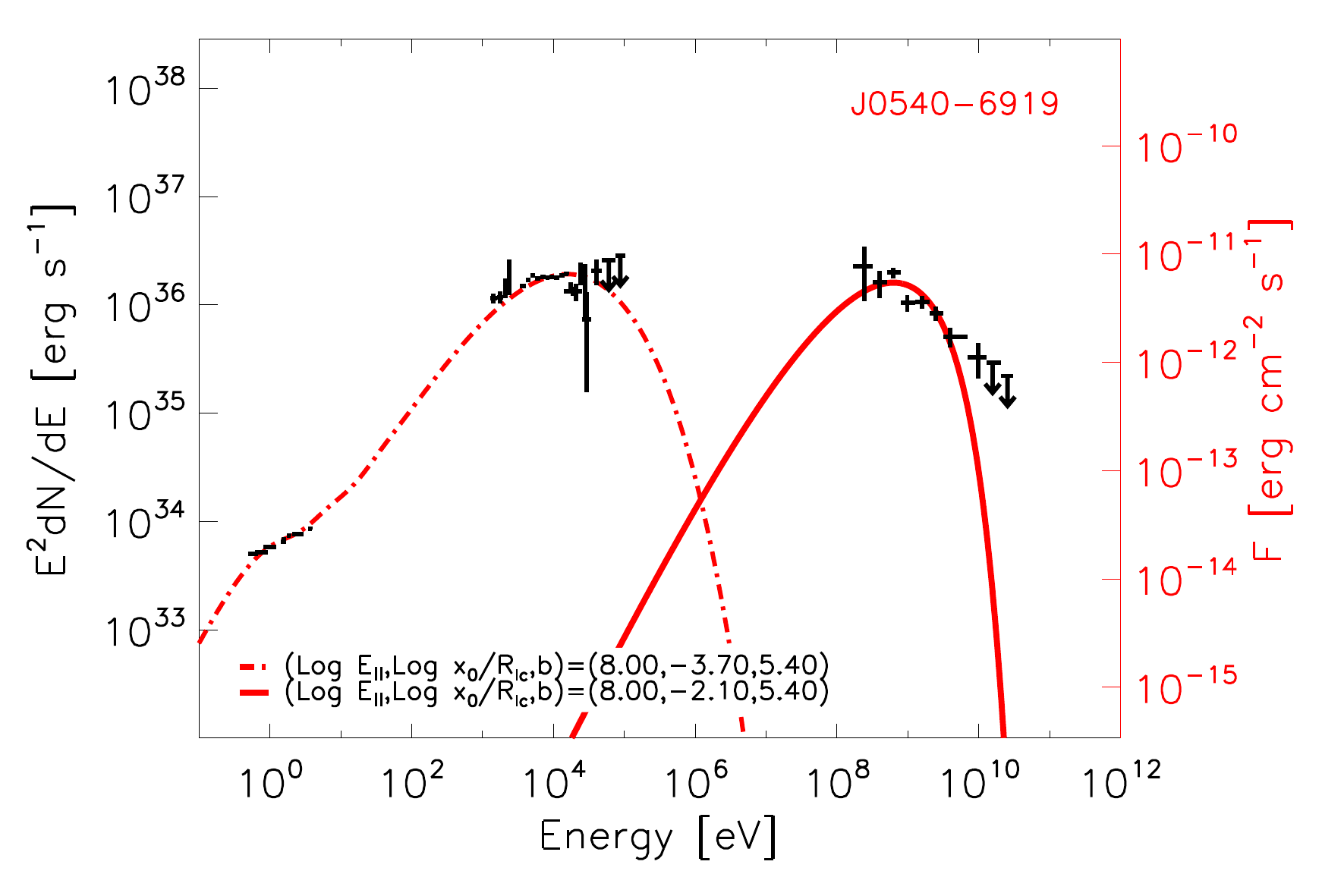}
\includegraphics[width=0.34\textwidth]{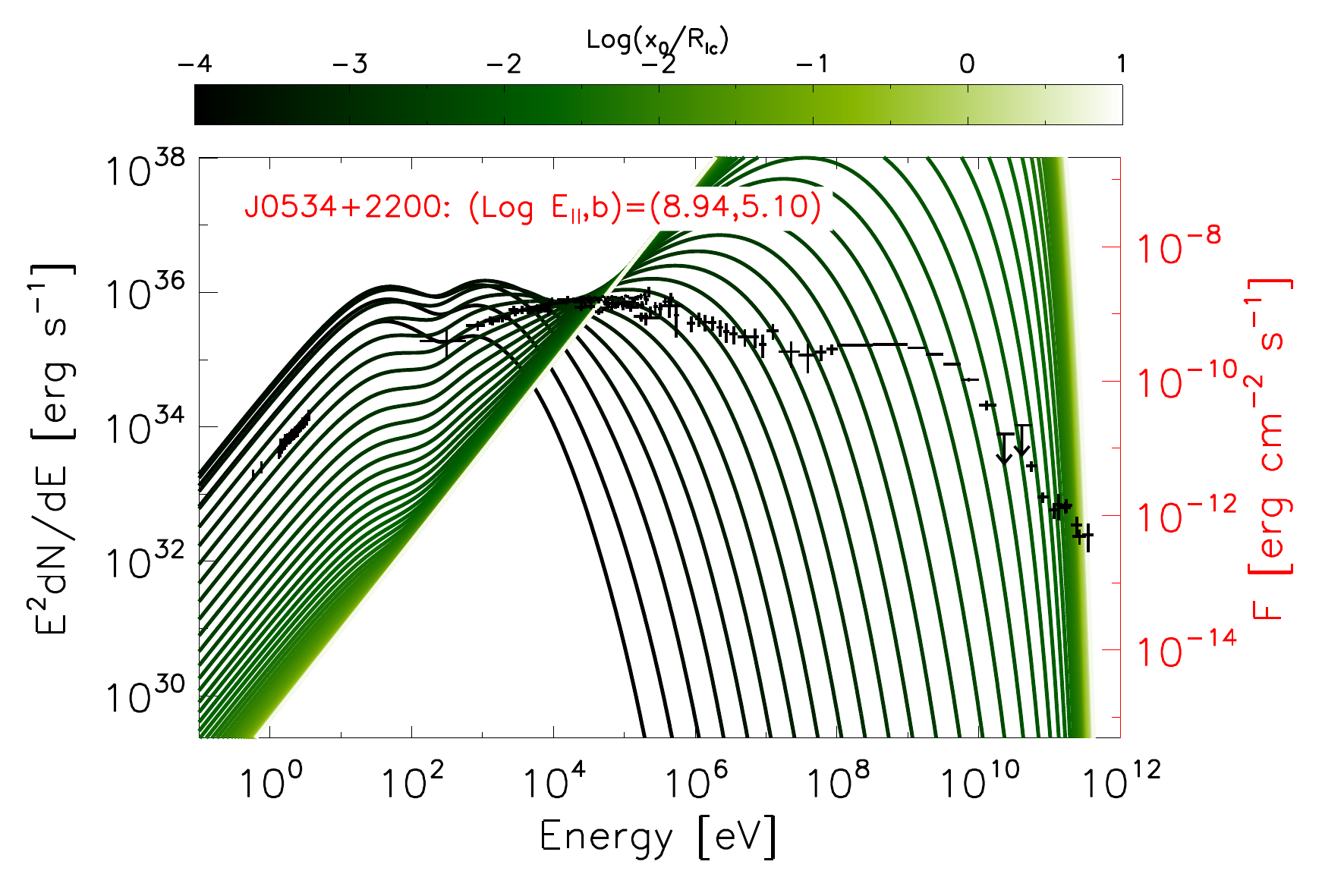}
\includegraphics[width=0.34\textwidth]{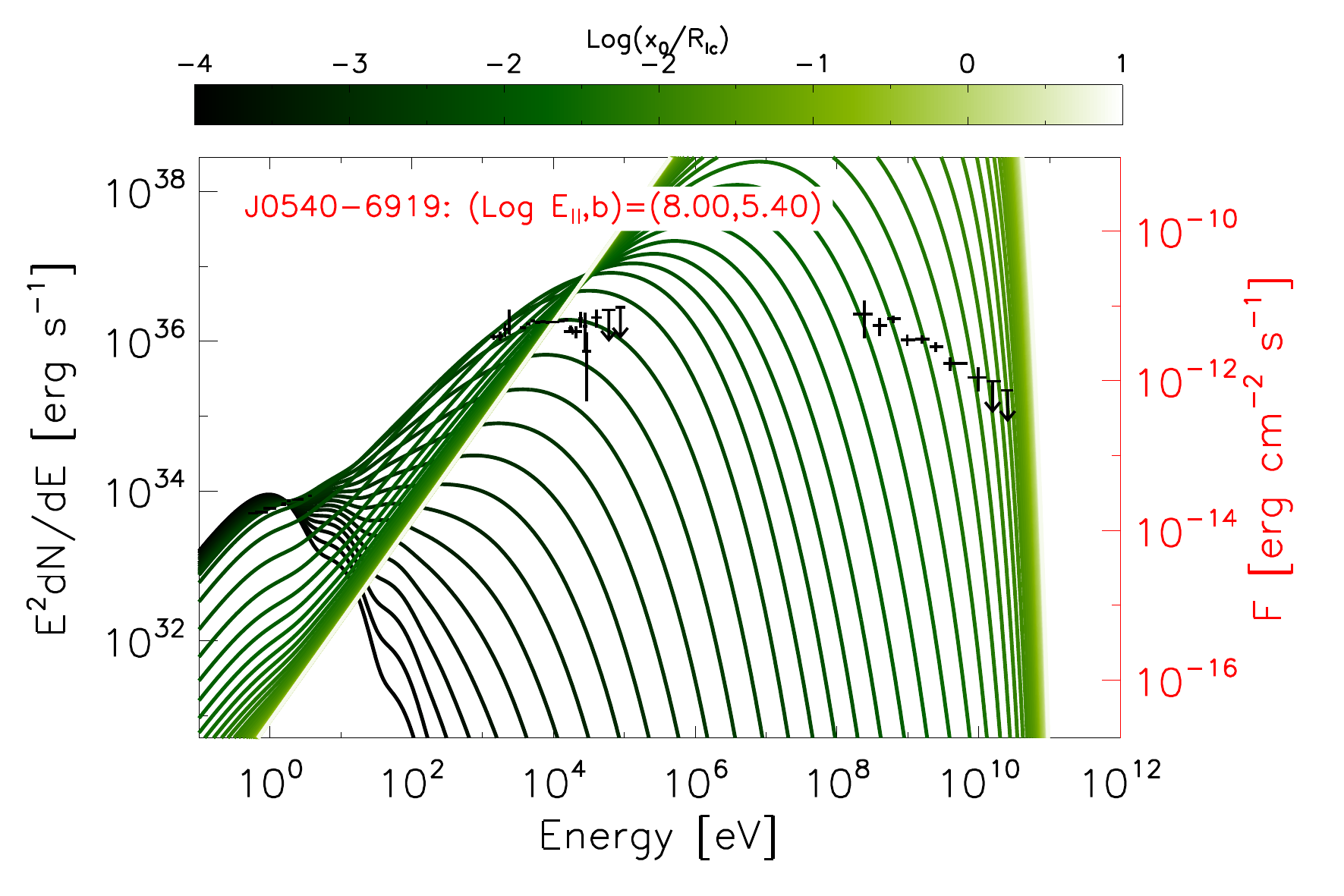}
\includegraphics[width=0.34\textwidth]{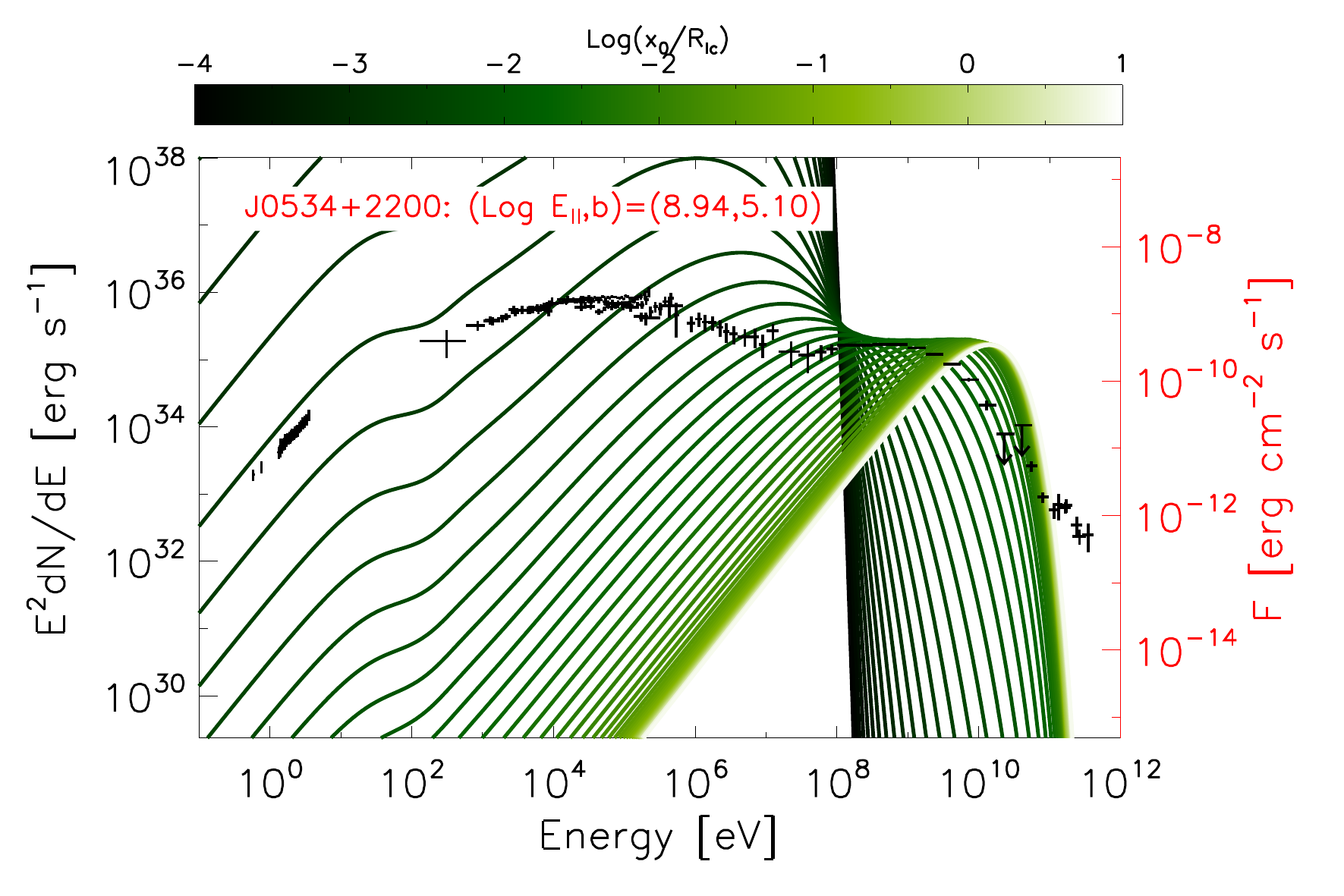}
\includegraphics[width=0.34\textwidth]{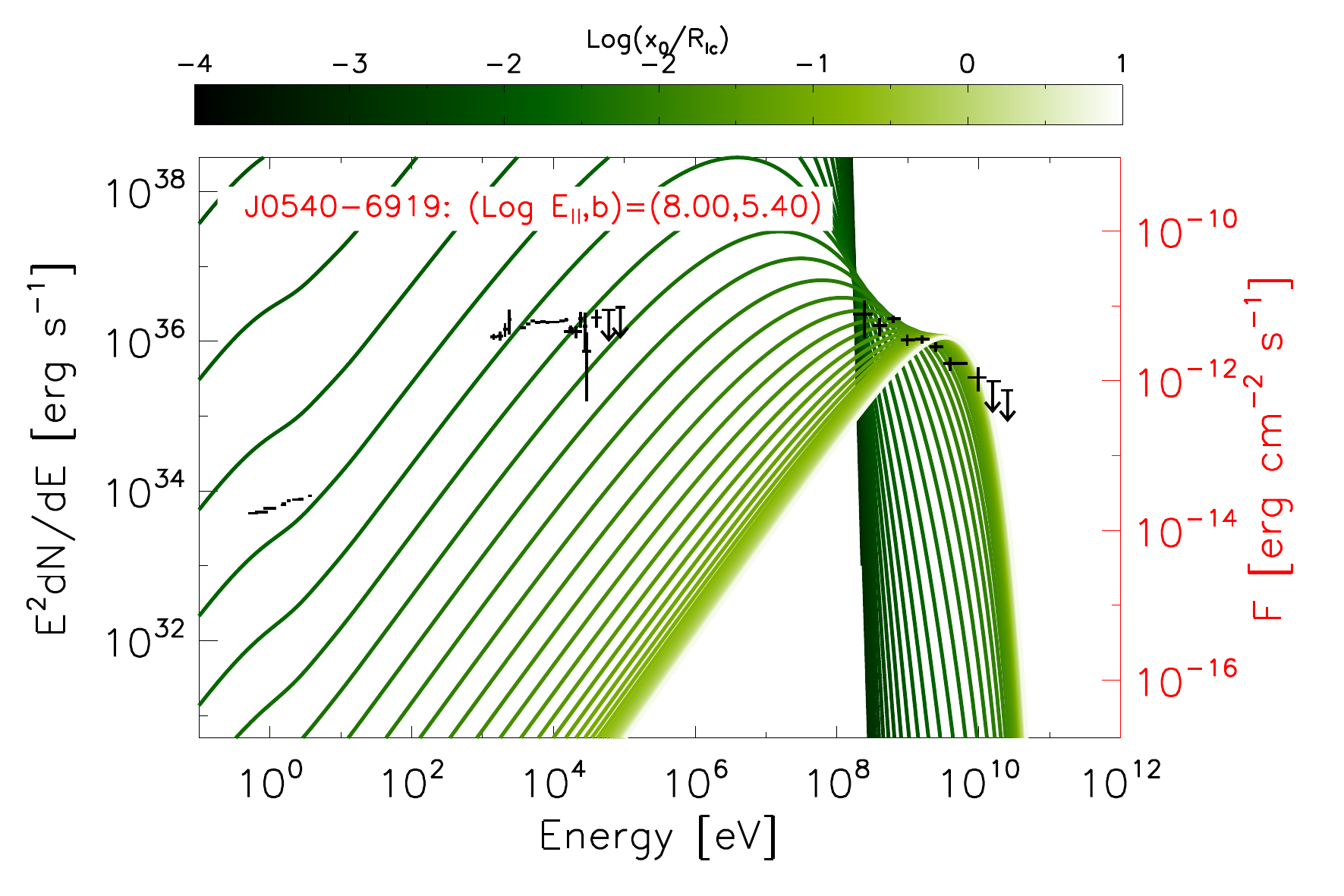}
\includegraphics[width=0.34\textwidth]{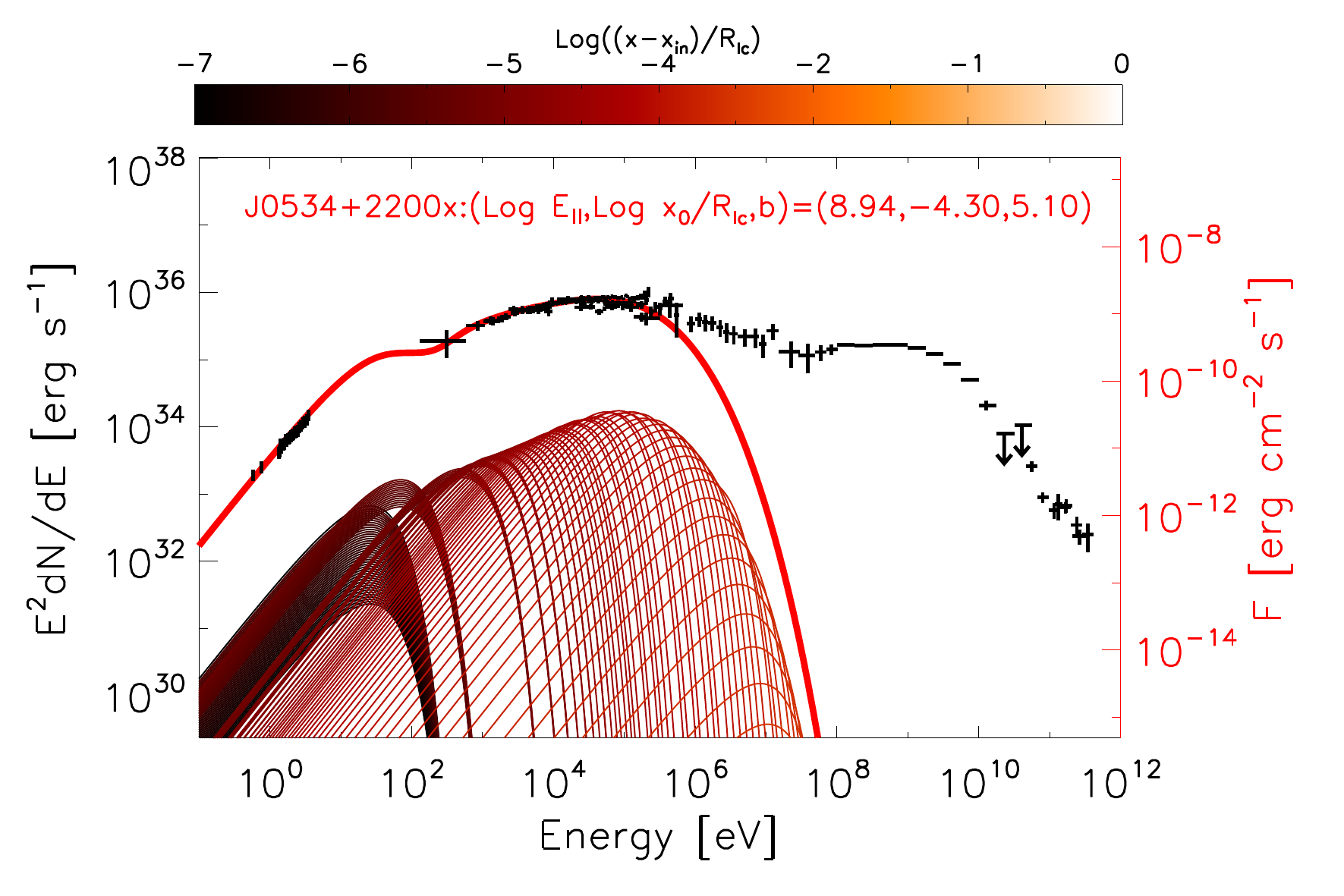}
\includegraphics[width=0.34\textwidth]{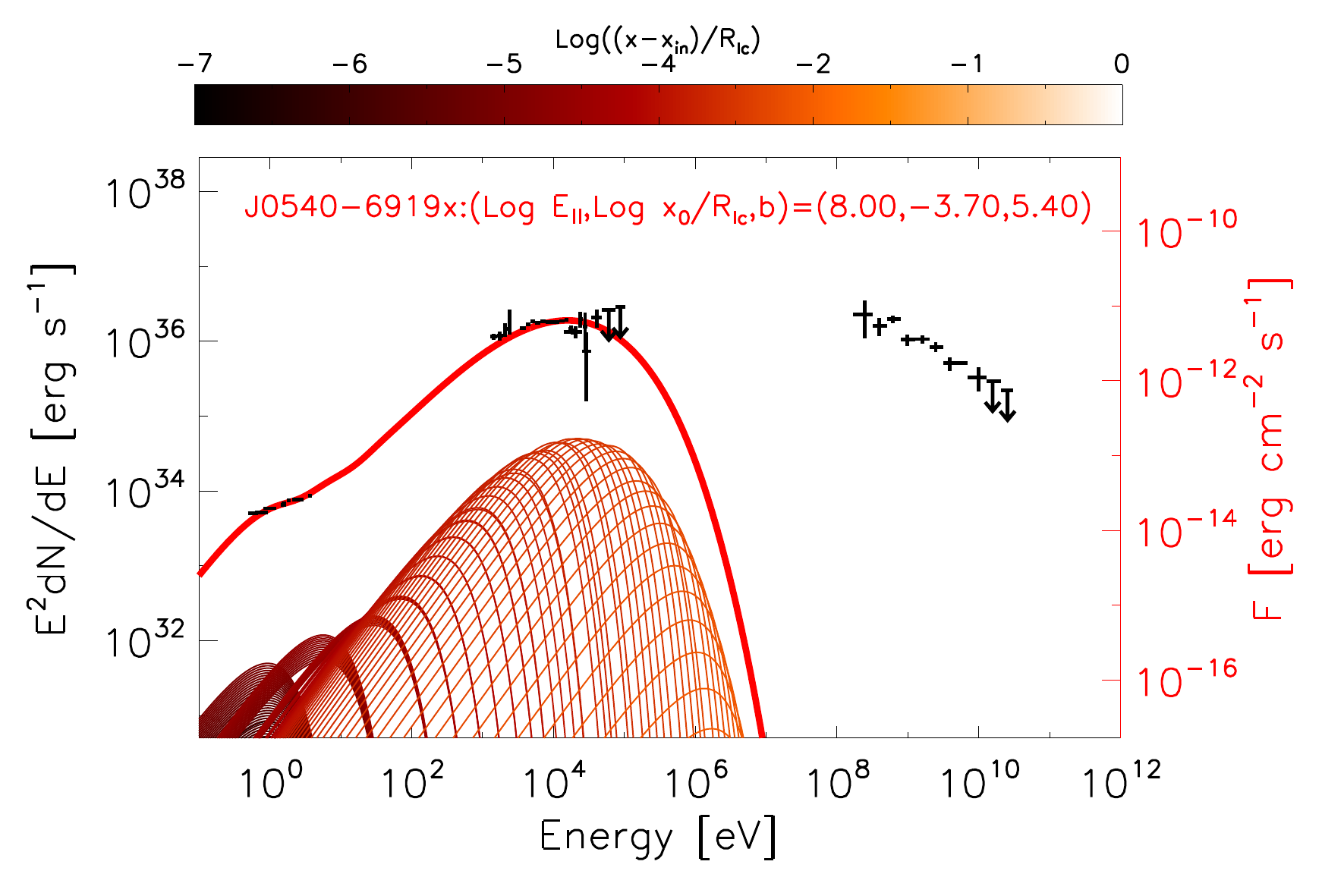}
\includegraphics[width=0.34\textwidth]{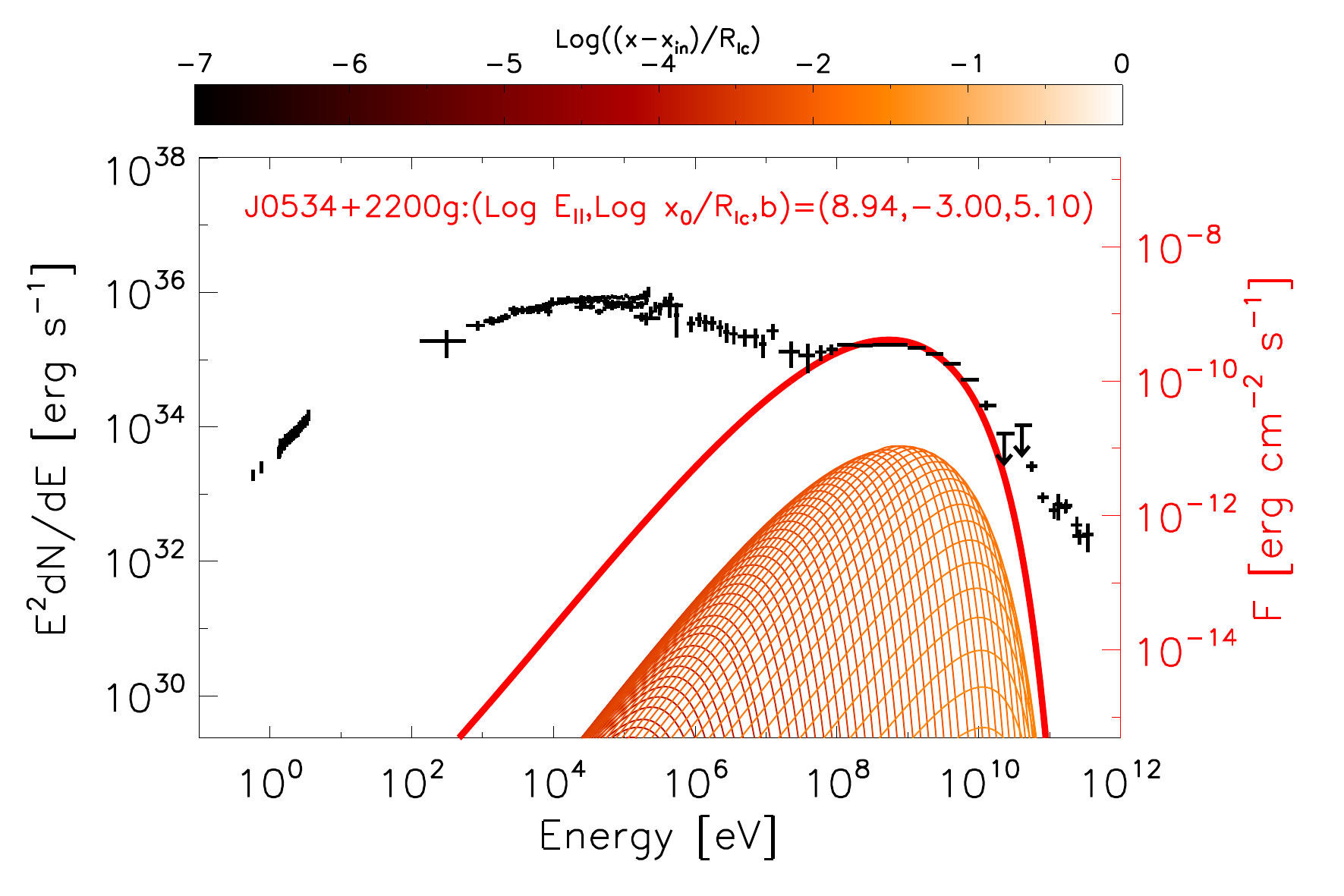}
\includegraphics[width=0.34\textwidth]{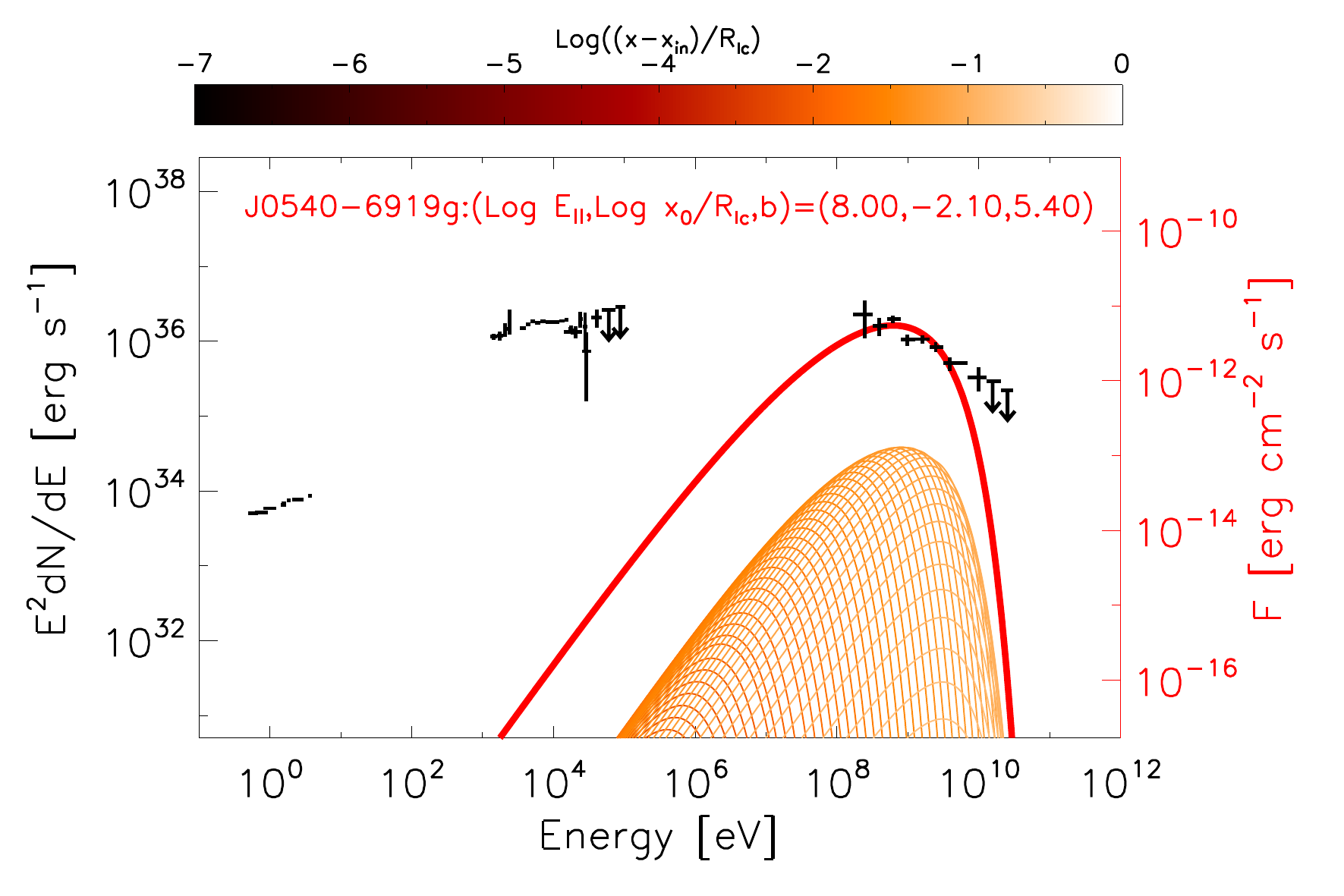}
\caption{Analysis of the model results for Crab (left panels) and the Crab Twin (right panels), located in the Large Magellanic Cloud. See text for discussion.  } 
\label{crab}
\end{center}
\end{figure*}

The best-fit SED coming from our model is shown in Figure~\ref{crab}. The values of the best-fit parameters are given in the last panel of Table \ref{uncommon-fits}.
We fitted the {\it Fermi} gamma-ray data at high energies (and got a qualitatively good fit also for the lower gamma-ray dataset as measured by EGRET) and the optical and keV RXTE data at lower energies.
Especially the lower energy fit is qualitatively good, although we find that the flatness of the spectrum in the X-ray regime requires a relatively large value of $b$, close to those found above for MSPs.
In comparison, an individual (not iterative) fitting of the gamma-ray spectrum would rather prefer a lower value of $b$, although such values cannot deal at all 
with the X-ray data. 
Instead, a larger value of $b$ can provide a good match to the X-ray / optical data set, and still be reasonably close to the high energy portion of the spectrum.

The best solution found under the same scheme  used for the rest of the pulsars in this paper is shown in the left panels of Figure~\ref{crab}.
There, we show the best-fit together with the influence of the contrast for each portion of the solution (the first three panels). 
The solution found in the first (top, left) panel comes from the best fitting components of the green plots representing the contrast influence.
Note that that difference between the curves plotted in the second and third panels is just in the normalization, the corresponding (color-coded) curves have the same $(E_{||}, b)$ and also $x_0/R_{lc}$ values.
The normalization, instead, is automatically chosen for each of these curves in Figure~\ref{crab} in order to provide the best fit --for that particular set of parameters-- to either the X-ray (second from top) or the gamma-ray (third from top) emission. 
Also here, the latter plots are useful to understand why a single solution cannot fit both sets of data (this is similar to the previously studied cases).
Finally, the last two left column plots show the spectral contributions to the spectrum of the low (mostly synchrotron) and high energy (mostly synchro-curvature to curvature) radiation.

\begin{figure*}
\begin{center}
\includegraphics[width=0.34\textwidth]{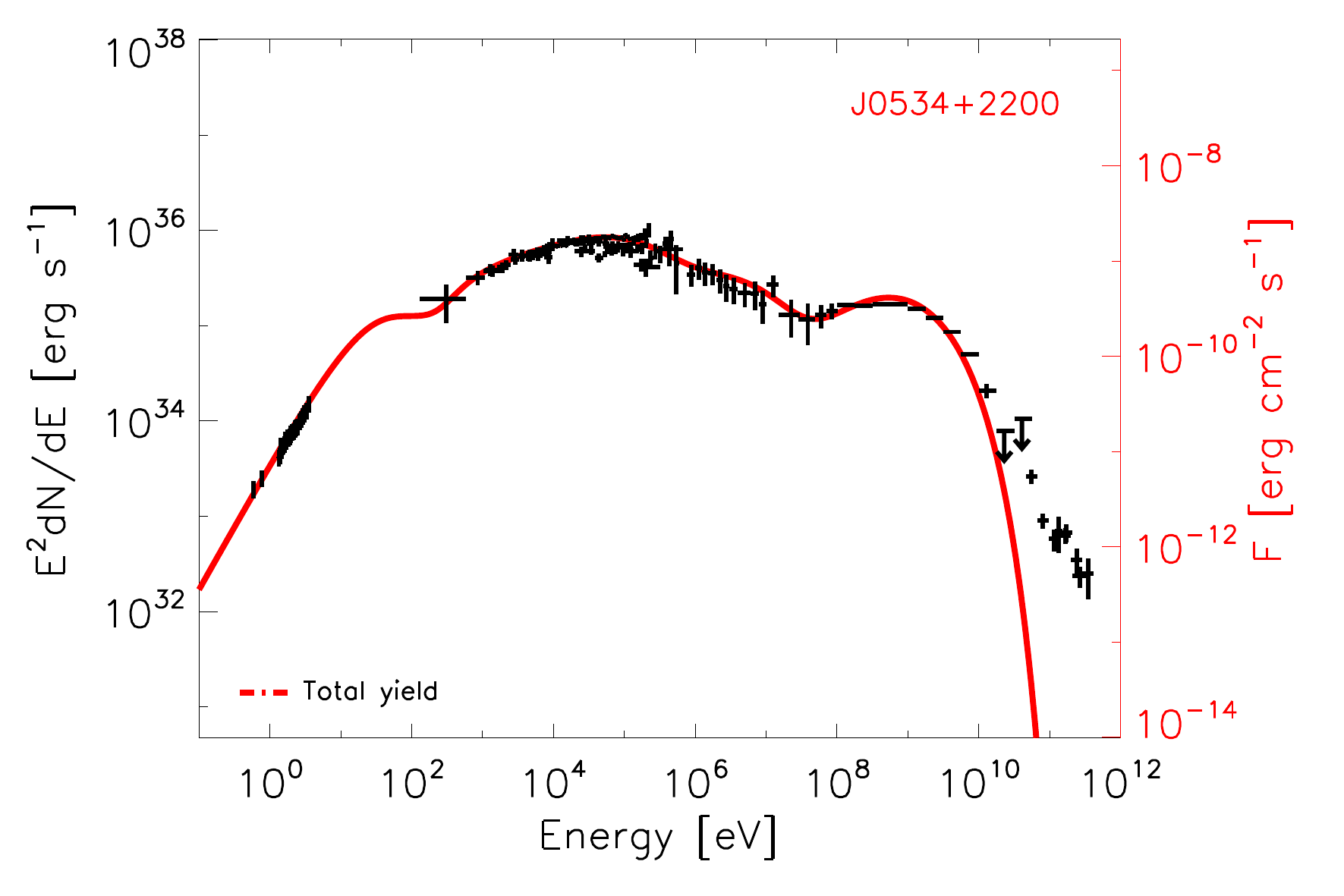}
\caption{Crab fitting: An additional set of $(x_0,N_0)$ is here assumed (for which the values are given in the text) to emphasize that the representation of the 
particle population might simply be what is behind the MeV underprediction shown in Figure~\ref{crab}. 
For a discussion of the TeV data, see text.
} 
\label{crab2}
\end{center}
\end{figure*}

We note that the MeV data is underpredicted in this model (summing the two components there is a underprediction in the MeV range).
As the Crab light curve is correlated at different energy regimes, the interpretation described in the right panel of Figure~\ref{concept} would justify the use of two sets of $(x_0,N_0)$ as an approximation
to the real population of particles emitting radiation towards us.
Then, we pose that the failure to reproduce well the MeV data may just indicate that this representation is still  not enough for a good accounting of the particle population.
Obviously, the transition from one set of geometrically-related parameters to the other should be a continuous process. 
This may become more visible for nearby pulsar with a large rotational power, which generate more particles flowing in the magnetosphere, becoming increasingly more difficult to represent
the particle distribution. 
It might be natural to think that 
to represent the particle population the approximation of 
two sets is also not enough. 
The quality of the data set for Crab would surface this.
A better representation of the particle distribution (still with the same $(E_{||}, b)$) may thus overcome this caveat. 
In fact, just to emphasize this possibility we notice that using the same  $(E_{||}, b)$ as given in Table \ref{uncommon-fits} to search for a fit 
to the MeV data with (i.e., with all parameters fixed except for $x_0$ --and $N_0$--) we find a match for a value 
a value of $(\log(x_0/R_{lc}),\log N_0)=(-3.80,36.65)$, intermediate between that needed for the low and high-energy parts of the SED. 
This contribution would naturally arise in a continuous representation of the distribution of particles emitting towards the observer, and the fit would not interfere with the prediction at hard X-rays, or X-rays, nor with that at higher gamma-ray energies, which would continue being produced as described. 
The total yield of such a model is represented in Figure 
\ref{crab2}.
Formally, it contains 7 free parameters, the two intrinsically related pulsar properties $(E_{||}, b)$, and three sets of $(x_0,N_0)$ used to represent the particle population --one of the $N_0$ can be considered fixed, since one can fit one of the SED portions only in its shape and normalize the rest referencing to it.
The  parameters used to represent an energy dependent $dN/dx$ in a piecewise energy-independent manner can probably be reduced if using a better representation of the particle distribution.
This remains to be studied, and represents a challenge. 
The increased number of parameters is to be put in context to those needed in other models: for instance, \cite{Lyutikov2013} uses 9 parameters to determine the particle population spectrum and its location to generate the Crab spectrum 
via a cyclotron self-Compton model. The latter model correctly describes the Crab's TeV emission (see next) 
but underpredicts the optical and IR yield.

We note that our Crab model underpredicts the TeV data. 
Could this be modelled similarly to the MeV part of the spectrum (i.e., using the same $(E_{||}, b)$ pair) just discussed?
We find it is indeed possible to represent the TeV component with a small population of particles emitting towards
us, but requiring a larger accelerating field, $\log E_{||} \gsim 9.84$.  
In this scheme, the TeV emission would be curvature, and it thus not ruled out that curvature can still be behind the largest photon energies if an enhancement of $E_{||}$ is possible.
However, such value of $E_{||}$ would in turn be inadequate for representing the X-ray to gamma-ray spectrum.
The need of a different $E_{||}$ cannot be encompassed in the interpretation posed above, where just two parameters define its intrinsic physics.
We tested whether this conclusion would be affected by a different assumption on $x_{in}$ or $x_{out}$, and the size of the accelerating region, finding that this is not the case.
Thus, these data could be produced somewhere else still via curvature with an enhanced $E_{||} $, or by a different process altogether, e.g. see \cite{Aharonian2012,Lyutikov2013,Osmanov2017,Harding2018}. 

Tests regarding the changing of $x_{in}$ or $x_{out}$, and the size of the accelerating region were also run to see whether the overall spectrum from optical to gamma-rays could be better fitted,
eventually by just a single set of parameters.
We find either these changes were innocuous or that they produced worse spectral representations.
Another tests we run regarded the possibility of having an evolution of $b$ across the accelerating region (from a high to a low value).
This is motivated by the fact that a lower value of $b$ assumed when fitting the gamma-ray data only would also produce a SED closer to the MeV data (the same problems with the underprediction of the TeV data would be maintained, though). 
But of course, the value of $b$ that is most influential in the SED is the one valid at the synchrotron-dominated part of the spectrum, and once this is fixed, little change is found admitting further variations along the trajectory. 

\subsection{Crab Twin}

For the Crab twin in the Large Magellanic Cloud only a less constraining data set is available: 
we use the gamma-ray data from \cite{Ackermann2015} reduced by 25\% assuming -as these authors say-- that about 25\% of the flux is contributed by a diffuse environment, 
as well as the compilation by \cite{kuiper15}, and optical to infrared (IR) data from \cite{Mignani2012b}. %
The fact that the optical flux density is characterized by a power-law spectrum and the luminosities 
correlate with the rotational power, like in Crab and other pulsars, prompted \cite{Mignani2012b} to suggest that this emission is non-thermal.
Given the not-fully understood jump in flux density just recently found in near ultra-violet (NUV), and under the assumption that it might 
be related with local absorption or an increased continuum contribution due to a non-magnetospheric component, or other effects, see Figure 3 in \cite{Mignani2019} for the data and the associated discussion regarding these issues, we choose not to include the two UV data points in our modelling.
We note in any case that only the NUV point is problematic (as it implies a jump of about 1 order of magnitude form adjacent data points in a $\nu f_\nu$ diagram, what makes it impossible for any model based on a single process with a continuous population of particles). 
Instead, the far UV data point in \cite{Mignani2019} would be in agreement with our model, despite we are not fitting against it.

It is important to note that the difference between the gamma-ray data sample of Crab and its twin is significant: the Crab data in gamma-rays have errors smaller than 1\%, whereas the errors for the Crab Twin 
range 10--50\%, in addition of the systematic uncertainty mentioned.
This will have an impact in the modelling. In fact, when the gamma-ray data set is modelled by itself only a broad range of 
admissible values of $E_{||}$ are obtained. 
Because of this, when searching for a separate fit using the same values of $(E_{||},b)$ we shall model first the optical to X-ray data (featuring extremely small error bars) 
and use the defined values for the pair to see whether a fit 
to the gamma-ray data is obtained. 
This is why the gamma-ray fit parameters described in the last panel of Table \ref{uncommon-fits} has no errors.
The results of the separate fitting with the same physical parameters of the accelerating region are shown in the right panels of Figure~\ref{crab}.
The similarities with the Crab fitting are striking, and visible along all plots. 
They are also similar regarding the fitting parameters.
Also a high value of $b$ is required for the Crab Twin fit.
These values of $b$ are the highest we found when fitting the pulsars with $P > 10$ ms in our sample (see Table 1), and promote the possibility that they could relate with the rotational power, or with the magnetic field at the light cylinder.
%

\section{Conclusions}

This paper presents a systematic theoretical analysis of the SED of non-thermally emitting pulsars for which the SED is sampled from X-rays (in some cases, also from IR/optical)
to gamma-rays. 
We used the model introduced by \cite{torres18}, which was found able to describe well the emission of several pulsars using only 3 physical parameters and 
a global (normalization) scale. 
These four parameters clearly distinguish themselves in two groups: the intrinsic ones (the accelerating electric field $E_{||}$, and the magnetic gradient $b$) related to the pulsar itself, 
and those related to the geometry of the pulsar with respect to the observer 
(the contrast $(x_0/R_{lc})^{-1}$, and the normalization $N_0$ for the number of particles flowing along the accelerating region that are emitting radiation towards us). 
Here, our theoretically scrutinised sample contains 36 of the 40 cases compiled by \cite{CotiZelati2019}, see also \cite{kuiper15} and references therein, and includes 32 normal pulsars and 4 MSPs having broad-band non-thermal SEDs.
We have only left out of our analysis a normal pulsar for which only a few data points could be derived in total, making fits less relevant  (J2043+2740),
and 3 MSPs (B1821-24, B1937+21, and J0218+4232) requiring additional {\it Fermi}-LAT observational analysis before fits can proceed.
For the latter, though, we have already mentioned  preliminary results.
This collection of pulsars has to be compared with the original reduced sample of 8 sources studies by
\cite{torres18}.

A comparison with such a large data set is a tour-de-force for any model, but particularly for one hosting such a reduced number of free variables.
However, we found that for  half of the pulsars herein considered (18 out of 36), the model 
is able to qualitatively describe well the observational data across seven orders of magnitude in energy.
This is explicitly shown in Figure~\ref{sed1}.
In addition, we have identified a group of 8 pulsars for which the model can still be argued to provide the basic features of the SED, although with caveats (see Figure~\ref{arguably-good});
and a different group of 10 pulsars (including Crab and the Crab twin) for which the model fails --in its original incarnation, i.e., with the 3 parameters and the normalization scale-- to correctly describe the SEDs (see Figure~\ref{no-good-fits}).
We studied how this inability happens, realizing of a simple extension of the model that can lead to good fits in all cases:
with the same pair of values $(E_{||}, b$) and two different pairs of values ($x_0/R_{lc}$, $N_0$),
the model can fit well all SEDs, both the X-ray and the gamma-ray data. 
This approach separates the model input from related to orientation and geometry from the properties intrinsic to the pulsar, as described above. 
We have provided 
two different interpretations (see Figure~\ref{concept} and Section 4) that explain why this extension is to be naturally expected.
It would be the most probable outcome if two regions of acceleration are visible to the observer and/or when the visibility of one accelerating region is favourable enough so that 
a more complex representation of the particle distribution is needed. 
Such model extension, then, does not change any of the model basic assumptions at all, and only adds two parameters.
Overall, it still makes for an austere model, particularly when the parameters being added are of the same kind and bears full resemblance with the ones previously existing.
The fact that with the inclusion of this extension the SEDs for all pulsars analyzed can be described with the utmost detail 
reinforces the conclusions obtained earlier regarding the utility of their free variables as order parameters in the SEDs complexity. 
As an interesting aside, our interpretation of the SEDs in these cases allows for a qualitative explanation of the X-ray to gamma-ray light curve phase shifts, 
what up to now have remained elusive.

We also remarked how well the model can act as a predictor for those pulsars that can be detectable in X-rays or lower energies, if knowing their high energy yield.
In all cases where a single set of parameters $(E_{||}, b, x_0/R_{lc}, N_0$) will be enough to describe the SED, the matching between the fit to the gamma-ray data only (assumed to be the only data set known) and the fit to the total X-ray and gamma-ray data is surprising.
This happens in essentially all the 18 such cases shown in Figure~\ref{sed1}.
However, we pointed the caveat that when a single set of parameters is not enough, the gamma-ray yield modelling may miss detectable pulsars, judging them as non-detectable.
Luckily, however, this does not act the other way around.
The predictive power may thus be incomplete, but if a gamma-ray fit predicts a detectable source, we find there is no risk in searching.
\cite{li18} have already shown this model can be a successful searching tool for pulsar detections, and more may yet to come with future surveys.
With a population of observed non-thermal X-ray pulsars that is currently >5 times smaller than that in gamma-rays, such predictive power is welcomed.

The analysis of this larger sample also confirms other previous findings. 
The appearance of sub-exponential cutoffs in the {\it Fermi}-LAT energy range is, as we advanced in \cite{torres18}, also a natural consequence of synchro-curvature-dominated losses.
We continue to find the great importance that synchrotron and synchro-curvature emission (i.e., the full process without approximations, whether neither synchrotron nor curvature dominates)
have for the spectral formation of all the pulsars analyzed. 
We can conclude that fully curvature-dominated emission plays a significant role only in a few cases, but not in most of the gamma-ray emitting pulsars. 
This is particularly visible in the explicit cases for which we show the pulsar SED formation (see Figure~\ref{formation}, \ref{formation2} and \ref{crab}).
Such Figures are consistent with the properties derived from the solution of the equations of motion (see e.g., Figure~\ref{traj1}). 
Regions of the trajectory between $\sim 10^{-5}$ and $\sim 10^{-2} R_{lc}$, where both synchrotron and curvature are relevant, are responsible for most of the observed emission. 
All in all, we continue to find that the relevant scale for the production of the pulsar's SED are small in comparison with the light cylinder radius.

Whereas for most cases, values of the magnetic gradient around $b=3$ are needed to describe 
the pulsar SED correctly, we are finding that relative larger magnetic gradients appear in the fits of a few pulsars.
In particular, in comparison with the radial dipole magnetic field line value of 3, we find that 
for the young and energetic Crab and Crab Twin, and especially for the MSPs, magnetic gradients of up to $b=6$ are needed.
This may indicate that the  local conditions of the magnetic 
field where the small accelerating region is located dominates the whole emission. 
This gradient may be different to the overall behavior of the field at larger scales, which may continue to be (for instance) a dipole. 
May the relatively large gradients in these cases, together with the smallness of the relevant emission region, point towards a region of acceleration close to the null point (after which
the separatrix goes along the equatorial plane)?
In such region, large numerical resolution works (see Figure 11 of \citet{Timokhin2006}, and references therein for earlier theoretical work)
have shown that the field is subject to a large variation, where magnetic gradients can be very high.

Another aspect that we find confirmed with our larger sample is the correlation between $E_{||}$ and $x_0$.
When using the single set fittings together with the separate fits in gamma-rays (i.e., with the $x_0$ corresponding to the one obtained in the gamma-ray fit) we see that 
the generally reduced error bars of our parameters confirm the gamma-ray trend shown in  Figure 5 of \cite{torres18} using {\it Fermi} 2PC data only.
Plotting $E_{||}$ versus the $x_0$ resulting from the separate X-ray fitting the trend is still conserved,  displaced to lower values of $x_0$.
We do not find such clear correlations between direct pulsar properties and model parameters, but only hints in
$b$ vs $E_{||}$,   or the spin-down power $E_{sd}$ vs $E_{||}$, and the light cylinder field $B_{lc}$ vs $E_{||}$, especially for the separate fittings.
In particular, we do not see a clear correlation between $b$ and $B_{lc}$ or $R_{lc}$ or $B_s$.
It is not ruled out that 
correlations between model and pulsar intrinsic parameters, derived from ($P, \dot P$), 
may appear not one-to-one but using combinations of the latter (i.e., with combinations of $B_{lc}$, $E_{sd}$, etc.).

It is interesting to note that there are a few other non-thermal X-ray pulsars with 
no gamma-ray emission yet detected (see  \cite{kuiper15}, and Figure 2 of \cite{torres18}). 
In the context of our findings, these can be a priori produced by a low(er) $E_{||}$, or by a particular geometrical orientation, where, for instance, only the 
region of acceleration with the geometry leading to the smaller contrast is visible.
However, in \cite{torres18} the fits were already done with $E_{||}$ varying at least from 
10$^4$ to 10$^{10}$ V/m, 
and most of the pulsars with no gamma-ray emission were found to prefer
a low value of $E_{||}$ with a relatively high value of $x_0/R_{lc}$, instead of a relatively high value of $E_{||}$ with a low value of $x_0/R_{lc}$, as it happens here.
This may imply that we are indeed seeing in these pulsars the tail of the population (towards lower  $E_{||}$).

We would like to conclude with a note about Crab and the Crab Twin.
Despite the difference in data quality of the data sets for the two pulsars, 
our analysis unveils some similarities:
they are both showcased having large magnetic gradients in the accelerating regions generating the radiation we see.
We find that the Crab Twin current data set, and most of the Crab data set can be well described with the model as explained, 
and have explained how the Crab MeV energy regime that is missed
can actually be encompassed in a more complete (less bumpy) representation of the particle distribution.
As an aside, we note that whereas the very high energy data at TeV energies cannot be encompassed in our model if the acceleration regime has the same vale of 
$E_{||}$ as the one needed to produce the X-ray and gamma-ray emission, it is possible for it to arise via curvature radiation
from a different and more localized magnetospheric extent, without invoking additional processes, such as inverse Compton.

\section*{Acknowledgements}

DFT and FCZ acknowledge support from the Spanish Ministry of Economy, Industry and Competitiveness grants PGC2018-095512-B-I00, SGR2017-1383, and AYA2017- 92402-EXP. DV acknowledges support from the Spanish Ministry of Economy, Industry and Competitiveness grants AYA2016-80289-P and AYA2017-82089-ERC (AEI/FEDER, UE). FCZ is supported by a Juan de la Cierva fellowship. DFT acknowledges the participants of the Workshop on Magnetospheres of Neutron Stars and Black Holes (June 2019, Goddard Space Flight Center) for comments.

\begin{table*}
\small
\begin{center}
\caption{Results of the broadband spectral modelling with a single set of $(\log(E_{||}),  b, \log(x_0/R_{lc}), \log(N_0))$ parameters. 
The table shows the pulsar type, name, distance used in the conversion of flux to luminosities, the best fit values and their
1$\sigma$ uncertainty in the form  
$(\log(E_{||}),  b, \log(x_0/R_{lc}), \log(N_0))$. 
The horizontal lines divide the pulsars whose SEDs are well represented by a single set of parameters (Figure~\ref{sed1}),
those that are arguably good (Figure~\ref{arguably-good}), and 
those for which one set of parameters is clearly not enough, suggesting we may be seeing significant contribution from two regions of acceleration or  that the geometrical
configuration is such that a single distribution is not valid at all energies (Figure~\ref{no-good-fits}).
} 
\label{common-fits}

\begin{tabular}{l@{$\quad$} l@{$\quad$} l@{$\quad$} c@{$\quad$}  c@{$\quad$}  c@{$\quad$}  c@{$\quad$}  }
\hline
\hline
Type & Pulsar & $D$      & $\log E_\parallel$  &  $b$   & $\log ({x_0}/{R_{\rm lc}})$ & $\log N_0$ \\
     &        & [kpc]  & [V/m]               &        &                             &            \\
\hline
PSR &  J0007+7303 &   1.40 & $ 8.36^{+0.10}_{-0.02}$ & $ 2.80^{+0.10}_{-0.10}$ & $-2.80^{+0.10}_{-0.10}$ & $32.22^{+0.01}_{-0.01}$ \\
PSR &  J0205+6449 &   3.20 & $ 9.52^{+0.02}_{-0.02}$ & $ 2.80^{+0.10}_{-0.10}$ & $-3.60^{+0.10}_{-0.10}$ & $30.67^{+0.01}_{-0.01}$ \\
PSR &  J0633+1746 &   0.19 & $ 7.80^{+0.02}_{-0.02}$ & $ 2.50^{+0.10}_{-0.10}$ & $-2.10^{+0.10}_{-0.10}$ & $31.15^{+0.01}_{-0.01}$ \\
PSR &  J0659+1414 &   0.29 & $ 8.54^{+0.10}_{-0.10}$ & $ 2.80^{+0.10}_{-0.10}$ & $-3.60^{+0.11}_{-0.10}$ & $31.98^{+0.01}_{-0.01}$ \\
PSR &  J0835$-$4510 &   0.28 & $ 8.31^{+0.01}_{-0.01}$ & $ 3.25^{+0.05}_{-0.05}$ & $-2.50^{+0.10}_{-0.10}$ & $32.05^{+0.01}_{-0.04}$ \\
PSR &  J1420$-$6048 &   5.60 & $ 8.68^{+0.26}_{-0.48}$ & $ 2.70^{+0.10}_{-0.10}$ & $-2.70^{+0.50}_{-0.20}$ & $31.95^{+0.30}_{-0.37}$ \\
PSR &  J1357$-$6429 &   3.10 & $ 8.14^{+0.98}_{-0.32}$ & $ 3.10^{+0.10}_{-0.10}$ & $-2.80^{+0.30}_{-0.80}$ & $32.96^{+0.10}_{-1.05}$ \\
PSR &  J1718$-$3825 &   3.60 & $ 8.24^{+0.12}_{-0.22}$ & $ 3.00^{+0.10}_{-0.10}$ & $-2.50^{+0.20}_{-0.10}$ & $32.65^{+0.15}_{-0.10}$ \\
PSR &  J1747$-$2958 &   2.52 & $ 8.58^{+0.12}_{-0.02}$ & $ 3.00^{+0.10}_{-0.10}$ & $-3.00^{+0.10}_{-0.10}$ & $33.13^{+0.01}_{-0.09}$ \\
PSR &  J1801$-$2451 &   3.80 & $ 8.72^{+1.02}_{-0.70}$ & $ 3.00^{+0.10}_{-0.10}$ & $-3.20^{+0.60}_{-0.80}$ & $33.24^{+0.49}_{-0.78}$ \\
PSR &  J1809$-$2332 &   0.88 & $ 8.42^{+0.02}_{-0.26}$ & $ 3.00^{+0.10}_{-0.10}$ & $-2.80^{+0.30}_{-0.10}$ & $32.10^{+0.01}_{-0.14}$ \\
PSR &  J1813$-$1246 &   2.63 & $ 9.22^{+0.02}_{-0.12}$ & $ 2.65^{+0.05}_{-0.05}$ & $-3.20^{+0.10}_{-0.10}$ & $31.17^{+0.18}_{-0.01}$ \\
PSR &  J1838$-$0537 &   1.80 & $ 8.36^{+0.36}_{-0.34}$ & $ 3.10^{+0.10}_{-0.30}$ & $-2.70^{+0.50}_{-0.40}$ & $32.15^{+0.16}_{-0.58}$ \\
PSR &  J1846$-$0258 &   5.80 & $ 6.14^{+0.06}_{-0.08}$ & $ 2.80^{+0.10}_{-0.10}$ & $-1.20^{+0.10}_{-0.10}$ & $34.57^{+0.05}_{-0.06}$ \\
PSR &  J2021+3651 &   1.80 & $ 8.40^{+0.02}_{-0.02}$ & $ 3.30^{+0.10}_{-0.10}$ & $-2.70^{+0.10}_{-0.10}$ & $32.74^{+0.01}_{-0.01}$ \\
PSR &  J2021+4026 &   2.15 & $ 8.00^{+0.18}_{-0.02}$ & $ 3.10^{+0.10}_{-0.10}$ & $-2.60^{+0.10}_{-0.20}$ & $33.61^{+0.07}_{-0.01}$ \\
PSR &  J2022+3842 &  10.00 & $ 9.00^{+0.24}_{-0.36}$ & $ 3.10^{+0.10}_{-0.10}$ & $-3.20^{+0.30}_{-0.20}$ & $32.59^{+0.53}_{-0.19}$ \\
PSR &  J2229+6114 &   3.00 & $ 8.92^{+0.02}_{-0.02}$ & $ 2.60^{+0.10}_{-0.10}$ & $-2.80^{+0.10}_{-0.10}$ & $31.19^{+0.01}_{-0.01}$ \\
\hline
PSR &  J0357+3205 &   0.83 & $ 6.92^{+0.08}_{-0.12}$ & $ 2.20^{+0.10}_{-0.10}$ & $-1.30^{+0.20}_{-0.20}$ & $31.04^{+0.17}_{-0.03}$ \\
PSR &  J1513$-$5908 &   4.40 & $ 6.23^{+0.01}_{-0.01}$ & $ 2.75^{+0.05}_{-0.05}$ & $-1.10^{+0.10}_{-0.10}$ & $35.01^{+0.01}_{-0.01}$ \\
PSR &  J1709$-$4429 &   2.60 & $ 8.30^{+0.02}_{-0.02}$ & $ 3.10^{+0.10}_{-0.10}$ & $-2.50^{+0.10}_{-0.10}$ & $33.11^{+0.01}_{-0.01}$ \\
PSR &  J1826$-$1256 &   1.55 & $ 8.26^{+0.02}_{-0.02}$ & $ 3.10^{+0.10}_{-0.10}$ & $-2.60^{+0.10}_{-0.10}$ & $32.58^{+0.01}_{-0.01}$ \\
PSR &  J1833$-$1034 &   4.10 & $ 9.86^{+0.12}_{-0.12}$ & $ 3.30^{+0.10}_{-0.10}$ & $-4.10^{+0.10}_{-0.12}$ & $32.79^{+0.08}_{-0.09}$ \\
PSR &  J1836+5925 &   0.30 & $ 7.76^{+0.02}_{-0.02}$ & $ 2.60^{+0.10}_{-0.10}$ & $-2.00^{+0.10}_{-0.10}$ & $30.86^{+0.01}_{-0.01}$ \\
PSR &  J2030+4415 &   0.88 & $ 9.24^{+0.10}_{-0.78}$ & $ 2.90^{+0.10}_{-0.10}$ & $-3.90^{+0.80}_{-0.10}$ & $31.99^{+0.01}_{-0.11}$ \\
PSR &  J2055+2539 &   0.62 & $ 9.43^{+0.01}_{-0.01}$ & $ 2.70^{+0.10}_{-0.10}$ & $-4.10^{+0.10}_{-0.10}$ & $31.39^{+0.01}_{-0.03}$ \\
\hline
PSR &  J1057$-$5226 &   0.30 & $ 7.40^{+0.02}_{-0.02}$ & $ 5.00^{+0.10}_{-0.10}$ & $-1.50^{+0.10}_{-0.10}$ & $30.39^{+0.01}_{-0.01}$ \\
PSR &  J1124$-$5916 &   5.00 & $ 8.20^{+0.02}_{-0.02}$ & $ 3.20^{+0.10}_{-0.10}$ & $-2.90^{+0.10}_{-0.10}$ & $34.17^{+0.01}_{-0.01}$ \\
PSR &  J1741$-$2054 &   0.30 & $ 9.10^{+0.02}_{-0.02}$ & $ 2.65^{+0.05}_{-0.05}$ & $-3.96^{+0.10}_{-0.10}$ & $31.59^{+0.01}_{-0.01}$ \\
PSR &  J1952+3252 &   3.00 & $ 9.64^{+0.02}_{-0.02}$ & $ 3.30^{+0.10}_{-0.10}$ & $-4.00^{+0.10}_{-0.10}$ & $33.37^{+0.01}_{-0.01}$ \\
MSP &  J0437$-$4715 &   0.16 & $11.04^{+0.02}_{-0.14}$ & $ 2.90^{+0.10}_{-0.10}$ & $-4.52^{+0.10}_{-0.10}$ & $29.17^{+0.08}_{-0.01}$ \\
MSP &  J0614$-$3329 &   0.62 & $11.80^{+0.02}_{-0.02}$ & $ 4.10^{+0.10}_{-0.10}$ & $-5.00^{+0.10}_{-0.10}$ & $29.84^{+0.01}_{-0.01}$ \\
MSP &  J0751+1807 &   1.10 & $11.50^{+0.02}_{-0.02}$ & $ 3.30^{+0.10}_{-0.10}$ & $-4.70^{+0.10}_{-0.10}$ & $29.60^{+0.02}_{-0.01}$ \\
MSP &  J1231$-$1411 &   0.42 & $ 8.80^{+0.02}_{-0.02}$ & $ 2.50^{+2.50}_{-0.10}$ & $-1.60^{+0.10}_{-0.10}$ & $28.57^{+0.20}_{-0.01}$ \\

\hline
\hline
\end{tabular}

\end{center}
\end{table*}

\begin{table*}
\small
\begin{center}
\caption{Results of piecewise spectral modelling (x (g) refers to fitting only the X-ray (gamma-ray) data set). 
The panels correspond to Figure~\ref{composite_fits}, 
\ref{arguablydouble}, and \ref{crab}, respectively. In general, the gamma-ray fit is used to determine $E_{||}$, whereas
the X-ray fit is used to determine $b$. Exceptions occur (e.g., 0540-6919g, or 0357+3205x) when such individual datasets are not currently constraining per se, and 
are discussed in the text.
} 
\label{uncommon-fits}

\begin{tabular}{l@{$\quad$} l@{$\quad$} l@{$\quad$} l@{$\quad$}  l@{$\quad$}  l@{$\quad$}  l@{$\quad$}  }
\hline
\hline
Type & Pulsar & $D$      & $\log E_\parallel$  &  $b$   & $\log ({x_0}/{R_{\rm lc}})$ & $\log N_0$ \\
     &        & [kpc]  & [V/m]               &        &                             &            \\
\hline
PSR & J1057$-$5226x &   0.30 & $ 7.38^{+\ldots}_{-\ldots}$ & $ 3.10^{+2.40}_{-0.10}$ & $-5.00^{+0.90}_{-0.10}$ & $31.74^{+5.19}_{-0.78}$ \\
PSR & J1057$-$5226g &   0.30 & $ 7.38^{+0.04}_{-0.06}$ & $ 3.10^{+\ldots}_{-\ldots}$ & $-1.50^{+0.10}_{-0.10}$ & $30.43^{+0.10}_{-0.03}$ \\
PSR & J1124$-$5916x &   5.00 & $ 7.98^{+\ldots}_{-\ldots}$ & $ 4.30^{+0.10}_{-0.10}$ & $-4.00^{+0.10}_{-0.10}$ & $37.47^{+0.01}_{-0.41}$ \\
PSR & J1124$-$5916g &   5.00 & $ 7.98^{+0.10}_{-0.46}$ & $ 4.30^{+\ldots}_{-\ldots}$ & $-2.40^{+0.60}_{-0.10}$ & $33.02^{+0.01}_{-0.37}$ \\
PSR & J1741$-$2054x &   0.30 & $ 6.84^{+\ldots}_{-\ldots}$ & $ 3.60^{+1.90}_{-0.90}$ & $-3.60^{+0.10}_{-1.40}$ & $36.37^{+0.23}_{-5.83}$ \\
PSR & J1741$-$2054g &   0.30 & $ 6.84^{+0.16}_{-0.10}$ & $ 3.60^{+\ldots}_{-\ldots}$ & $-1.10^{+0.30}_{-0.40}$ & $30.33^{+0.37}_{-0.16}$ \\
PSR & J1952+3252x &   3.00 & $ 8.34^{+\ldots}_{-\ldots}$ & $ 3.90^{+0.60}_{-0.10}$ & $-4.20^{+0.50}_{-0.10}$ & $34.44^{+1.99}_{-0.01}$ \\
PSR & J1952+3252g &   3.00 & $ 8.34^{+0.02}_{-0.08}$ & $ 3.90^{+\ldots}_{-\ldots}$ & $-2.30^{+0.10}_{-0.10}$ & $32.22^{+0.01}_{-0.08}$ \\
MSP & J0437$-$4715x &   0.16 & $ 8.32^{+\ldots}_{-\ldots}$ & $ 3.50^{+0.10}_{-0.10}$ & $-3.70^{+0.10}_{-0.10}$ & $30.58^{+0.01}_{-0.21}$ \\
MSP & J0437$-$4715g &   0.16 & $ 8.32^{+0.22}_{-0.22}$ & $ 3.50^{+\ldots}_{-\ldots}$ & $-1.70^{+0.40}_{-0.30}$ & $28.31^{+0.18}_{-0.33}$ \\
MSP & J0614$-$3329x &   0.62 & $ 9.52^{+\ldots}_{-\ldots}$ & $ 6.50^{+0.10}_{-0.10}$ & $-5.00^{+0.10}_{-0.10}$ & $34.68^{+0.01}_{-0.01}$ \\
MSP & J0614$-$3329g &   0.62 & $ 9.52^{+0.02}_{-0.08}$ & $ 6.50^{+\ldots}_{-\ldots}$ & $-2.60^{+0.10}_{-0.10}$ & $29.39^{+0.01}_{-0.04}$ \\
MSP & J0751+1807x &   1.10 & $ 9.34^{+\ldots}_{-\ldots}$ & $ 6.10^{+0.40}_{-0.80}$ & $-4.70^{+0.10}_{-0.30}$ & $34.49^{+0.90}_{-1.59}$ \\
MSP & J0751+1807g &   1.10 & $ 9.34^{+2.16}_{-0.58}$ & $ 6.10^{+\ldots}_{-\ldots}$ & $-2.50^{+1.00}_{-2.30}$ & $29.28^{+0.46}_{-0.74}$ \\
MSP & J1231$-$1411x &   0.42 & $ 8.86^{+\ldots}_{-\ldots}$ & $ 6.40^{+0.10}_{-0.10}$ & $-4.60^{+0.10}_{-0.10}$ & $36.08^{+0.27}_{-0.31}$ \\
MSP & J1231$-$1411g &   0.42 & $ 8.86^{+0.06}_{-0.04}$ & $ 6.40^{+\ldots}_{-\ldots}$ & $-1.80^{+0.10}_{-0.10}$ & $28.82^{+0.05}_{-0.10}$ \\
\hline
PSR & J0357+3205x &   0.83 & $ 6.86^{+\ldots}_{-\ldots}$& $ 3.00^{+\ldots}_{-\ldots}$ & $-3.30^{+0.10}_{-0.10}$ & $32.48^{+0.01}_{-0.01}$ \\
PSR & J0357+3205g &   0.83 & $ 6.86^{+0.12}_{-0.14}$ & $ 3.00^{+\ldots}_{-\ldots}$ & $-1.30^{+0.40}_{-0.20}$ & $31.25^{+0.10}_{-0.31}$ \\
PSR & J1513$-$5908x &   4.40 & $ 7.00^{+\ldots}_{-\ldots}$ & $ 3.10^{+0.10}_{-0.10}$ & $-3.40^{+0.10}_{-0.10}$ & $33.91^{+0.01}_{-0.01}$ \\
PSR & J1513$-$5908g &   4.40 & $ 7.00^{+0.10}_{-0.02}$ & $ 3.10^{+\ldots}_{-\ldots}$ & $-1.90^{+0.10}_{-0.10}$ & $34.34^{+0.01}_{-0.08}$ \\
PSR & J1709$-$4429x &   2.60 & $ 8.54^{+\ldots}_{-\ldots}$ & $ 3.50^{+0.60}_{-0.50}$ & $-5.00^{+1.70}_{-0.10}$ & $33.44^{+1.91}_{-1.50}$ \\
PSR & J1709$-$4429g &   2.60 & $ 8.54^{+0.02}_{-0.02}$ & $ 3.50^{+\ldots}_{-\ldots}$ & $-2.80^{+0.10}_{-0.10}$ & $33.34^{+0.01}_{-0.01}$ \\
PSR & J1826$-$1256x &   1.55 & $ 8.08^{+\ldots}_{-\ldots}$ & $ 3.80^{+\ldots}_{-\ldots}$ & $-3.20^{+0.10}_{-0.10}$ & $34.24^{+0.01}_{-0.01}$ \\
PSR & J1826$-$1256g &   1.55 & $ 8.08^{+0.18}_{-0.16}$ & $ 3.80^{+\ldots}_{-\ldots}$ & $-2.40^{+0.20}_{-0.20}$ & $32.60^{+0.05}_{-0.12}$ \\
PSR & J1833$-$1034x &   4.10 & $ 8.04^{+\ldots}_{-\ldots}$ & $ 3.40^{+0.10}_{-0.10}$ & $-3.70^{+2.40}_{-0.10}$ & $33.32^{+2.57}_{-0.01}$ \\
PSR & J1833$-$1034g &   4.10 & $ 8.04^{+2.14}_{-0.48}$ & $ 3.40^{+\ldots}_{-\ldots}$ & $-2.20^{+0.70}_{-2.20}$ & $32.55^{+0.32}_{-0.43}$ \\
PSR & J1836+5925x &   0.30 & $ 7.76^{+\ldots}_{-\ldots}$ & $ 3.30^{+1.70}_{-0.20}$ & $-3.90^{+0.10}_{-0.59}$ & $32.01^{+2.70}_{-0.93}$ \\
PSR & J1836+5925g &   0.30 & $ 7.76^{+0.02}_{-0.02}$ & $ 3.30^{+\ldots}_{-\ldots}$ & $-2.00^{+0.10}_{-0.10}$ & $30.88^{+0.01}_{-0.01}$ \\
PSR & J2030+4415x &   0.88 & $ 7.46^{+\ldots}_{-\ldots}$ & $ 3.70^{+0.10}_{-0.10}$ & $-3.80^{+0.10}_{-0.10}$ & $34.78^{+0.01}_{-0.01}$ \\
PSR & J2030+4415g &   0.88 & $ 7.46^{+\ldots}_{-0.02}$ & $ 3.70^{+\ldots}_{-\ldots}$ & $-2.00^{+0.10}_{-0.10}$ & $31.65^{+0.01}_{-0.01}$ \\
PSR & J2055+2539x &   0.62 & $ 7.10^{+\ldots}_{-\ldots}$ & $ 2.80^{+\ldots}_{-\ldots}$ & $-4.70^{+0.30}_{-0.10}$ & $30.84^{+2.33}_{-0.63}$ \\
PSR & J2055+2539g &   0.62 & $ 7.10^{+0.14}_{-0.18}$ & $ 2.80^{+\ldots}_{-\ldots}$ & $-1.30^{+0.60}_{-0.30}$ & $30.46^{+0.22}_{-0.32}$ \\
\hline
PSR & J0534+2200x &   2.00 & $ 8.94^{+\ldots}_{-\ldots}$ & $ 5.10^{+0.10}_{-0.10}$ & $-4.30^{+0.10}_{-0.10}$ & $38.89^{+0.01}_{-0.01}$ \\
PSR & J0534+2200g &   2.00 & $ 8.94^{+0.02}_{-0.02}$ & $ 5.10^{+\ldots}_{-\ldots}$ & $-3.00^{+0.10}_{-0.10}$ & $33.29^{+0.01}_{-0.01}$ \\
PSR & J0540$-$6919x &  49.70 & $ 8.00^{+0.02}_{-0.02}$ & $ 5.40^{+0.10}_{-0.10}$ & $-3.70^{+0.10}_{-0.10}$ & $40.48^{+0.01}_{-0.01}$ \\
PSR & J0540$-$6919g &  49.70 & $ 8.00^{+\ldots}_{-\ldots}$ & $ 5.40^{+\ldots}_{-\ldots}$ & $-2.10^{+\ldots}_{-\ldots}$ & $34.08^{+\ldots}_{-\ldots}$ \\

\hline
\hline
\end{tabular}

\end{center}
\end{table*}


\bibliographystyle{mnras} 

\label{lastpage}
\end{document}